\documentclass[a4paper,12pt,oneside]{report}
\usepackage[titletoc]{appendix}
\usepackage[left=3cm, right=2cm]{geometry}
\usepackage[utf8]{inputenc}
\usepackage[sc]{mathpazo} 
\usepackage{setspace}
\usepackage{relsize}
\usepackage{tikz} 
\usetikzlibrary{shapes,arrows,positioning,automata,backgrounds,calc,er,patterns}
\usepackage{tikz-feynman}
\tikzfeynmanset{compat=1.0.0}
\usetikzlibrary{decorations.pathmorphing, patterns,shapes}
\usepackage{color}
\usepackage{amsmath}
\usepackage[export]{adjustbox}
\usepackage{footnote}
\usepackage{braket}
\usepackage{flexisym}
\usepackage{graphicx}
\usepackage{dcolumn}% Align table columns on decimal point
\graphicspath{{figures/}}
\usepackage{fancyhdr}

\usepackage{slashed}
\usepackage{calrsfs}
\usepackage{cite}
\usepackage{amssymb}    

\usepackage{booktabs}

\usepackage[backref=page]{hyperref}
\renewcommand*{\backref}[1]{}
\renewcommand*{\backrefalt}[4]{%
	\ifcase #1 (Not cited.)%
	\or        (Cited on page~#2.)%
	\else      (Cited on pages~#2.)%
	\fi}

\newcommand{\drv}{{\rm d}}

\newcommand{\DY}{\Delta Y}

\newcommand{\tref}[1]{~\ref{#1}}

\newcommand{\tarr}{
	\begin{array}}
	\newcommand{\earr}{\end{array}}

\begin{document}
 
\begin{titlepage}
	\vspace*{-3cm}
	%	\begin{spacing}{1.5}
	\center 
	\begin{figure}[hbtp]
		\centering
		{\includegraphics[scale=0.35]{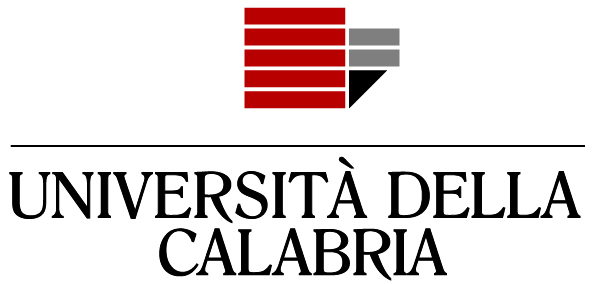}}
	\end{figure} 
	%\textsc{ \LARGE UNIVERSITÀ DELLA CALABRIA}\\[0.2cm]
	\vspace{1.0cm}
	{ Dipartimento di Fisica}\\
	\vspace{1.0cm} 
	{\bfseries Dottorato di Ricerca in\\
		\vspace{0.2cm} Scienze e Tecnologie Fisiche, Chimiche e dei Materiali} \\
	\vspace{1.0cm}

	\textsc{\small  \bf{Ciclo\\ \vspace{0.2cm}
			XXXIV}}\\[1.5cm]
	% \vspace{0.3cm}
	{\LARGE \bfseries Hunting stabilization effects of the high-energy resummation at the LHC\par}\vspace{1.0cm}
 
	{\bfseries Settore Scientifico Disciplinare\\
		FIS/02 - Fisica Teorica, Modelli e Metodi Matematici}
	\vspace{1.5cm}
	\begin{flushleft} 
		\emph{\bf Coordinatore:}
		\textcolor{black}{Prof.ssa Gabriella \textsc{Cipparrone}}\\
		\vspace{0.2cm}
	%	\hspace{2.8cm}\emph{Firma}
	%	\noindent\rule{5cm}{0.4pt}\\
		\hspace{2.6cm}	\includegraphics[scale=0.35]{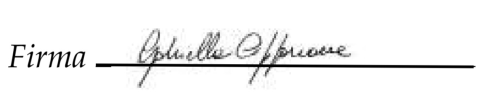}\\
		%\vspace{1.0cm}
		\emph{\bf Supervisore:}\hspace{0.2cm}
		\textcolor{black}{Prof. Alessandro \textsc{Papa}}
		\vspace{0.2cm}\\
	%	\hspace{2.8cm}\emph{Firma}
	%	\noindent\rule{5cm}{0.4pt}	
	  \hspace{2.6cm} \includegraphics[scale=0.35]{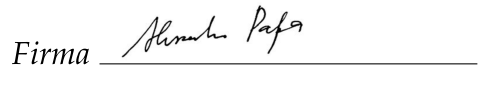}\\	
\end{flushleft}
	%\end{minipage}~
	\vspace{0.4cm}
	\begin{flushright}
		Dottorando:~Dott.
		{Mohammed Maher Abdelrahim \textsc{Mohammed}}\\
		\vspace{0.2cm}
		%\emph{Firma}\noindent\rule{7.55cm}{0.4pt}
			\includegraphics[scale=0.3]{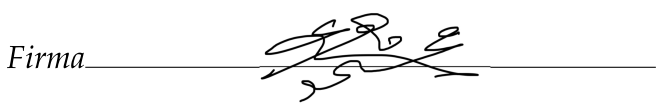}\\
	\end{flushright}
	
	%	 \HRule\\
	%	\textbf{Anno Accademico 2020/2021} 
	%	\vfill 
	%\end{spacing}
\end{titlepage}
\setcounter{secnumdepth}{-2} 
\pagenumbering{roman}
\chapter{Abstract}
\begin{spacing}{1.5}
Studying semi-hard processes in the large center-of-mass
energy limit gives us an opportunity to further test perturbative QCD
in an unexplored kinematical configuration, contributing to a better understanding of the dynamics of strong interactions.
For semi-hard reactions in kinematics at large center-of-mass energy $\sqrt{s}$, the BFKL resummation of energy logarithms comes into play, since large energy logarithms compensate the smallness of QCD coupling $\alpha_s$ and must therefore be accounted for to all perturbative orders. 

Tracing the path toward performing precision calculations via BFKL resummation of high-energy logarithms, in this thesis we present phenomenological analyses for distinct inclusive processes, highlighting the recognized problem of instabilities under higher-order corrections and energy-scales variations, that would abort any possibility to investigate semi-hard reactions with high-precision at natural energy-scales. At the same time, we present new reactions that seem to act as fair stabilizers of the high-energy series. 

First, the inclusive production at the LHC of a charged light hadron and of a jet, featuring a wide separation in rapidity, is presented making use of optimization methods to fix energy-scale. We report some predictions, tailored on the CMS and CASTOR acceptances, for the cross section averaged over the azimuthal angle between the identified jet and hadron and for azimuthal correlations. Then, we propose as a novel probe channel for the manifestation of the BFKL dynamics, the inclusive hadroproduction of a Higgs boson and of a jet, featuring large transverse momenta and well separated in rapidity. We present predictions for azimuthal Higgs-jet correlations and other observables, to be possibly compared with typical experimental
analyses at the LHC. Finlay, we propose the inclusive semi-hard production, in proton-proton collisions, of two bottom-flavored hadrons, as well as of a single bottom-flavored hadron accompanied by a light jet, as novel channels for targeting stabilization effects of the high-energy resummation under higher-order corrections. Moreover, we propose the study of double differential distributions in the transverse momenta of the two final-state particles as a common basis to investigate the interplay of different resummation approaches.

\end{spacing}

\chapter{Sintesi in lingua italiana}
\begin{spacing}{1.5}
	Lo studio dei processi semiduri nel limite di alta energia nel centro di massa
	ci offre l'opportunità di testare ulteriormente la QCD perturbativa in una
	configurazione cinematica finora inesplorata, contribuendo a una migliore
	comprensione della dinamica delle interazioni forti.
	Per le reazioni semidure in cinematica a grande energia del centro di massa
	$\sqrt{s}$, entra in gioco la risommazione BFKL dei logaritmi di energia,
	poiché i logaritmi di grande energia compensano la piccolezza
	dell'accoppiamento QCD $\alpha_s$ e devono quindi essere calcolati a
	tutti gli ordini perturbativi.
	
	Tracciando il percorso verso l'esecuzione di calcoli di precisione tramite
	la risommazione BFKL dei logaritmi ad alta energia, in questa tesi
	presentiamo analisi fenomenologiche per diversi processi inclusivi,
	evidenziando il noto problema delle instabilità sotto correzioni di
	ordine superiore e variazioni delle scale energetiche, che annullerebbero ogni
	possibilità per studiare reazioni semi-dure con alta precisione alle scale
	energetiche naturali per quei processi. Allo stesso tempo, presentiamo nuove
	reazioni che sembrano fungere da corretti stabilizzatori della serie ad
	alta energia.
	
	In primo luogo, viene presentata la produzione inclusiva ad LHC di un adrone
	leggero carico e di un getto, caratterizzati da un'ampia separazione in
	rapidità, utilizzando metodi di ottimizzazione per fissare la scala
	energetica. Riportiamo alcune previsioni, adattate alle accettanze di CMS e
	CASTOR, per la sezione trasversale mediata sull'angolo azimutale tra getto e
	adrone identificati e per le correlazioni azimutali. Quindi,
	proponiamo come nuovo canale sonda per la manifestazione della dinamica BFKL,
	l'adroproduzione inclusiva di un bosone di Higgs e di un getto,
	caratterizzati da grandi momenti trasversali e ben separati in rapidità.
	Presentiamo previsioni per le correlazioni azimutali di Higgs-jet e altre
	osservabili, da confrontare eventualmente con analisi sperimentali ad LHC.
	Infine, proponiamo la produzione semi-dura inclusiva, nelle collisioni
	protone-protone, di due adroni con sapore ``bottom'', nonché di un singolo
	adrone con sapore ``bottom'' accompagnato da un getto, come nuovi canali per
	studiare gli effetti di stabilizzazione ad alta energia sotto correzioni di
	ordine superiore. Proponiamo anche lo studio delle doppie distribuzioni
	differenziali nei momenti trasversi delle due particelle di stato finale
	come base comune per investigare l'interazione tra diversi approcci alla
	risommazione.
\end{spacing}
 
%\end{spacing}
\tableofcontents
\listoffigures

\setcounter{secnumdepth}{-2} 
\chapter{Introduction}
\pagenumbering{arabic}
\begin{spacing}{1.5}
The standard model (SM) of particle physics represents a unified description of the fundamental interactions that govern the dynamical behavior of the elementary particles at sub-atomic scale. The validity of SM has been leaned on a long history of theoretical speculations and experimental measurements, crowned with the discovery of the Higgs boson~\cite{ATLAS:2012yve,CMS:2012qbp}, at the Large Hadron Collider (LHC). 
%After this discovery the particle physics community has entered a new chapter of high precise experimental measurements, at the same time the equivalence theoretical calculations have been made a long years effort in computing the cross sections and related observables at high accuracy. 
With more data to be collected at the LHC, the study of semi-hard processes\footnote{Semi-hard processes~\cite{Gribov:1983ivg} are those with center-of-mass energy squared $s$ is substantially larger than one or more hard scales $Q_{i}^{2}$, $s\geqslant Q_{i}^{2}\geqslant \Lambda_{QCD}^{2}$, with $\Lambda_{QCD}$ is the QCD scale.} in the large center-of-mass energy limit allows us to further test perturbative QCD (pQCD) in unexplored kinematical configurations thus contributing to a better understanding of the dynamics of strong interactions. Within pQCD computations, reducing theoretical uncertainties coming from higher-order corrections is required to have a reliable estimate of the production rate. At high energies, the validity of the perturbative expansion, truncated at a certain order in the strong coupling $\alpha_s$, is spoiled. This is due to the appearance of large logarithms of the center-of-mass energy squared, $s$, associated with the perturbative calculations and it is needed to resum them to all orders in $\alpha_{s}$. The most powerful framework to perform this resummation is the Balitsky–Fadin–Kuraev–Lipatov (BFKL)~\cite{Lipatov:1985uk,Fadin:1975cb,kuraev1976multi,Kuraev:1977fs,Lipatov:1976zz,Balitsky:1978ic} approach, initially developed at the so-called leading logarithmic approximation (LLA), where it prescribes how to resum all terms proportional to $(\alpha_s \ln s )^n$. To improve the obtained results at LLA, the so-called next-to-leading logarithmic approximation (NLA) was considered~\cite{Fadin:1998py,Ciafaloni:1998gs}, where also all terms proportional to $\alpha_s(\alpha_s \ln s)^n$, were resumed. A significant question for collider phenomenology is highlighting at which energies the BFKL dynamics becomes significant and cannot be overlooked. Typical BFKL observables that can be studied at the LHC are the azimuthal coefficients of the Fourier expansion of the cross section differential in the variables of the tagged objects over the relative azimuthal-angle. They take a certain factorization form given as the convolution of a universal BFKL Green’s function with process-dependent impact factors, the latter describing the transition from each colliding proton to the respective final-state identified object. The BFKL Green's function obeys an integral equation, whose kernel is known at the next-to-leading order (NLO)~\cite{Fadin:1998py,Ciafaloni:1998gs,Fadin:1998jv,Fadin:2004zq,Fadin:2005zj}. In order to confront the theoretical prediction with the observed data, we need the NLO corrections to impact factors, so as to treat different processes with consistent accuracy. So far the available impact factors calculated at NLO order are: 1) colliding-parton (quarks and gluons) impact factors~\cite{Fadin:1999de,Fadin:1999df}, which represents the basis for constructing the 2) forward-jet impact factor~\cite{Bartels:2001ge,Bartels:2002yj,Caporale:2011cc,Caporale:2012ih,Ivanov:2012ms,Colferai:2015zfa} and 3) forward charged-light hadron one~\cite{Ivanov:2012iv}, 4) the impact factor describing the $\gamma^*$ to light-vector-meson (LVM) leading twist transition~\cite{Ivanov:2004pp}, 5) the one detailing the $\gamma^* \rightarrow \gamma^*$ subprocess~\cite{Bartels:2000gt,Bartels:2001mv,Bartels:2002uz,Bartels:2004bi,Fadin:2001ap,Balitsky:2012bs}, and 6) the one for the production of a forward Higgs boson in the infinite top-mass limit~\cite{Hentschinski:2020tbi,Nefedov:2019mrg}. 

Over last years, pursuing the goal of identifying observables that fit the data where BFKL approach is distinct, and other possible resummations approaches fail, a number of reactions have been proposed for different collider environments: the exclusive diffractive leptoproduction of two light vector mesons~\cite{Pire:2005ic,Segond:2007fj,Enberg:2005eq,Ivanov:2005gn,Ivanov:2006gt}, the inclusive hadroproduction of two jets featuring large transverse momenta and well separated in rapidity (Mueller-–Navelet channel~\cite{Mueller:1986ey}), for which several phenomenological studies have appeared so far~(see, \emph{e.g.},~Refs.~\cite{Colferai:2010wu,Caporale:2012ih,Ducloue:2013hia,Ducloue:2013bva,Caporale:2013uva,Caporale:2014gpa,Colferai:2015zfa,Caporale:2015uva,Ducloue:2015jba,Celiberto:2015yba,Celiberto:2015mpa,Celiberto:2016ygs,Celiberto:2016vva,Caporale:2018qnm}), the inclusive detection of two light-charged rapidity-separated hadrons~\cite{Celiberto:2016hae,Celiberto:2016zgb,Celiberto:2017ptm}, three- and four-jet hadroproduction~\cite{Caporale:2015vya,Caporale:2015int,Caporale:2016soq,Caporale:2016vxt,Caporale:2016xku,Celiberto:2016vhn,Caporale:2016djm,Caporale:2016lnh,Caporale:2016zkc}, $J/\Psi$-plus-jet~\cite{Boussarie:2017oae}, hadron-plus-jet~\cite{Bolognino:2019cac}, Higgs-plus-jet~\cite{Celiberto:2020tmb,Celiberto:2021fjf}, heavy-light dijet system~\cite{Bolognino:2021mrc,Bolognino:2021hxx} and forward Drell–Yan dilepton production with a possible backward-jet tag~\cite{Golec-Biernat:2018kem}. 

In phenomenological analyses of inclusive processes with a theoretical setup based on the BFKL resummation, a significant issue was recognized that NLA corrections both to the BFKL Green’s function and impact factors turn out to be of the same size and with opposite sign of pure LLA contributions, that making the perturbative series highly unstable.
This behavior manifests as a large uncertainty arising from the renormalization (and factorization) scale choice, which by its turn calls for some optimization procedure to make reliable predictions. 

Recently, a new direction toward restoring the stability of the BFKL series at natural scales was initiated, via the analysis of semi-hard reactions with detected objects featuring large transverse masses, for example, the inclusive production of a Higgs in association with a jet~\cite{Celiberto:2020tmb,Celiberto:2021fjf}, which bring evidence that high-energy resummed distributions in rapidity and transverse momentum exhibit solid stability under higher-order corrections. A relevant step in this direction was taken in~\cite{Bolognino:2021mrc,Bolognino:2021hxx,Celiberto:2021dzy}, where a hybrid formalism combining the BFKL resummation and collinear factorization was adopted. 

Aware of the importance to pursue the phenomenological paths toward looking for signals of stabilization of the high-energy resummation under higher-order corrections and energy-scale variation, we will study and discuss two distinct classes of semi-hard reactions, according to the necessity for stabilizing the (scale optimization) procedure of the perturbative series. The first type of processes we are interested in, are the reactions that show sensitivity to the renormalization/factorization scale choice, and cannot be studied at a "natural" scale, while the second processes are characterized by natural energy scales provided by the nature of the large transverse masses and kinematics configurations of the final detected particles, where there is no need to use scale optimization procedures. 
\begin{center}
\noindent\rule{8cm}{0.4pt}
\end{center}
The thesis is organized as follows: In Chapter~\ref{chapter1}, we present briefly the basic building blocks of BFKL formalism, such as the gluon Reggeization, the BFKL factorization structure of the amplitudes at LLA and NLA, and the definition of impact factors at both LLA and NLA. In Chapter~\ref{chapter2}, we will sketch a general framework used to describe inclusive processes at the hand of BFKL resummation, and investigate the Hadron-jet production channel, by addressing the previously mentioned issues of the unsuitability effects under choosing different energy-scales values. In Chapter~\ref{chapter3}, we present studies on Higgs-plus-jet and Bottom-flavored inclusive emissions as stabilizers of the high-energy resummation. In Chapter~\ref{chapter4}, we will make general conclusions and point to possible future outlooks.

The results presented in this thesis, specifically in chapter~\ref{chapter3} and section~\ref{sec:Hadron-jet} of chapter~\ref{chapter2}, are the subject of three publications:\\
\begin{itemize}
	\item 
	A.D.~Bolognino, F.G.~Celiberto, D.Yu.~Ivanov, M.M.A.~Mohammed, A.~Papa,
	%  \emph{Hadron-jet correlations in high-energy hadronic collisions at the LHC},
	Eur.\ Phys.\ J.\ C {\bf 78} (2018) no.9,  772
	%doi:10.1140/epjc/s10052-018-6253-7}
	[arXiv:1808.05483 [hep-ph]].
	%%CITATION = doi:10.1140/epjc/s10052-018-6253-7;%%
	%1 citations counted in INSPIRE as of 25 Jan 2019
	
	\item 
	F.G.~Celiberto, D.Yu.~Ivanov, M.M.A.~Mohammed, A.~Papa,
	%  \emph{Hadron-jet correlations in high-energy hadronic collisions at the LHC},
	Eur.\ Phys.\ J.\ C {\bf 81} (2021), 293
	%doi:10.1140/epjc/s10052-018-6253-7}
	[arXiv:2008.00501 [hep-ph]].
	
	\item
	F.G.~Celiberto, M.~Fucilla, D.Yu.~Ivanov, M.M.A.~Mohammed, A.~Papa,
	%  \emph{Hadron-jet correlations in high-energy hadronic collisions at the LHC},
	Phys.\ Rev.\ D. {\bf 104} (2021), 114007
	[arXiv:2109.11875 [hep-ph]].
	 
\end{itemize}
\end{spacing}
\setcounter{secnumdepth}{2}

\chapter{Basics of the BFKL approach}\label{chapter1}
\begin{spacing}{1.5}

The BFKL approach goes back to the middle of the 70-th of the last century, as an attempt to answer the question, what is the structure of the Pomeron in the gauge field theories, in particular, what is the perturbative QCD (pQCD) Pomeron. In pQCD, for scattering of particles at high center-of-mass (c.o.m) energy $\sqrt{s}$, and fixed transfer momentum $t$ (i.e. not growing with $s$), the amplitude is dominated by Reggeized gluons exchanges in the $t$-channel, and goes like $s^{j(t)}$ within BFKL approach.
This chapter is devoted to showing the main building blocks used in the derivation of BFKL equation which determines the behavior pQCD amplitudes at high-energies in which vacuum quantum numbers are exchanged in the $t$-channel, by resumming order by order terms proportional to powers of large energy logarithms that appear in the perturbative series. A pedagogical derivation of the BFKL equation in momentum space can be found in the textbooks~\cite{barone2002high, forshaw1997quantum, ioffe2010quantum}, our presentation focuses on those points which will be relevant for the rest of this thesis. In the first section, we discuss Gluon Reggeization, the second and third sections are dedicated to presenting the BFKL approach in both leading (LLA) and next-to-leading (NLA) approximations, respectively. Finally, the derivation of the BFKL amplitude in multi-Regge kinematics, and the definition of the impact factors are presented.

%============================================================
\section{Gluon Reggeization}\label{Gluon_Reggeization}
%============================================================

The cornerstone of the BFKL approach is the gluon Reggeization, the notion of Reggei-zation of elementary particles in perturbation theory dates back to the work of Gell-Mann and his collaborators~\cite{Gell-Mann:1962xxa, Gell-Mann:1963zxa, Gell-Mann:1964ncb}.
% In terms of the relativistic partial wave amplitude $A_j(t)$, analytically continued to complex $j$ values. 
The Reggeization of a given elementary particle of spin $j$ and mass $m$ usually means that the amplitude of a scattering process with exchange of the quantum numbers of that particle in the $t$-channel goes like $s^{j(t)}$, in the limit $s \gg |t|$ (Regge limit). The function $j(t)$ is called the Regge trajectory of the given particle and takes the value of the spin of that particle when $t$ is equal to its squared mass. In perturbative QCD, the notion of gluon Reggeization means that for a scattering process in a dynamical regime with
large $\sqrt{s}$ and fixed transfer momentum $t$ (i.e. not growing with $s$), the leading contribution in each order of perturbation theory to the process amplitude is given by the Reggeized gluon. The Reggeization of the gauge boson was checked in both QED and QCD, where in the former case only the electron does Reggeize in perturbation theory~\cite{Gell-Mann:1964ncb}, but the photon remains elementary~\cite{Mandelstam:1965zz}, and in the latter case all elementary particles of the theory are Reggeize (eg.gluons~\cite{Fadin:1975cb,kuraev1976multi},\cite{Grisaru:1973vw,grisaru1973reggeization,Lipatov:1976zz}, as well as quark~\cite{fadin1976fermion, Fadin:1977jr, Bogdan:2002sr}). 
The gluon Reggeization is very important at high-energy processes since the exchanges of the gluon in the $t$-channel provide nondecrease of cross sections with energy, where the Reggeized gluon is a gluon with a modified propagator of the form
\begin{equation}\label{eq:reggeized_gluon_propagator}
D_{\mu\nu}(s,q^{2})=-i\frac{q_{\mu\nu}}{q^{2}}\left(\frac{s}{s_{0}}\right)^{\alpha_{g}(q^{2})-1},
\end{equation}
where $\alpha_{g}(q^{2})=1+\omega(t)$ is the gluon trajectory. 

\begin{figure}[h]
	\centering
	\includegraphics[scale=0.55]{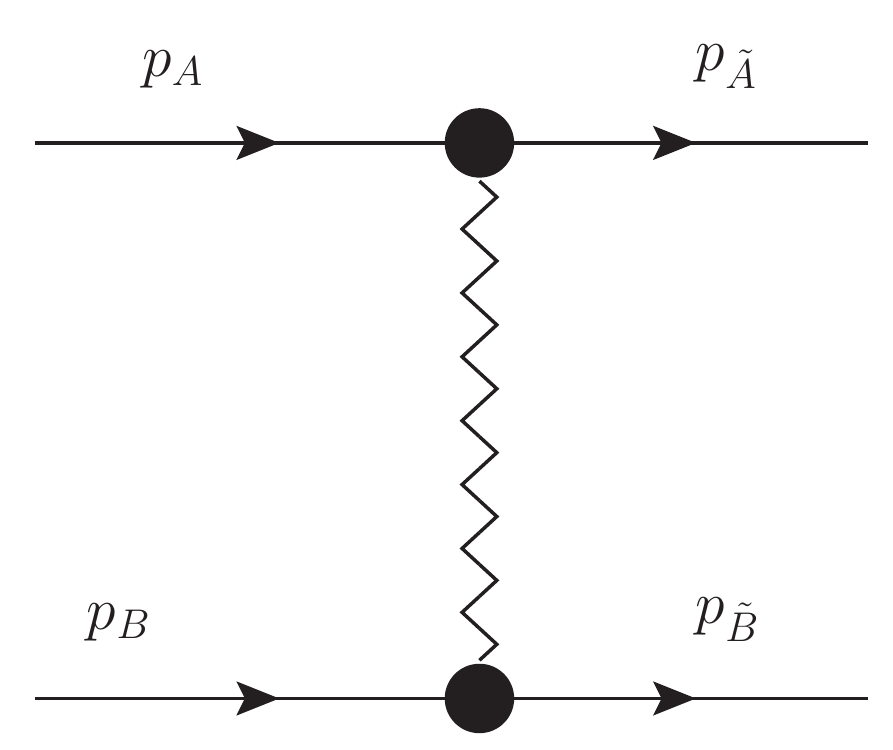}
	\caption{\textbf{The process $A+B\rightarrow A^{\prime}+B^{\prime}$ with colour octet in the $t$ channel and negative signature. The zigzag line represent Reggeized gluon exchange.}}
	\label{fig1:elastic}
\end{figure}

The simplest realization of the gluon Reggeization is in the elastic process $A+B\rightarrow A^{\prime}+B^{\prime}$ (Fig~\ref{fig1:elastic}), with an exchange of gluon quantum numbers in the $t$-channel. Gluon Reggeization means that in the Regge kinematical region $s\simeq-u\rightarrow \infty$, $t$ fixed (i.e. not growing with s), the amplitude of this process takes the Regge form
\begin{equation}\label{eq:Regge_form_amplitude}
\left(A_{8}\right)^{A^{\prime}B^{\prime}}_{AB}=\Gamma^{c}_{AA^{\prime}}\frac{s}{t}\bigg[\left(\frac{s}{-t}\right)^{\omega(t)}+\left(\frac{-s}{-t}\right)^{\omega(t)}\bigg]\Gamma^{c}_{BB^{\prime}},
\end{equation}
which is valid in the LLA as well as in the NLA. Here, $c$ is a colour index, and $\Gamma^{c}_{pp^{\prime}}$ are the particle-particle-Reggeon (PPR) vertices\footnote{At leading-order PPR vertices are determined by the Born amplitudes}, not
depending on $s$, and $\omega(t)$ is the Reggeized gluon trajectory, at 1-loop is determined by the $s$-channel discontinuity of the amplitudes. It reads
\begin{equation}\label{eq:LLA_gluon_trajectory}
\omega^{(1)}(t)= \frac{g^{2}N_c t}{2(2\pi)^{D-1}}\int\frac{d^{D-2}q_1}{\vec{q_1}^{\:2}(\vec{q}-\vec{q_1})^{2}}\simeq -\frac{g^2N_c \Gamma(1-\epsilon)}{(4\pi)^{2+\epsilon}}\frac{2}{\epsilon}\left(\vec{q}^{\:2}\right)^{\epsilon}.
\end{equation}
The space-time dimension $D=4+2\epsilon$ is taken different from 4 to regularize infrared
divergencies.
\begin{figure}[h]
	\centering
	\includegraphics[scale=0.35]{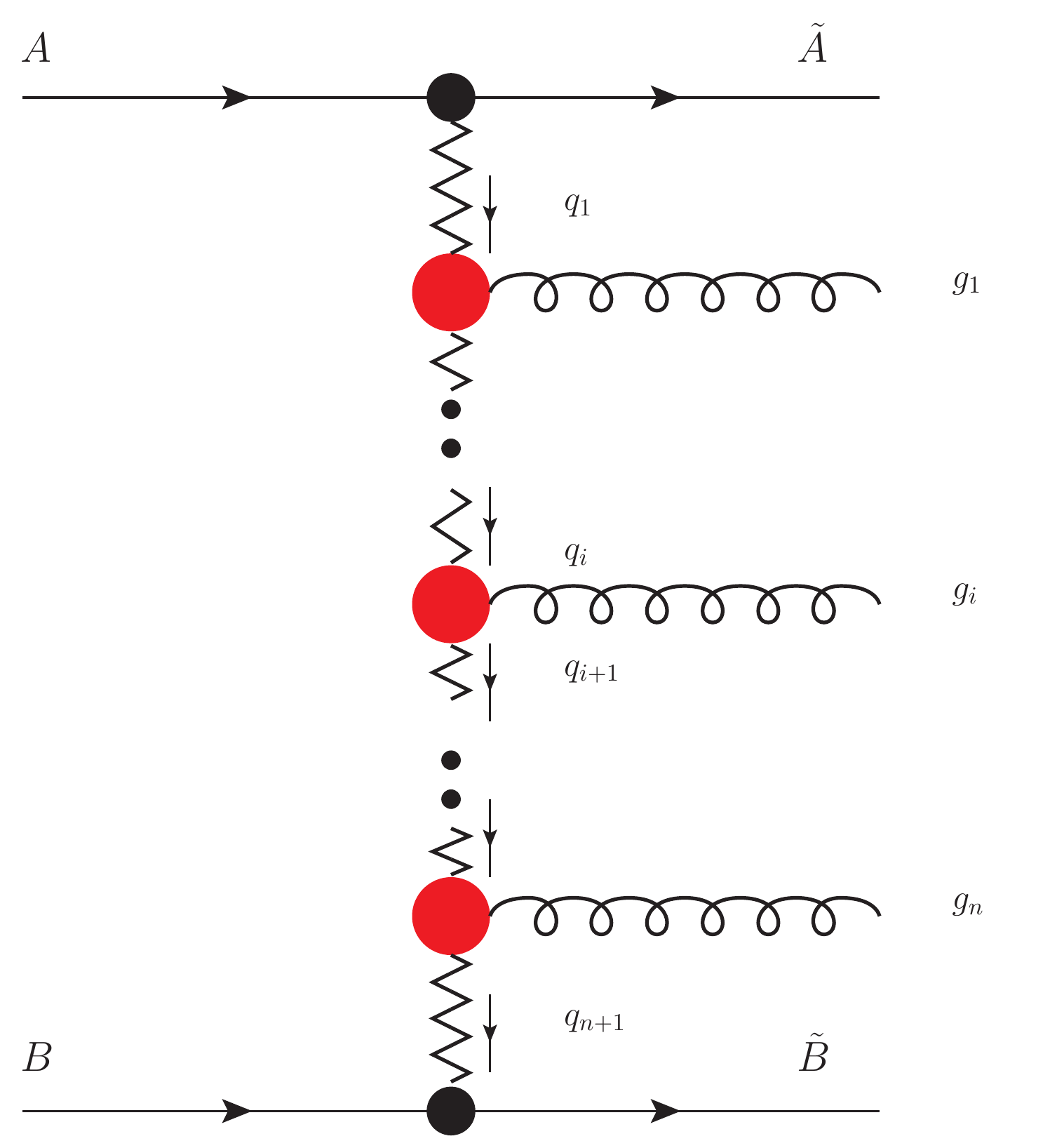}
	\caption{\textbf{Diagrammatical representation of the production amplitude $A^{A^{\prime}B^{\prime}+n}_{AB}$ in LLA}.}
	\label{fig2:inelastic}
\end{figure}
%============================================================
\section{BFKL in LLA}\label{LLA_BFKL}
%============================================================
If we consider inelastic amplitude for the process $A+B\rightarrow A^{\prime}+B^{\prime}+n$ (see Fig~\ref{fig2:inelastic}), where $n$ is a system of produced particles with momenta $k_i$  $(i=1,...,n)$.
In the LLA the main contributions to the unitarity relations come from the multi-Regge kinematics (MRK)\footnote{When rapidities of the produced particles are strongly ordered and their transverse momenta do not grow with $s$.}, where the real part of the production amplitude can be written in the factorized form
\begin{equation}\label{eq:lla_regge_amplitude}
\left(A_{8}\right)^{A^{\prime}B^{\prime}+n}_{AB}=2s\Gamma^{c_1}_{AA^{\prime}}
\bigg[\prod_{i=1}^{n}\gamma^{P_i}_{c_ic_i+1}\left(q_i,q_{i+1}\right)\frac{1}{t_i}\left(\frac{s_i}{s_R}\right)^{\omega_i}\bigg]\frac{1}{t_{n+1}}\left(\frac{s_{n+1}}{s_R}\right)^{\omega_{n+1}}\Gamma^{c_{n+1}}_{BB^{\prime}},
\end{equation}
where $\gamma^{P_i}_{c_ic_i+1}\left(q_i,q_{i+1}\right)$ are the gauge invariant effective vertices of two $t$-channel gluons (Reggeon) with color indices $c_{i,i+1}$ and momenta $q_{i,i+1}$, to a produced $s$-channel gluons of momenta $k_i=q_i-q_{i+1}$, $q_0\equiv p_A$, $q_{n+1}\equiv -p_B$, $s_i=(k_{i-1}+k_i)^2$, $k_0\equiv p_{\tilde{A}}$, $k_{n+1}\equiv p_{\tilde{B}}$, and $\omega_i$ stand for $\omega(t_i)$, with $t_i=q^2_i$. The elastic amplitude for the process $A+B\rightarrow A^{\prime}+B^{\prime}$ at high-energies can be obtained from the previous expression for the production amplitudes by using $s$-channel unitarity, and reads
\begin{multline}\label{lla_amplitude}
A_{AB}^{A^{\prime}B^{\prime}}= \frac{is}{(2\pi)^{D-2}}\int \frac{d^{D-2}q_{1}}{\vec{q^2}_{1}\vec{q}^{\prime 2}_{1}}\frac{d^{D-2}q_{2}}{\vec{q^2}_{2}\vec{q}^{\prime 2}_{2}}\int_{\delta-i\infty}^{\delta+i\infty}\frac{d\omega}{\sin(\pi\omega)}\sum_{R,\nu}\Phi^{(R,\nu)}_{AA^{\prime}}(\vec{q}_{1};\vec{q};s_0)
\\
\times\left[\left(\frac{s}{s_{0}}\right)^{\omega}-\tau\left(\frac{-s}{s_{0}}\right)^{\omega}\right] G^{(R)}_{\omega}(\vec{q}_{1},\vec{q}_{2};\vec{q})\Phi^{(R,\nu)}_{BB^{\prime}}(-\vec{q}_{2};-\vec{q};s_{0}).
\end{multline}
The quantities $\Phi^{(R,\nu)}_{PP^{\prime}}$ are the so-called impact factors in the $t$-channel color state $(R,\nu)$, they are process dependent parts, describing the transition $A\rightarrow A^{\prime}$ and $B\rightarrow B^{\prime}$, the summation is over the color group irreducible representations $R$, which are contained in the product of two adjoint representations and over its states $\nu$. The function $G^{(R)}_\omega$ represents the Mellin transform of the Green’s functions for Reggeon-Reggeon scattering and obeys the following iterated equation:
\begin{equation}\label{bfkl_green_fun}
\omega G^{(R)}_{\omega}(\vec{q}_{1},\vec{q}_{2};\vec{q})= \vec{q}^2_{1}\vec{q}^{2\prime}_{1} \delta^{D-2}(\vec{q}_{1}-\vec{q}_{2})+\int \frac{d^{D-2}\vec{q}_r}{\vec{q}^2_{r}\vec{q}^{2\prime}_{r}} K^{(R)}(\vec{q}_{1},\vec{q}_{r};\vec{q})G^{(R)}_{\omega}(\vec{q}_{r},\vec{q}_{2};\vec{q}),
\end{equation}
whose integral kernel is
\begin{equation}\label{bfkl_lla_kernel}
K^{(R)}(\vec{q}_{1},\vec{q}_{2};\vec{q})= \left[\omega(-\vec{q}^2_1)+\omega(-\vec{q}^{2\prime}_{1})\right] \delta^{D-2}(\vec{q}_{1}-\vec{q}_{2}) + K^{(R)}_r(\vec{q}_{1},\vec{q}_{2};\vec{q}),
\end{equation}
where the first term related to the gluon trajectory corresponds to the virtual corrections, while the second one corresponds to the real ones. The BFKL equation is the integral equation given in Eq~(\ref{bfkl_green_fun}), and is valid for $t = 0$
and singlet ($R=0$) quantum numbers in the $t$-channel.
%============================================================
\section{BFKL in NLA}\label{NLA_BFKL}
%============================================================
The ingredients used to compute BFKL equation at NLA~\cite{fadin1998bfkl}, were the 2-loop gluon Regge trajectory~\cite{fadin1995gluon,fadin1996gluon}, the 1-loop vertex for production of one gluon in the collision of two Reggeons (the Lipatov effective vertex~\cite{fadin1996next}),the -Born level- vertices for two gluon emission~\cite{fadin1997one,kotsky1998two} and for the production of a quark-antiquark
pair~\cite{fadin1998quark}.

In the NLA the unitarity relations must be modified from two perspectives: the first one due to the 1-loop corrections in effective
vertex for the production of the particles $P_i$, and the particle-particle-Reggeon vertices, these set of corrections is realized by performing one of the following replacements in the production amplitudes~\ref{eq:lla_regge_amplitude}:
$$\omega^{(1)}\rightarrow \omega^{(2)}, \quad \Gamma^{C(Born)}_{P^{\prime}P} \rightarrow \Gamma^{C(1-loop)}_{P^{\prime}P},  \gamma^{G_i(Born)}_{c_ic_{i+1}} \rightarrow \gamma^{G_i(1-loop)}_{c_ic_{i+1}}.$$
 
The second set of modifications is consists of allowing the emission of one pair of particles with rapidities of the same order of magnitude, both in the central or in the fragmentation region, this means performing one of the following replacements in the production amplitudes~\ref{eq:lla_regge_amplitude}:
$$\Gamma^{C(Born)}_{P^{\prime}P} \rightarrow \Gamma^{C(Born)}_{\{f\}P},  \gamma^{G_i(Born)}_{c_ic_{i+1}} \rightarrow \gamma^{Q\bar{Q}(Born)}_{c_ic_{i+1}},\quad \gamma^{G_i(Born)}_{c_ic_{i+1}} \rightarrow \gamma^{GG(Born)}_{c_ic_{i+1}},$$
where $\Gamma_{\{f\}P}$ stands for the all intermediate states in the fragmentation region of particle $P$ scattering off Reggeon, and  $\gamma^{Q\bar{Q}}, \gamma^{GG}$ are the effective vertices for the production of a quark-antiquark pair and of a two-gluon pair, respectively.

At NLA the terms proportional to $\beta_0$, that account for the inclusion of the running of the coupling, introduce a logarithmic dependence in momentum space representation, thus the positiveness of cross sections is not always ensured. Moreover, the NLL corrections to the BFKL Green's function are known to be very large and negative~\cite{ross1998effect}, and all that led to many ways to improve the convergence of the perturbative series~\cite{PhysRevD.68.114003,PhysRevD.60.054031,brodsky1999qcd,levin1999bfkl,salam1998resummation,ciafaloni2003extending}.  
%============================================================
\section{The BFKL amplitude}\label{BFKL_Amplitude}
%============================================================
To draw a general scheme about how to study different processes within NLA BFKL approach (which depends on the available NLO impact factors), let us consider the following process which involves the exchange of the vacuum quantum numbers in the $t$-channel: $A+B\rightarrow A^{\prime}+B^{\prime}$.
\begin{figure}[h]
	\centering
	\includegraphics[scale=0.45]{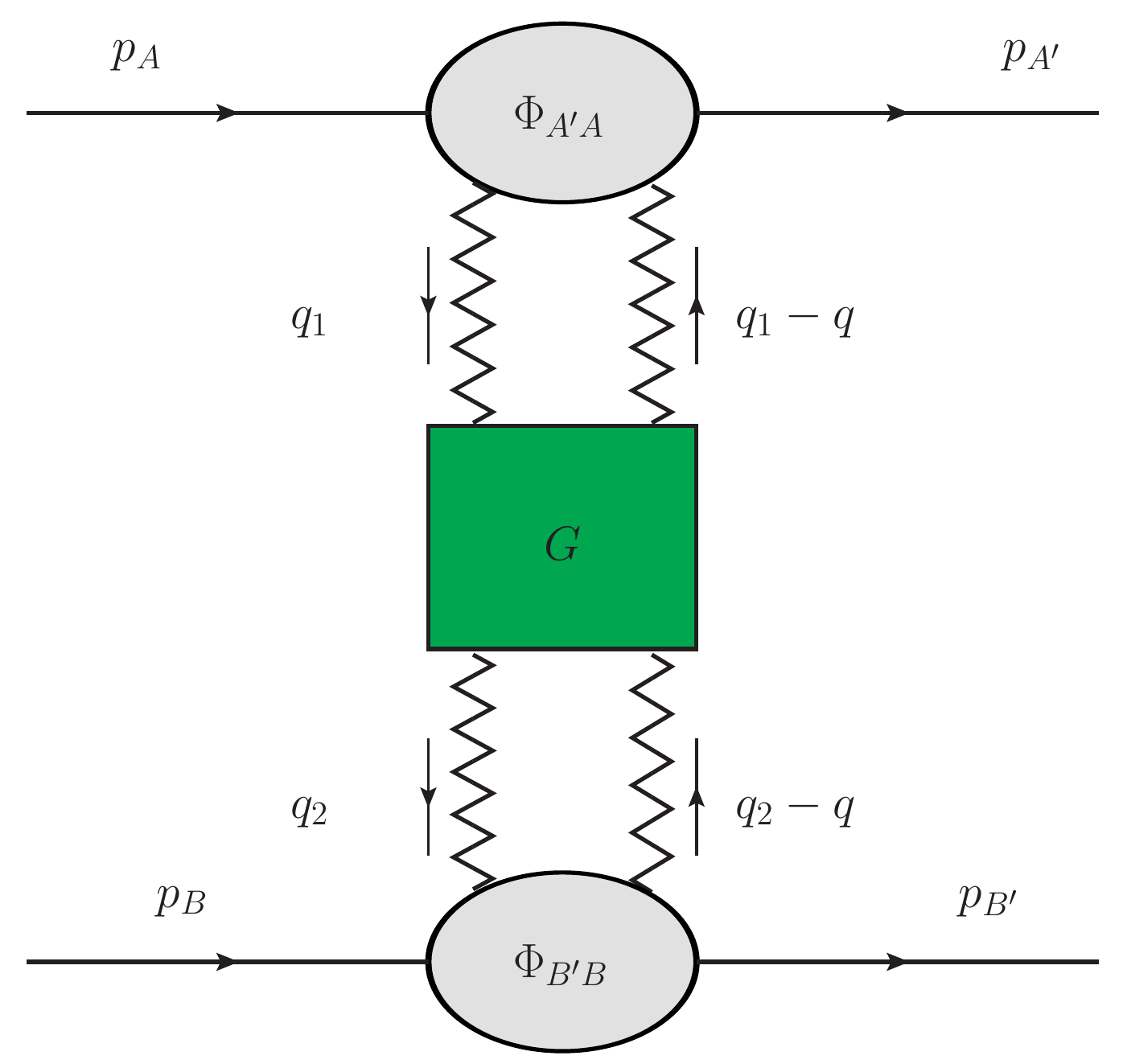}
	\caption{\textbf{Diagrammatic representation of the imaginary part of amplitude for the process} $A+B\rightarrow A^{\prime}+B^{\prime} $.}
	\label{factorized_amplitude}
\end{figure}\\
Using the optical theorem \cite{donnachie2002pomeron,PhysRevD.2.2963,PhysRevD.6.2906}, the total cross-section of this process reads
\begin{equation}\label{2.26}
\sigma_{AB}(s)=\frac{\text{Im}_{s}A_{AB}^{A^{\prime}B^{\prime}}}{s}.
\end{equation}
Within BFKL approach, the imaginary part expression of the amplitude can be factorized \cite{fadin1998bfkl} as a convolution of the Green’s function of two interacting Reggeized gluons with the impact factors\footnote{Which are practically the couplings of the pomeron to the hadrons.} (vertices $\Phi_{A\rightarrow A^{\prime}}$ and $\Phi_{B\rightarrow B^{\prime}}$) of the colliding particles [$\Phi_{A\rightarrow A^{\prime}}\otimes G \otimes \Phi_{B\rightarrow B^{\prime}}$] , and it is given as
\begin{multline}\label{imag_amplitude}
\text{Im}_{s}A_{AB}^{A^{\prime}B^{\prime}}= \frac{s}{(2\pi)^{D-2}}\int \frac{d^{D-2}q_{1}}{\overrightarrow{q}_{1}}\Phi_{A}(\overrightarrow{q}_{1},s_{0})\int_{\delta-i\infty}^{\delta+i\infty}\frac{d\omega}{2\pi i}\left(\frac{s}{s_{0}}\right)^{\omega}G_{\omega}(\overrightarrow{q}_{1},\overrightarrow{q}_{2})
\\
\times\int\frac{d^{D-2}q_{2}}{\overrightarrow{q}_{2}}\Phi_{B}(-\overrightarrow{q}_{2},s_{0}), 
\end{multline}
where $\overrightarrow{q}_{1}$, $\overrightarrow{q}_{2}$  are the transverse momenta of the Reggeized gluons, $s$ the squared center-of-mass energy and $s_{0}$ are arbitrary energy scale parameter.
The Green's function $G_{\omega}(\overrightarrow{q}_{1},\overrightarrow{q}_{2})$ is process independent and obeys the BFKL equation,
\begin{equation}\label{amp_green_fun}
\omega G_{\omega}(\overrightarrow{q}_{1},\overrightarrow{q}_{2})= \delta^{D-2}(\overrightarrow{q}_{1}-\overrightarrow{q}_{2})+\int d^{D-2}\overrightarrow{q} K(\overrightarrow{q}_{1},\overrightarrow{q})G_{\omega}(\overrightarrow{q},\overrightarrow{q}_{2}),
\end{equation}
with $ K(\overrightarrow{q}_{1}, \overrightarrow{q})$ the BFKL kernel. The BFKL kernel determines the energy dependence of scattering amplitudes.
The vertices (impact factors $\Phi_{A\rightarrow A^{\prime}},\Phi_{B\rightarrow B^{\prime}}$) depend on the particular process under analysis, they are known in NLA for a few processes, in particular, the impact factors for colliding-parton (quarks and gluons) impact
factors~\cite{Fadin:1999de,Fadin:1999df}, which represent the common
basis for the calculation of the forward-jet impact
factor~\cite{Bartels:2001ge,Bartels:2002yj,Caporale:2011cc,Caporale:2012ih,Ivanov:2012ms,Colferai:2015zfa}, and of
the forward light-charged hadron one~\cite{Ivanov:2012iv}; the impact
factor describing the $\gamma^*$ to light-vector-meson leading twist
transition~\cite{Ivanov:2004pp}, and the $\gamma^*$ to $\gamma^*$
transition~\cite{Bartels:2000gt,Bartels:2001mv,Bartels:2002uz,Bartels:2004bi,Fadin:2001ap,Balitsky:2012bs}, and recently, the one for the production of a forward Higgs boson in the infinite top-mass limit~\cite{Hentschinski:2020tbi,Nefedov:2019mrg}.

Now to derive the general form for the amplitude in the so-called $(\nu, n)$-representation, it is favorable to work within the transverse momentum representation, which is defined as,
$$\hat{\overrightarrow{q}}\ket{\overrightarrow{q}_{i}}=\overrightarrow{q}_{i}\ket{\overrightarrow{q}_{i}},\quad \braket{\overrightarrow{q}_{1}|\overrightarrow{q}_{2}}=\delta^{(2)}(\overrightarrow{q}_{1}-\overrightarrow{q}_{2}),$$
\begin{equation}\label{pT_reprs}
\braket{A|B}=\braket{A|\overrightarrow{k}}\braket{\overrightarrow{k}|B}=\int d^{2}kA(\overrightarrow{k})B(\overrightarrow{k}).
\end{equation}
The BFKL kernel $ K(\overrightarrow{q}_{1},\overrightarrow{q}_{2})$ becomes 
\begin{equation}\label{pT_kernel}
K(\overrightarrow{q}_{1},\overrightarrow{q}_{2})=\bra{\overrightarrow{q}_{1}}\hat{K}\ket{\overrightarrow{q}_{2}},
\end{equation}
and the equation (\ref{amp_green_fun}) reads
$$\omega\bra{\overrightarrow{q}_{1}}\hat{G}_{\omega}\ket{\overrightarrow{q}_{2}}=\braket{\overrightarrow{q}_{1}|\overrightarrow{q}_{2}}+\int d^{2}\overrightarrow{q}\bra{\overrightarrow{q}_{1}}\hat{K}\ket{\overrightarrow{q}}\bra{\overrightarrow{q}}\hat{G}_{\omega}\ket{\overrightarrow{q}_{2}},$$
$$\omega\bra{\overrightarrow{q}_{1}}\hat{G}_{\omega}\ket{\overrightarrow{q}_{2}}-\bra{\overrightarrow{q}_{1}}\hat{K}\hat{G}_{\omega}\ket{\overrightarrow{q}_{2}}=\braket{\overrightarrow{q}_{1}|\overrightarrow{q}_{2}},$$
\begin{equation}\label{pT_green_fun}
\bra{\overrightarrow{q}_{1}}(\omega-\hat{k})\hat{G}_{\omega}\ket{\overrightarrow{q}_{2}}=\braket{\overrightarrow{q}_{1}|\overrightarrow{q}_{2}},
\end{equation}
which has the solution 
\begin{equation}\label{hat_green_fun}
\hat{G}_{\omega}=(\omega-\hat{k})^{-1}.
\end{equation}
The kernel expanded in terms of the strong coupling, given as 
\begin{equation}\label{hat_kernel}
\hat{K}=\bar{\alpha}_{S}\hat{K}^{0}+\bar{\alpha}_{S}^{2}\hat{K}^{1},
\end{equation} 
with $\bar{\alpha}_{S}=\alpha_{S}N_{c}/\pi$, and $N_{c}$ the number of colours, $\hat{K}^{0}$ and $\hat{K}^{1}$ are BFKL kernel in LLA and NLA accuracy, respectively. Using this kernel expansion we can rewrite the Green function operator $\hat{G}_{\omega}$ with NLA accuracy as
\begin{equation}\label{xpnd_hat_green_fun}
\hat{G}_{\omega}=(\omega-\bar{\alpha}_{S}\hat{K}^{0})^{-1}+(\omega-\bar{\alpha}_{S}\hat{K}^{0})^{-1}(\bar{\alpha}_{s}^{2}\hat{K}^{1})(\omega-\bar{\alpha}_{S}\hat{K}^{0})^{-1}+\mathcal{O}\left[(\bar{\alpha}_{s}^{2}\hat{K}^{1})^{2}\right].
\end{equation} 
In this representation the amplitude (\ref{imag_amplitude}) takes the form
\begin{equation}\label{pT_imag_amplitude}
Im_{s}A_{AB}^{A^{\prime}B^{\prime}}= \frac{s}{(2\pi)^{D-2}}\int_{\delta-i\infty}^{\delta+i\infty}\frac{d\omega}{2\pi i}\left(\frac{s}{s_{0}}\right)^{\omega}\bra{\frac{\Phi_{A}}{\overrightarrow{q}_{1}}}\hat{G}_{\omega}\ket{\frac{\Phi_{B}}{\overrightarrow{q}_{2}}}.
\end{equation}
The basis of the eigenfunction of $\hat{K}^{0}$, with $\chi(\nu, n)$ LLA BFKL trajectory,
\begin{equation}\label{lo_kernel_act}
\hat{K}^{0}\ket{\nu, n}=\chi(\nu, n)\ket{\nu, n}, \quad \chi(\nu, n)=2\psi(1)-\psi(\frac{n}{2}+\frac{1}{2}+i\nu)-\psi(\frac{n}{2}+\frac{1}{2}-i\nu),
\end{equation}
with $\psi(x)=\Gamma^{\prime}(x)/\Gamma(x)$, is the logarithmic derivative of the Euler gamma function; is given by the set of functions bellow
\begin{equation}\label{lo_kernel_eigvl}
\braket{\overrightarrow{q}|\nu, n}=\frac{1}{\pi\sqrt{2}}(\overrightarrow{q}^{2})^{i\nu-1/2}e^{in\phi},
\end{equation}
where $\phi$ is the azimuthal angle of $\overrightarrow{q}$. So, the orthonormality and completeness conditions, respectively read
\begin{equation}\label{norm_lo_kernel_eigvl}
\braket{\nu^{\prime}, n^{\prime}|\nu, n}=\int\frac{d^2q}{2\pi^{2}}(\overrightarrow{q}^{2})^{i\nu-i\nu^{\prime}-1}e^{i(n-n^{\prime})\phi}=\delta(\nu-\nu^{\prime})\delta_{nn^{\prime}}.
\end{equation} 
\begin{equation}\label{orth_lo_kernel_eigvl}
\hat{1}=\sum_{n=-\infty}^{+\infty}\int_{-\infty}^{+\infty}d\nu \ket{\nu, n}\bra{\nu, n}.
\end{equation}
Using (\ref{orth_lo_kernel_eigvl}) twice in (\ref{pT_imag_amplitude}) we will end up with
\begin{equation}\nonumber
Im_{s}A_{AB}^{A^{\prime}B^{\prime}}= \frac{s}{(2\pi)^{D-2}}\sum_{n=-\infty}^{+\infty}\int_{-\infty}^{+\infty}d\nu\sum_{n^{\prime}=-\infty}^{+\infty}\int_{-\infty}^{+\infty}d\nu^{\prime}\int_{\delta-i\infty}^{\delta+i\infty}\frac{d\omega}{2\pi i}\left(\frac{s}{s_{0}}\right)^{\omega}
\end{equation}
\begin{equation}\label{nu_n_imag_amplitude}
\times \braket{\frac{\Phi_{A}}{\overrightarrow{q}_{1}}|\nu, n}\bra{\nu, n}\hat{G}_{\omega}\ket{\nu^{\prime}, n^{\prime}}\braket{\nu^{\prime}, n^{\prime}|{\frac{\Phi_{B}}{\overrightarrow{q}_{2}}}},
\end{equation}
where $\braket{\frac{\Phi_{1}}{\overrightarrow{q}_{1}}|\nu, n}$ and $\braket{\nu^{\prime}, n^{\prime}|{\frac{\Phi_{2}}{\overrightarrow{q}_{2}}}}$ are projections of the impact factors onto the eigenfunctions of LO BFKL, and are given as
\begin{equation}\label{projected_A_IF}
\braket{\frac{\Phi_{A}}{\overrightarrow{q}_{1}}|\nu, n}\equiv \int d^{2}q_{1}\frac{\Phi_{A}(\overrightarrow{q}_{1})}{\overrightarrow{q}_{1}^{2}}\frac{1}{\pi\sqrt{2}}(\overrightarrow{q}_{1}^{2})^{i\nu-\frac{1}{2}}e^{in\phi_{1}}.
\end{equation}
\begin{equation}\label{projected_B_IF}
\braket{\nu, n|\frac{\Phi_{B}}{\overrightarrow{q}_{2}}}\equiv \int d^{2}q_{2}\frac{\Phi_{B}(\overrightarrow{-q}_{2})}{\overrightarrow{q}_{2}^{2}}\frac{1}{\pi\sqrt{2}}(\overrightarrow{q}_{2}^{2})^{-i\nu-\frac{1}{2}}e^{-in\phi_{2}}.
\end{equation}
The action of the full NLO BFKL kernel on the set of functions Eq~(\ref{lo_kernel_eigvl}) is expressed as follows:
\begin{equation}\nonumber
\hat{K}\ket{\nu, n}=\bar{\alpha}_{s}(\mu_{R})\chi(\nu, n)\ket{\nu, n}+\bar{\alpha}^{2}_{s}(\mu_{R})\bigg(\chi^{(1)}(\nu, n)
+\frac{\beta_{0}}{4N_{c}}\chi(\nu, n)\ln(\mu^{2}_{R})\bigg)\ket{\nu, n}
\end{equation}
\begin{equation}\label{nlo_kernel_act}
+\bar{\alpha}^{2}_{s}\frac{\beta_{0}}{4N_{c}}\chi(\nu, n)\bigg(i\frac{\partial}{\partial\nu}\bigg)\ket{\nu, n},
\end{equation}
where $\mu_{R}$ is the renormalization scale; the first term represents the action of LO kernel, while the second and the third ones stand for the diagonal and the non-diagonal parts of the NLO kernel,
\begin{equation}\label{qcd_beta_0}
\beta_0=\frac{11}{3} N_c - \frac{2}{3}n_f, 
\end{equation}
is the first coefficient of the QCD $\beta$-function, where $n_f$ is the number
of active flavors.

Knowing the action of the kernel (as in Eq. (\ref{lo_kernel_act})), we can use Eq. (\ref{xpnd_hat_green_fun}) to write the matrix element $\bra{\nu, n}\hat{G}_{\omega}\ket{\nu^{\prime}, n^{\prime}}$ of the BFKL Green’s function in NLA,
\begin{equation}\nonumber
\bra{n,\nu}\hat{G}_{\omega}\ket{n,\nu^{\prime}}=\delta_{nn^\prime}\delta(\nu-\nu^{\prime})\bigg[\frac{1}{\omega-\overline{\alpha}_{s}\chi(n,\nu)}+\frac{1}{(\omega-\overline{\alpha}_{s}\chi(n,\nu))^{2}}
\end{equation} 
\begin{equation}\nonumber
\times\bigg(\overline{\alpha}^{2}_{s}\bigg(\chi^{(1)}(n,\nu)+\frac{\beta_{0}}{4N_{c}}\chi(n,\nu)\ln(\mu_{R})^{2}\bigg)\bigg)\bigg]
\end{equation}
\begin{equation}\label{braket_hat_green_fun}
+\frac{\overline{\alpha}^{2}_{s}\frac{\beta_{0}}{4N_{c}}\chi(n,\nu^{\prime})\delta_{nn^\prime}}{(\omega-\overline{\alpha}_{s}\chi(n,\nu))(\omega-\overline{\alpha}_{s}\chi(n,\nu^{\prime}))}\bigg(i\frac{\partial}{\partial\nu^{\prime}}\delta(\nu-\nu^{\prime})\bigg),
\end{equation}
where the first part is LLA contribution, while other remaining terms are due to the diagonal and the non-diagonal parts of NLO kernel $\hat{K}^{1}$,
the function $\chi^{(1)}(\nu, n)$ for general conformal spin $n$,has been calculated in Ref. \cite{Kotikov:2000pm}, and conveniently given by
\begin{equation}\label{chi1_function}
\chi^{(1)}(\nu, n)=-\frac{\beta_{0}}{8N_{c}}\bigg(\chi^{2}(\nu, n)-\frac{10}{3}\chi(\nu, n)-i\chi^{\prime}(\nu, n)\bigg)+\bar{\chi}(\nu, n),
\end{equation}
where
\begin{equation}\nonumber
\bar{\chi}(\nu, n) =-\frac{1}{4}\bigg[\frac{\pi^{2}-4}{3}\chi(\nu, n)-6\zeta(3)-\chi^{\prime\prime}(\nu, n)+2\phi(\nu, n)+2\phi(-\nu, n)
\end{equation}
\begin{equation}\label{chiBar}
+\frac{\pi^{2}\sinh(\pi\nu)}{2\nu\cosh^{2}(\pi\nu)}\bigg(\bigg(3+\bigg(1+\frac{n_{f}}{N^{3}_{c}}\bigg)\frac{11+12\nu^{2}}{16(1+\nu^{2})}\bigg)\delta_{n0}-\bigg(1+\frac{n_{f}}{N^{3}_{c}}\bigg)\frac{1+4\nu^{2}}{32(1+\nu^{2})}\delta_{n2}\bigg)\bigg],
\end{equation}
\begin{equation}\nonumber
 \phi(\nu, n)=-\int_{0}^{1}dz\frac{x^{-1/2+i\nu+n/2}}{1+x}\bigg[\frac{1}{2}\bigg(\psi^{\prime}\big(\frac{n+1}{2}\big)-\zeta(2)\bigg)+Li_{2}(x)+Li_{2}(-x)
\end{equation}
\begin{equation}\nonumber
+\ln x\bigg(\psi(n+1)-\psi(1)+\ln(1+x)+\sum_{k=1}^{\infty}\frac{(-x)^{k}}{k+n}\bigg)+\sum_{k=1}^{\infty}\frac{x^{k}}{(k+n)^{2}}(1-(-1)^{k})\bigg]
\end{equation}
\begin{equation}\nonumber
=\sum_{k=1}^{\infty}\frac{(-1)^{k+1}}{k+(n+1)/2+i\nu}\bigg[\psi^{\prime}(k+n+1)-\psi^{\prime}(k+1)+(-1)^{k+1}(\beta^{\prime}(k+n+1)
\end{equation}
\begin{equation}\label{phi_function}
+\beta^{\prime}(k+1)-\frac{1}{k+(n+1)/2+i\nu}(\psi(k+n+1)-\psi(k+1))\bigg]
\end{equation}
with
\begin{equation}\label{beta_prime_Li_fun}
\beta^{\prime}(z)=\frac{1}{4}\bigg[\psi^{\prime}\bigg(\frac{z+1}{2}\bigg)-\psi^{\prime}\bigg(\frac{z}{2}\bigg)\bigg], \qquad Li_{2}(x)=-\int_{0}^{x}dt\frac{\ln(1-t)}{t}.
\end{equation}
The impact factors in Eqs. (\ref{projected_A_IF}) and (\ref{projected_B_IF}) expanded in terms of $\alpha_{s}$, read
\begin{equation}\label{xpnd_AB_IF}
\Phi_{1,2}=\alpha_{s}(\mu_{R})\Bigg(c_{1,2}(\nu, n)+\alpha_{s}(\mu_{R})c^{(1)}_{1,2}(\nu,n)\Bigg),
\end{equation}
where $c_{1,2}(\nu, n)$ and $c^{(1)}_{1,2}(\nu,n)$ are the LO and NLO impact factors in the ($\nu,n$)-re-presentation, respectively.

Substituting Eq~(\ref{chi1_function}) in Eq~(\ref{braket_hat_green_fun}), we can write the matrix element as
\begin{equation}\nonumber
\bra{n,\nu}\hat{G}_{\omega}\ket{n,\nu^{\prime}}=\delta_{nn^\prime}\delta(\nu-\nu^{\prime})\bigg[\frac{1}{\omega-\overline{\alpha}_{s}\chi(n,\nu)}+\frac{\overline{\alpha}^{2}_{s}}{(\omega-\overline{\alpha}_{s}\chi(n,\nu))^{2}}
\end{equation}
\begin{equation}\nonumber
\times\bigg\{\overline{\chi}(n,\nu)+\frac{\beta_{0}}{4N_{c}}\chi(n,\nu)\bigg(-\chi(n,\nu)+\frac{10}{3}+2\ln(\mu_{R})^{2}+i\frac{\partial}{\partial\nu}\chi(n,\nu)\bigg)\bigg\}\bigg]
\end{equation}
\begin{equation}\label{finl_braket_hat_green_fun}
+\frac{\overline{\alpha}^{2}_{s}\frac{\beta_{0}}{4N_{c}}\chi(n,\nu^{\prime})\delta_{nn^\prime}}{(\omega-\overline{\alpha}_{s}\chi(n,\nu))(\omega-\overline{\alpha}_{s}\chi(n,\nu^{\prime}))}\bigg(i\frac{\partial}{\partial \nu^{\prime}}\delta(\nu-\nu^{\prime})\bigg).
\end{equation}
It will be useful to write the following formulas, which appear latter on when we carry the integration over $\omega$ and one of the $\nu$'s variable in Eq~(\ref{nu_n_imag_amplitude})
\begin{equation}\label{usfl_int1}
\int_{\delta-i\infty}^{\delta+i\infty}\frac{d\omega}{2\pi i} \frac{\left(\frac{s}{s_{0}}\right)^{\omega}}{(\omega-\overline{\alpha}_{s}\chi(n,\nu))^{2}}=\ln\left(\frac{s}{s_{0}}\right)\left(\frac{s}{s_{0}}\right)^{\overline{\alpha}_{s}\chi(n,\nu)},
\end{equation}
\begin{equation}\label{usfl_int2}
\int_{-\infty}^{+\infty}d\nu^{\prime}\Phi_{A}(n,\nu)\bigg(i\frac{\partial}{\partial \nu^{\prime}}\delta(\nu-\nu^{\prime})\bigg)\Phi_{B}(n,\nu^{\prime})=\Phi_{A}(n,\nu)\bigg(-i\frac{\partial\Phi_{B}(n,\nu)}{\partial \nu}\bigg).
\end{equation}
Using formulas (\ref{xpnd_AB_IF}),(\ref{finl_braket_hat_green_fun}),(\ref{usfl_int1}), and (\ref{usfl_int2}) into (\ref{nu_n_imag_amplitude}), and after a few algebraic steps, we end up with the following final representation of the BFKL amplitude at NLA  
\begin{equation}\nonumber
Im_{s}A_{AB}^{A^{\prime}B^{\prime}}= \frac{s}{(2\pi)^{D-2}}\sum_{n=-\infty}^{+\infty}
\int_{-\infty}^{+\infty}d\nu\left(\frac{\hat{s}}{s_{0}}\right)^{\bar{\alpha}_{s}(\mu_{R})\chi(n,\nu)}\alpha_{s}^{2}(\mu_{R})c_{1}(n,\nu)c_{2}(n,\nu)
\end{equation}
\begin{equation}\nonumber
\times\bigg[1+\bar{\alpha}_{s}(\mu_{R})\left(\frac{c_{2}^{(1)}(n,\nu)}{c_{2}(n,\nu)}\right)
+\bar{\alpha}_{s}^{2}(\mu_{R})\ln\frac{\hat{s}}{s_{0}}
\bigg\{\bar{\chi}(n,\nu)+\frac{\beta_{0}}{8N_{c}}\chi(n,\nu)
\end{equation}
\begin{equation}
\times\bigg(-\chi(n,\nu)+\frac{10}{3}+2\ln\mu_{R}^{2}+i\frac{d\ln(\frac{c_{1}(n,\nu)}{c_{2}(n,\nu)})}{d\nu}\bigg)\bigg\}\bigg].
\end{equation}
 
%============================================================
\section{Impact factors} \label{Impact_factors_defs}
%============================================================
The notion of the Impact factors has been introduced to account for the coupling of the Pomeron to the hadrons.
We need NLO impact factos in NLA, which take contributions from virtual corrections (one-particle intermediate states) and real particle production 
(two-particle intermediate states). Although impact factors are process dependent, they all share the following universal property
$$\Phi(k,q)|^{k-q\rightarrow 0}_{k\rightarrow 0}\quad \rightarrow 0,$$
which guarantees the infra-red finiteness of the amplitude.

From the previous section~\ref{BFKL_Amplitude}, we can define the LO impact factor, which describes the transition $A\rightarrow A^\prime$, as follows (see, {\it e.g.},
Ref.~\cite{Fadin:1998fv}):
\begin{equation}\label{LLA_IF_def}
\Phi^{(0)}_{A^\prime A}(\vec{q}_R,\vec{q})=\frac{1}{\sqrt{N^2_c-1}}\sum_{\left\{f\right\}}\int  \frac{dM_{a}}{2\pi} \Gamma^{(0)}_{\{a\}A}(q_R)\Gamma^{(0)}_{\{a\}A^\prime}(q^\prime_R)^* d\rho_a,
\end{equation} 
where $\Gamma^{(0)}$ are vertices evaluated in Born approximation, the integration here is over the standard phase space $d\rho_a$ of an intermediate state $\{a\}$ as well as over its squared invariant mass $M^2_{a}$. 

The phase space of a state $a$ consisting of particles with momenta $l_n$ is
\begin{equation}
d\rho_a=(2\pi)^D\delta\left(p_A-q_R-\sum_{n\in a}l_n\right)\prod_{n \in a}\frac{d^{D-1}l_n}{(2\pi)^{D-1}2E_n},
\end{equation}
while the squared invariant mass of the state $a$
$$M^2_a=(p_A-q_R)^2=(p^\prime_A-q^\prime_R)^2.$$

At NLA the simple expression for the LLA impact factor must be modified from two perspectives. The first one due to the modifications in production vertices after including the radiative corrections, and the second is to account for the gluon emission in the central rapidity region. Therefore, the NLA definition of the impact factor reads
\begin{equation}\nonumber
\Phi^{(1)}_{A^\prime A}(\vec{q}_R,\vec{q}ls_0)= \frac{1}{\sqrt{N^2_c-1}}\sum_{\left\{f\right\}}\int  \frac{dM^2_{a}}{2\pi} \Gamma^{(0)}_{\{a\}A}(q_R)\Gamma^{(0)}_{\{a\}A^\prime}(q^\prime_R)^* d\rho_a\theta(s_{\Lambda}-M^2_a)
\end{equation}
\begin{equation}\label{NLA_IF_def}
\begin{split}
-\frac{1}{2}\int\frac{d^{D-2}q_{R^\prime}}{\vec{q}^2\vec{q}^{\prime 2}_{R^\prime}}\Phi^{(Born)}_{A^{\prime}A}(\vec{q}_{R^\prime},\vec{q})\left[K^{(Born)}_r(\vec{q}_{R^\prime},\vec{q}_R;\vec{q})\ln\left(\frac{s^2_\Lambda}{(\vec{q}_{R^\prime}-\vec{q}_R)^2s_0}\right) \right.\\
\left.+\vec{q}^{\prime 2}_{R^\prime}\vec{q}^{2}_{R^\prime}\delta(\vec{q}_R-\vec{q}_{R^\prime})\left\{\omega(t_R)\ln\left(\frac{\vec{q}^2_R}{s_0}\right)+\omega(t^\prime_R)\ln\left(\frac{\vec{q}^{\prime 2}_R}{s_0}\right)\right\}\right],
\end{split}
\end{equation}
where the intermediate parameter $s_\Lambda$ should go to infinity, and the dependence on this parameter vanishes due to the cancellation between the first and second terms; the trajectory function $\omega(t)$ is the one defined at 1-loop in Eq~(\ref{eq:LLA_gluon_trajectory}).
\end{spacing}
 
\chapter{Hadron-jet production channel}\label{chapter2}
%========================================
%\section{Jet and Dihadron production} 
\section{General remarks} 
%========================================
\begin{spacing}{1.5}
	
Inclusive processes at high-energies with tagged objects in the final state featuring large rapidity separation have been considered as a promising testfield for the search of BFKL dynamics in current and future colliders. The more the tagged objects are separated in rapidity, the deeper into the small-$x$ region of collinear PDFs we enter. In this region, and because of the conservation of momentum, the cross section decreases as the rapidity gap increases. Recently, a number of probes for BFKL signals have been proposed for different collider environments: the diffractive leptoproduction of two light vector
mesons~\cite{Ivanov_2004,Ivanov_2006,Ivanov_2007,Enberg:2005eq},
the total cross section of two highly-virtual photons~\cite{Ivanov:2014hpa},
the inclusive hadroproduction of two jets with large transverse momenta
and well separated in rapidity (Mueller-Navelet channel~\cite{Mueller:1986ey}),
for which several phenomenological studies have carried out so
far (for more details see~\cite{Celiberto:2020wpk} and references therein),
the inclusive detection of two light-charged hadrons~\cite{Celiberto_2016,Celiberto_2017,Celiberto_2017s}, three- and four-jet
hadroproduction~\cite{Caporale:2015vya,Caporale:2015int,Caporale:2016soq,Caporale:2016xku,Caporale:2016lnh,Caporale:2016zkc}, $J/\Psi$-jet~\cite{Boussarie_2018},
hadron-jet~\cite{Bolognino_2018,bolognino2019inclusive,bolognino2019highenergy}, the inclusive production of rapidity-separated $\Lambda$-$\Lambda$ or $\Lambda$-jet pairs~\cite{Celiberto:2020rxb}, and recently, double $\Lambda_{c}$ or of a $\Lambda_{c}$ plus a light-flavored jet system~\cite{Celiberto:2021dzy}, Drell-Yan--jet~\cite{Golec_Biernat_2018,Deak_2019} and heavy-quark pair
photo- and hadroproduction~\cite{Celiberto:2017nyx,Bolognino:2019ouc,Bolognino:2019yls}. 

This chapter is dedicated to give a description of inclusive semi-hard reactions with full NLA accuracy, and to give the full NLA BFKL analysis for the cross-section and azimuthal-angle for the inclusive production at the LHC of a charged
light hadron and of a jet, featuring a wide separation in rapidity, as suggested probe process for the investigation of the BFKL formalism of resummation of high-energy logarithms in the pQCD series. Moreover, we make use of optimization methods to set the values of the renormalization scale ($\mu_{R}$) and study the effect of choosing different values for the factorization scale ($\mu_{F})$, entering the theoretical description of this process. 

\begin{figure}[hbtp]
	\centering
	\includegraphics[scale=0.45]{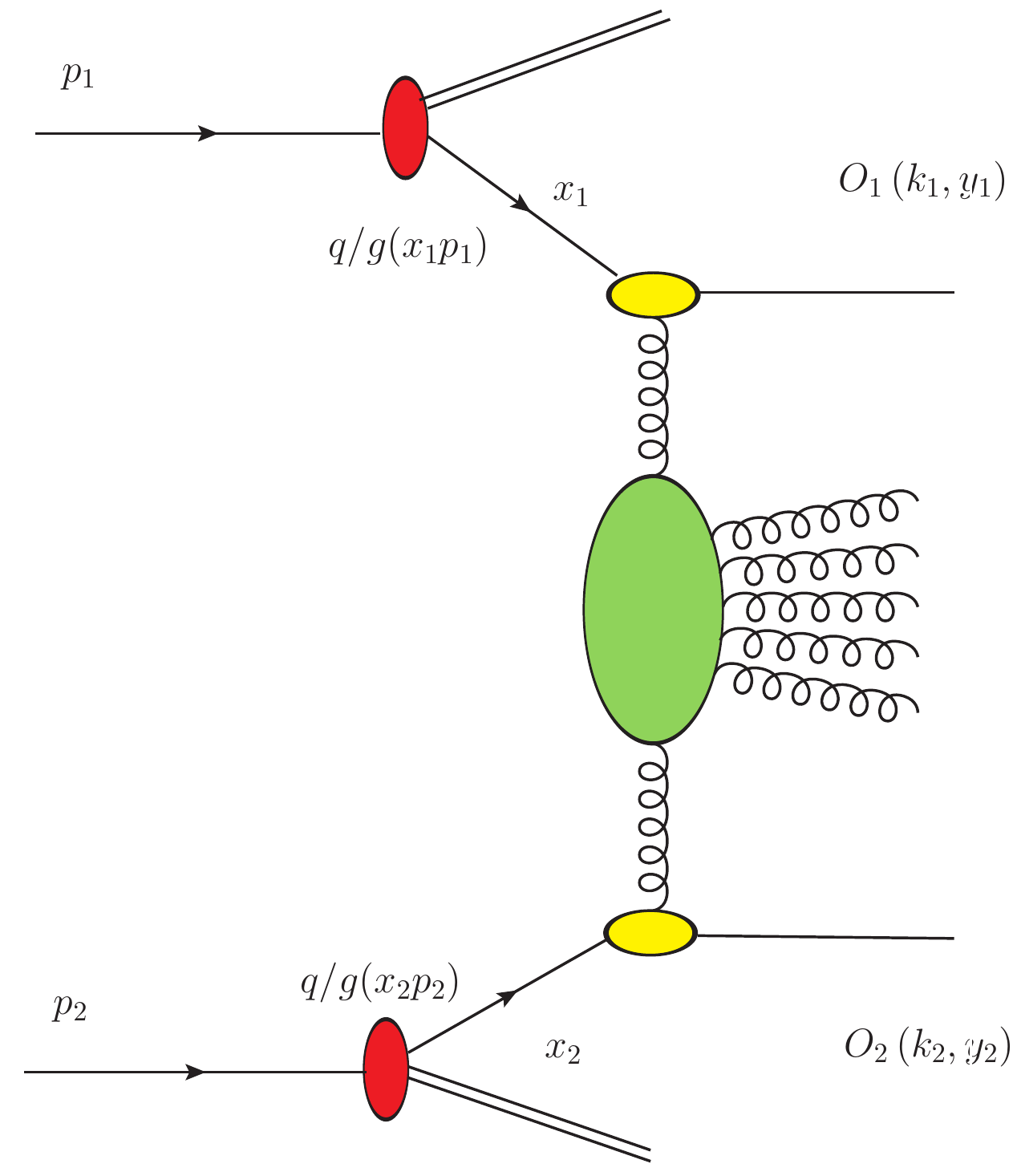}
	\caption{\textbf{Schematic representation of a generic semi-hard process}.}
	\label{generic_semi_hard_process}
\end{figure} 

%========================================
%\section{Mueller–Navelet jets}\label{ch3.1}
\subsection{Theoretical setup} \label{th: general-setup}
%========================================
A generic expression of the inclusive hadroproduction processes of our considerations~(see Fig~\ref{generic_semi_hard_process}):
\begin{equation}\label{semi_hard_process}
{\rm proton}(p_1) \ + \ {\rm proton}(p_2) \ \to \ O_1(\vec k_1, y_1) \ + \ {\rm X} \ + \ {\rm O_2}(\vec k_2, y_2),
\end{equation}
where $ O_{1,2}$ are emitted objects with high transverse momenta of similar sizes $\vec k_{1,2}^2$, and wide rapidity separation $\Delta Y = |y_1-y_2|$. The symbol $X$ is stands for an undetected system of hadrons.
The colliding protons' momenta $P_1$ and $P_2$ are taken as Sudakov basis vectors satisfying
$P^2_1= P^2_2=0$ and $2 (P_1\cdot P_2) = s$,  so that the momenta of detected
objects can be decomposed as
\begin{equation}\label{g_sudakov}
k_{1,2}= x_{1,2} P_{1,2}+ \frac{\vec p_{1,2}^2}{x_{1,2} s}P_{2,1}+k_{1,2\perp} \ , \quad
k_{1,2\perp}^2=-\vec k_{1,2}^2
\end{equation}
The longitudinal-momentum fractions $x_{1,2}$ are related to their
rapidities in the c.o.m. frame through the relation
$y_{1,2}=\pm\frac{1}{2}\ln\frac{x_{1,2}^2 s}
{\vec k_{1,2}^2}$, 
so that $dy_{1,2}=\pm\frac{dx_{1,2}}{x_{1,2}}$,
and $\Delta Y=y_1-y_2=\ln\frac{x_1x_2 s}{|\vec k_1||\vec k_2|}$, here the
spatial part of the four-vector $k_{1\parallel}$ being taken positive.

In QCD collinear factorization, the processes can be viewed as
initiated by two protons each emitting one parton, according to its parton distribution function
(PDF) and ended with the detected object in the final state. In the case of hadrons in the final state, they are formulated according to their fragmentation functions (FFs). Within collinear factorization, a general formula for the processes of our interest, reads 
\begin{equation}\label{collinear_fact_}
\frac{d\sigma}{dx_{1}dx_{2}d^2 k_{1}d^2 k_{2}}
= \sum_{i,j=q,\bar q,g}\int\limits^1_0 dx_a \int\limits^1_0 dx_b\, 
f_i(x_a,\mu_F) f_j(x_b,\mu_F) \frac{d\hat \sigma_{i,j}(\hat s,\mu_F)}
{dx_{1}dx_{2}d^2 k_{1}d^2  k_{2}},
\end{equation}
where the $i, j$ indices specify the parton types 
(quarks $q = u, d, s, c, b$;
antiquarks $\bar q = \bar u, \bar d, \bar s, \bar c, \bar b$; 
or gluon $g$), $f_{i,j}\left(x_{a,b}, \mu_F \right)$ represent the initial proton PDFs and the; 
$x_{a,b}$ are the longitudinal fractions of the partons involved in the hard
subprocess, while $\mu_F$ is the factorization scale;
$d\hat\sigma_{i,j}\left(\hat s \right)$ is the partonic cross section and $\hat s \equiv x_ax_bs$ is the squared center-of-mass energy of the parton-parton collision subprocess.

In the BFKL approach the cross section of the hard subprocesses can be presented as the Fourier sum of the azimuthal coefficients ${\cal C}_n$, 
having so
\begin{equation}\label{BFKL_series_fact}
\frac{d\sigma}
{dy_1dy_2\, d|\vec k_1| \, d|\vec k_2|d\phi_1 d\phi_2}
=\frac{1}{(2\pi)^2}\left[{\cal C}_0+\sum_{n=1}^\infty  2\cos (n\phi )\,
{\cal C}_n\right]\, ,
\end{equation}
where $\phi=\phi_1-\phi_2-\pi$, with $\phi_{1,2}$ the emitted objects $O_{1,2}$ 
azimuthal angles, while $y_{1,2}$ and $\vec p_{1,2}$ are their
rapidities and transverse momenta, respectively. 
The $\phi$-averaged cross section ${\cal C}_0$ 
and the other coefficients ${\cal C}_{n\neq 0}$ are given
by
\begin{equation}\nonumber
{\cal C}_n \equiv \int_0^{2\pi}d\phi_1\int_0^{2\pi}d\phi_2\,
\cos[n(\phi_1-\phi_2-\pi)] \,
\frac{d\sigma}{dy_1dy_2\, d|\vec p_1| \, d|\vec p_2|d\phi_1 d\phi_2}\;
\end{equation}
\begin{equation}\nonumber
= \frac{e^{\Delta Y}}{s}
\int_{-\infty}^{+\infty} d\nu \, \left(\frac{x_1 x_2 s}{s_0}
\right)^{\bar \alpha_s(\mu_R)\left\{\chi(n,\nu)+\bar\alpha_s(\mu_R)
	\left[\bar\chi(n,\nu)+\frac{\beta_0}{8 N_c}\chi(n,\nu)\left[-\chi(n,\nu)
	+\frac{10}{3}+2\ln\left(\frac{\mu_R^2}{\sqrt{\vec k_1^2\vec k_2^2}}\right)\right]\right]\right\}}
\end{equation}
\begin{equation}\nonumber
\times \alpha_s^2(\mu_R) c_1(n,\nu,|\vec k_1|, x_1)
[c_2(n,\nu,|\vec k_2|,x_2)]^*\,
\end{equation}
\begin{equation}\label{cn-coefficients}%\nonumber
\times \left\{1
+\alpha_s(\mu_R)\left[\frac{c_1^{(1)}(n,\nu,|\vec k_1|,
	x_1)}{c_1(n,\nu,|\vec k_1|, x_1)}
+\left[\frac{c_2^{(1)}(n,\nu,|\vec k_2|, x_2)}{c_2(n,\nu,|\vec k_2|,
	x_2)}\right]^*\right]\right.
\end{equation}
\begin{equation}\nonumber
\left. + \bar\alpha_s^2(\mu_R) \ln\left(\frac{x_1 x_2 s}{s_0}\right)
\frac{\beta_0}{4 N_c}\chi(n,\nu)f(\nu)\right\}\;.
\end{equation}
\\
Here $\bar \alpha_s(\mu_R) \equiv \alpha_s(\mu_R) N_c/\pi$, with
$N_c$ the number of colors, $\chi\left(n,\nu\right)$ is the LO BFKL characteristic function given in Eq~\ref{lo_kernel_eigvl},
$c_{1,2}^{(1)}(n,\nu,|\vec k_{1,2}|,x_{1,2})$ are the LO forward /backward objects impact factors in the $(n,\nu)$-repre\-sen\-ta\-tion, given as an integral in the partons fractions $x_{1,2}$, containing the PDFs of the gluon and of the different quark /antiquark flavors
in the proton, and the FFs of the detected hadrons. The LO impact factor for the possible cases of our consideration, can be given in compact form as follows 
%\begin{equation}\nonumber
%c_{H_b}(n,\nu,|\vec p|,x) \;\;=\;\;2 \sqrt{\frac{C_F}{C_A}}
%(\vec p^2)^{i\nu-1/2}\,\int_{x}^1\frac{d\beta}{\beta}
%\left( \frac{\beta}{x}\right)^{2 i\nu-1}
%\end{equation}
%\begin{equation}
%\;\;\times\;\;\left[\frac{C_A}{C_F}f_g(x)D_g^{H_b}\left(\frac{x}{\beta}\right)
%+\sum_{r=q,\bar q}f_r(x)D_r^\Lambda\left(\frac{x}{\beta}\right)\right] \;,
%\end{equation}
% 
%$c_J(n,\nu)$ is the LO forward jet vertex in the $\nu$-repre\-sen\-ta\-tion,
%\begin{equation}
%\label{cJ}
%c_J(n,\nu,|\vec p|,x)=2\sqrt{\frac{C_F}{C_A}}
%(\vec p^{\,2})^{i\nu-1/2}\,\left(\frac{C_A}{C_F}f_g(x)
%+\sum_{s=q,\bar q}f_s(x)\right)
%\end{equation}
\begin{equation}\nonumber
c_{i}(n,\nu,|\vec k|,x) \;\;=\;\;2 \sqrt{\frac{C_F}{C_A}}
(\vec k^2)^{i\nu-1/2}\,\int_{x}^1d\beta
\left( \frac{\beta}{x}\right)^{2 i\nu-1}
\end{equation}
\begin{equation}
\;\;\times\;\;\left[\frac{C_A}{C_F}f_g(\beta) \mathcal{S}_g(x,\beta)
+\sum_{r=q,\bar q}f_r(\beta) \mathcal{S}_r(x,\beta)\right] \;,
\end{equation}
where
\begin{equation}
\mathcal{S}_{g,r}(x,\beta) = \left\{
\begin{aligned}
&  \frac{1}{\beta} D_{g,r}^{H_b}(x/\beta) \; , 
\qquad & i\equiv \text{hadron} \; ; \\ 
& \delta(\beta-x) \; ,
\qquad & i\equiv\text{jet}  \; . 
\end{aligned}
\right.
\end{equation}
and the $f(\nu)$ function is defined by
\begin{equation}
\label{fnu}
i\frac{d}{d\nu}\ln\left(\frac{c_1}{[c_2]^*}\right)=2\left[f(\nu)
-\ln\left(\sqrt{|\vec k_1| |\vec k_2|}\right)\right] \;.
\end{equation}
The remaining objects are the NLO impact factors $c_{1,2}^{(1)}(n,\nu,|\vec k_{1,2}|, x_{1,2})$, for the hadron case, its expression given in Ref.~\cite{Ivanov:2012iv}, while the jet one is defined in Ref.~\cite{Caporale:2012ih}. 
%========================================
%\section{Mueller–Navelet jets}\label{ch3.1}
\subsection{BLM optimization scale setting} 
\label{BLM}
%========================================
It is broadly perceived that the NLO corrections of both, impact factors and the BFKL kernel, in the MS renormalization scheme, are very large in absolute value, thus making the perturbative series highly unstable. For this reason we need some optimization procedure to make reliable predictions. To fix the  arbitrarily chosen renormalization scale $\mu_R$, we follow  Ref.~\cite{Caporale:2015uva}, which showed that, using Brodsky-Lepage-Mackenzie (BLM) optimization method, one can reduce the large uncertainties in the $\mu_R$  setting, by absorbing the non-conformal $\beta_0$-terms (which are present not only in the NLA BFKL kernel, but also in the expressions for the NLA impact factor) into the running coupling. This leads to a non-universality of the BLM scale and to its dependence on the energy of the process.

 We first perform a finite renormalization from the $\overline{\rm MS}$ to
 the physical MOM scheme, whose definition is related to the 3-gluon vertex
 being a key ingredient of the BFKL approach and get
 \begin{equation}
  \alpha_s^{\overline{\rm MS}}=\alpha_s^{\rm MOM}\left(1+\frac{\alpha_s^{\rm MOM}}{\pi}T
 \right)\;,
 \end{equation}
 with $T=T^{\beta}+T^{\rm conf}$,
 \begin{equation}
  T^{\beta}=-\frac{\beta_0}{2}\left( 1+\frac{2}{3}I \right)\, ,
 \end{equation}
 \[ 
 T^{\rm conf}= \frac{3}{8}\left[ \frac{17}{2}I +\frac{3}{2}\left(I-1\right)\xi
 +\left( 1-\frac{1}{3}I\right)\xi^2-\frac{1}{6}\xi^3 \right] \;,
 \]
 where $I=-2\int_0^1dx\frac{\ln\left(x\right)}{x^2-x+1}\simeq2.3439$ and $\xi$
 is the gauge parameter of the MOM scheme, fixed at zero in the following.
 Then, the ``optimal'' BLM scale $\mu_R^{\rm BLM}$ is the value of $\mu_R$
 that makes the $\beta_0$-dependent part in the expression for the observable
 of interest vanish.  
 
Finally, the optimal scale  $\mu^{\rm BLM}_R$
is the value of $\mu_R$ satisfies the condition
\[
C^{\beta_0}_n
\propto \!\!
\int_{y^{\rm min}_1}^{y^{\rm max}_1}dy_1
\int_{y^{\rm min}_2}^{y^{\rm max}_2}dy_2\int_{k^{\rm min}_1}^{k^{\rm max}_1}dk_1
\int_{k^{\rm min}_2}^{k^{\rm max}_2}dk_2
\!\! 
\int\limits^{\infty}_{-\infty} \!\!d\nu\,e^{Y \bar \alpha^{\rm MOM}_s(\mu^{\rm BLM}_R)\chi(n,\nu)}
%\left(\alpha^{\rm MOM}_s (\mu^{\rm BLM}_R)\right)^3
c_1(n,\nu)[c_2(n,\nu)]^*
%\frac{\beta_0}{2 N_c}
\]
\[
\left[\frac{5}{3}
+\ln \frac{(\mu^{\rm BLM}_R)^2}{|\vec k_1|
	|\vec k_2|} +f(\nu)-2\left( 1+\frac{2}{3}I \right)
\right.
\]
\begin{equation}
\label{blmb0}
\left.
+\bar \alpha^{\rm MOM}_s(\mu^{\rm BLM}_R) Y \: \frac{\chi(n,\nu)}{2}
\left(-\frac{\chi(n,\nu)}{2}+\frac{5}{3}+\ln \frac{(\mu^{\rm BLM}_R)^2}{|\vec k_1|
	|\vec k_2|}
+f(\nu)-2\left( 1+\frac{2}{3}I \right)\right)\right]=0 \, .
\end{equation}
%\end{widetext}

The first term in the r.h.s. of~Eq.~(\ref{blmb0}) originates from the NLA
corrections to the hadron/jet vertices and the second one (proportional to
$\alpha^{\rm MOM}_s$) from the NLA part of the kernel. 
We finally plug these scales into our expression for the integrated
coefficients in the BLM scheme (for the derivation see
Ref.~\cite{Caporale:2015uva}):
\begin{equation}\label{Cn_int_blm}
C_n =
\int_{y^{\rm min}_1}^{y^{\rm max}_1}dy_1
\int_{y^{\rm min}_2}^{y^{\rm max}_2}dy_2\int_{k^{\rm min}_1}^{k^{\rm max}_1}dk_1
\int_{k^{\rm min}_2}^{k^{\rm max}_2}dk_2
\!\! 
\int\limits^{\infty}_{-\infty} \!\!d\nu
\end{equation}
\begin{equation}\nonumber
\frac{e^Y}{s}\,
e^{Y \bar \alpha^{\rm MOM}_s(\mu^{\rm BLM}_R)\left[\chi(n,\nu)
	+\bar \alpha^{\rm MOM}_s(\mu^{\rm BLM}_R)\left(\bar \chi(n,\nu) +\frac{T^{\rm conf}}
	{3}\chi(n,\nu)\right)\right]}
\left(\alpha^{\rm MOM}_s (\mu^{\rm BLM}_R)\right)^2 
\end{equation}
\[
\times c_1(n,\nu)[c_2(n,\nu)]^*
\left\{1+\bar \alpha^{\rm MOM}_s(\mu^{\rm BLM}_R)\left[\frac{\bar c^{(1)}_1(n,\nu)}
{c_1(n,\nu)}+\left[\frac{\bar c^{(1)}_2(n,\nu)}{c_2(n,\nu)}\right]^*
+\frac{2T^{\rm conf}}{3} \right] \right\} \, .
\]
%========================================
%\section{Mueller–Navelet jets}\label{ch3.1}
%\section{Phenomenology} 
%========================================
\subsection{BFKL-sensitive observables}
%========================================
To explore different kinematic configurations, based on realistic one used at the LHC, we integrate the coefficients~(\ref{cn-coefficients}) over the phase space for the two
emitted objects, $ O_{1,2}(\vec k_{1,2} , y_{1,2} )$, while their rapidity distance $\Delta Y$, is kept fixed
\begin{equation}\label{Integrated_coefficients}
C_n(\Delta Y,s) =
\int_{k^{\rm min}_1}^{{k^{\rm max}_1}}d|\vec k_1|
\int_{k^{\rm min}_2}^{{k^{\rm max}_2}}d|\vec k_2|
\int_{y^{\rm min}_1}^{y^{\rm max}_1}dy_1
\int_{y^{\rm min}_2}^{y^{\rm max}_2}dy_2
\delta \left( y_1 - y_2 - \Delta Y \right){\cal C}_n
\end{equation}
Since that the typical BFKL observables at the LHC are the azimuthal angle $\phi$ correlations of tagged particles in the final state, which are separated in rapidity, a specific attention has been drawn to the behaviour of the so-called azimuthal correlation. In attempt to to find, less inclusive, observable to test high-energy logs resummation, we consider the $\phi$-summed cross section, $C_0$, and the dependence on the relative azimuthal angle between the tagged particles in the final state, where the latter is defined to be
\begin{equation}
\langle \cos\left[ n\left(\phi_{1}-\phi_{2}-\pi \right)\right]
\rangle\ \equiv\ \frac{{\cal C}_n}{{\cal C}_0}\, , \, \, \, 
\text{with} \, \, \, n=1,2,3 \, .
\end{equation}
The emission of multiple gluons manifests as a fast decrease of $\langle \cos\left[ n\left(\phi_{1}-\phi_{2}-\pi \right)\right]\rangle$ with the difference rapidity between the final observed objects. However, even this observable suffer from a large collinear logs contamination, since the $n=0$ moment is very sensitive to collinear dynamics, thus we need observable has less influence of the collinear region. A further step in this direction was to eliminate the $n=0$ dependence, by proposing the so-called "conformal ratios" $R_{nm}$~\cite{Vera:2006un,Vera:2007kn}:
\begin{equation}
	\frac{\langle 
	\cos \left[ 2 \left(\phi_{1}-\phi_{2}-\pi\right)\right]
	\rangle}{\langle \cos \left(\phi_{1}-\phi_{2}-\pi \right)
	\rangle} \equiv \frac{{\cal C}_2}{{\cal C}_1} \equiv R_{21}
\ ,\
\frac{\langle 
	\cos \left[ 3 \left(\phi_{1}-\phi_{2}-\pi\right)\right]
	\rangle}{\langle \cos \left[ 2\left(\phi_{1}-\phi_{2}-\pi \right)
	\right]\rangle} \equiv \frac{{\cal C}_3}{{\cal C}_2} \equiv R_{32}.
\, 
\end{equation}
Further generalization for the conformal ratios has been suggested for recent proposed processes involving three and four-jets~\cite{Caporale:2015vya,Caporale:2016soq,Caporale:2016lnh,Caporale:2016zkc,Caporale:2015int,Caporale:2016xku,Caporale:2016vxt,Caporale:2016djm} in the final states, as special Mueller-Navelet jets with central tagged jets.
\section{Hadron-jet process}
\label{sec:Hadron-jet}
\begin{figure}[hbtp]
	\centering
	\includegraphics[scale=0.45]{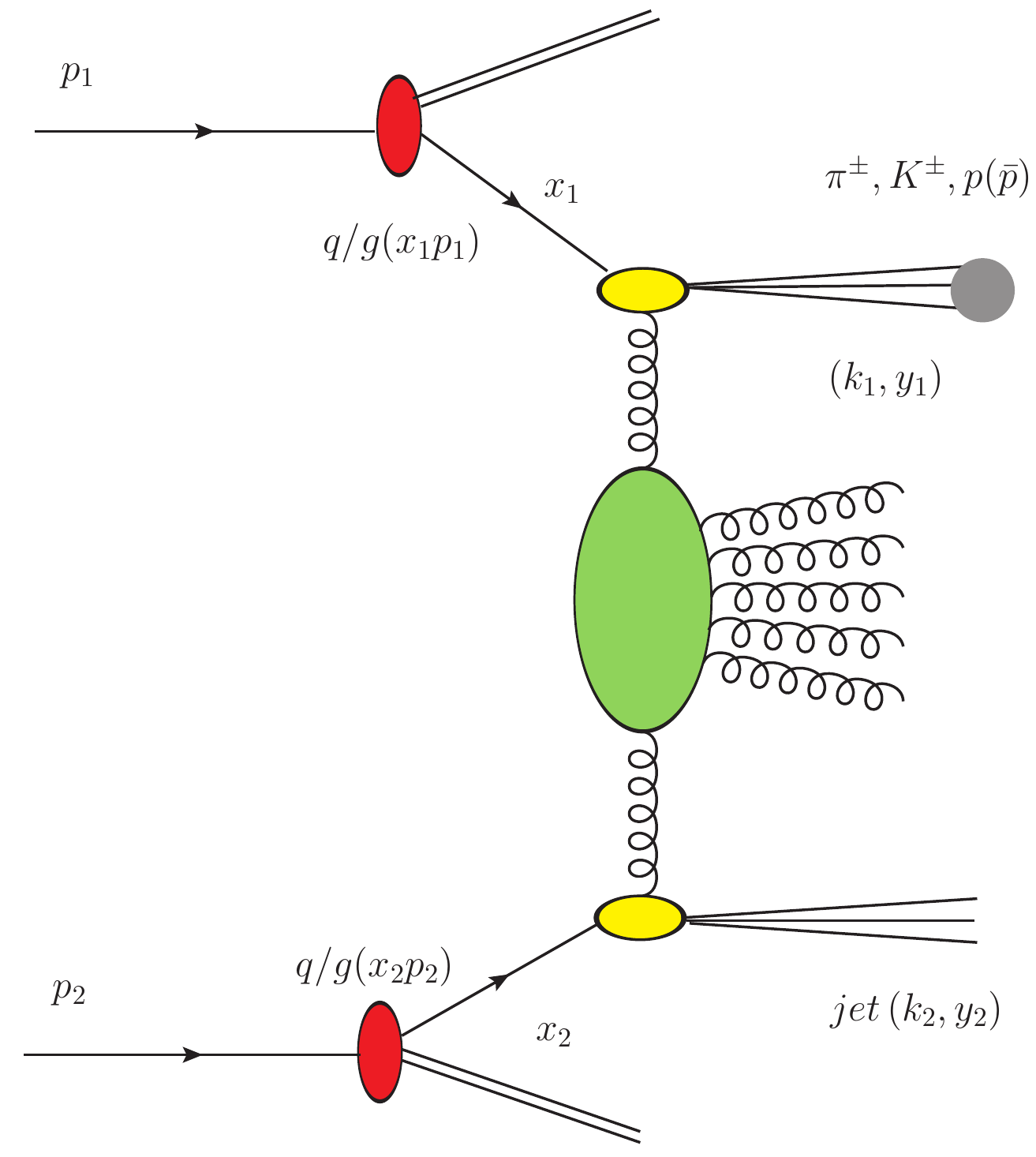}
	\caption{\textbf{Schematic presentation for the Hadron-jet process.}} 
	\label{hadron_jet}
\end{figure}

%===================================
In this section we present, the inclusive production at the LHC of a charged light hadron and of a jet~\cite{Bolognino:2018oth,Bolognino:2019yqj,Bolognino:2019cac}, with  a wide separation in rapidity,
\begin{eqnarray}
\label{process:Hadron-jet}
{\rm proton}(p_1) + {\rm proton}(p_2) 
\to 
{\rm hadron}(k_1, y_1) + {\rm X} + {\rm jet}(k_2, y_2) \;,
\end{eqnarray}
as suggested probe channel for the investigation of the BFKL mechanism of resummation of high-energy logarithms in the pQCD series. We present a numerical predictions, tailored, as in previous section, for both the CMS and CASTOR detectors, for the cross section averaged over the azimuthal angle between the tagged jet and hadron and for azimuthal correlations coefficients. Here, a charged light hadron: $\pi^{\pm}, K^{\pm}, p \left(\bar p\right)$ and a
jet with high transverse momenta, separated by a large interval of rapidity,
are produced together with an undetected hadronic system X
(see Fig.~\ref{hadron_jet}). 
The process~(\ref{process}) has many common features with the inclusive
$J/\Psi$-meson plus backward jet production, considered in
Ref.~\cite{Boussarie:2017oae}. From the theoretical point of view, $J/\Psi$-meson process, has
more uncertainties in comparison to the Hadron-jet one. First the impact factor for
$J/\Psi$-meson production was considered in LO. Instead, impact factor for
the light hadron is known up to NLO accuracy. Previous studies with BFKL calculations for various processes at LHC show that the account of NLO corrections to the impact
factors reduce the theoretical uncertainties.
%The theoretical task to build predictions for cross section and azimuthal
%correlations for our process is embarrassingly simple: one should simply
%replace one of the two jet impact factor entering the Mueller-Navelet formulas
%with the vertex for the proton-to-hadron transition. From the theoretical point
%of view, this process is definitely an easy target, since all the needed
%building blocks are available, with NLO accuracy.
The motivations for building a numerical predictions for this process are summarized in the following points:
\begin{itemize}

	\item  Hadron-jet process is fully describable within NLO BFKL, thus it can be used to test the specific BFKL factorization structure experimentally.
		
%	\item In Refs.~\cite{Caporale:2014gpa,Celiberto:2015yba,Celiberto:2015mpa}
%	it was discussed, in the context of Mueller-Navelet jet production, that
%	using {\em asymmetric} cuts for the transverse momenta of the tagged jets
%	suppresses the Born term, present only for back-to-back jets, thus enhancing
%	the effects of the additional undetected hard gluon radiation and 
%	making therefore more visible the impact of the BFKL resummation, with
%	respect to the fixed-order (DGLAP) contribution. For the process we
%	are considering here this kind asymmetry would be naturally imposed by the
%	completely different nature of the two tagged objects: the identified jet
%	should have transverse momentum not smaller than 20~GeV or so, whereas the
%	minimum hadron transverse momentum can be as small as 5~GeV.

    \item The process we are considering here is naturally imposed asymmetry cuts due to the different nature of the two tagged objects.
    
%	\item For the process under consideration only one hadron in the final state
%	should be identified, instead of two as in the hadron-hadron inclusive
%	production, the other identified object being a jet with a typically much
%	larger transverse momentum. This should facilitate the mining of these
%	events out of the minimum-bias ones.
    \item  The different nature of the two detected objects should simplify the detection of these events out of the minimum-bias ones.
	
	\item From the theoretical point of view one can use this process to compare and constrain collinear FFs, handling the linear expressions as in the di-hadron process. And/or to compare different models for jet algorithms  which are not quadratic as it would be in the Mueller-Navelet jet case.
	
\end{itemize}

%==============================================================
\subsection{Theoretical framework} \label{th:HdJt-framework}
%==============================================================
In the considered process~(\ref{hadron_jet}), both identified hadron and jet have high transverse momenta, $\vec{k}_{J}^{2} \sim \vec{k}_{H}^{2}\geq \Lambda_{QCD}^{2}$ and large separation in rapidity $Y = y_{1}-y_{2}$, which makes the BFKL re-summation come into play.
The momenta of hadron and jet can be written using the Sudakov vector decomposition form: 
\begin{equation}\nonumber
k_H= x_H p_1+ \frac{\vec k_H^2}{x_H s}p_2+k_{H\perp} \ , \quad
k_{H\perp}^2=-\vec k_H^2 \ ,
\end{equation}

\begin{equation}\label{4.2}
k_J= x_J p_2+ \frac{\vec k_J^2}{x_J s}p_1+k_{J\perp} \ , \quad
k_{J\perp}^2=-\vec k_J^2 \ .
\end{equation}
where $x_{H}$,$x_{J}$ are the longitudinal momentum fractions for hadron and jet respectively and the connections between them and the rapidities are given through the relations:
\begin{equation}\label{4.3}
y_H=\frac{1}{2}\ln\frac{x_H^2 s}
{\vec k_H^2}, \quad y_J=\frac{1}{2}\ln\frac{\vec k_J^2}{x_J^2 s}.
\end{equation}

%\begin{figure}[hbtp]
%	\centering
%	\includegraphics[scale=0.25]{jet-hadron.png}
%	\caption{\textbf{schematic presentation for the process $\text{Proton}(p_{1}) + \text{Proton}(p_{2}) \rightarrow \text{hadron}(k_{1}) + \text{jet}(k_{2})+ X$}} 
%\end{figure} 
In this process the vertices describing the dynamics in the proton fragmentation region is combination of proton -to- identified hadron vertex and proton-to-jet vertex.
The LO impact factor in the $\nu$ -representation for the identified hadron, written in expression contains the PDFs of the gluon and of the different quark/anti-quark flavors in the proton, and the FFs of the detected hadron, reads:
\begin{equation}\label{4.4}
c_H(n,\nu,|\vec k_H|,x_H) = 2 \sqrt{\frac{C_F}{C_A}}
(\vec k_H^2)^{i\nu-1/2}\,\int_{x_H}^1\frac{dx}{x}
\left( \frac{x}{x_H}\right)
% ^{2 i\nu-1+\bar\alpha_s(\mu_R)\chi(n,\nu)}
^{2 i\nu-1} 
\end{equation}
\begin{equation}\label{cH}
\times\left[\frac{C_A}{C_F}f_g(x)D_g^h\left(\frac{x_H}{x}\right)
+\sum_{r=q,\bar q}f_r(x)D_r^h\left(\frac{x_H}{x}\right)\right] \;,
\end{equation}
and $c_J(n,\nu)$ is the LO forward jet vertex in the $\nu$-repre\-sen\-ta\-tion,
\begin{equation}
c_J(n,\nu,|\vec k_J|,x_J)=2\sqrt{\frac{C_F}{C_A}}
(\vec k_J^{\,2})^{i\nu-1/2}\,\left(\frac{C_A}{C_F}f_g(x_J)
+\sum_{s=q,\bar q}f_s(x_J)\right).
\end{equation}
 
%For the jet vertex, both LO and NLO correction to the forward jet impact factor in the small-cone limit $c_{2}^{(1)}(n,\nu,|\overrightarrow{k_{2}}|,x_{2})$, are given explicitly in Appendix (\ref{A.1}).
 
The differential cross-section can be presented as Fourier sum of the azimuthal coefficients $C_{n}$, as below:
\begin{equation}\label{4.9}
\frac{d\sigma}
{dy_Hdy_J\, d|\vec k_H| \, d|\vec k_J|d\phi_H d\phi_J}
=\frac{1}{(2\pi)^2}\left[C_0+\sum_{n=1}^\infty  2\cos (n\phi )\,
C_n\right]\, ,
\end{equation} 
where $\phi_{1,2}$ are hadron/jet azimuthal angles, and $C_{0}$ the differential cross-section integrated over $\phi_{1,2}$. The $C_{n}$ coefficients are $\phi_{1,2}$ independent, they depend just on the transverse momenta, rapidities and the re-normalization, factorization and energy scale parameters. \\
It is useful to introduce the variable
\begin{equation}\label{4.10}
Y_{0}=\ln\bigg(\frac{s_{0}}{|\overrightarrow{k_{H}}||\overrightarrow{k_{J}}|}\bigg),
\end{equation} 
in such a way that
\begin{equation}\label{4.11}
Y-Y_{0}=\ln\bigg(\frac{x_{H}x_{J}s}{s_{0}}\bigg),
\end{equation}
where $Y=y_{1}-y_{2}=\ln\bigg(\frac{x_{H}x_{J}s}{|\overrightarrow{k_{H}}||\overrightarrow{k_{J}}|}\bigg)$, is the rapidity gap.\\
Using all formulas above, the coefficients $C_{n}$ in the so-called exponentiated representation are given by
\begin{equation}\nonumber
C_n \equiv \int_0^{2\pi}d\phi_H\int_0^{2\pi}d\phi_J\,
\cos[n(\phi_H-\phi_J-\pi)] \,
\frac{d\sigma}{dy_Hdy_J\, d|\vec k_H| \, d|\vec k_J|d\phi_H d\phi_J}\;
\end{equation}
\begin{equation}\nonumber
= \frac{e^Y}{s}
\int_{-\infty}^{+\infty} d\nu \, \left(\frac{x_H x_J s}{s_0}
\right)^{\bar \alpha_s(\mu_R)\left\{\chi(n,\nu)+\bar\alpha_s(\mu_R)
	\left[\bar\chi(n,\nu)+\frac{\beta_0}{8 N_c}\chi(n,\nu)\left[-\chi(n,\nu)
	+\frac{10}{3}+2\ln\left(\frac{\mu_R^2}{\sqrt{\vec k_H^2\vec k_J^2}}\right)\right]\right]\right\}}
\end{equation}
\begin{equation}\nonumber
\times \alpha_s^2(\mu_R) c_H(n,\nu,|\vec k_H|, x_H)
[c_J(n,\nu,|\vec k_J|,x_J)]^*\,
\end{equation}
\begin{equation}\label{eq:HdJt-Cn-coefficients}%\nonumber
\times \left\{1
+\alpha_s(\mu_R)\left[\frac{c_H^{(1)}(n,\nu,|\vec k_H|,
	x_H)}{c_H(n,\nu,|\vec k_H|, x_H)}
+\left[\frac{c_J^{(1)}(n,\nu,|\vec k_J|, x_J)}{c_J(n,\nu,|\vec k_J|,
	x_J)}\right]^*\right]\right.
\end{equation}
\begin{equation}\nonumber
\nonumber
\left. + \bar\alpha_s^2(\mu_R) \ln\left(\frac{x_H x_J s}{s_0}\right)
\frac{\beta_0}{4 N_c}\chi(n,\nu)f(\nu)\right\}\;.
\end{equation}
Here $\bar{\alpha}_{s}^{2}(\mu_{R})\equiv \alpha_{s}^{2}(\mu_{R})N_{c}/\pi$ 
,with $N_{c}$ the number of colours, $\beta_{0}$ is the first coefficient of the QCD $\beta$-function, and $c_{H,J}^{(1)}(n,\nu,|\vec k_{H,J}|, x_{H,J})$ are the hadron/jet NLO impact factor corrections in $(\nu, n)$-representation, their
expressions being given in Appendix (\ref{A}).\\

%===================================
\subsection{Phenomenological analysis}
%===================================
With the theoretical setup described in~\ref{th:HdJt-framework}, in this section we give an  interpretation to the numerical analysis results presented at the end of the section.
In order to match the realistic kinematic cuts at LHC, we integrate the coefficients
over the phase space for two final-state objects and keep fixed the rapidity
interval, $Y$, between the hadron and the jet: 
\begin{equation}
\label{Hdjt-Cn_int}
C_n= 
\int_{y^{\rm min}_H}^{y^{\rm max}_H}dy_H
\int_{y^{\rm min}_J}^{y^{\rm max}_J}dy_J\int_{k^{\rm min}_H}^{k^{\rm max}_H}dk_H
\int_{k^{\rm min}_J}^{{k^{\rm max}_J}}dk_J
\, \delta \left( y_H - y_J - Y \right)
\, {\cal C}_n \left(y_H,y_J,k_H,k_J \right)\, .
\end{equation}
We consider two distinct ranges for the final-state objects:
\begin{itemize}
	\item \textit{\bf CMS-jet}: both the hadron and the jet tagged by the CMS detector
	in their typical kinematic configurations, \emph{i.e.}: 5 GeV $< k_H <$
	21.5 GeV, 35 GeV $< k_J <$ 60 GeV, $|y_H| \leq$ 2.4, $|y_H| \leq$ 4.7;
	
	\item \textit{\bf CASTOR-jet}: a hadron always tagged inside CMS, together with a very backward jet detected by CASTOR in the range: 5 GeV $< k_J\lesssim $  17.68 GeV, -6.6 $< y_J <$ -5.2.
\end{itemize}

%===================================
\subsection{Results and discussion}
\begin{figure}[p]
	\centering
	\includegraphics[scale=0.33,clip]{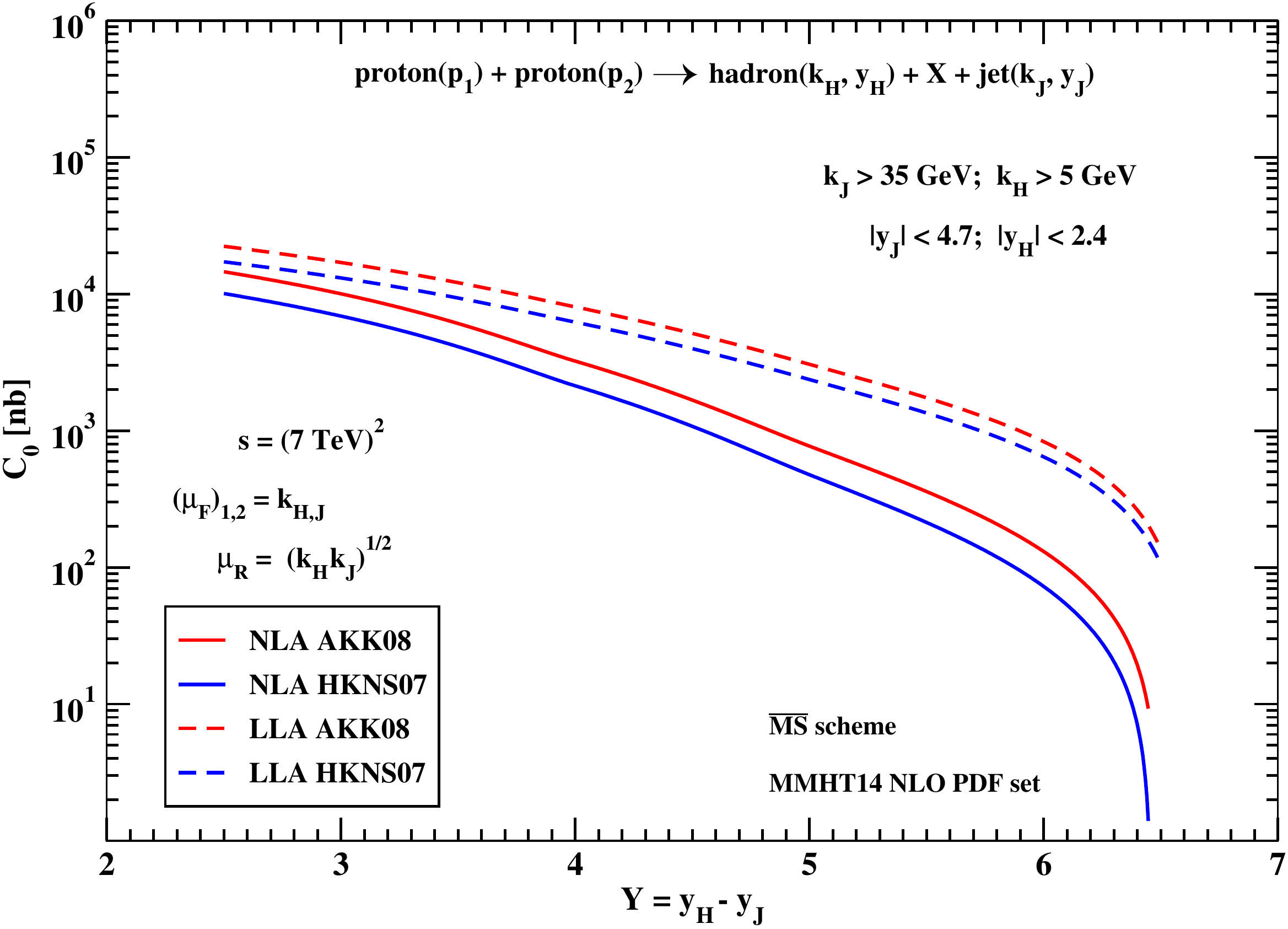}
	\includegraphics[scale=0.33,clip]{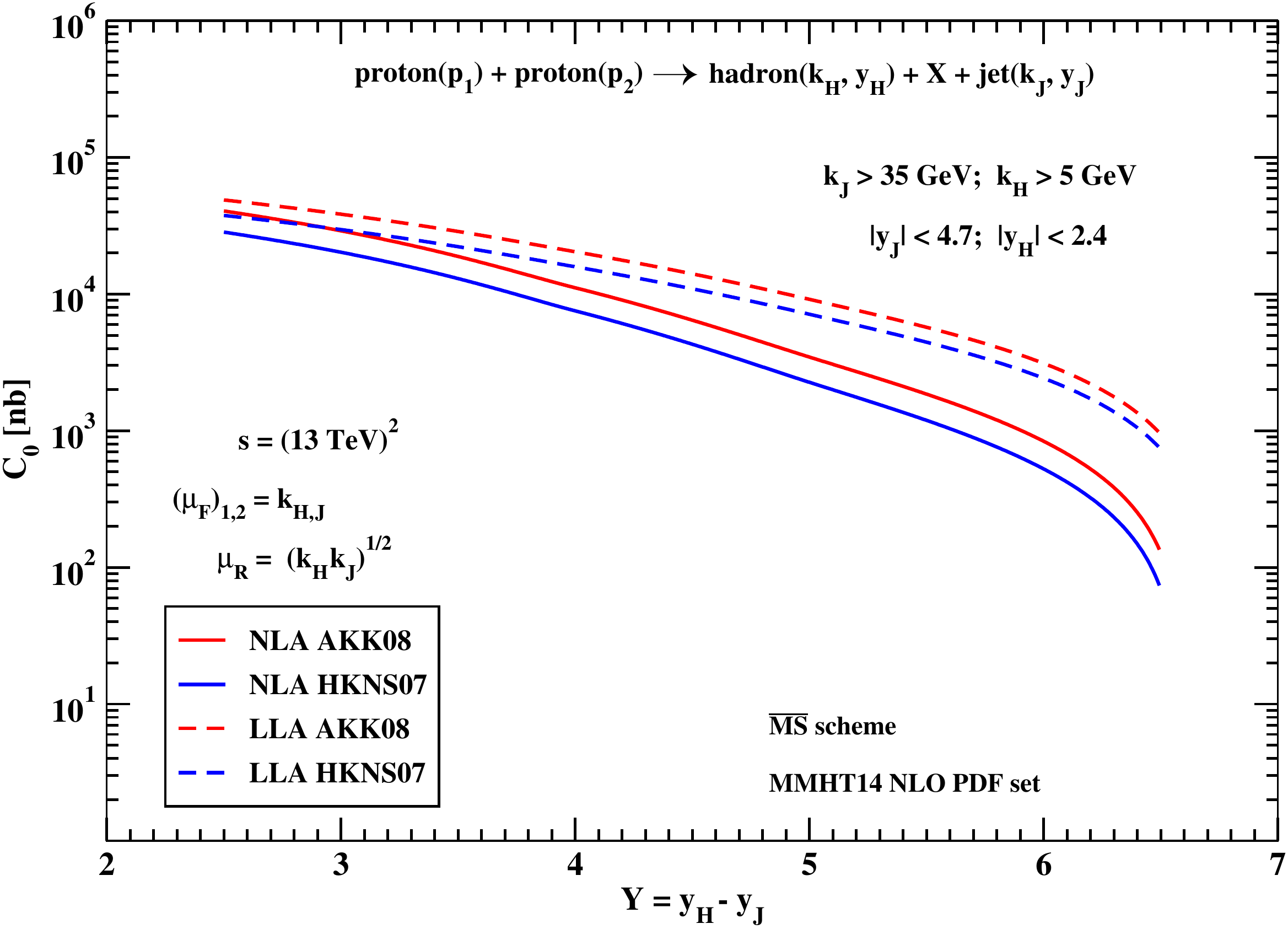}
	\caption{$Y$-dependence of $C_0$ for $\mu_R = \mu_N =
		\sqrt{|\vec k_H||\vec k_J|}$, $(\mu_F)_{1,2} = |\vec k_{H,J}|$, for
		$\sqrt{s}= 7$ TeV (left) and $\sqrt{s} = 13$ TeV (right), and
		$Y \leq 7.1$ ({\it CMS-jet}
		configuration).}	
	\label{fig:C0_MSb_NS_CMS}	
\end{figure}

%===================================
In Fig.~\ref{fig:C0_MSb_NS_CMS} we present results  for the $Y$-dependence of $\phi$-summed cross section $C_{0}$ in $\overline{\text{MS}}$ scheme at $\sqrt{s}=13,7$ TeV in the {\it CMS-jet} kinematic configuration; in this case we choose for the factorization scale $(\mu_{F})_{H,J}=|\overrightarrow{k}_{H,J}|$, and keep fixing the natural values for the renormalization scale $\mu_{R}=\mu_{N}=\sqrt{|\overrightarrow{k}_{H}||\overrightarrow{k}_{J}|}$. We can see that the expected BFKL pattern is manifest, where the NLO corrections become larger with the increasing $Y$. As a matter of fact, these corrections are very large in absolute value, thus making
the perturbative series highly unstable. For this reason we need some optimization
procedure to make reliable predictions.

\begin{figure}[p]
	\centering
	\includegraphics[scale=0.33,clip]{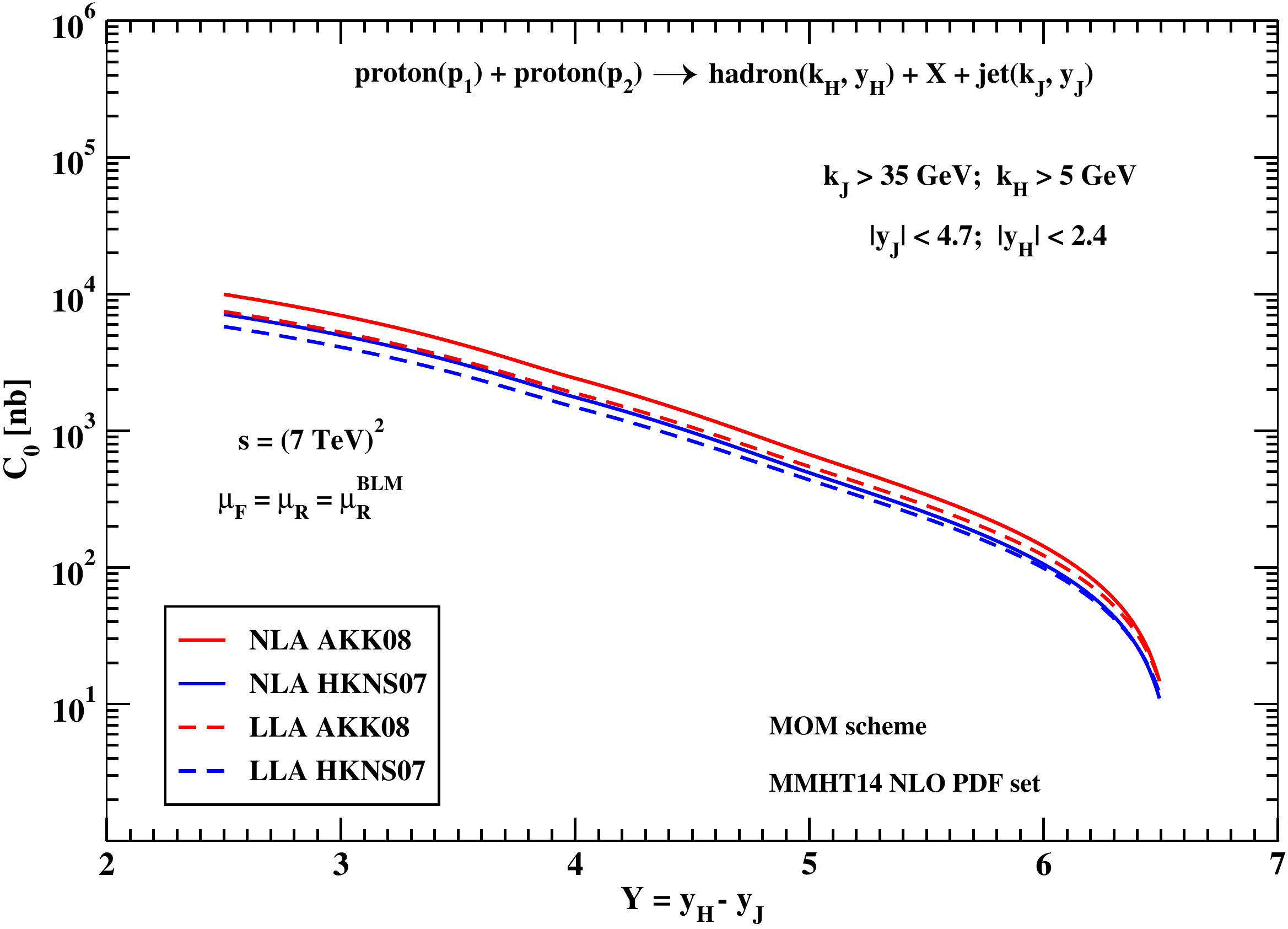}
	\includegraphics[scale=0.33,clip]{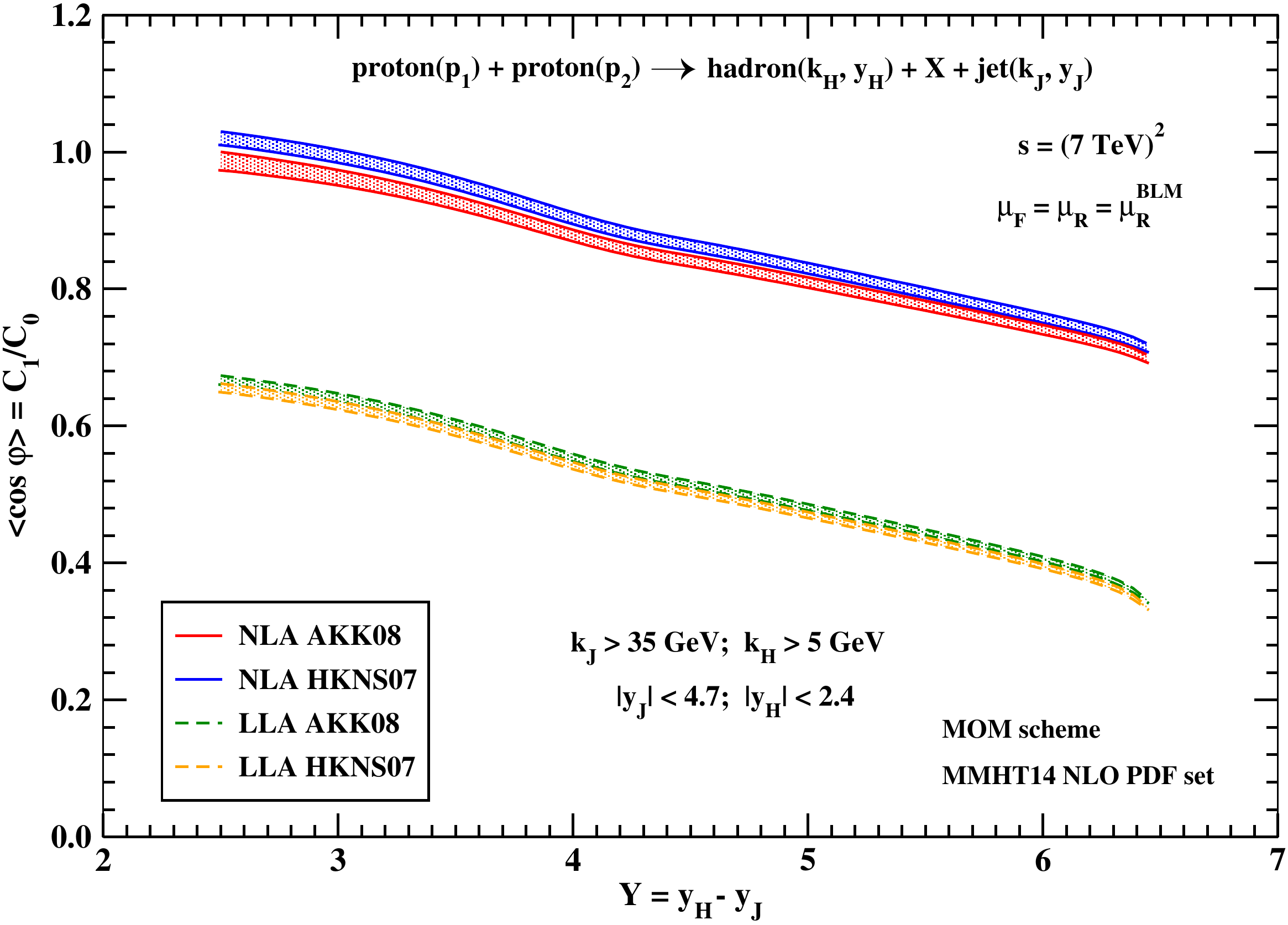}
	
	\includegraphics[scale=0.33,clip]{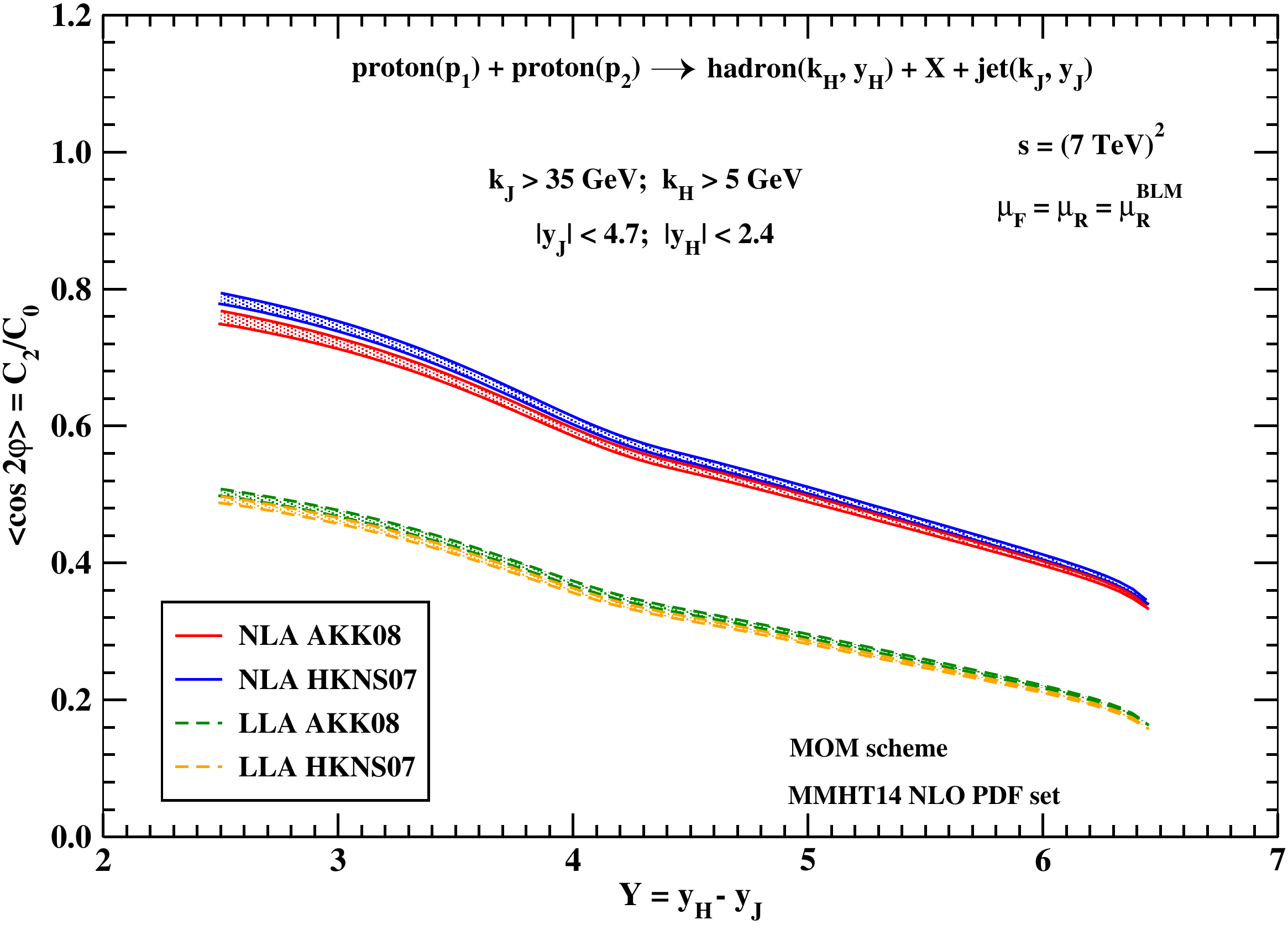}
	\includegraphics[scale=0.33,clip]{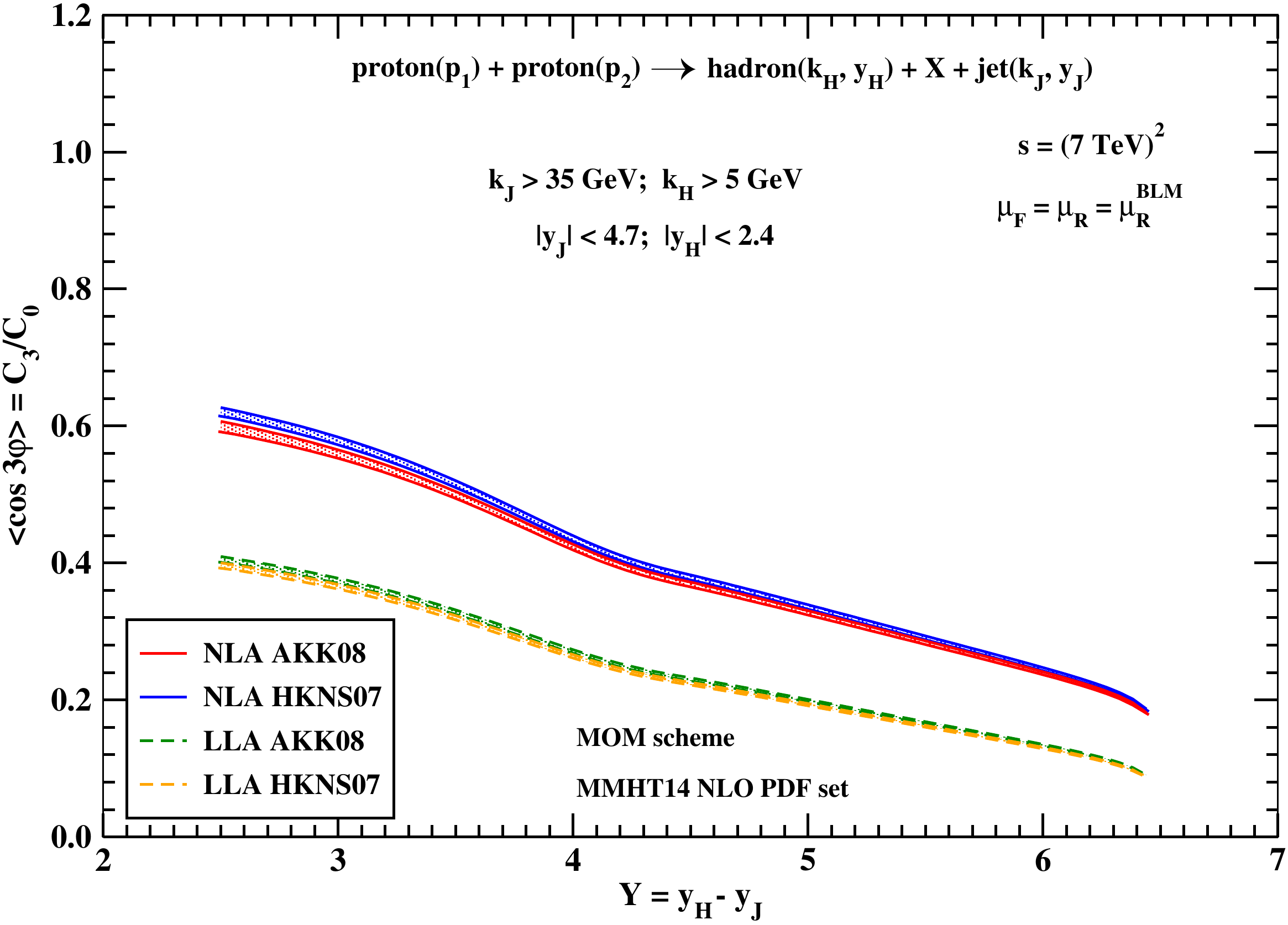}
	
	\includegraphics[scale=0.33,clip]{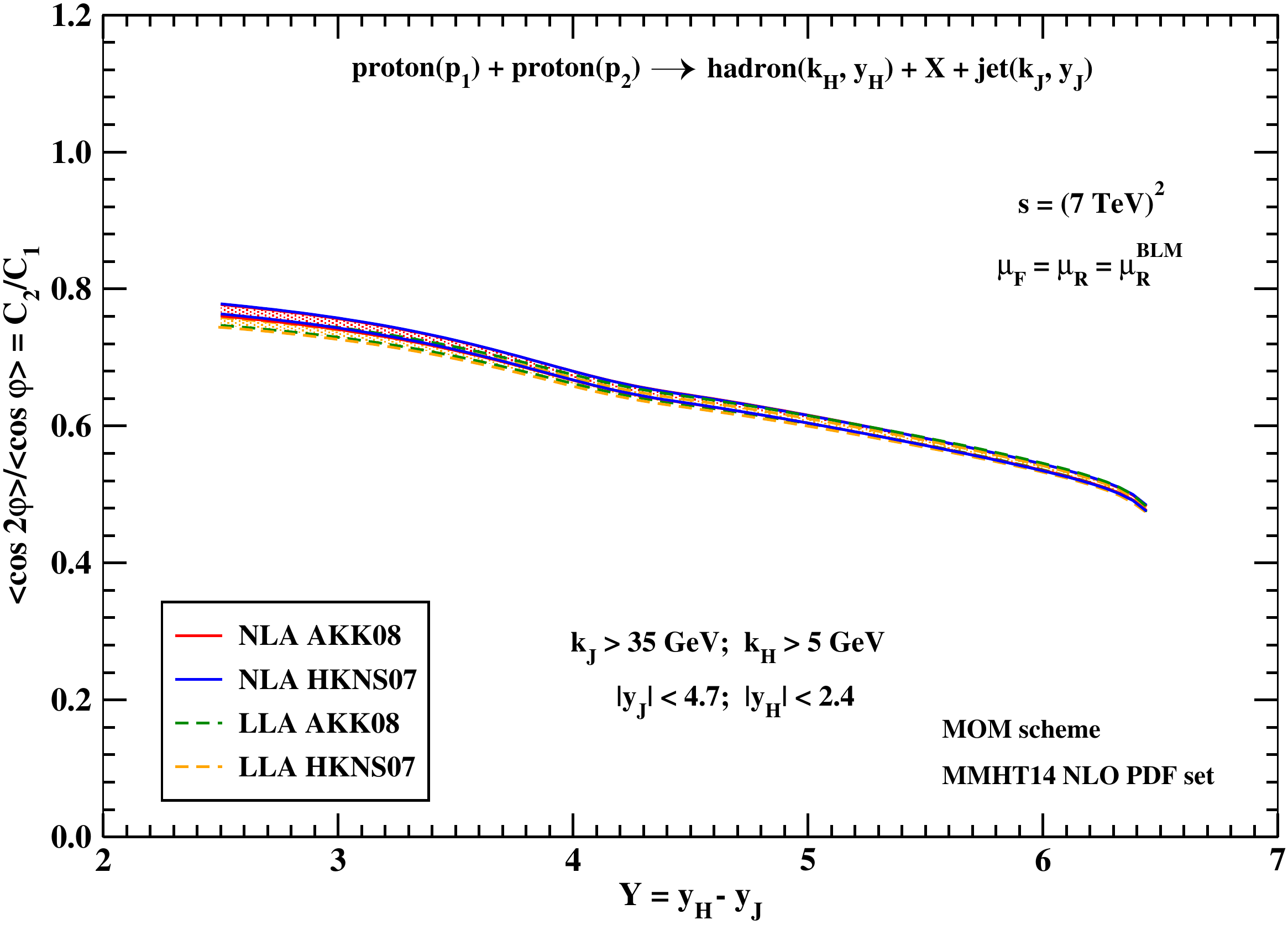}
	\includegraphics[scale=0.33,clip]{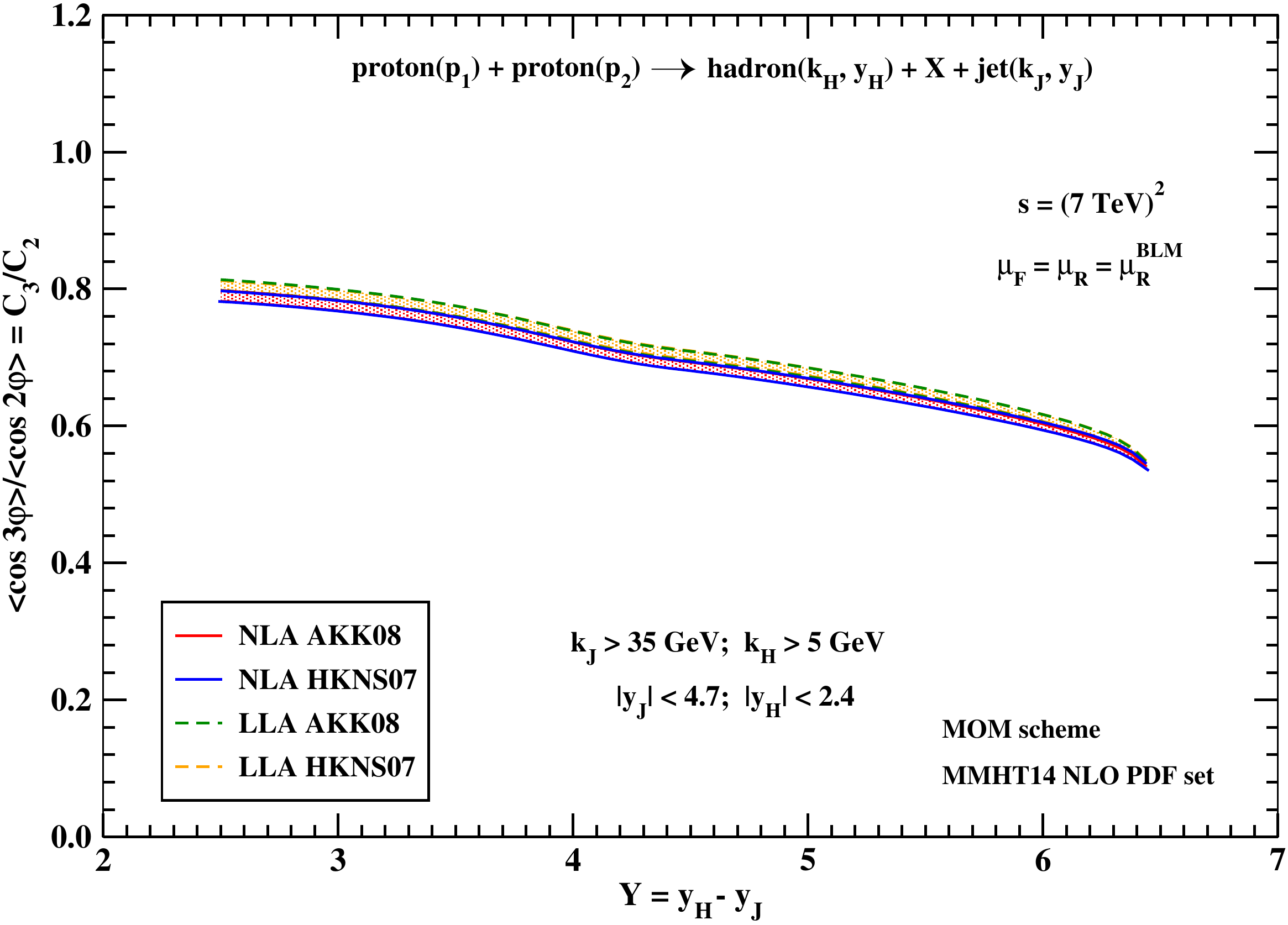}
	\caption{$Y$-dependence of $C_0$ and of several ratios $C_m/C_n$ for 
		$\mu_F = \mu_R^{\rm BLM}$, $\sqrt{s} = 7$ TeV, and $Y \leq 7.1$
		({\it CMS-jet} configuration).}
	\label{fig:Cn_MOM_BLM_CMS_7}
\end{figure}

\begin{figure}[p]
	\centering
	\includegraphics[scale=0.33,clip]{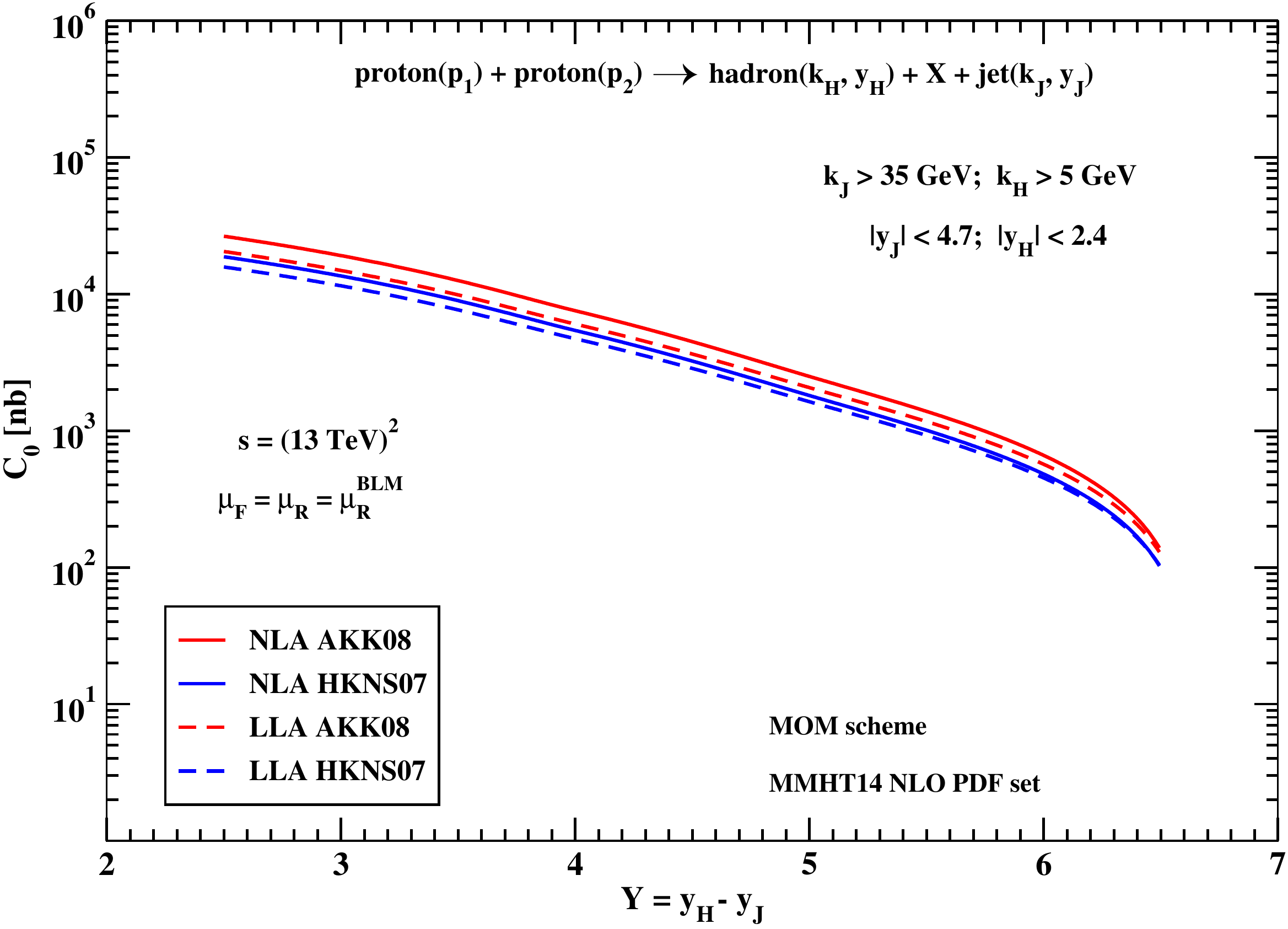}
	\includegraphics[scale=0.33,clip]{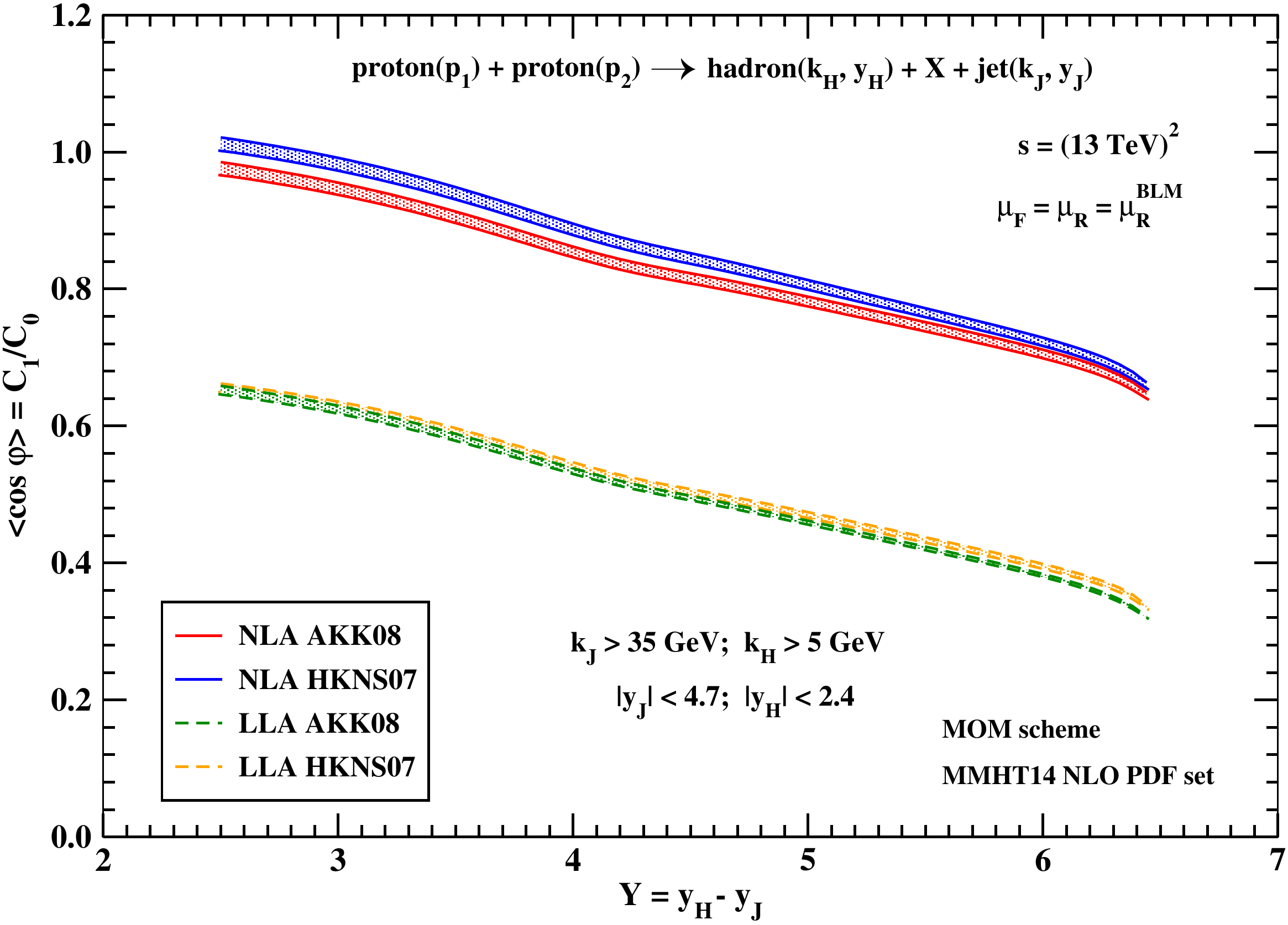}
	
	\includegraphics[scale=0.33,clip]{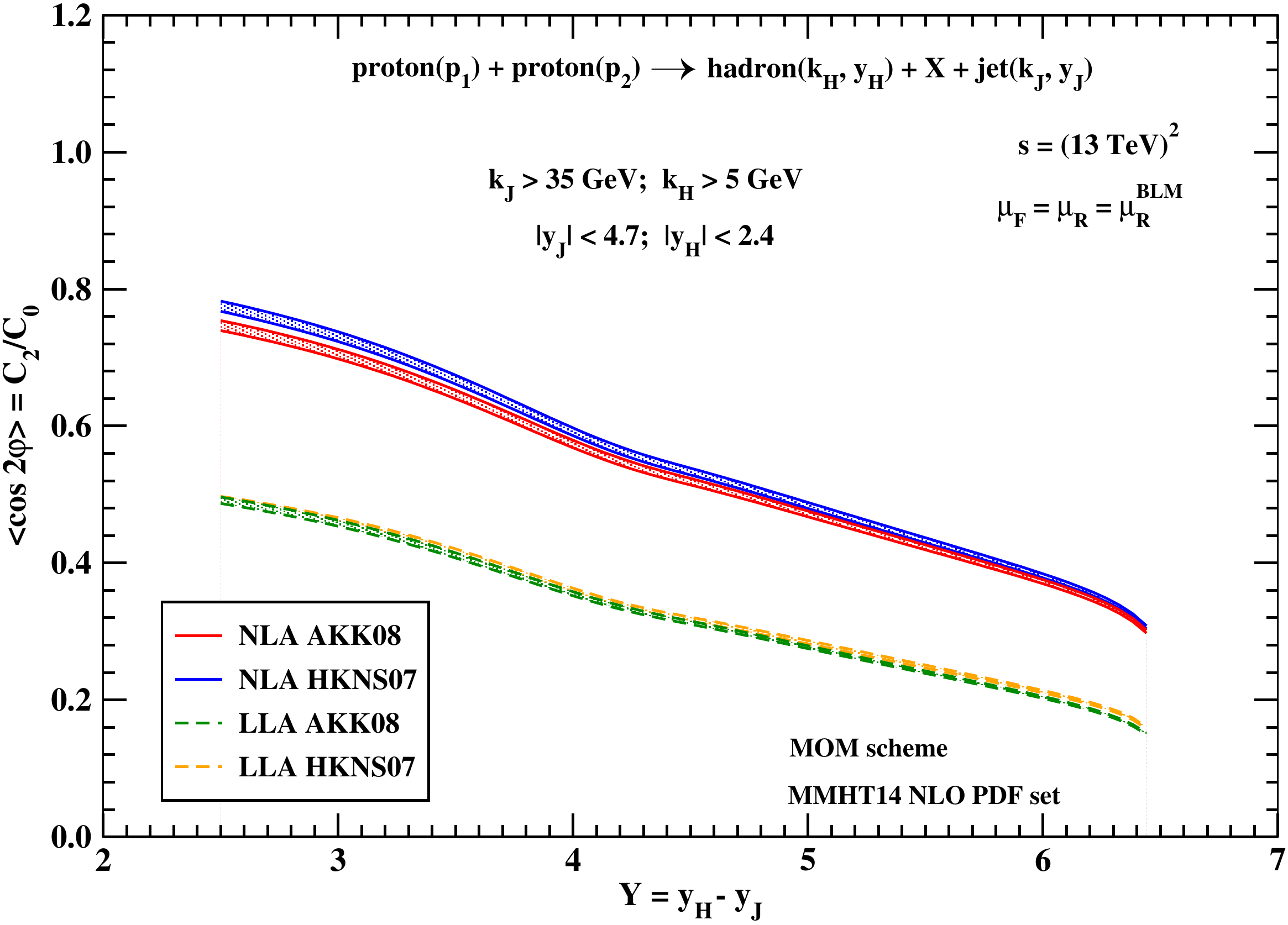}
	\includegraphics[scale=0.33,clip]{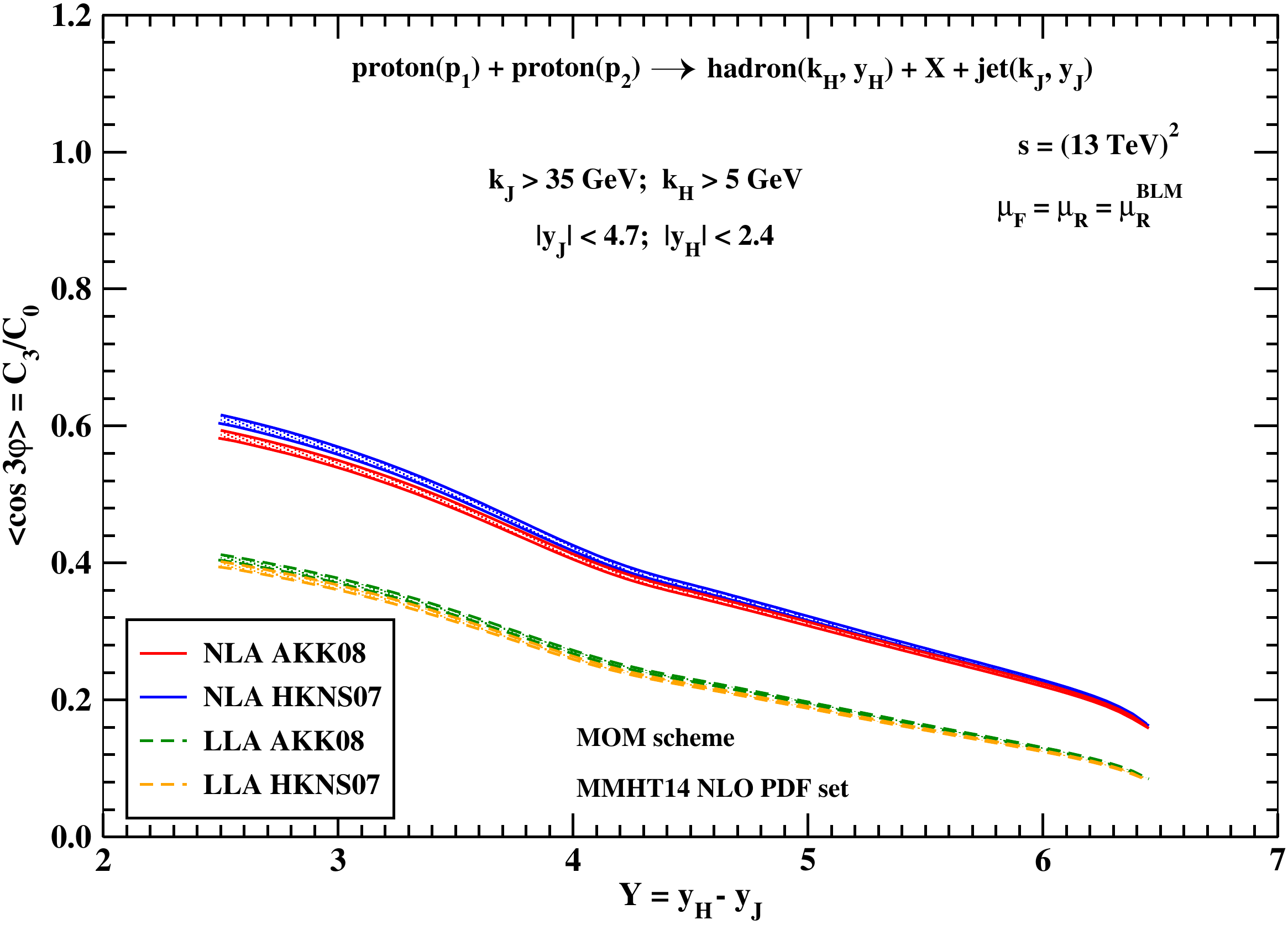}
	
	\includegraphics[scale=0.33,clip]{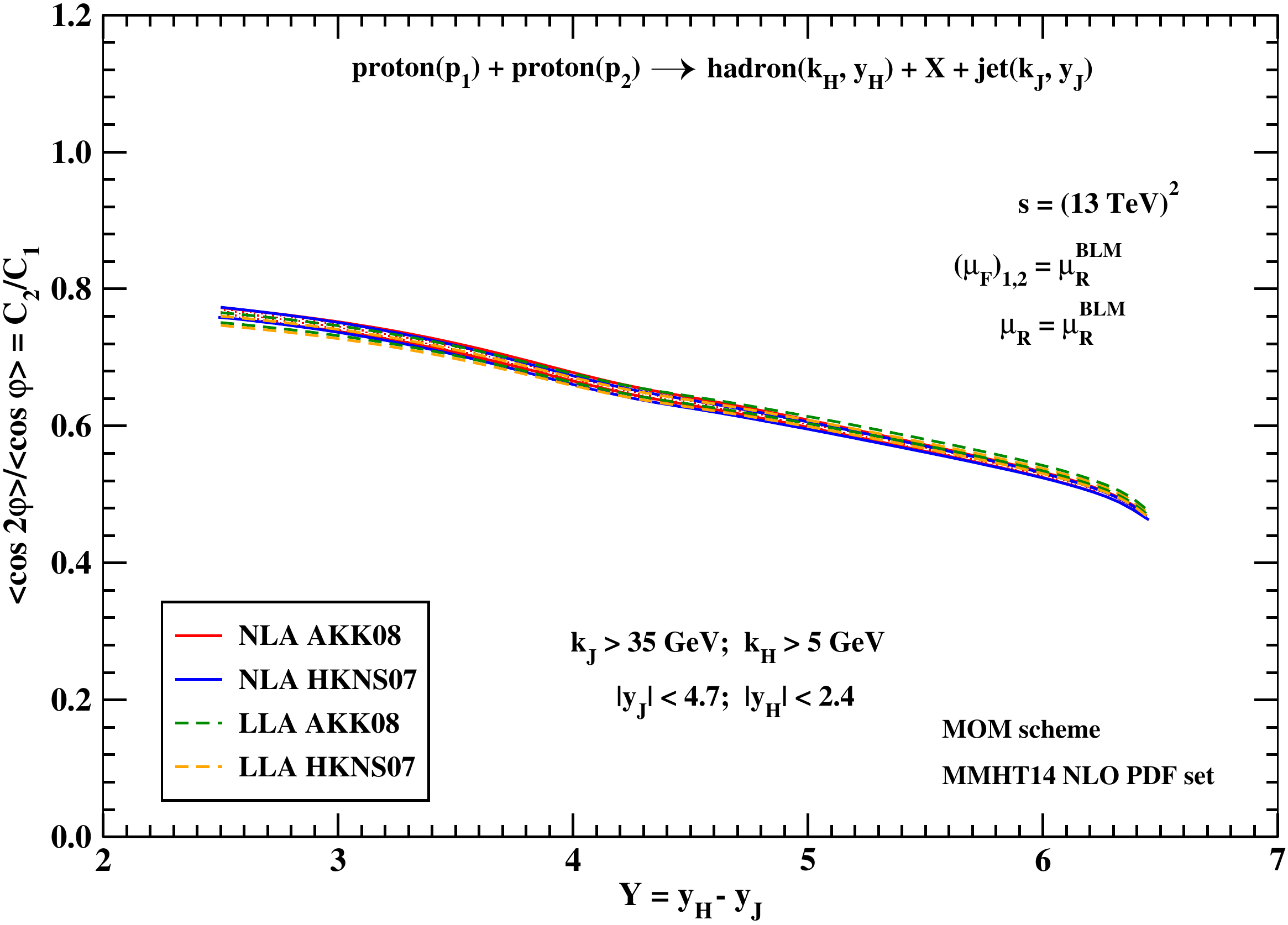}
	\includegraphics[scale=0.33,clip]{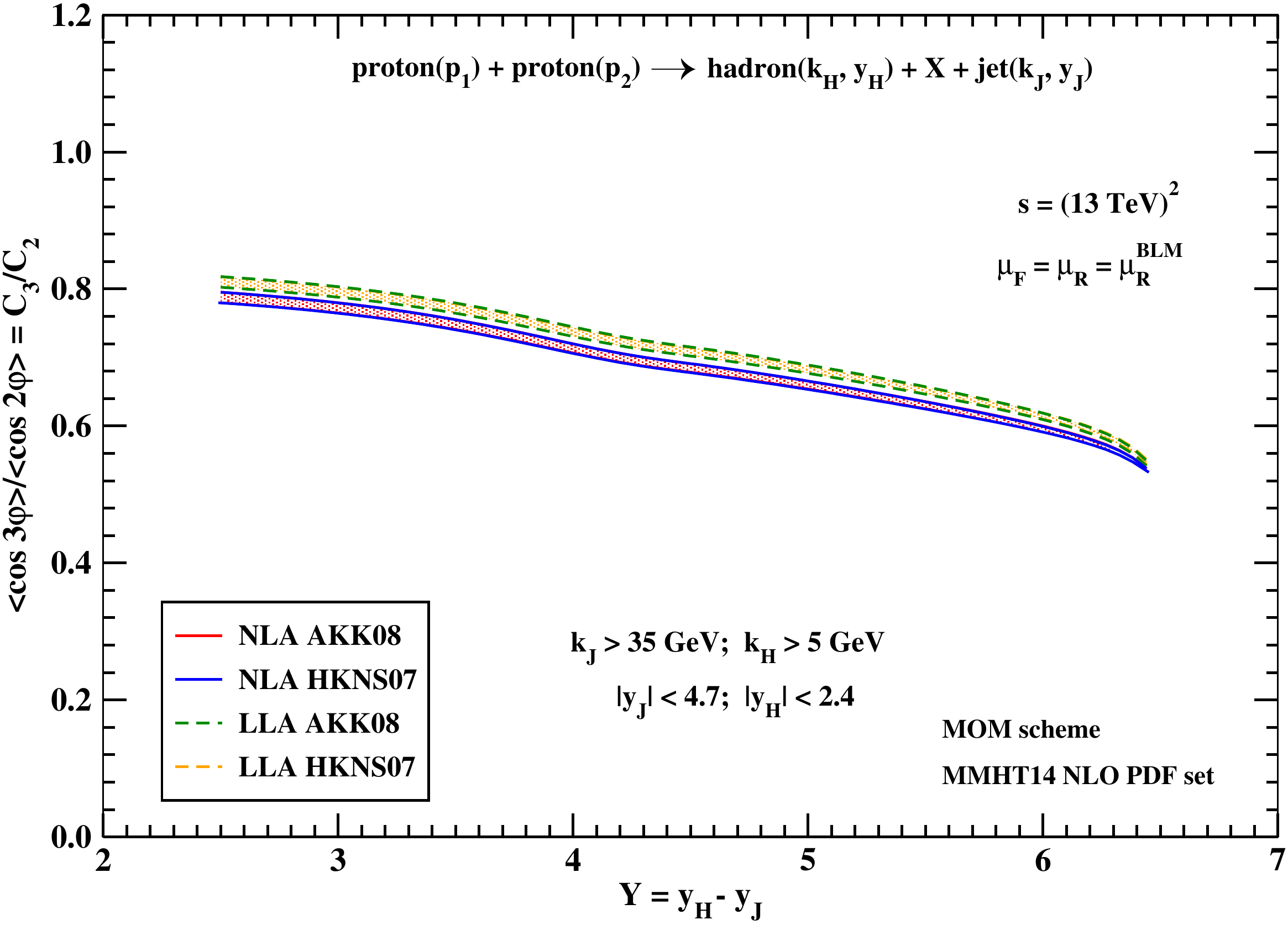}
	\caption{$Y$-dependence of $C_0$ and of several ratios $C_m/C_n$ for 
		$\mu_F = \mu_R^{\rm BLM}$, $\sqrt{s} = 13$ TeV, and $Y \leq 7.1$
		({\it CMS-jet} configuration).}
	\label{fig:Cn_MOM_BLM_CMS_13}
\end{figure}

In Figs.~\ref{fig:Cn_MOM_BLM_CMS_7} and~\ref{fig:Cn_MOM_BLM_CMS_13}, we report
predictions with the BLM scale optimization for $C_0$ and
several $R_{nm}$ ratios with the jet tagged inside the CMS
detector for $\sqrt{s} = 7$ and 13~TeV, respectively.
Here, we can see that, as a result of using BLM optimization, the LLA and NLA predictions for $C_0$ are now comparable, which a sign of stability in the BFKL series. For the ratios $R_{n0}$, the NLA predictions appear to be above the LLA ones, as observed in Mueller-Navelet jets and in the hadron-hadron~\cite{Celiberto_2016,Celiberto_2017,Celiberto_2017s} case. The ratios $R_{21}$ and $R_{32}$, show insensitivity to the NLO corrections.

\begin{figure}[p]
	\centering
	\includegraphics[scale=0.33,clip]{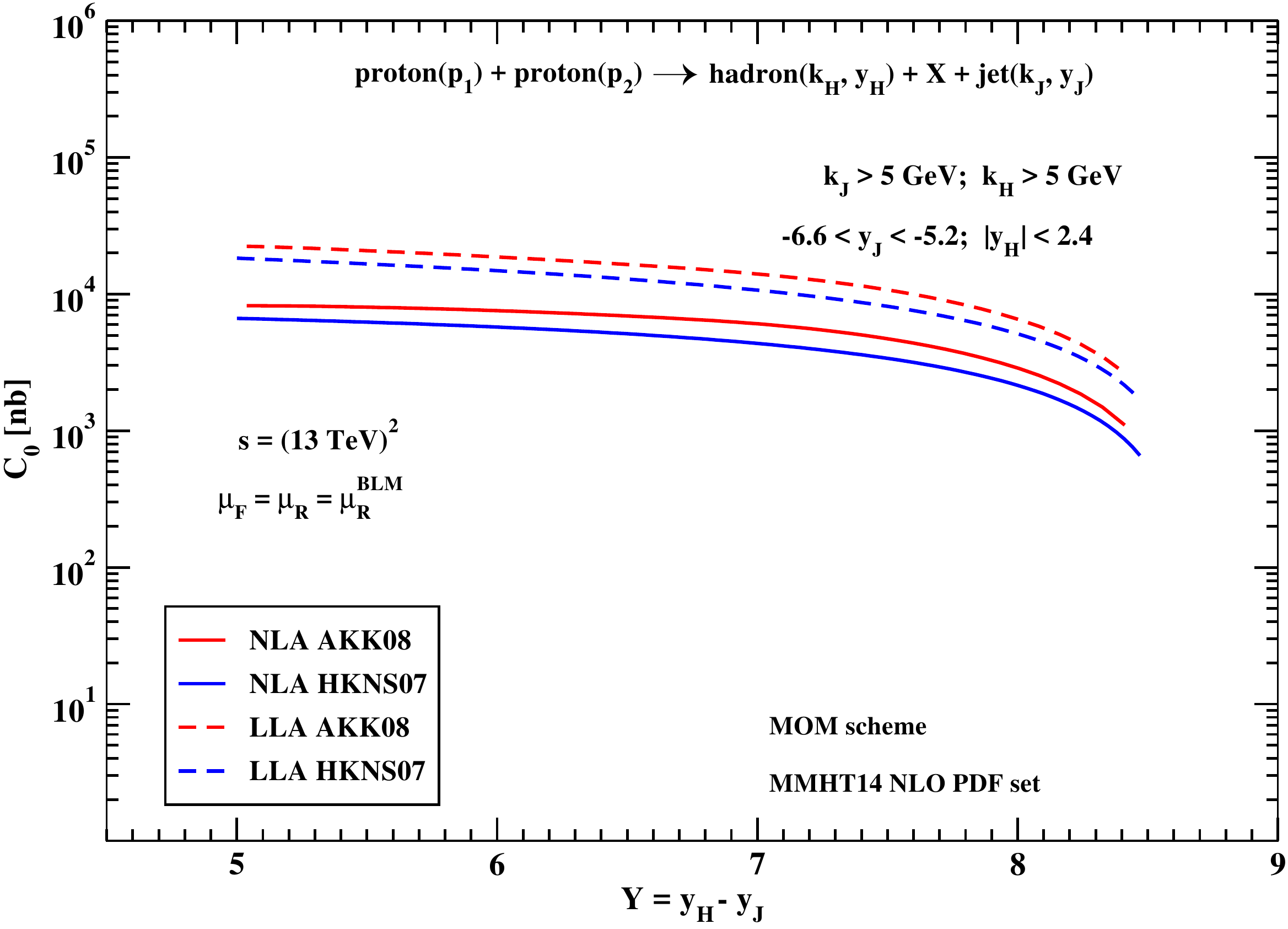}
	\includegraphics[scale=0.33,clip]{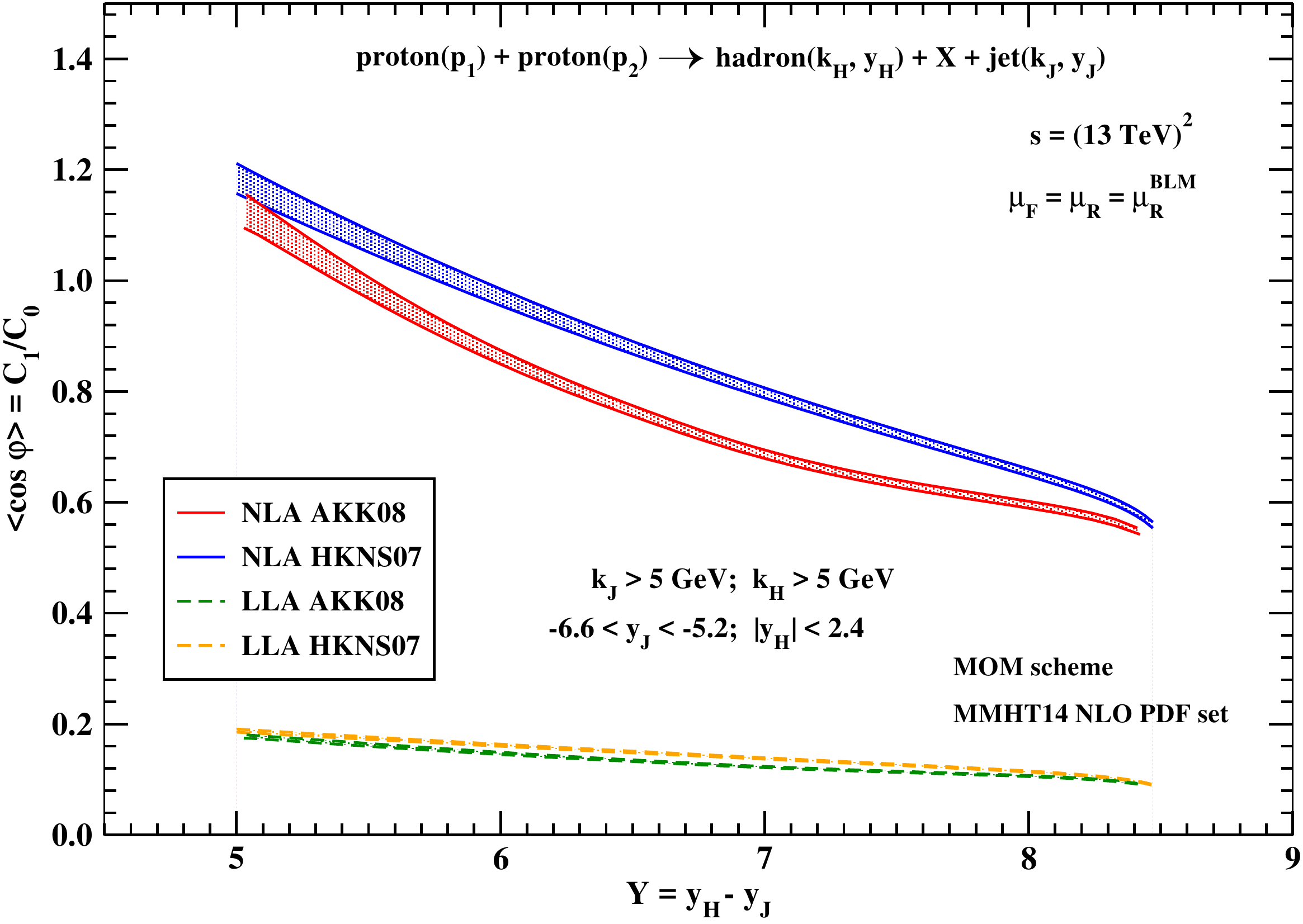}
	
	\includegraphics[scale=0.33,clip]{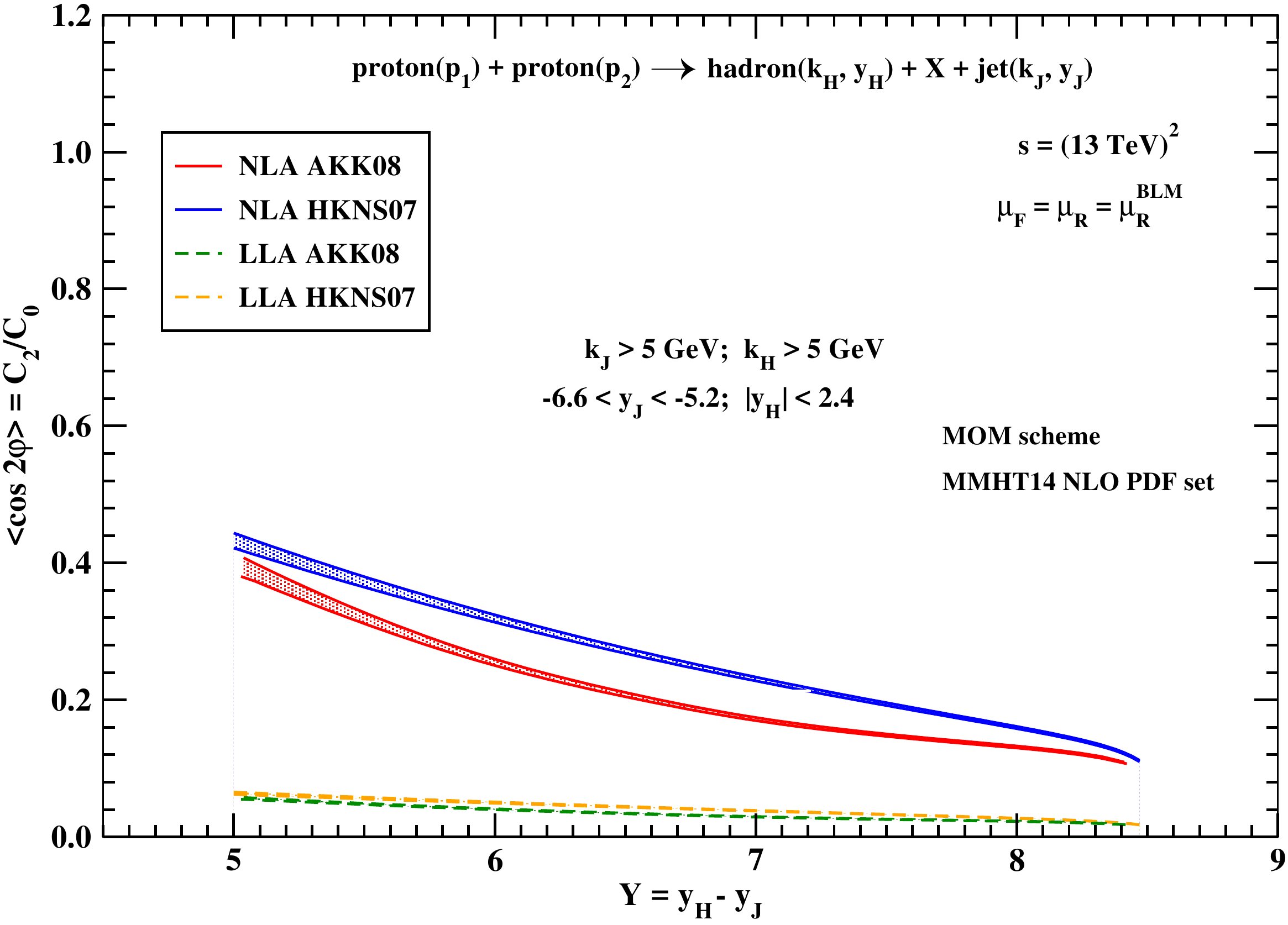}
	\includegraphics[scale=0.33,clip]{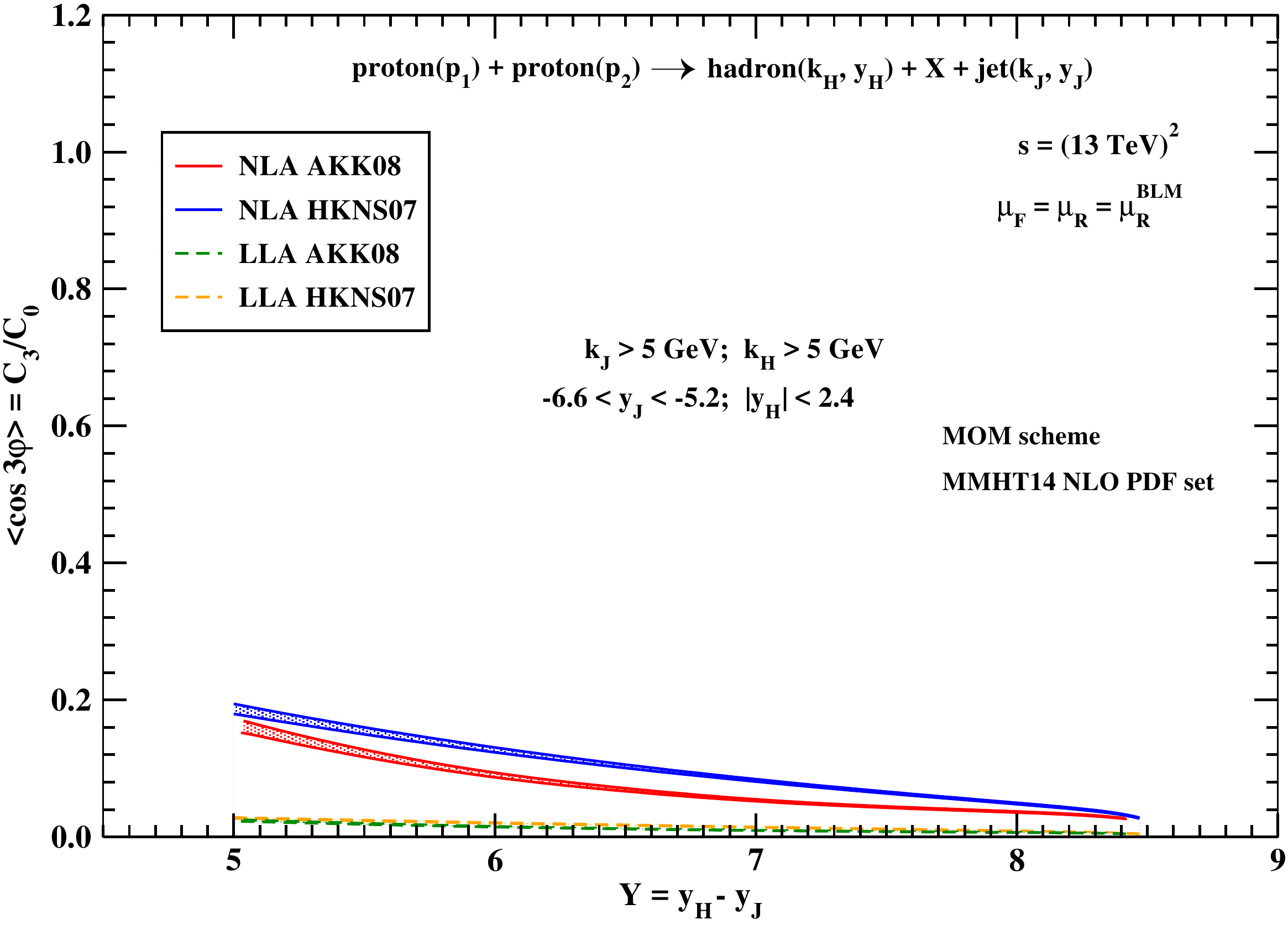}
	
	\includegraphics[scale=0.33,clip]{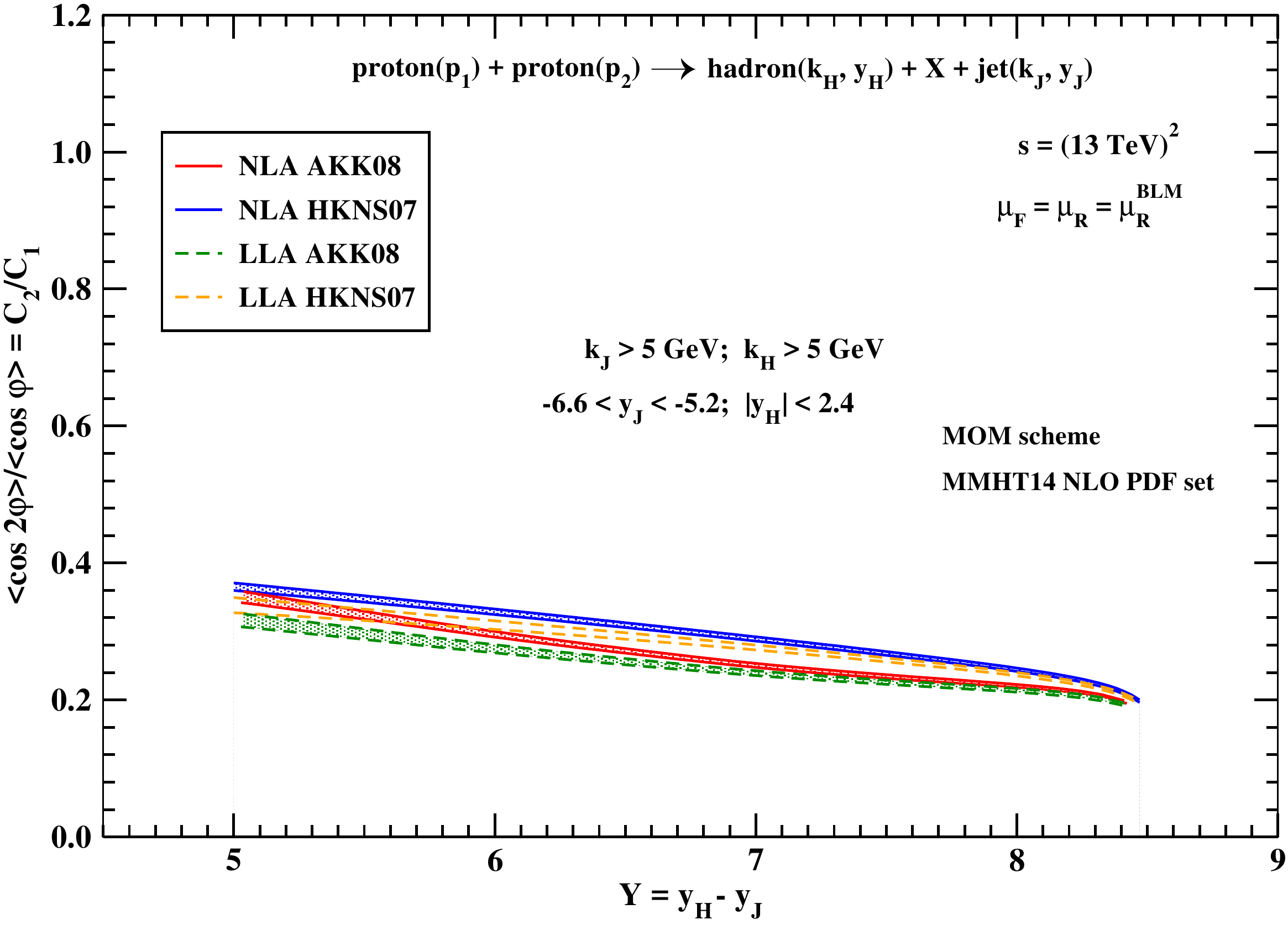}
	\includegraphics[scale=0.33,clip]{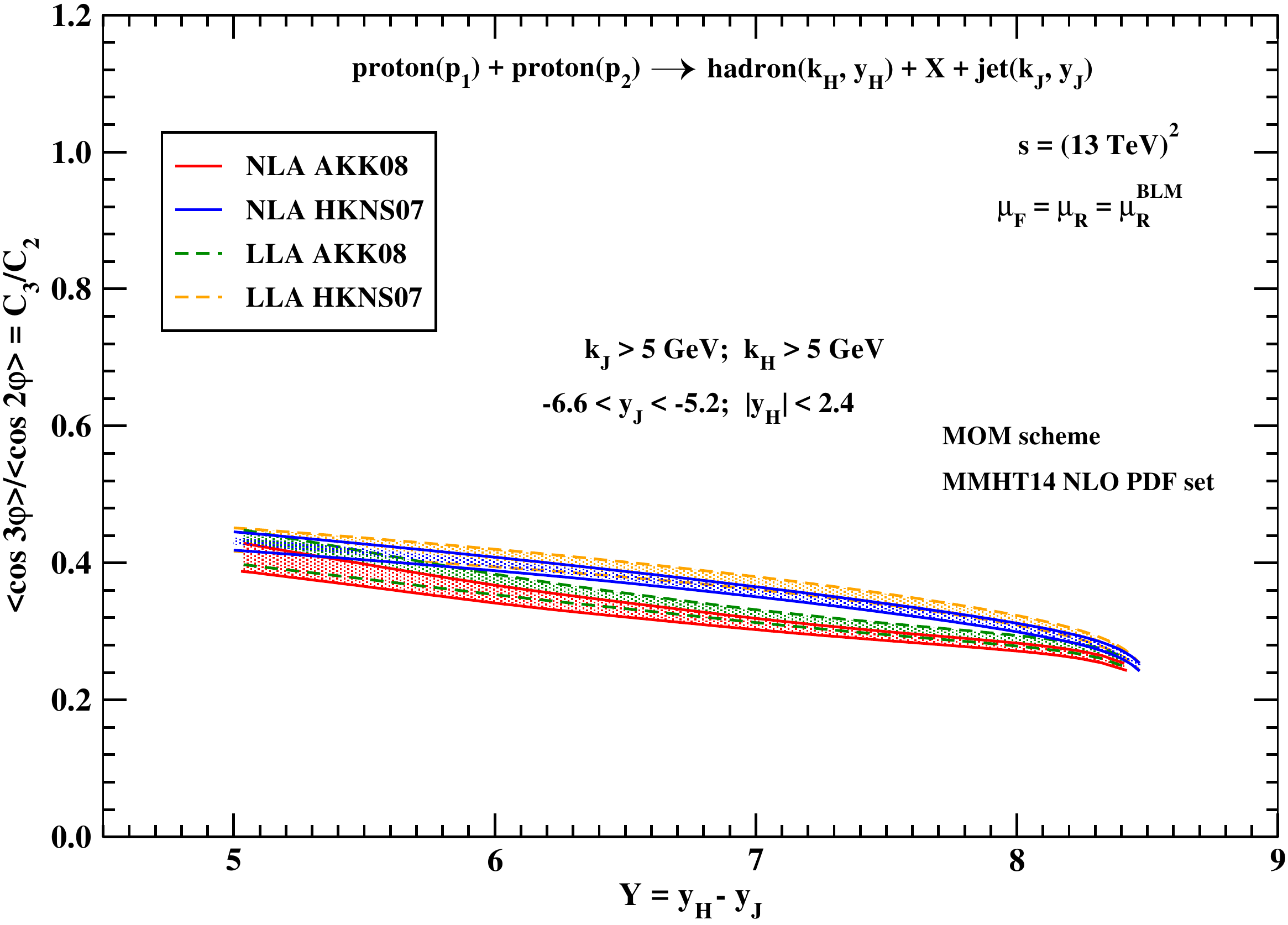}
	\caption{$Y$-dependence of $C_0$ and of several ratios $C_m/C_n$ for 
		$\mu_F = \mu_R^{\rm BLM}$, $\sqrt{s} = 13$ TeV, and $Y \leq 9$
		({\it CASTOR-jet} configuration).}
	\label{fig:Cn_MOM_BLM_CASTOR_13}
\end{figure}

%Panels in Fig.~\ref{fig:Cn_MOM_BLM_CASTOR_13} show results with BLM scale
%optimization for $C_0$ and several $R_{nm}$ ratios in the {\it CASTOR-jet}
%configuration at $\sqrt{s} = 13$ TeV. They exhibit some new and, to
%some extent, unexpected features: (i) the two parametrizations for the FFs
%lead to clearly distinct predictions, (ii) $\langle \cos \phi\rangle$ exceeds
%one at the smaller values for $Y$, a clearly unphysical effect. The
%reason for these phenomena could reside in the fact that, the lower values
%for $Y$ in the {\it CASTOR-jet} case are obtained for negative values of the
%hadron rapidity, {\it i.e.} in final-state configurations where both
%jet and hadron are backward.
Panels in Fig.~\ref{fig:Cn_MOM_BLM_CASTOR_13} show results for $C_0$ and several $R_{nm}$ ratios in the {\it CASTOR-jet} configuration at $\sqrt{s} = 13$ TeV. Here the following behaviour is observed: (i) the two parametrizations for the FFs lead to clearly distinct predictions, (ii) $\langle \cos \phi\rangle$ exceeds  one at the smaller values for $Y$. The reason for these phenomena could be, due to the lower values for rapidity range provided by CASTOR-jet configuration, where both jet and hadron are backward.

\begin{figure}[p]
	\centering
	\includegraphics[scale=0.33,clip]{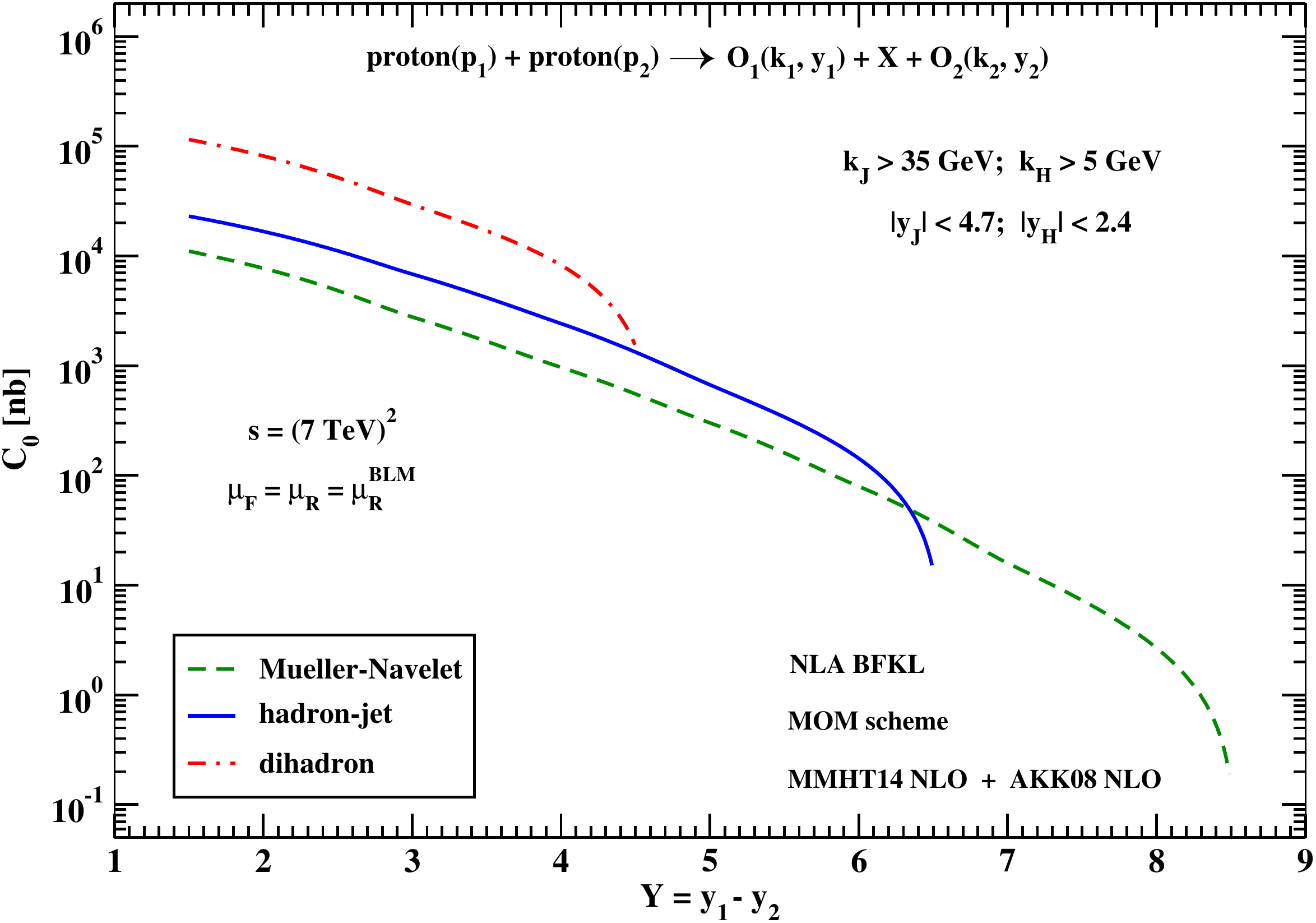}
	\includegraphics[scale=0.33,clip]{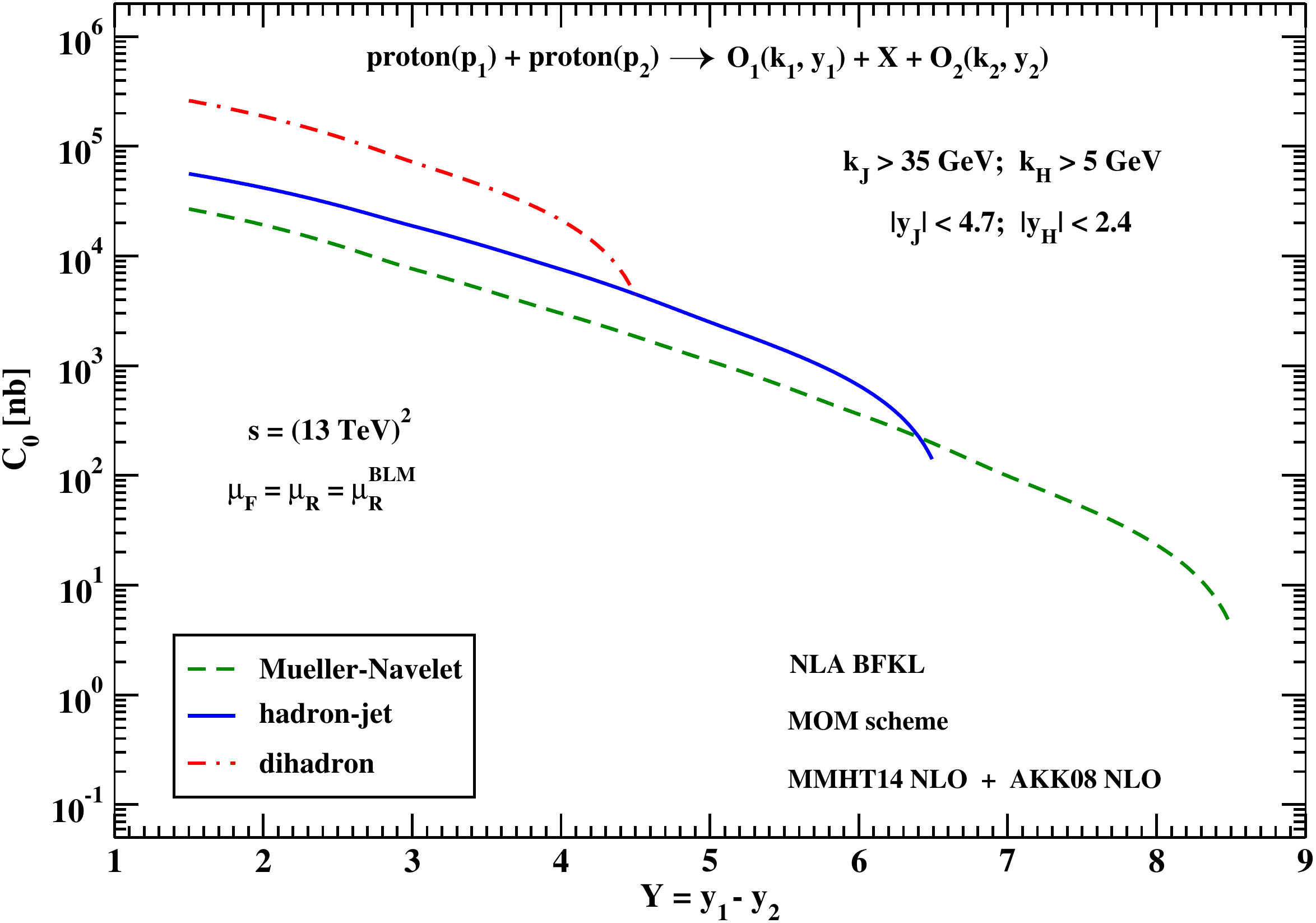}	
	\caption{Comparison of the $\phi$-averaged cross section $C_0$ in different
		NLA BFKL processes: Mueller-Navelet jet, hadron-jet and dihadron production,
		for $\mu_F = \mu_R^{\rm BLM}$, $\sqrt{s} = 7$ and 13 TeV, and $Y \leq 7.1$
		({\it CMS-jet} configuration).}
	\label{fig:C0_comp_NLA_BLM_CMS}
\end{figure}

Finally, for the sake of completeness, in Fig.~\ref{fig:C0_comp_NLA_BLM_CMS} we compare the $\phi$-summed cross section $C_0$ in different NLA BFKL processes: Mueller-Navelet jet,
hadron-jet and hadron-hadron production, for $\mu_F = \mu_R^{\rm BLM}$, at 
$\sqrt{s} = 7$ and 13 TeV, and $Y \leq 7.1$ in the {\it CMS-jet} case.
The hadron-hadron cross section, with the kinematical cuts adopted,
dominates over the jet-jet one by an order of magnitude, with the hadron-jet
cross section lying, as expected, in-between.

 %===================================================
% \subsection{The prescription on energy scales}
 %===================================================
 The above $C_{n}$ expression \ref{eq:HdJt-Cn-coefficients} contains two sources of uncertainties, originated from the NLO corrections to both, the impact factors and the BFKL kernel. 
 Using the same BLM optimization discussed in the previous section we will eliminate these ambiguities. Performing a finite renormalization from the modified minimal subtraction scheme ($\overline{\text{MS}}$) to the physical  momentum subtraction scheme (MOM), which is related to the 3-gluon vertex by definition.
 In order to find the values of the BLM scales, we introduce the ratios $m_R=\mu_R^{\rm BLM}/\mu_N$, where $\mu_N=\sqrt{|\vec k_H||\vec k_J|}$ is the ``natural'' scale suggested by the kinematic of the process. We look here for the values of $m_R$ that solve Eq.~(\ref{blmb0}),
 the final expression for the integrated coefficients in the BLM scheme:
 \begin{equation}
 \label{hadjet_Cn_int_blm}
 C_n =
 \int_{y^{\rm min}_H}^{y^{\rm max}_H}dy_H
 \int_{y^{\rm min}_J}^{y^{\rm max}_J}dy_J\int_{k^{\rm min}_H}^{k^{\rm max}_H}dk_H
 \int_{k^{\rm min}_J}^{k^{\rm max}_J}dk_J
 \,
 \int\limits^{\infty}_{-\infty} d\nu 
 \end{equation}
 \begin{equation}\nonumber
 \frac{e^Y}{s}\,
 e^{Y \bar \alpha^{\rm MOM}_s(\mu^{\rm BLM}_R)\left[\chi(n,\nu)
 	+\bar \alpha^{\rm MOM}_s(\mu^{\rm BLM}_R)\left(\bar \chi(n,\nu) +\frac{T^{\rm conf}}
 	{3}\chi(n,\nu)\right)\right]}
 \left(\alpha^{\rm MOM}_s (\mu^{\rm BLM}_R)\right)^2 
 \end{equation}
 \[
 \times c_H(n,\nu)[c_J(n,\nu)]^*
 \left\{1+\bar \alpha^{\rm MOM}_s(\mu^{\rm BLM}_R)\left[\frac{\bar c^{(1)}_H(n,\nu)}
 {c_H(n,\nu)}+\left[\frac{\bar c^{(1)}_J(n,\nu)}{c_J(n,\nu)}\right]^*
 +\frac{2T^{\rm conf}}{3} \right] \right\} \, .
 \]
 The coefficient $C_0$ gives the $\phi$-summed cross section, while the ratios
 $R_{n0} \equiv C_n/C_0 = \langle\cos(n\phi)\rangle$ determine the values of the
 mean cosines, or azimuthal correlations, of the produced hadron and jet.
% In Eq.~(\ref{Cn_int_blm}), $\bar \chi(n,\nu)$ is the eigenvalue of NLA BFKL
% kernel~\cite{Kotikov:2000pm} and its expression is given, {\it e.g.} in
% Eq.~(23) of Ref.~\cite{Caporale:2012ih}, whereas $\bar c^{(1)}_{H,J}$ are the
% NLA parts of the hadron/jet vertices (see Ref.~\cite{Caporale:2015uva}).
 
 We set the factorization scale $\mu_F$ equal to the renormalization scale
 $\mu_R$, as assumed by the MMHT~2014 PDF.
 
 All calculations are done in the MOM scheme. For comparison, we present results
 for the $\phi$-averaged cross section $C_0$ in the $\overline{\rm MS}$ scheme.
 In the latter case, we choose natural values for $\mu_R$, {\it i.e.} 
 $\mu_R = \mu_N \equiv \sqrt{|\vec k_H||\vec k_J|}$, and two different values of
 the factorization scale, $(\mu_F)_{1,2} = |\vec k_{H,J}|$, depending on which of
 the two vertices is considered. We checked that the effect of using natural
 values also for $\mu_F$, {\it i.e.} $\mu_F = \mu_N$, is negligible with respect
 to our two-value choice.
 
%===================================
\section{Used tools}
All numerical calculations were done using {\tt JETHAD}~\cite{Celiberto:2020wpk}, a promising standard software recently developed, suited for the computation of cross sections and
related observables for semi-hard reactions. 
A two-loop running coupling setup with $\alpha_s\left(M_Z\right)=0.11707$ and five quark flavors was chosen. All PDF sets and the NNFF1.0 FF parametri-zation were used via the Les Houches Accord PDF Interface (LHAPDF) 6.2.1~\cite{Buckley:2014ana}. We selected the MMHT~2014~PDF set, together with the AKK~2008~\cite{Albino:2008fy} and HKNS~2007 ~\cite{Hirai:2007cx} FF interfaces.

%===================================
\section{Closing statements} 
%===================================
The lower cutoff for the hadron's transverse momentum $k_{H}$, allows to
explore additional kinematic range with respect to jet case. In Hadron-jet production process we observe that, just as in the case of Mueller-Navelet jets decorrelation reduced with the increasing rapidity intervals, by the inclusion of NLO correction. And the predicted cross-sections for Hadron-jet process, as expected, are laying in the middle with respect to the Mueller-Navelet jets and investigated dihadron production\cite{Celiberto_2016,Celiberto_2017,Celiberto_2017s}, since that, this process is a hybridization between these two investigated processes, which are sharing the same theoretical description.  
\end{spacing}

\chapter{Proposed stabilizers of the high-energy resummation}\label{chapter3}
 
%===============================
\section{Inclusive Higgs-plus-jet production}
\label{sec:higgs-jet}
%===============================
\begin{spacing}{1.5}
With more data to be collected at the LHC, studying the Higgs boson plus multijet processes will allow us to further test the perturbative QCD on the Higgs boson production, contributing to a better understanding the dynamic of strong interactions in the high-energy limit.
In this section, we present recent BFKL phenomenological results for the inclusive production of a Higgs in association with a jet, as a possible testfield for the semi-hard regime of QCD.
We show how the large energy scales provided by the emission of a Higgs boson stabilize
the BFKL series, and discuss the possible extension of this work in the full NLA BFKL
analysis, by including the NLO jet impact factor, with a realistic implementation of the
jet selection function, and the NLO forward-Higgs impact factor.

\begin{figure}[t]
	\centering
	\includegraphics[width=0.45\textwidth]{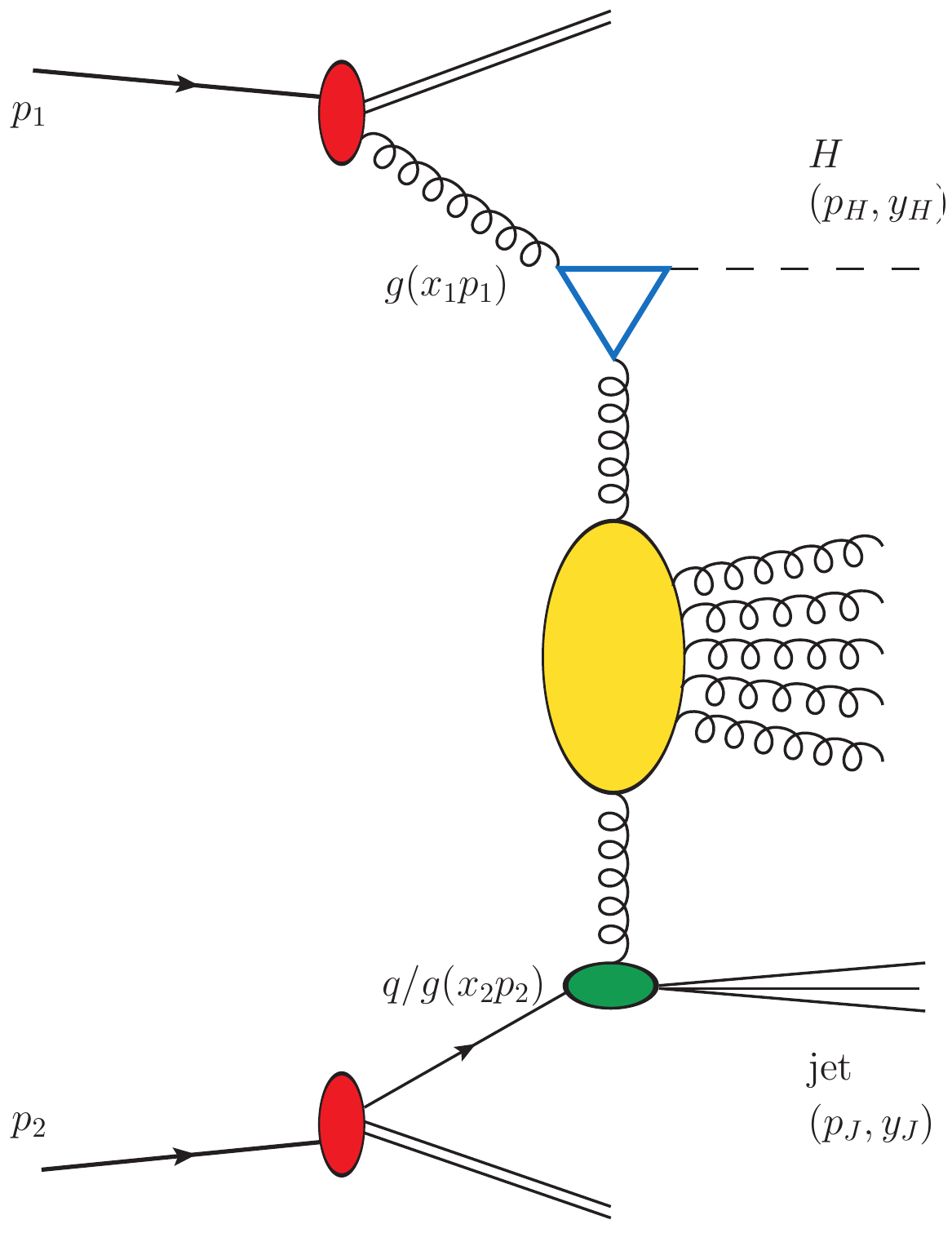}
	\caption{Schematic representation of the inclusive Higgs-jet hadroproduction.}
	\label{fig:Higgs-jet}
\end{figure}

In this work we introduce and study with NLA BFKL accuracy a novel semi-hard
reaction~\cite{Celiberto:2020tmb,Celiberto:2021fjf}, \emph{i.e.} the concurrent inclusive production of a Higgs boson
and a jet (depicted in Fig.~\ref{fig:Higgs-jet}):
\begin{equation}
\label{process}
{\rm proton}(p_1) \ + \ {\rm proton}(p_2) \ \to \ H(\vec p_H, y_H) \ + \ {\rm X} \ + \ {\rm jet}(\vec p_J, y_J) \;,
\end{equation}
emitted with large transverse momenta, $\vec p_{H,J} \gg \Lambda_{\rm QCD}$, and
separated by a large rapidity gap, $\Delta Y = y_H - y_J$, where the Higgs
boson and the jet are tagged in the forward and backward rapidity region, respectively. The tagged jet in the final state insure the existence of a large rapidity intervals,
$\Delta Y = y_H - y_J \simeq \ln(s/Q^2)$, with $Q^2$ a typical hard-scale value. In our proposal, following Ref.~\cite{DelDuca:1993ga}, we adopt a kinematics which strictly respects the semi-hard regime, with suitable cuts on transverse momenta to single out only the high-energy logarithms\footnote{However, when $p_{H,J}^2\equiv p_T^2 \ll M_H^2$, double logarithmic terms $\ln^2(M_H^2/p_T^2)$ appear, which are belong to back-to-back region of the phase space.}.

\subsection{Theoretical framework}
\label{th:Hgsjt-framework}
%===============================
%For the process under consideration (see Fig.~\ref{fig:process}) we plan to
%construct the cross section, differential in some of the kinematic variables of
%the Higgs and the jet, and some azimuthal correlations between them. In the
%BFKL approach the cross section takes the factorized form, diagrammatically
%represented in Fig.~\ref{fig:BFKL_factorization}, given by the convolution of
%the Higgs and jet impact factors with the BFKL gluon Green's function, $G$.

%\begin{figure}[t]
%	\centering
%	\includegraphics[width=0.60\textwidth]{./figures/H_jet_amplitude_lla.pdf}
%	\caption{Schematic representation of the BFKL factorization for the Higgs-jet
%		hadroproduction.}
%	\label{fig:BFKL_factorization}
%\end{figure}
%===============================
%\subsection{Cross section and azimuthal coefficients} 
%===============================
%\label{cross_section}

%For the sake of simplicity, we consider final-state configurations where the
%Higgs is always tagged in a more forward direction with respect to the jet,
%thus implying $\Delta Y \equiv y_H - y_J > 0$.
As anticipated, the Higgs and the jet are expected to feature large
transverse momenta,
$|\vec p_H|^2 \sim |\vec p_J|^2 \gg \Lambda^2_{\rm QCD}$.
The four-momenta of the parent protons, $p_{1,2}$, are taken as Sudakov vectors
satisfying $p^2_{1,2} = 0$ and $p_1 p_2 = s/2$, so that the final-state
transverse momenta can be decomposed in the following way:
\begin{eqnarray}
p_H &=& x_H p_1 + \frac{M_{H,\perp}^{\, 2}}{x_H s}p_2 + p_{H\perp} \ , \quad
p_{H\perp}^{\, 2} = - |\vec p_H|^2 \ , \nonumber \\
p_J &=& x_J p_2 + \frac{|\vec p_J|^2}{x_J s}p_1 + p_{J\perp} \ , \quad
p_{J\perp}^{\, 2} = - |\vec p_J|^2 \ ,
\label{higgs-Sudakov}
\end{eqnarray}
with the spacial part of the four-vector $p_{1\parallel}$ being taken positive;
$M_{H,\perp} = \sqrt{M_H^2+|\vec p_H|^2}$ is the Higgs-boson transverse mass.

The longitudinal-momentum fractions, $x_{H,J}$, for the Higgs and jet are
related to the corresponding rapidities in the center-of-mass frame via the
relations
\begin{equation}\label{rapidities}
y_H = \frac{1}{2}\ln\frac{x_H^2 s}{M_{H, \perp}^2} 
\; , \qquad 
y_J = \frac{1}{2}\ln\frac{|\vec p_J|^2}{x_J^2 s} 
\; , \qquad
dy_{H,J} = \pm \frac{dx_{H,J}}{x_{H,J}}
\; .
\end{equation}
As for the rapidity distance, one has
\begin{equation}\label{rapidity_interval}
\Delta Y = y_H - y_J = \ln \frac{x_H x_J s}{M_{H,\perp} |\vec p_J|} \;.
\end{equation}

%Using QCD collinear factorization to build the (differential) hadronic the
%cross section, one has 
%\begin{equation}
%\label{dsigma_QCD}
%\frac{d\sigma}{dx_H dx_J d^2\vec p_H d^2\vec p_J}
%=\sum_{i, j = q,{\bar q},g}
%\int_0^1 dx_1 \int_0^1 dx_2 \ 
%f_i \left( x_1, \mu_{F_1} \right) \
%f_j \left(x_2, \mu_{F_2} \right)
%\frac{d{\hat\sigma}_{i,j}
%	\left( \hat s, \mu_{F_{1,2}} \right)}
%{dx_H dx_J d^2\vec p_H d^2\vec p_J}\;,
%\end{equation}
%where the $i, j$ indices run over the parton kinds 
%(quarks $q = u, d, s, c, b$; antiquarks $\bar q = \bar u, \bar d, \bar s,
%\bar c, \bar b$;  or gluon $g$), $f_{i, j} \left(x, \mu_{F_{1,2}} \right)$ are
%the incoming-proton PDFs; 
%$x_{1, 2}$ denote the longitudinal fractions of the partons involved in the
%hard subprocess, whereas $\mu_{F_{1,2}}$ stand for the factorization scales
%characteristic of the two fragmentation regions of the incoming hadrons;
%$d\hat\sigma_{i, j}\left(\hat s \right)$ is
%the partonic cross section, with $\hat s \equiv x_1 x_2s$ the squared
%center-of-mass energy of the parton-parton collision subreaction.
%In the present case, the sum over the parton kinds $i$ restricts to the
%gluon contribution only, consistently with a LO treatment of the Higgs impact
%factor, as discussed in the previous section.
The BFKL cross section can be presented 
(see Ref.~\cite{Caporale:2012ih} for the derivation)
as the Fourier series of the so-called \emph{azimuthal coefficients},
${\cal C}_n$,
\begin{equation}
\frac{d\sigma}
{dy_H dy_J\, d|\vec p_H| \, d|\vec p_J|d\varphi_H d\varphi_J}
=\frac{1}{(2\pi)^2}\left[{\cal C}_0 + \sum_{n=1}^\infty  2\cos (n\varphi )\,
{\cal C}_n\right]\; ,
\end{equation}
where $\varphi=\varphi_H-\varphi_J-\pi$, with $\varphi_{H,J}$ the Higgs and the
jet azimuthal angles. A comprehensive formula for the $\varphi$-averaged
cross section, ${\cal C}_0$, and the other coefficients, ${\cal C}_{n > 0}$,
reads 
\begin{equation}\nonumber
{\cal C}_n \equiv \int_0^{2\pi} d\varphi_H \int_0^{2\pi} d\varphi_J\,
\cos(n \varphi) \,
\frac{d\sigma}{dy_H dy_J \, d|\vec p_H| \, d|\vec p_J| d\varphi_H d\varphi_J}
\end{equation}
\begin{equation}\nonumber
= \frac{e^{\Delta Y}}{s} \frac{M_{H,\perp}}{|\vec p_H|}
\end{equation}
\begin{equation}\nonumber
\times  \int_{-\infty}^{+\infty} d\nu \, \left(\frac{x_J x_H  s}{s_0}
\right)^{\bar \alpha_s(\mu_{R_c})\left\{\chi(n,\nu)+\bar\alpha_s(\mu_{R_c})
	\left[\bar\chi(n,\nu)+\frac{\beta_0}{8 N_c}\chi(n,\nu)\left[-\chi(n,\nu)
	+\frac{10}{3}+4\ln\left(\frac{\mu_{R_c}}{\sqrt{\vec p_H\vec p_J}}\right)\right]\right]\right\}}
\end{equation}
\begin{equation}\nonumber
\times \left\{ \alpha_s^2(\mu_{R_1}) c_H(n,\nu,|\vec p_H|, x_H) \right\}
\left\{ \alpha_s(\mu_{R_2}) [c_J(n,\nu,|\vec p_J|,x_J)]^* \right\} \,
\end{equation}
\begin{equation}\label{Cn_start_WIP}%\nonumber
\times \left\{1
+ \alpha_s(\mu_{R_1}) \frac{c_H^{(1)}(n,\nu,|\vec p_H|,
	x_H)}{c_H(n,\nu,|\vec p_H|, x_H)}
+ \alpha_s(\mu_{R_2}) \left[\frac{c_J^{(1)}(n,\nu,|\vec p_J|, x_J)}{c_J(n,\nu,|\vec p_J|,
	x_J)}\right]^*  \right\} \; , %\right.
\end{equation}
%  \begin{equation}\nonumber
%  \left. + \bar\alpha_s^2(\mu_R) \ln\left(\frac{x_H x_J s}{s_0}\right)
%  \frac{\beta_0}{4 N_c}\chi(n,\nu)f(\nu)\right\}\;.
%  \end{equation}
\\
where $\bar \alpha_s \equiv N_c/\pi \, \alpha_s$, with $N_c$ the QCD color
number, $\beta_0$
%\begin{equation}
%\beta_0=\frac{11}{3} N_c - \frac{2}{3}n_f
%\end{equation}
the first coefficient in the expansion of the QCD $\beta$-function ($n_f$ is the active-flavor number), and $\chi\left(n,\nu\right) $
%\begin{equation}
%\chi\left(n,\nu\right) = 2\psi\left(1\right) - \psi\left(\frac{n + 1}{2} + i\nu \right)-\psi\left(\frac{n + 1}{2} - i\nu \right)
%\end{equation}
the eigenvalue of the LO BFKL kernel defined in Eq.~\ref{lo_kernel_eigvl}, and $c_{H,J}(n,\nu)$ are the Higgs and the jet LO impact factors in the ($\nu,n$)-space, given by
\begin{equation}
\label{c-higgs}
c_H(n,\nu, |\vec p_H|, x_H) = \frac{1}{v^2} \frac{|\mathcal{F}(\vec p_H^{\: 2})|^2}
{128\pi^{3}\sqrt{2(N^2_{c}-1)}}
\left( \vec p_H^{\: 2} \right)^ {i\nu + 1/2} f_g(x_H,\mu_{F_1}) \; ,
\end{equation}
\begin{equation}
\label{c-jet}
c_J(n,\nu, |\vec p_{J}|, x_J) = 
2 \sqrt{\frac{C_F}{C_A}}
\left( \vec p_J^{\: 2} \right)^{i\nu-1/2} 
\left( \frac{C_A}{C_F} f_g(x_J,\mu_{F_2}) + \sum_{a = q,\bar{q}}f_a(x_J,\mu_{F_2})
\right) \;. 
\end{equation}
The energy-scale parameter, $s_0$, is arbitrary within NLA accuracy and will
be fixed in our analysis at $s_0 = M_{H,\perp} |\vec p_J|$.
The remaining quantities are the NLO impact-factor corrections,
$c_{H,J}^{(1)}(n,\nu,|\vec p_{H,J}|, x_{H,J})$. The expression for the Higgs NLO
impact factor at the time of studying this process was not available, recently it has been calculated in the infinite top-mass limit within Lipatov’s high-energy effective
action~\cite{Hentschinski:2020tbi,Nefedov:2019mrg}. It is possible, however, 
to include some ``universal'' NLO contributions to the Higgs impact factor,
which can be expressed through the corresponding LO impact factor, and 
are fixed by the requirement of stability within the NLO under variations
of the energy scale $s_0$, the renormalization scale $\mu_R$ and of the
factorization scale $\mu_F$, getting
\begin{equation}
\alpha_s c_H^{(1)}(n,\nu,|\vec p_H|, x_H) \to \bar \alpha_s \tilde
c_H^{(1)}(n,\nu,|\vec p_H|, x_H) \; ,
\end{equation}
with
\[
\tilde c_H^{(1)}(n,\nu,|\vec p_H|, x_H) =
c_H(n, \nu, |\vec p_H|, x_H) \left\{ \frac{\beta_0}{4 N_c} 
\left( 2 \ln \frac{\mu_{R_1}}{|\vec p_H|} + \frac{5}{3} \right)
+ \frac{\chi\left(n,\nu\right)}{2} \ln \left( \frac{s_0}{M_{H,\perp}^2} \right)
\right.
\]
\begin{equation}
\label{cH1}
+ \frac{\beta_0}{4 N_c} \left( 2 \ln \frac{\mu_{R_1}}{M_{H,\perp}}\right)
\end{equation}
\[
\left.
-\frac{1}{2N_c f_g(x_H,\mu_{F_1})} \ln \frac{\mu_{F_1}^2}{M_{H,\perp}^2} \int_{x_H}^1\frac{dz}{z}
\left[P_{gg}(z)f_g\left(\frac{x_H}{z},\mu_{F_1}\right)+\sum_{a={q,\bar q}} P_{ga}(z)
f_a\left(\frac{x_H}{z},\mu_{F_1}\right)\right]
\right\} \;.
\]
The jet impact factor is known at the
NLO~\cite{Caporale:2011cc,Ivanov:2012ms,Colferai:2015zfa}, nonetheless
we treated it on the same ground as the Higgs one, including only the NLO
corrections fixed by the renormalization group and leading to  
\begin{equation}
\label{cJ1}
\tilde c_J^{(1)}(n,\nu,|\vec p_J|, x_J) =
c_J(n, \nu, |\vec p_J|, x_J) \left\{ \frac{\beta_0}{4 N_c} 
\left( 2 \ln \frac{\mu_{R_2}}{|\vec p_J|} + \frac{5}{3} \right)
+ \frac{\chi\left(n,\nu\right)}{2} \ln \left( \frac{s_0}{|\vec p_J|^2} \right)
\right.
\end{equation}
\[
-\frac{1}{2N_c \left( \frac{C_A}{C_F} f_{g}(x_J,\mu_{F_2}) +
	\sum_{a = q,\bar{q}}f_{a}(x_J,\mu_{F_2}) \right) }
\ln \frac{\mu_{F_2}^2}{|\vec p_J|^2}
\]
\[
\times \biggl(
\frac{C_A}{C_F} 
\int_{x_J}^1\frac{dz}{z}
\left[P_{gg}(z)f_g\left(\frac{x_J}{z},\mu_{F_2}\right)+\sum_{a={q,\bar q}} P_{ga}(z)
f_a\left(\frac{x_J}{z},\mu_{F_2}\right)\right]\biggr.
\]
\[
\left.\biggl.
+ \sum_{a = q,\bar{q}} \int_{x_J}^1\frac{dz}{z}
\left[P_{ag}(z) f_g\left(\frac{x_J}{z},\mu_{F_2}\right)
+ P_{aa}(z)f_a\left(\frac{x_J}{z},\mu_{F_2}\right)\right] \biggr)
\right\} \; .
\]

Combining all the ingredients, we can write our master formula for the
azimuthal coefficients,
\begin{equation}\nonumber
{\cal C}_n = \frac{e^{\Delta Y}}{s} \frac{M_{H,\perp}}{|\vec p_H|}
\end{equation}
\begin{equation}\nonumber
\times  \int_{-\infty}^{+\infty} \!\!\! d\nu \left(\frac{x_J x_H s}{s_0}
\right)^{\bar \alpha_s(\mu_{R_c})\left\{\chi(n,\nu)+\bar\alpha_s(\mu_{R_c})
	\left[\bar\chi(n,\nu)+\frac{\beta_0}{8 N_c}\chi(n,\nu)\left[-\chi(n,\nu)
	+\frac{10}{3}+4\ln\left(\frac{\mu_{R_c}}{\sqrt{\vec p_H\vec p_J}}\right)\right]\right]\right\}}
\end{equation}
\begin{equation}%\nonumber
\times \left\{ \alpha_s^2(\mu_{R_1}) c_H(n,\nu,|\vec p_H|, x_H) \right\}
\left\{ \alpha_s(\mu_{R_2}) [c_J(n,\nu,|\vec p_J|,x_J)]^* \right\} \,
\end{equation}
\begin{equation}\label{Cn_master}\nonumber
\times \left\{1
+ \bar \alpha_s(\mu_{R_1}) \frac{\tilde c_H^{(1)}(n,\nu,|\vec p_H|,
	x_H)}{c_H(n,\nu,|\vec p_H|, x_H)}
+ \bar \alpha_s(\mu_{R_2}) \left[\frac{\tilde c_J^{(1)}(n,\nu,|\vec p_J|, x_J)}{c_J(n,\nu,|\vec p_J|,
	x_J)}\right]^*  \right\} \; . %\right.
\end{equation}
The renormalization scales ($\mu_{R_{1,2,c}}$) and the factorization ones
($\mu_{F_{1,2}}$) can, in principle, be chosen arbitrarily, since their variation
produces effects beyond the NLO. It is however advisable to relate them to the
physical hard scales of the process.
We chose to fix them differently from each other, depending on the subprocess
to which they are related: $\mu_{R_1} \equiv \mu_{F_1} = C_\mu M_{H,\perp}$,
$\mu_{R_2} \equiv \mu_{F_2} = C_\mu |\vec p_J|$,
$\mu_{R_c} = C_\mu \sqrt{M_{H,\perp} |\vec p_J|}$, where $C_\mu$ is a variation
parameter introduced to gauge the effect of a change of the scale (see the
discussion at the end of Section~\ref{results}).
%===============================
\subsection{Phenomenological analysis}
\label{phenomenology}
%===============================
In order to fit realistic kinematic cuts adopted by the current
experimental analyses at the LHC, we constrain the Higgs emission inside the
rapidity acceptances of the CMS barrel detector, \emph{i.e.} $|y_H| < 2.5$,
while we allow for a larger rapidity range of the
jet~\cite{Khachatryan:2016udy}, which can be detected also by the CMS endcaps,
namely $|y_J| < 4.7$.
Moreover, three distinct configurations for the final-state transverse momenta are
considered:
\begin{itemize}
	\item[a)]
	\textit{\textbf{symmetric}} configuration, fit to the pursuit of pure BFKL
	effects, where both the Higgs and the jet transverse momenta lie in the
	range: 20 GeV $< |\vec p_{H,J}| <$ 60 GeV; \,
	\item[b)]
	\textit{\textbf{asymmetric}} selection, typical of the ongoing LHC
	phenomenology, where the Higgs transverse momentum runs from 10 GeV to
	$2 M_t$, where the jet is tagged inside its typical CMS configuration, from
	20 to 60 GeV;
	\item[c)]
	\textit{\textbf{disjoint windows}}, which allows for the maximum
	exclusiveness in the final state (to possibly discriminate BFKL from other resummation approaches): 35 GeV $< |\vec p_J| <$ 60 GeV and
	60 GeV $< |\vec p_H| < 2 M_t$.
\end{itemize}
%===============================
%\subsection{Azimuthal correlations and $p_T$-distribution}
\subsubsection{Observables:}
\label{observables}
%===============================
\begin{itemize}
	\item The first observables of our consideration are the azimuthal-angle coefficients
	\emph{integrated} over the phase space for two final-state particles, while
	the rapidity interval, $\Delta Y$, between the Higgs boson and the jet is
	kept fixed:
	\begin{equation}
	\label{Cn_int}
	C_n(\Delta Y, s) =
	\int_{p^{\rm min}_H}^{p^{\rm max}_H}d|\vec p_H|
	\int_{p^{\rm min}_J}^{{p^{\rm max}_J}}d|\vec p_J|
	\int_{y^{\rm min}_H}^{y^{\rm max}_H}dy_H
	\int_{y^{\rm min}_J}^{y^{\rm max}_J}dy_J
	\, \delta \left( y_H - y_J - \Delta Y \right)
	\, {\cal C}_n %\left(|\vec p_H|, |\vec p_J|, y_H, y_J \right)
	\, .
	\end{equation}
	
	We study the $\varphi$-averaged cross section (\emph{alias} the
	$\Delta Y$-distribution), $C_0(\Delta Y, s)$, the azimu\-thal-correlation
	moments, $R_{n0}(\Delta Y, s) = C_{n}/C_{0} \equiv \langle \cos n \varphi \rangle$,
	and their ratios, $R_{nm} = C_{n}/C_{m}$~\cite{Vera:2006un,Vera:2007kn} as
	functions of the Higgs-jet rapidity distance, $\Delta Y$.
	
	\item The second observable of our interest is the $p_H$-distribution for a given
	value of $\Delta Y$:
	\begin{equation}
	\label{pT_distribution}
	\frac{d\sigma(|\vec p_H|, \Delta Y, s)}{d|\vec p_H| d\Delta Y} =
	\int_{p^{\rm min}_J}^{{p^{\rm max}_J}}d|\vec p_J|
	\int_{y^{\rm min}_H}^{y^{\rm max}_H}dy_H
	\int_{y^{\rm min}_J}^{y^{\rm max}_J}dy_J
	\, \delta \left( y_H - y_J - \Delta Y \right)
	\, {\cal C}_0 
	%\left(|\vec p_H|, |\vec p_J|, y_H, y_J \right)
	\, ,
	\end{equation}
	the Higgs and jet rapidity ranges being given above and 35 GeV $< |\vec p_J| <$
	60 GeV.
\end{itemize}

%===============================
\subsection{Results and discussion}
\label{results}
\begin{figure}[hbt!]
	\centering
	\includegraphics[scale=0.53,clip]{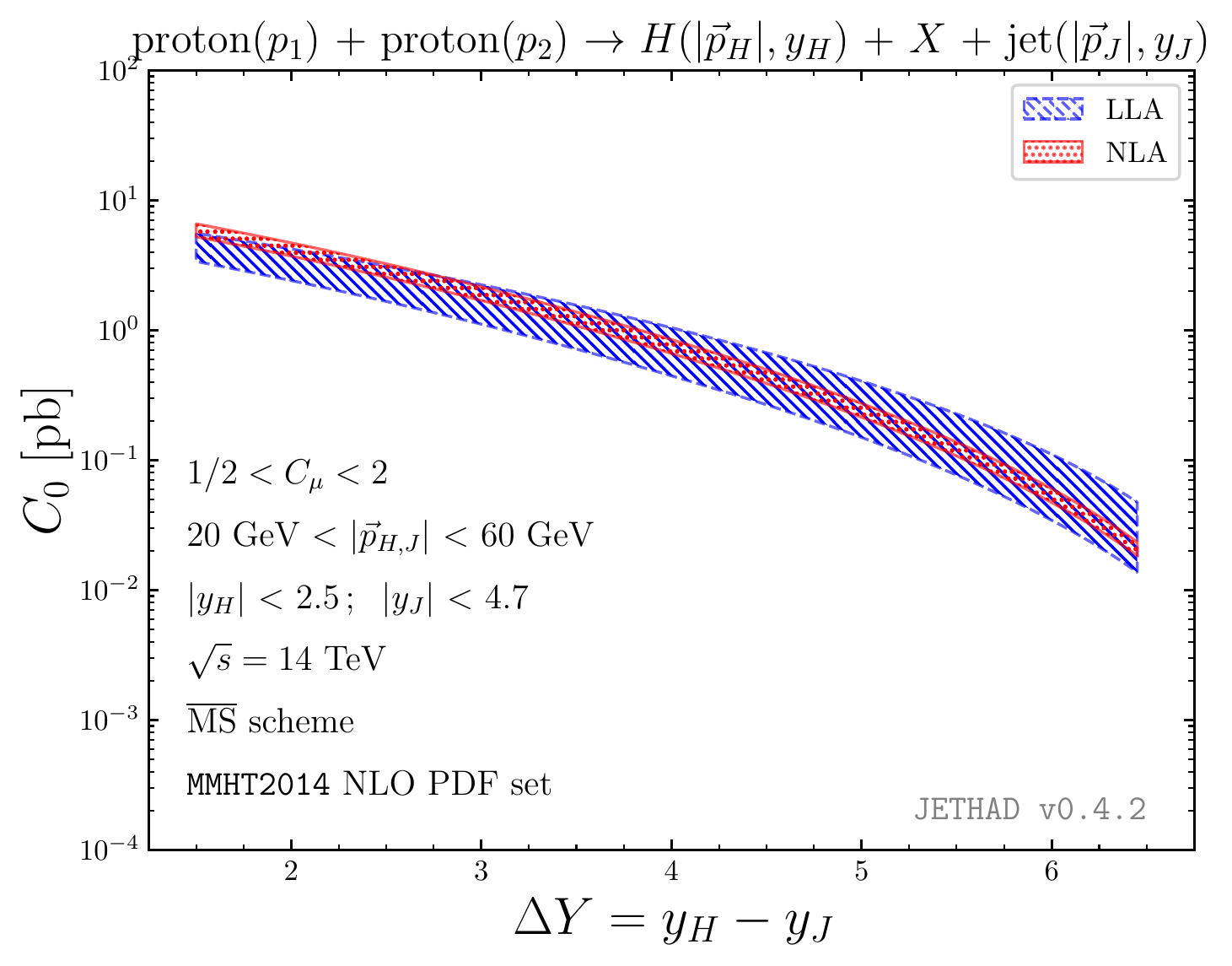}
	\includegraphics[scale=0.53,clip]{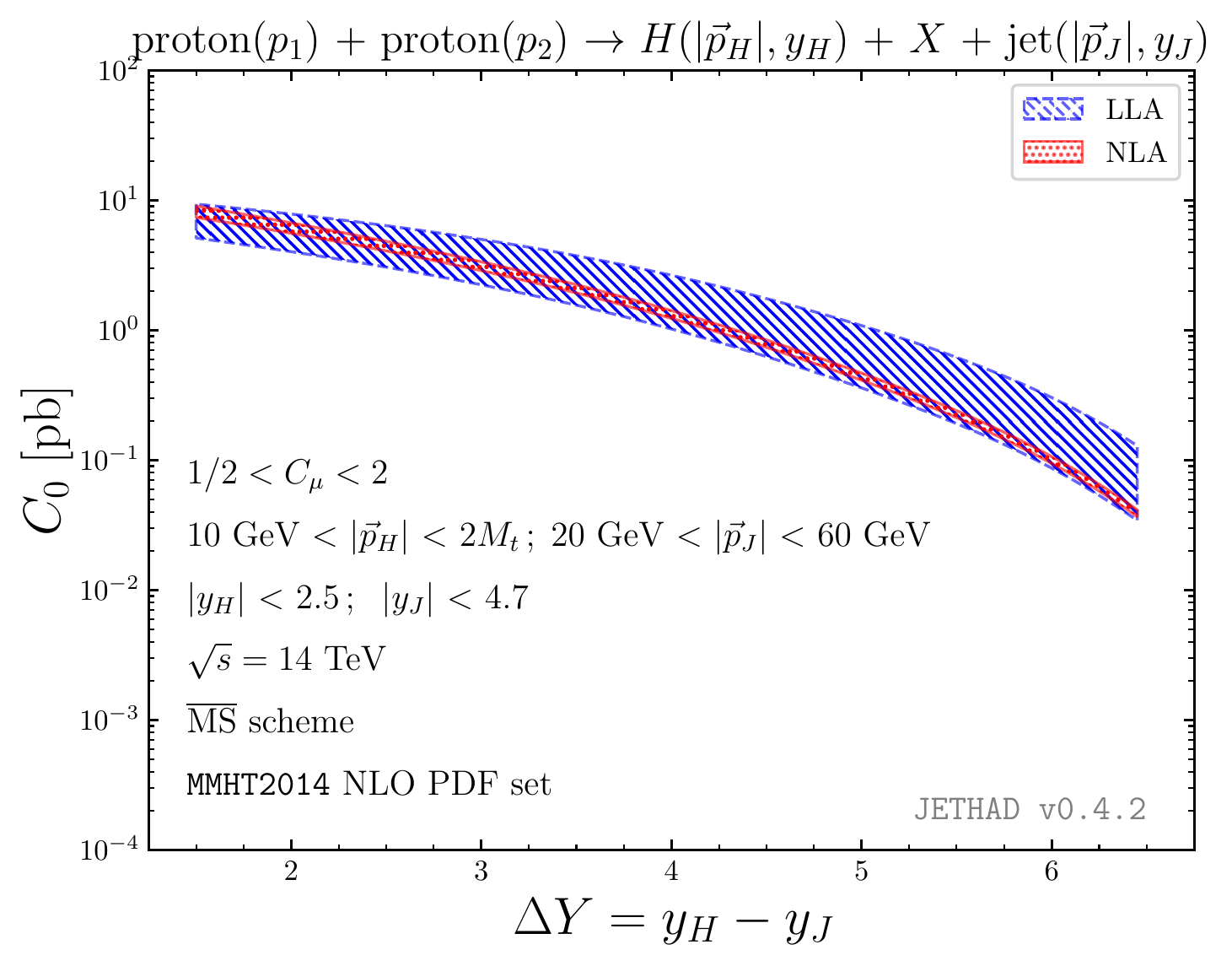}
	\includegraphics[scale=0.53,clip]{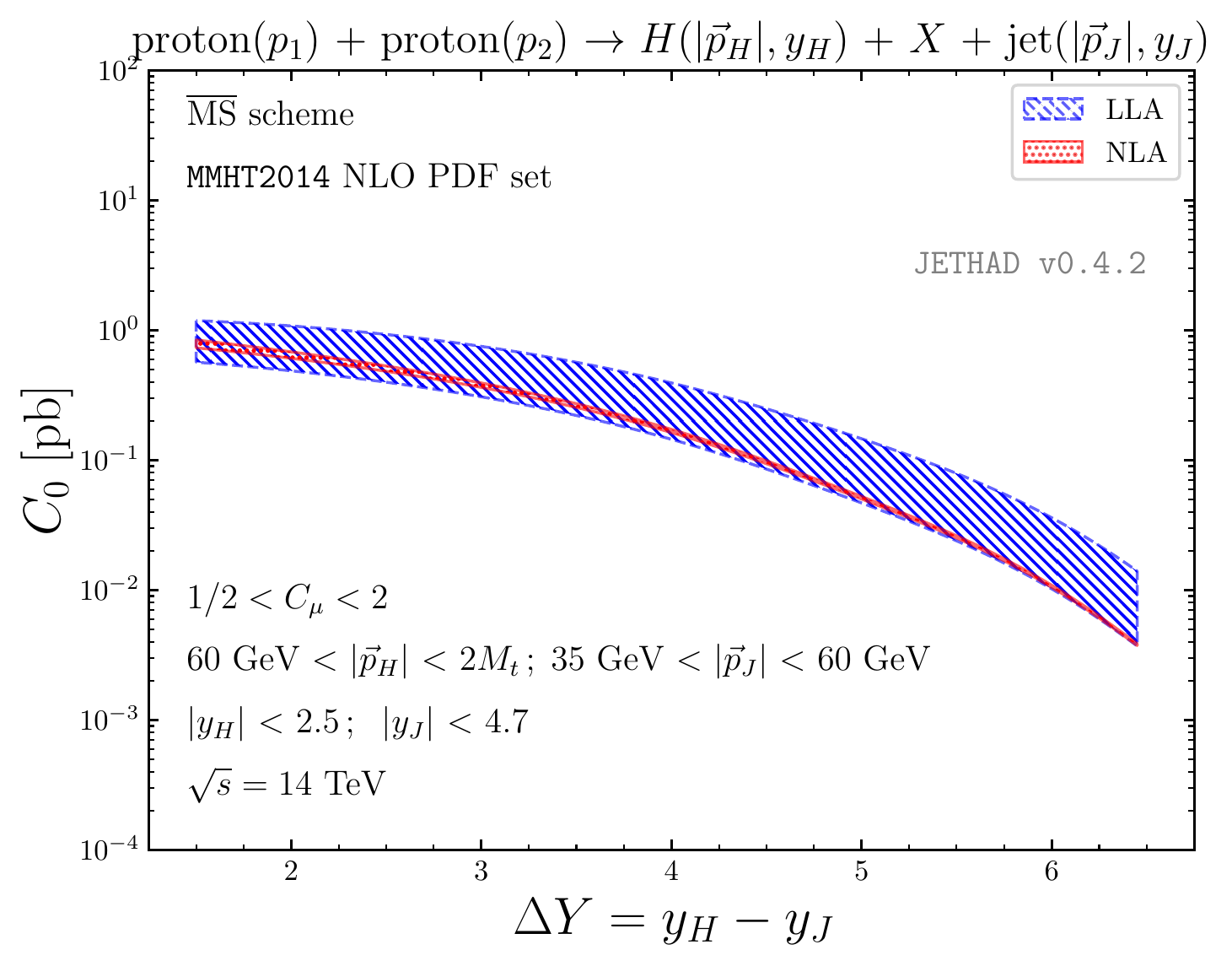}
	\caption{$\Delta Y$-dependence of the $\varphi$-averaged cross section, $C_0$, for the inclusive Higgs-jet hadroproduction in the three considered $p_T$-ranges and for $\sqrt{s} = 14$ TeV. %Shaded bands give the combined effect of the scale variation with the uncertainty coming from the phase-space numerical integration.
		.}
	\label{fig:C0_kt-asw}
\end{figure}

\begin{figure}[hpt]
	\centering
	\includegraphics[scale=0.54,clip]{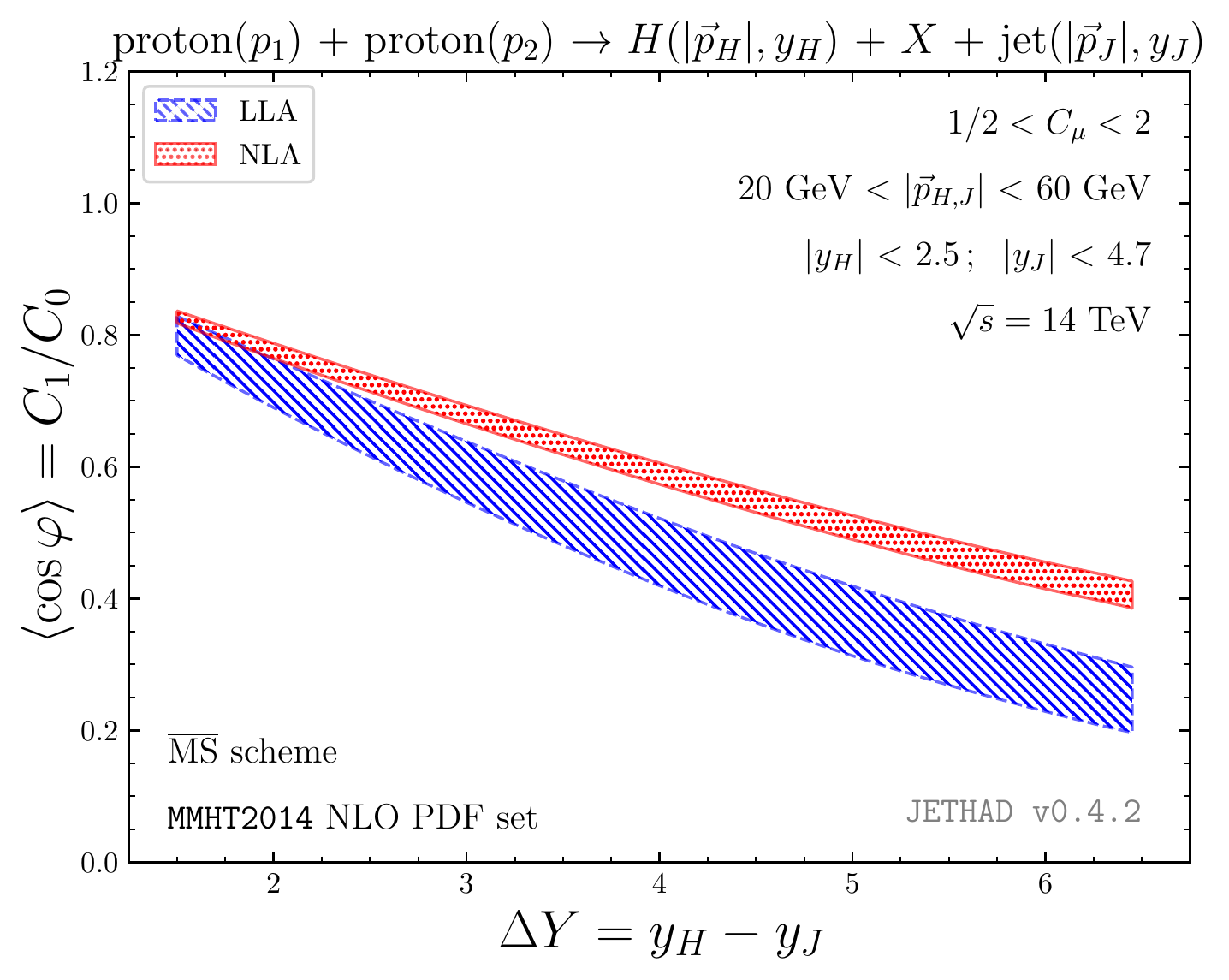}
	\includegraphics[scale=0.54,clip]{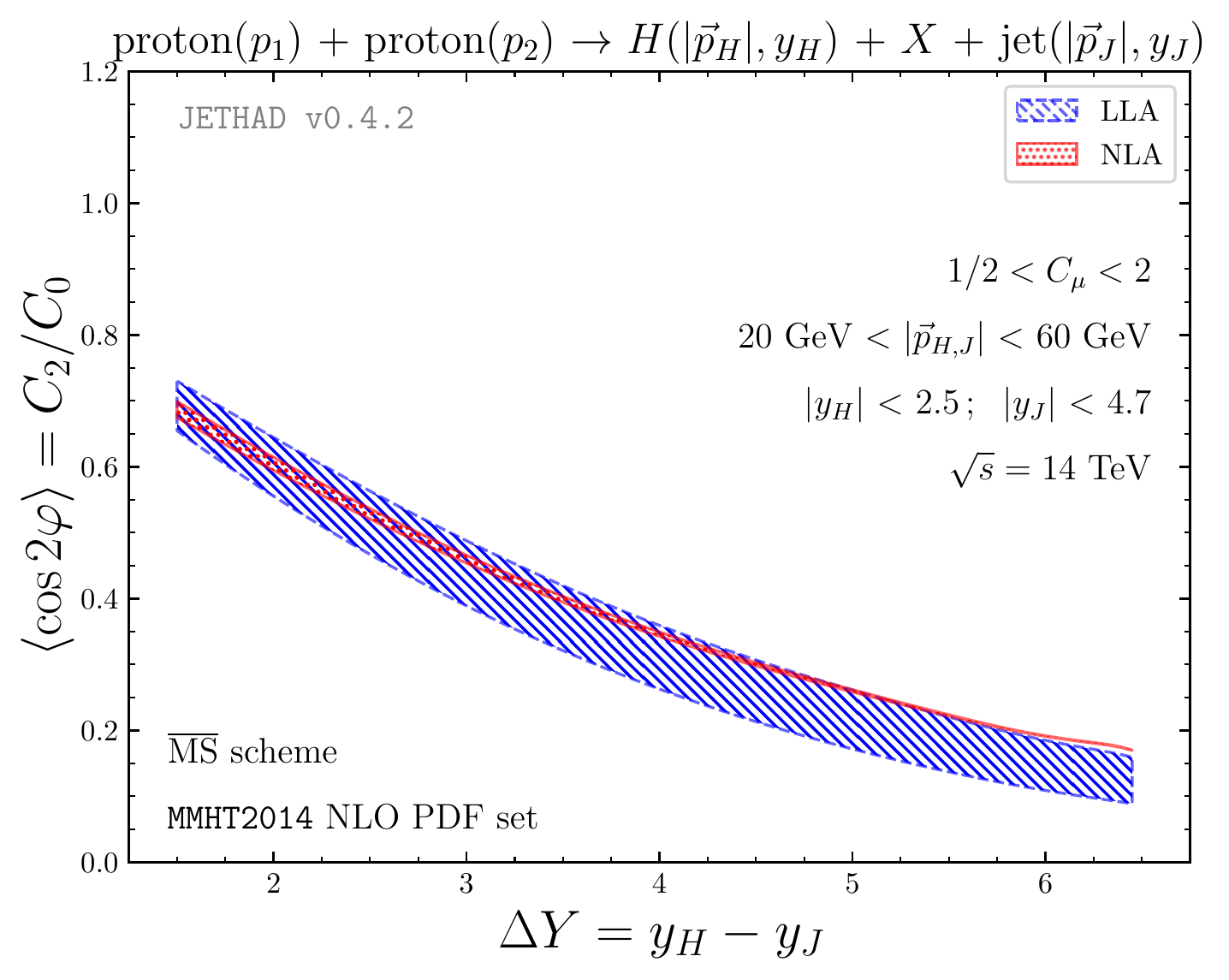}
	
	\includegraphics[scale=0.54,clip]{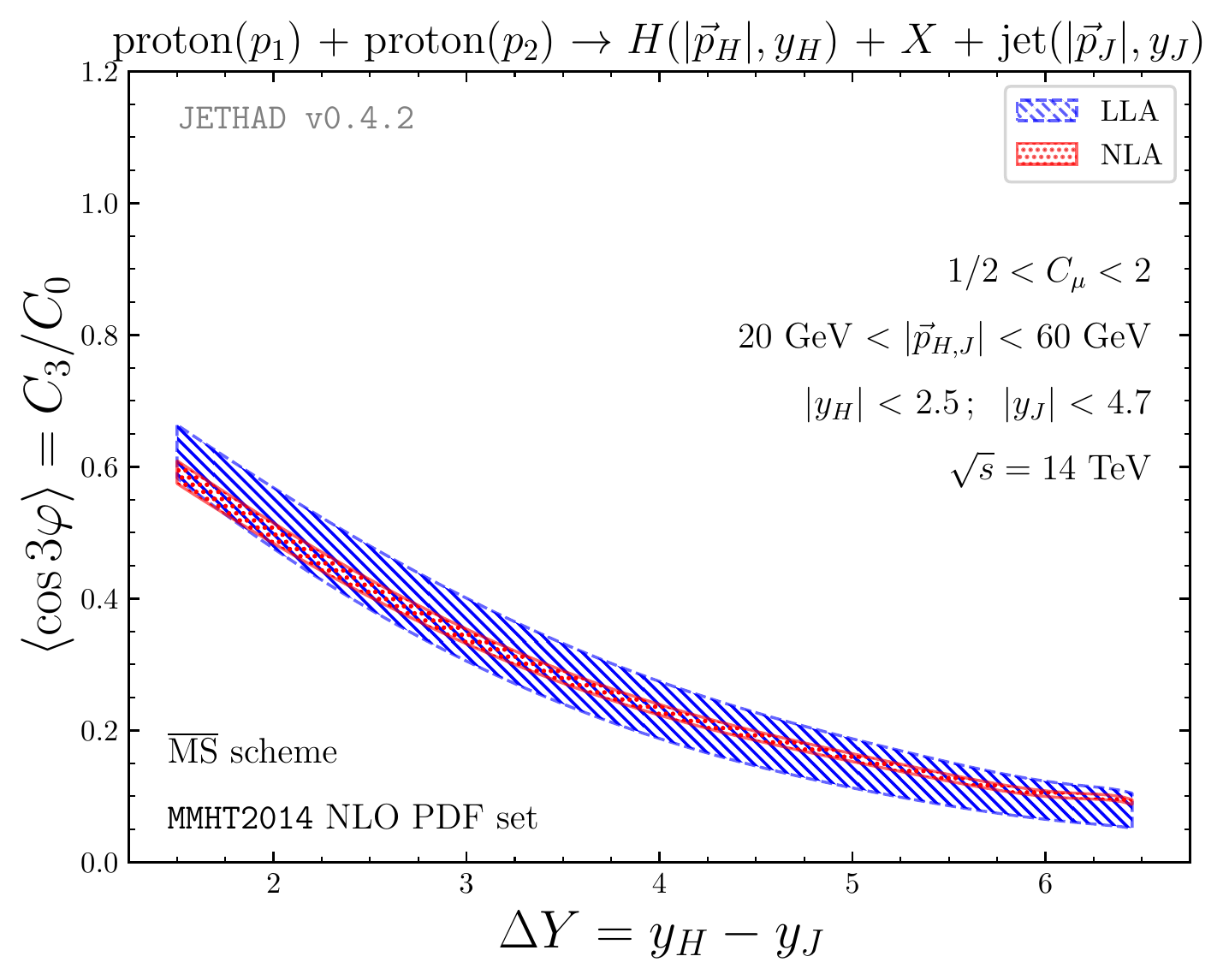}
	\includegraphics[scale=0.54,clip]{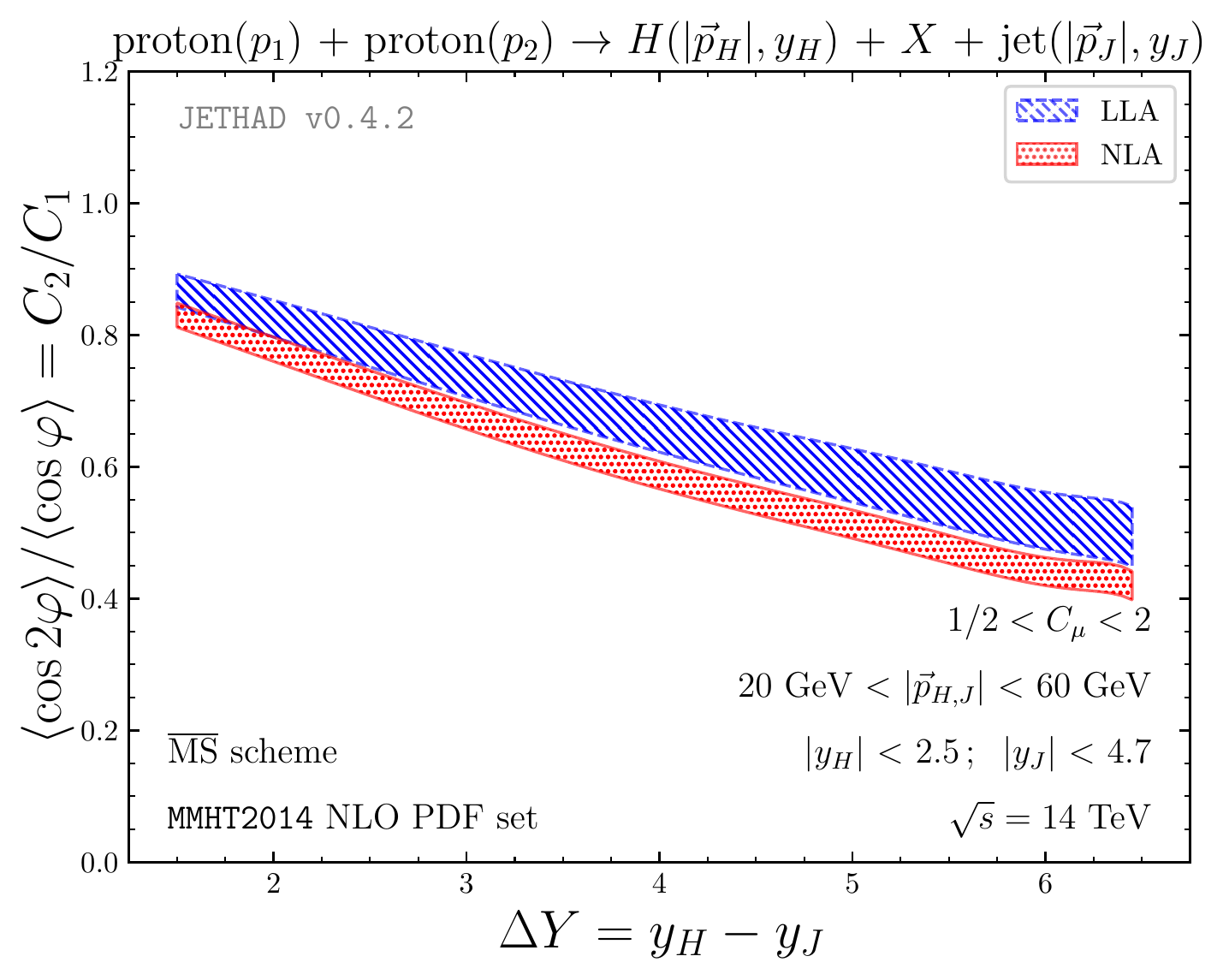}
	
	\includegraphics[scale=0.54,clip]{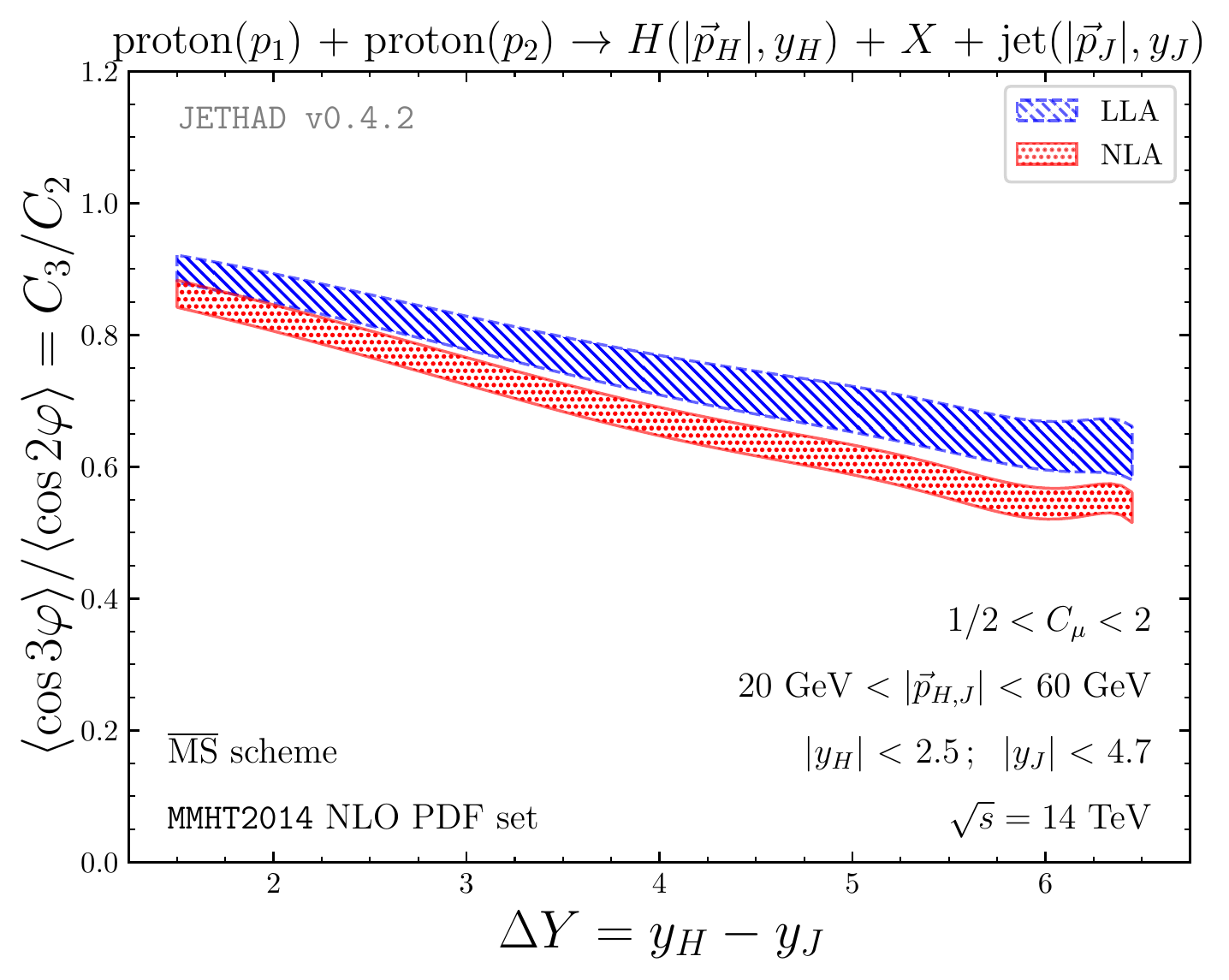}
	\caption{$\Delta Y$-dependence of several ratios $R_{nm} \equiv C_{n}/C_{m}$,
		for the inclusive Higgs-jet hadroproduction in the $p_T$-\emph{symmetric}
		configuration and for $\sqrt{s} = 14$ TeV.
		% Shaded bands give the combined effect of the scale variation with the uncertainty coming from the phase-space numerical integration.
		}
	\label{fig:Rnm_kt-s}
\end{figure}

%===============================
In Fig.~\ref{fig:C0_kt-asw} we present results for the $\Delta Y$-dependence of the
$\phi-summed$ cross section, $C_0$, in the three considered kinematic configurations.
Here, the usual onset of BFKL effects comes easily into play. The rise with energy of the purely partonic cross section is quenched by the convolution with PDFs, this leading to decrease the total cross section as $\Delta Y$ of hadronic distributions increases. 
Notably, NLA predictions (red) are almost entirely contained inside LLA uncertainty bands (blue), this could be an evidence of the stabilizing effects of the BFKL series under NLO corrections, due to the large energy scales provided by the emission of a Higgs boson.
For all the considered $R_{nm}$ cases (Figs.~\ref{fig:Rnm_kt-s}), NLA corrections show a milder discrepancy with respect to pure LLA ones. This reflects the fact that in the Higgs-jet hadroproduction process, the renormalization scale needs not to be too large as for other processes where BLM optimization had to be used~\cite{Ducloue:2013bva,Caporale:2014gpa,Caporale:2015uva}.

The predictions for the $p_H$-distribution are reported in Fig.~\ref{fig:pT_kt-w}, in the range 10 GeV $< |\vec p_H| < 2 M_t$, for two fixed values of $\Delta Y =$ 3, 5.
Here, the Born contribution (green), which describes the back-to-back emission of
the Higgs and of the jet with no additional gluon radiation, obtained by switching off the BFKL green function, reads
\begin{equation}\nonumber
\frac{d\sigma^{\rm Born}(|\vec p_H|, \Delta Y, s)}{d|\vec p_H| d\Delta Y} =
\pi \frac{e^{\Delta Y}}{s} M_{H,\perp}
\int_{y^{\rm min}_H}^{y^{\rm max}_H}dy_H
\int_{y^{\rm min}_J}^{y^{\rm max}_J}dy_J
\, \delta \left( y_H - y_J - \Delta Y \right)
%\left(|\vec p_H|, |\vec p_J|, y_H, y_J \right)
\end{equation}
\begin{equation} \label{pT_distribution_Born}
\times \alpha_s^2(\mu_{R_1}) \,
\frac{1}{v^2} \frac{|\mathcal{F}(\vec p_H^{\: 2})|^2}
{128\pi^{3}\sqrt{2(N^2_{c}-1)}} f_{g}(x_H,\mu_{F_1}) 
\end{equation}
\begin{equation}\nonumber
\times  \alpha_s(\mu_{R_2}) \, 2 \sqrt{\frac{C_F}{C_A}} 
\left( \frac{C_A}{C_F} f_{g}(x_J,\mu_{F_2}) + \sum_{a = q,\bar{q}}f_{a}(x_J,\mu_{F_2}) \right)
\, .
\end{equation}
%Our calculation in the Born limit at $\Delta Y = $ 3 (left panel of
%Fig.~\ref{fig:pT_kt-w}) is
%in fair agreement with the corresponding pattern in Ref.~\cite{DelDuca:2003ba}
%(solid line in the left panel of Fig.~2), up to a factor two, due to
%the fact that we restricted $\Delta Y$ to be positive, which means that
%the Higgs particle is always more forward than the jet\footnote{Note that
%	in Ref.~\cite{DelDuca:2003ba} the Higgs
%	mass is a free parameter. We compare our result with the corresponding one
%	at $M_H =$ 120 GeV.}. In our study, this calculation cannot exceed a given
%upper cut-off in the $|\vec p_H|$-range, say around 125 GeV. This is due to our
%choice for the final-state kinematic ranges, where consistency with
%experimental cuts in the rapidities of the detected objects would lead to
%$x_J > 1$ for sufficiently large jet transverse momenta.
In our study, an upper cut-off in the $|\vec p_H|$-range, say around 125 GeV, is introduced to avoid the $x_J > 1$ limits for large jet transverse momenta.
\begin{figure}[hpt]
	\centering
	\includegraphics[scale=0.54,clip]{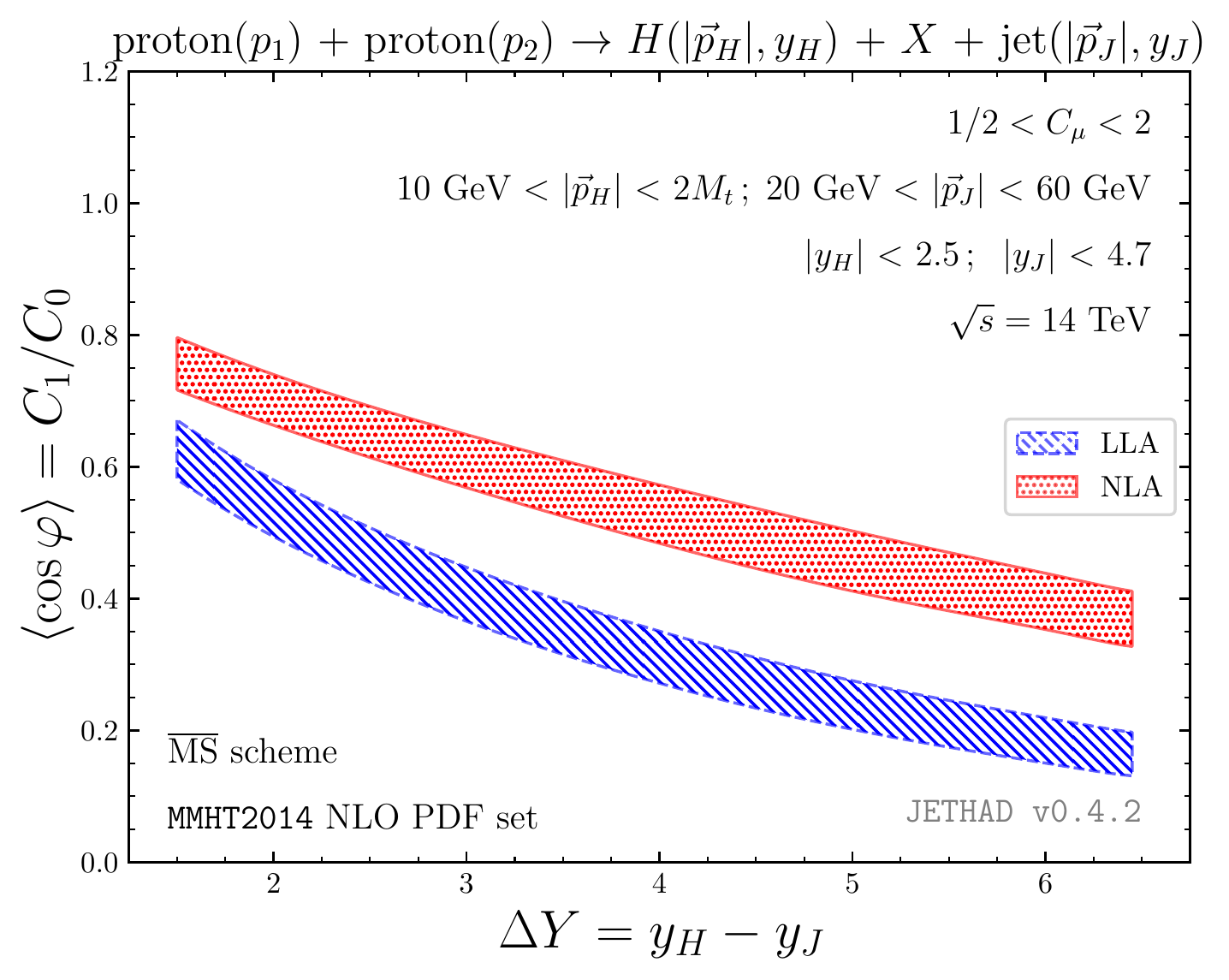}
	\includegraphics[scale=0.54,clip]{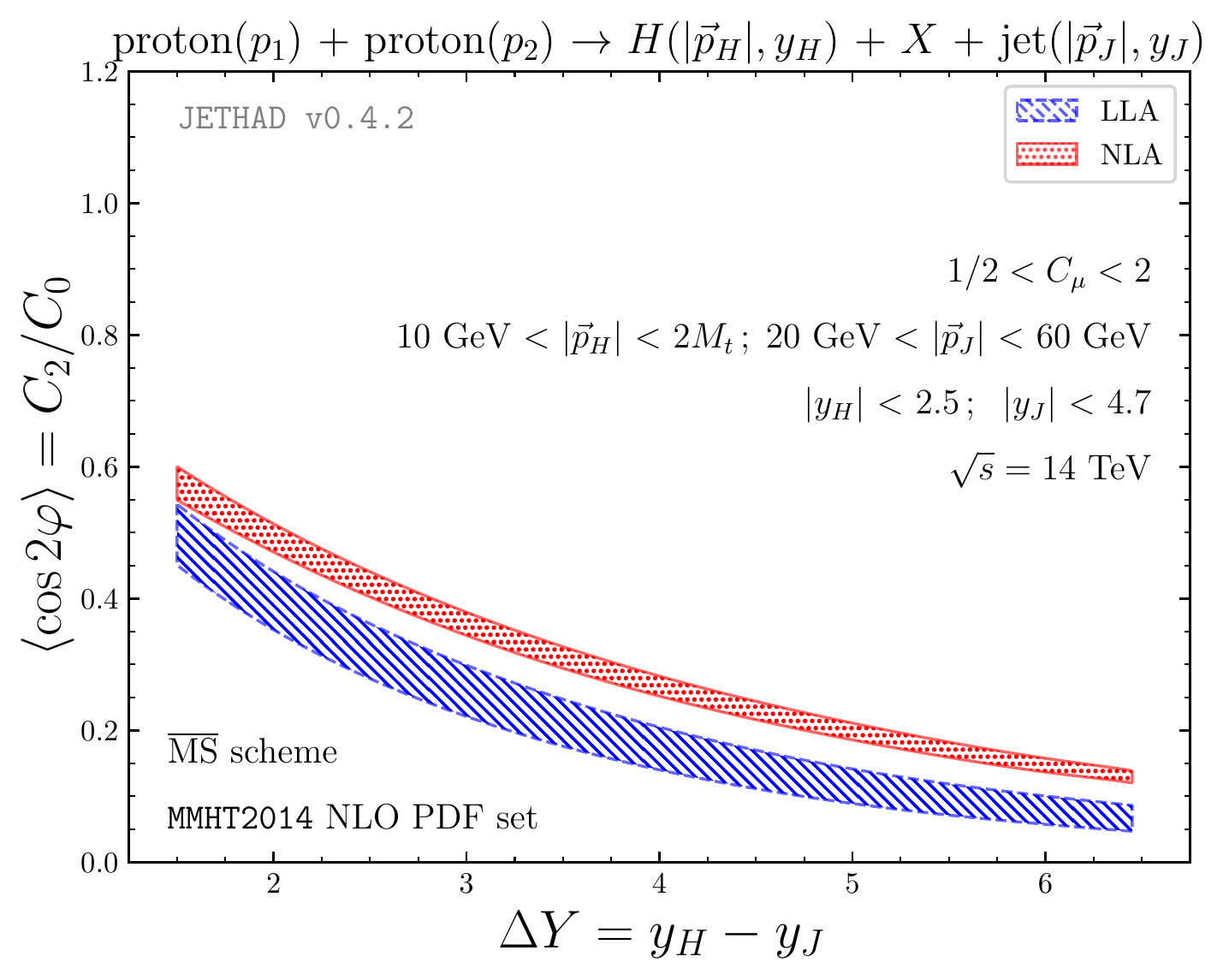}
	
	\includegraphics[scale=0.54,clip]{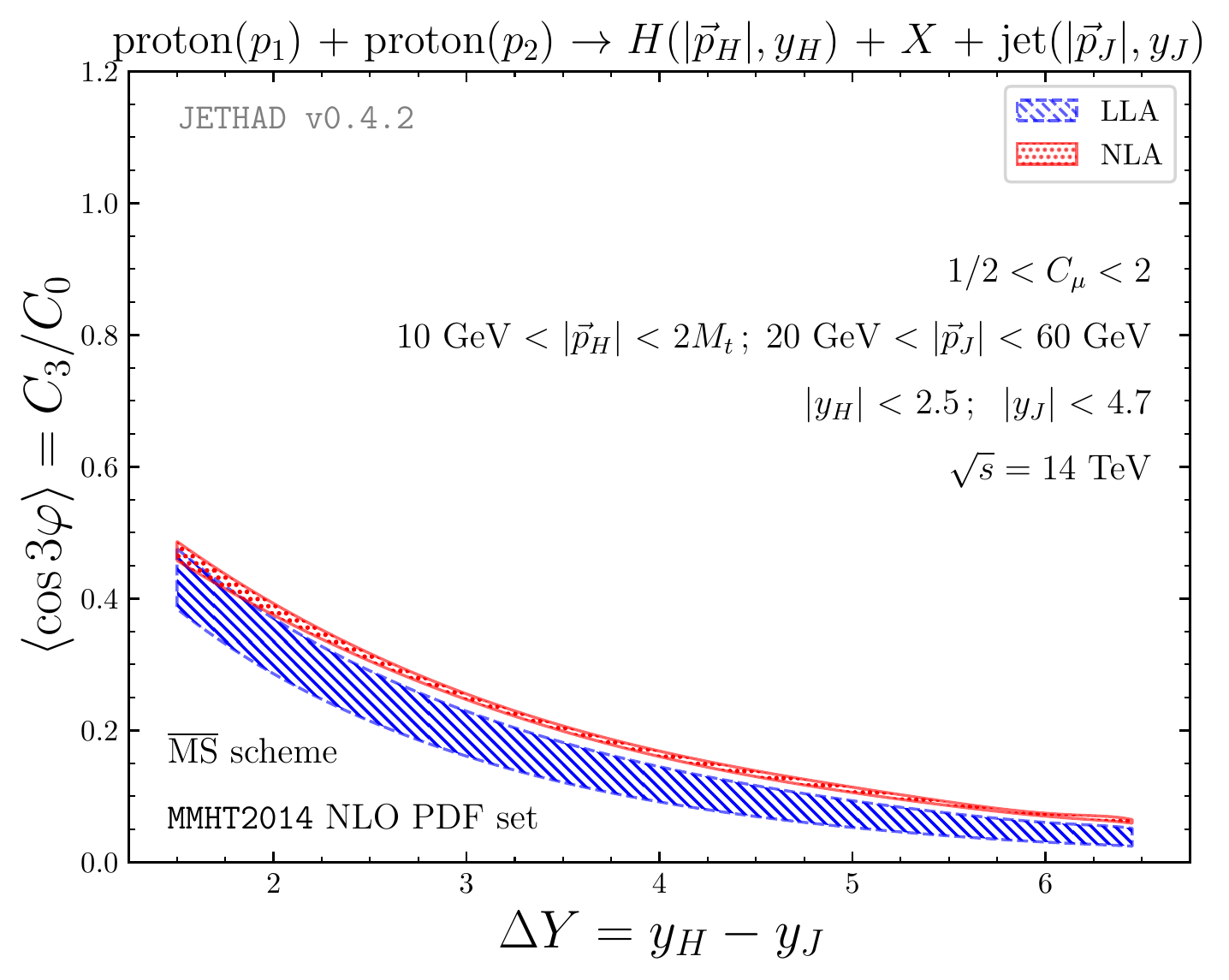}
	\includegraphics[scale=0.54,clip]{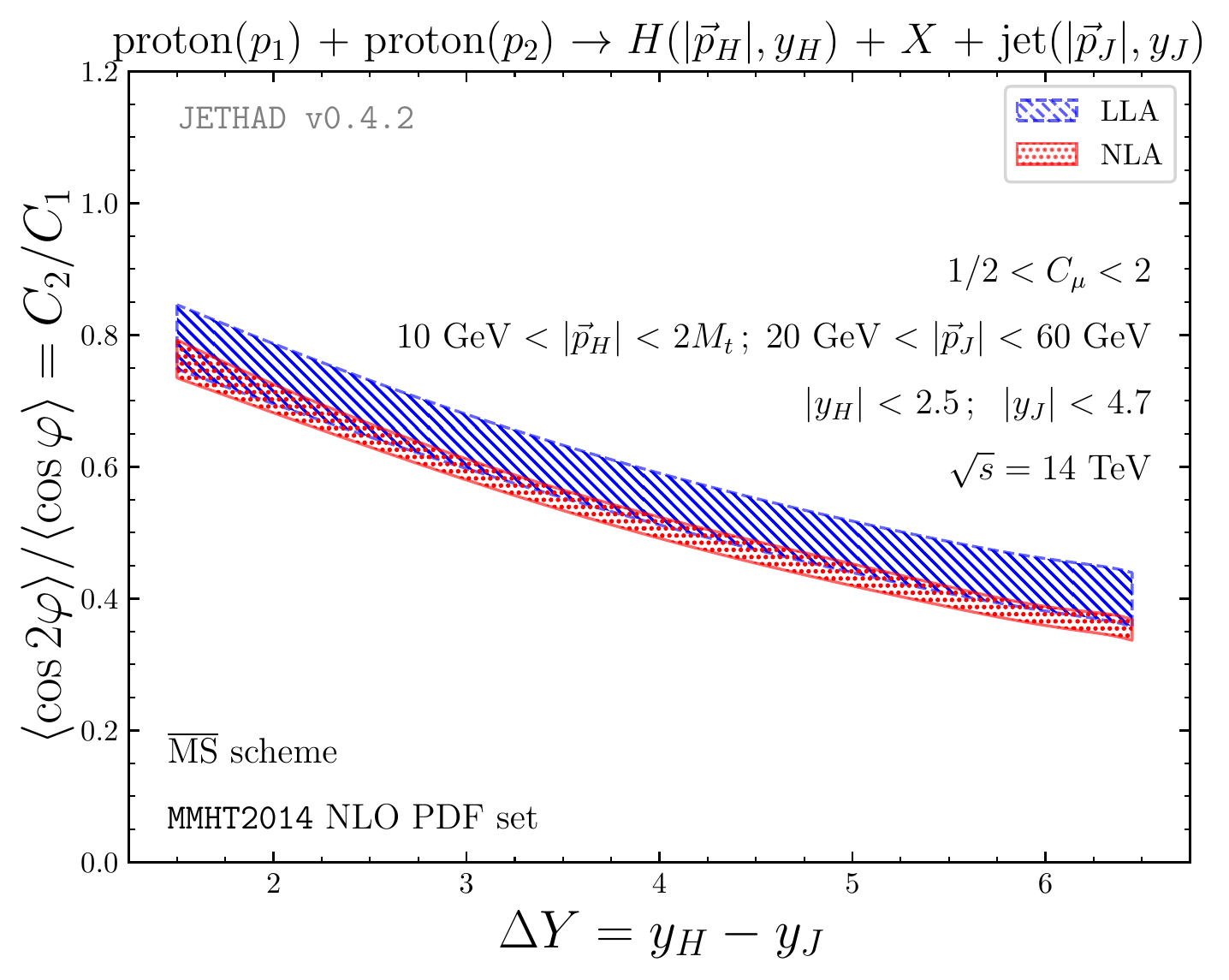}
	
	\includegraphics[scale=0.54,clip]{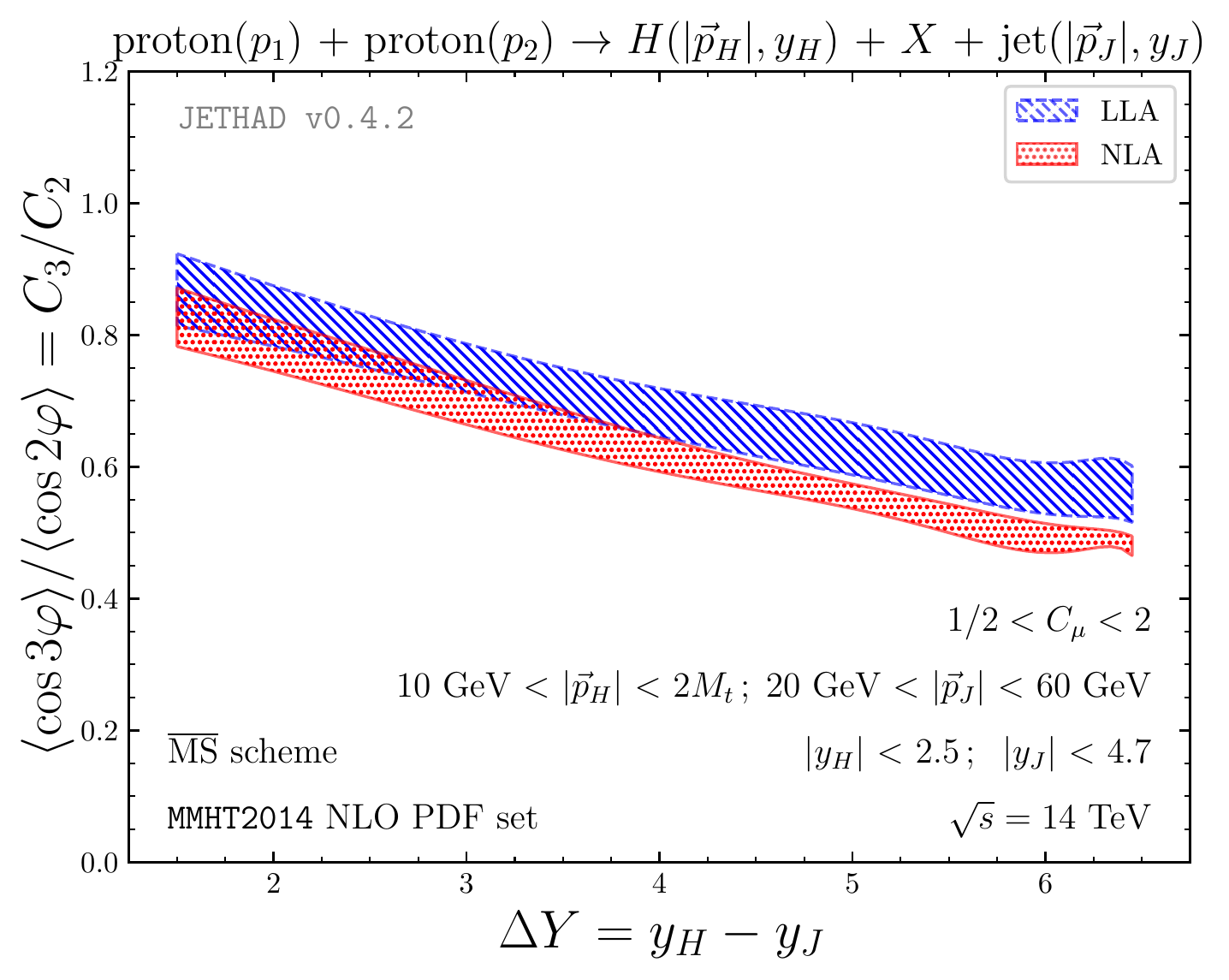}
	\caption{$\Delta Y$-dependence of several ratios $R_{nm} \equiv C_{n}/C_{m}$,
		for the inclusive Higgs-jet hadroproduction in the $p_T$-\emph{asymmetric}
		configuration and for $\sqrt{s} = 14$ TeV.
		%Shaded bands give the combined effect of the scale variation with the uncertainty coming from the phase-space numerical integration
		}
	\label{fig:Rnm_kt-a}
\end{figure}

\begin{figure}[hpt]
	\centering
	\includegraphics[scale=0.54,clip]{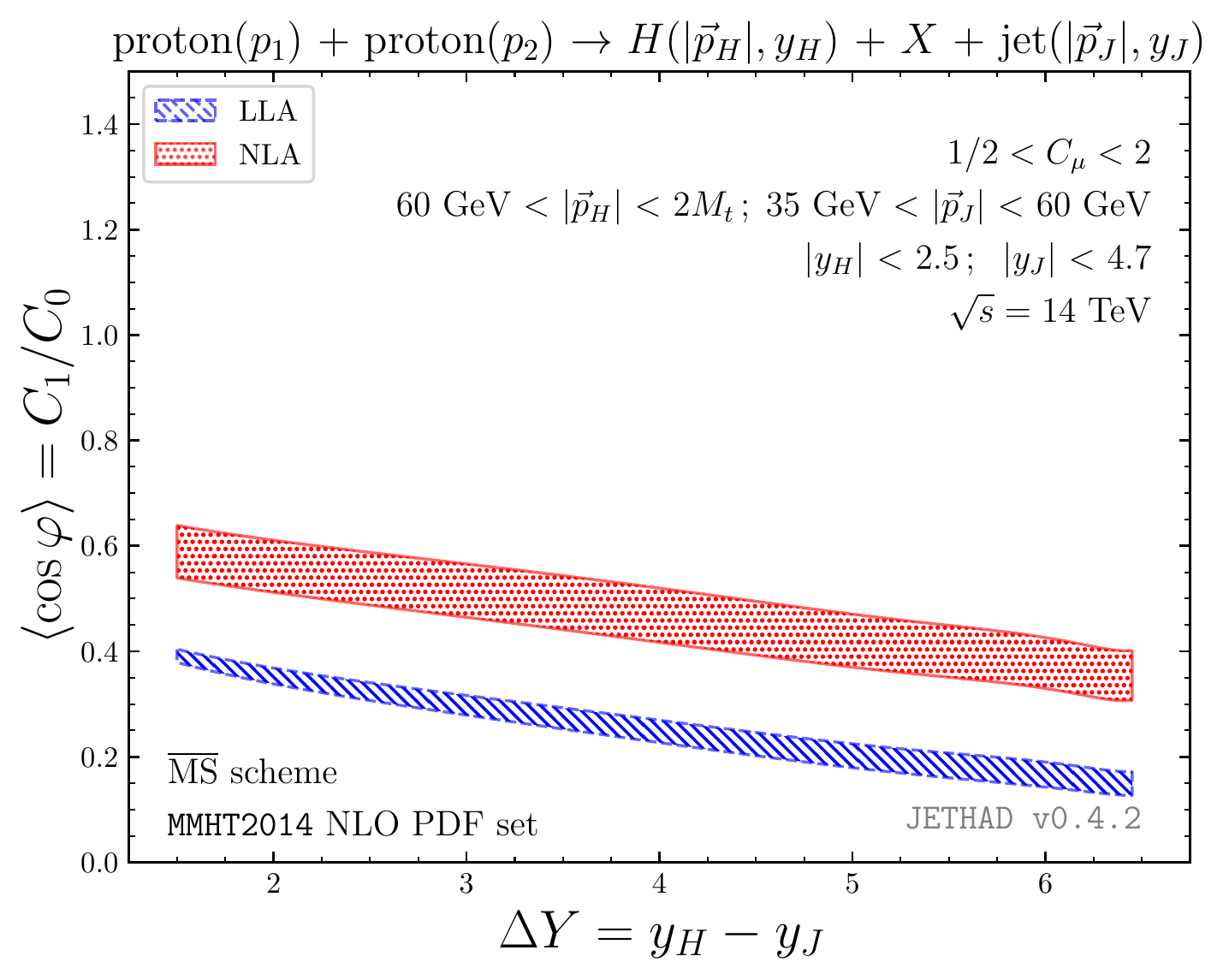}
	\includegraphics[scale=0.54,clip]{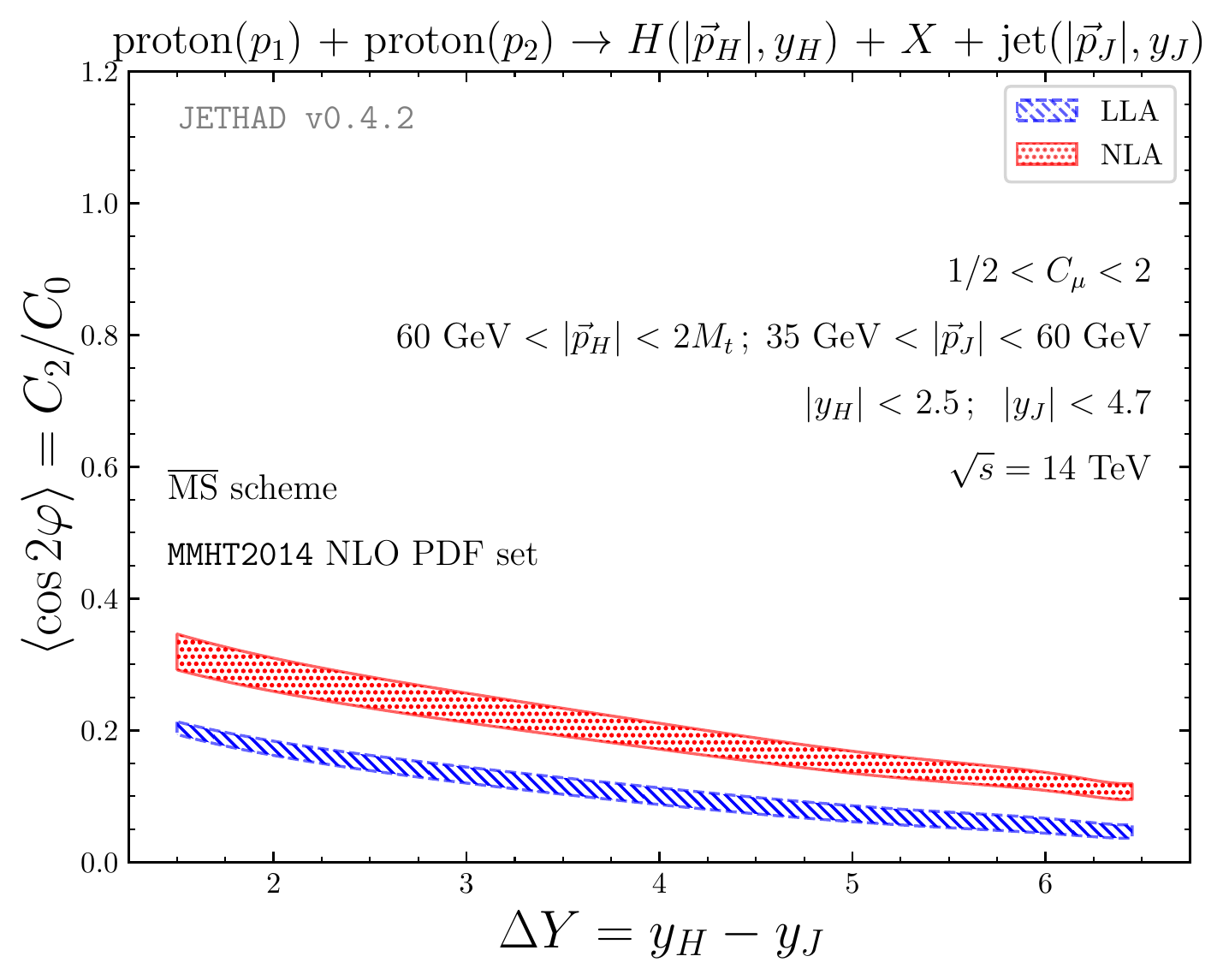}
	
	\includegraphics[scale=0.54,clip]{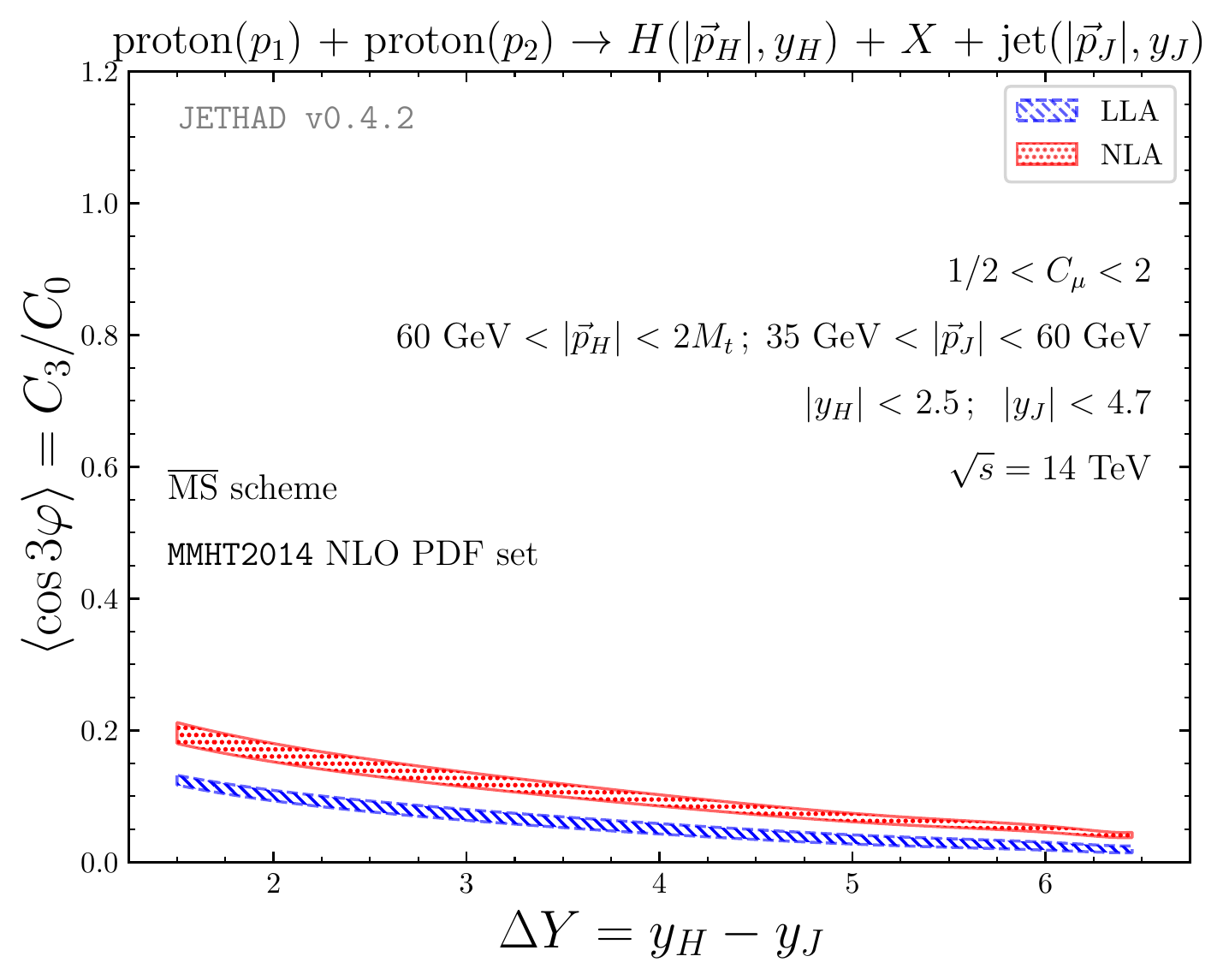}
	\includegraphics[scale=0.54,clip]{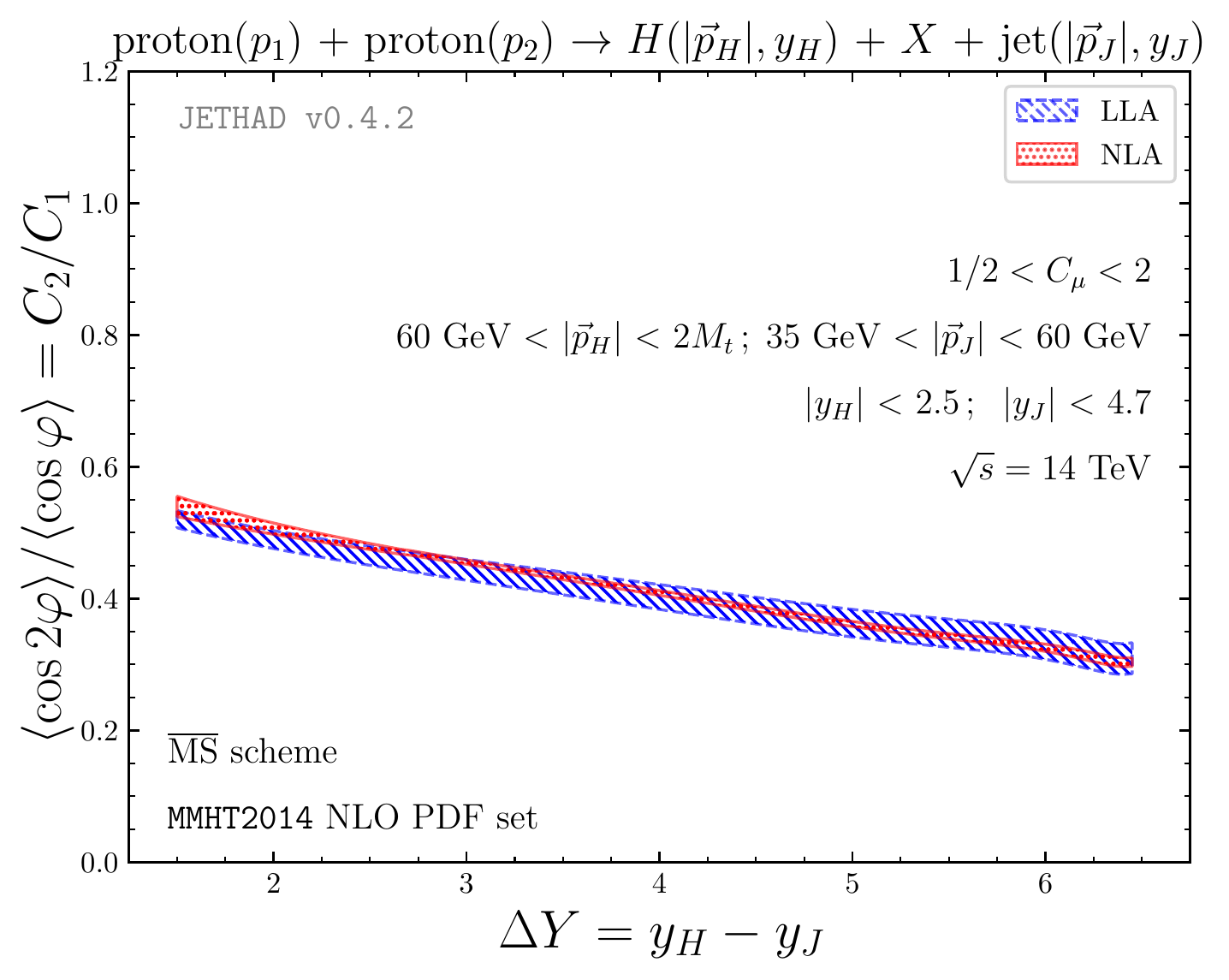}
	
	\includegraphics[scale=0.54,clip]{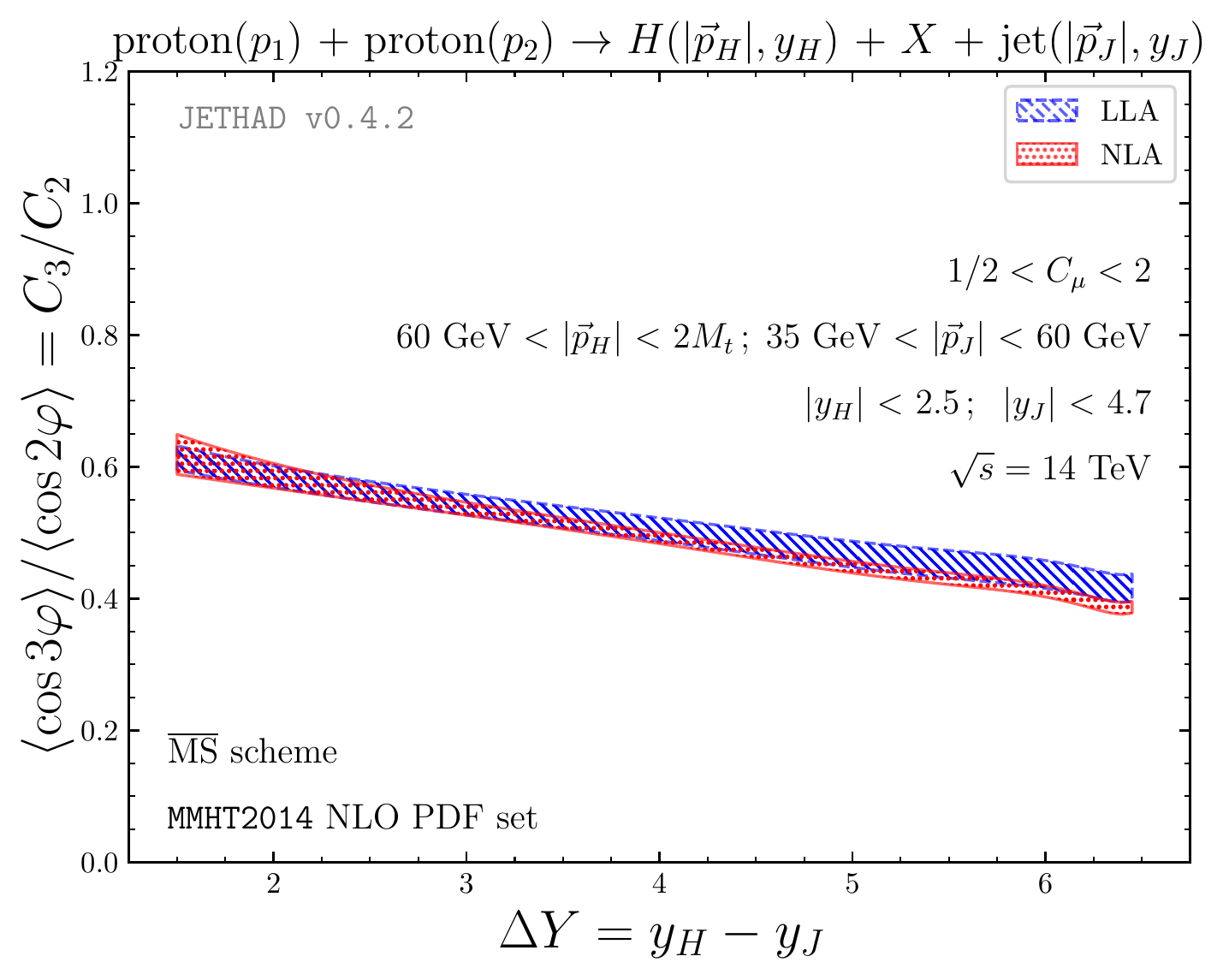}
	\caption{$\Delta Y$-dependence of several ratios $R_{nm} \equiv C_{n}/C_{m}$,
		for the inclusive Higgs-jet hadroproduction in the \emph{disjoint}
		$p_T$-windows configuration and for $\sqrt{s} = 14$ TeV. 
		%Shaded bands give the combined effect of the scale variation with the uncertainty coming from the phase-space numerical integration
		}
	\label{fig:Rnm_kt-w}
\end{figure}

In Fig.~\ref{fig:pT_kt-w}, both NLA and LLA corrections show a peak around $|\vec p_H|=$ 40 GeV for the two fixed values of $\Delta Y$, and a decreasing behavior at large $|\vec p_H|$. Here, we distinguish three kinematic ranges, the first is the low-$|\vec p_H|$ region, {\emph{i.e.}} $|\vec p_H| < 10$ GeV, which is dominated by large transverse-momentum logarithms, need resummation formalism not accounted by our approach. The second subregion is the intermediate-$|\vec p_H|$ region, where $|\vec p_H|$ =$|\vec p_J|$ and ranges from 35 to 60 GeV. It is essentially the peak region plus the first part of the decreasing tail, where NLA uncertainty bands are entirely contained inside the LLA ones.
Here, our description at the hand of the BFKL resummation is validated by the impressive stability behavior of the perturbative series. The third region is the large-$|\vec p_H|$ represented by the long tail, where the NLA distributions start to decouple from LLA ones and exhibit an increasing sensitivity to scale variation. Here, collinear logarithms together with
\emph{threshold} effects~\cite{Bonciani:2003nt} start to become relevant, thus the convergence of the high-energy series is not guaranteed.
With the aim of providing a comparison between fixed-order calculations and BFKL predictions, in Fig.~\ref{fig:pT_kt-w} we present also the $p_H$-distributions at $\Delta Y=3$ and$~5$, as obtained by a fixed-order NLO calculation implemented according to the POWHEG method~\cite{Nason:2004rx}, for this implementation we have adopted the subroutines dedicated to the inclusive Higgs plus jet final state~\cite{Campbell:2012am}. It is interesting to observe that the NLO fixed-order prediction is systematically lower in comparison to the LLA- and NLA-BFKL ones, and this is more visible at the larger $\Delta Y$, where the effect of resummation is expected to be more important.
Finally, we compared the distributions presented above with the corresponding ones obtained in the large top-mass limit, $M_t \to + \infty$. 
We noted that, when this limit is taken, cross sections become at most $5 \div 7 \, \%$ larger,
%(left panels of Fig.~\ref{fig:C0-Rnm_std-vs-ltop}) become at most $5 \div 7 \, \%$ larger,
whereas the effect on azimuthal correlations % ($R_{10}$ in the right panels of Fig.~\ref{fig:C0-Rnm_std-vs-ltop}) 
is very small or negligible\footnote{Here, the bands related to the large top-mass limit are hardly distinguishable from the ones with physical top mass, thus we do not show figures for this comparison.}. The impact on the $p_H$ -distribution reported in Fig.~\ref{fig:pT_kt-w_ltop}, which is also quite small in the $|p_H|\sim
|p_J|$ range, while it become more manifest at larger values of $|p_H|$.
\begin{figure}[hpt]
	\centering
	\includegraphics[scale=0.505,clip]{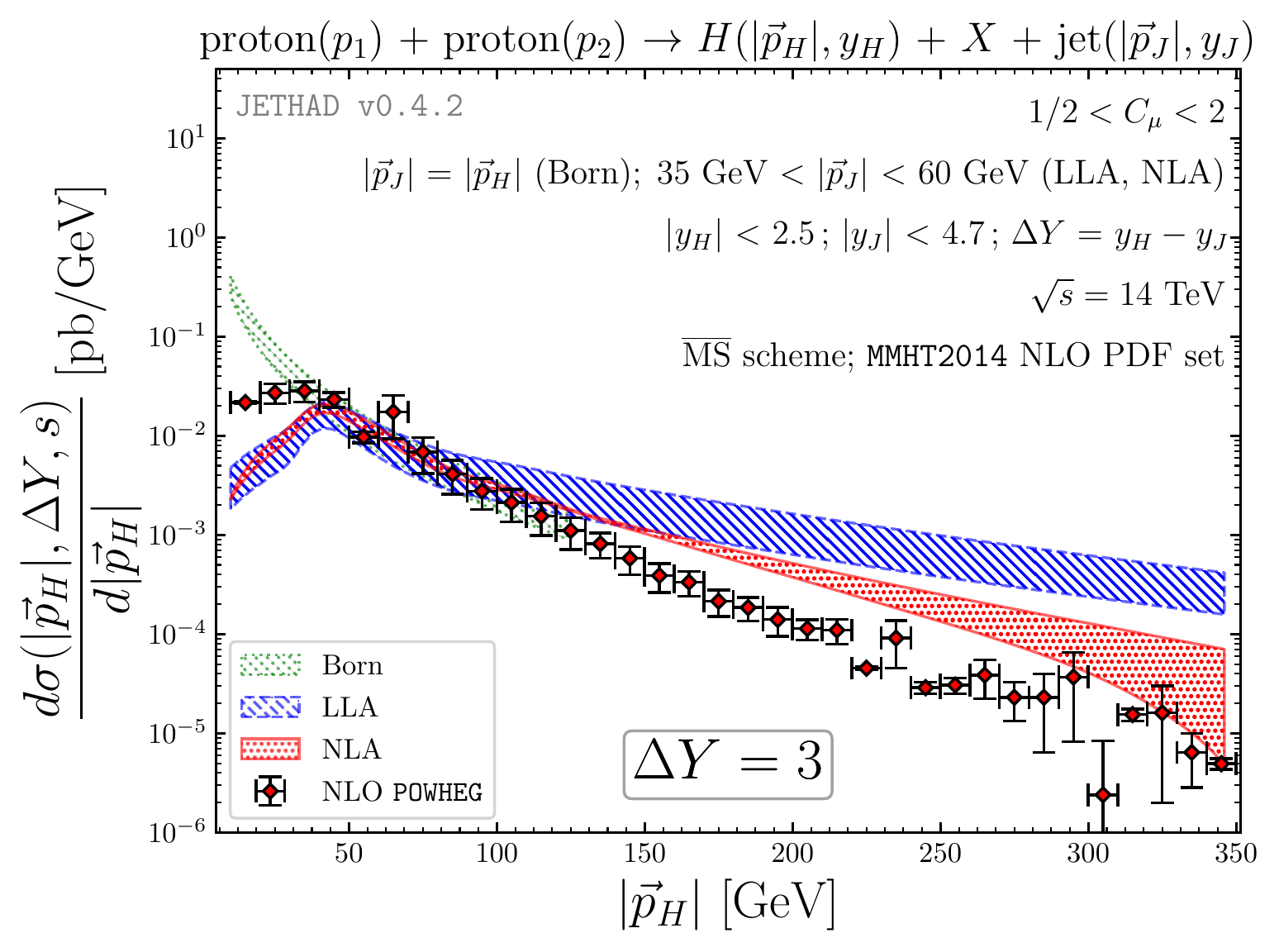}
	\includegraphics[scale=0.505,clip]{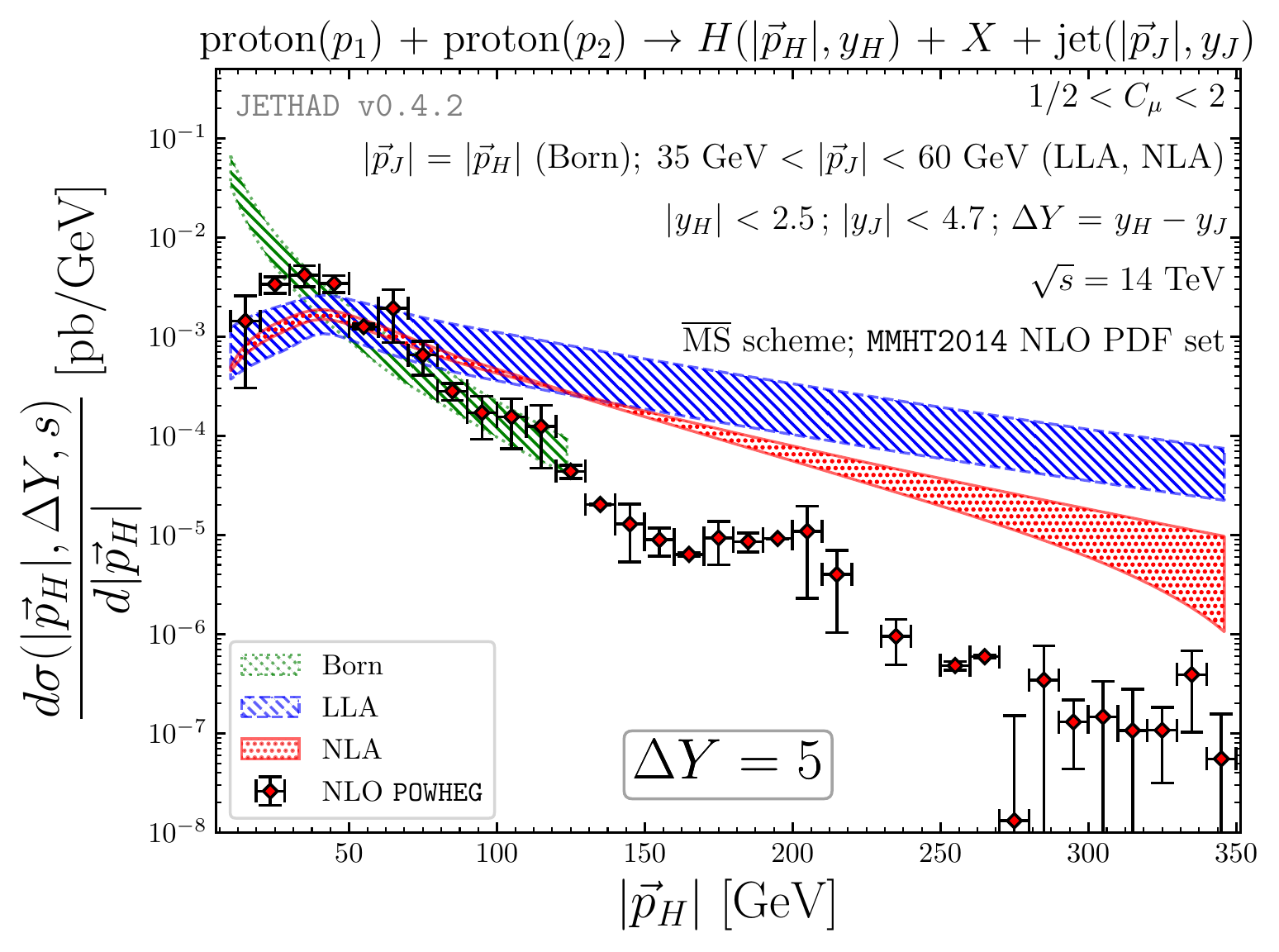}
	\caption{$p_T$-dependence of the cross section for the inclusive Higgs-jet
		hadroproduction for 35 GeV $< p_T <$ 60 GeV, $\sqrt{s} = 14$ TeV and for
		$\Delta Y = 3, 5$. 
		%Shaded bands give the combined effect of the scale variation with the uncertainty coming from the phase-space numerical integration
		}
	\label{fig:pT_kt-w}
\end{figure}

All these considerations brace the message that an a comprehensive investigation of the
$|\vec p_H|$-distribution  would depend on a brought together formalism where distinct
resummations are simultaneously exemplified. Specifically, the effect of the BFKL 
resummation could rely on the delicate interplay among the Higgs transverse
mass, the Higgs transverse momentum and the jet transverse momentum entering,
in logarithmic form, the expressions of partial NLO corrections to impact
factors (see Eqs.~(\ref{cH1}) and~(\ref{cJ1})). Future investigations including full
higher-order corrections will permit us to additional check the strength of our 
estimations.
%

%\begin{figure}[p]
%\centering
%\includegraphics[scale=0.53,clip]{C0_0J_sc_2019_ltop_kt-s_Tnvv_MSB_CMS14.pdf}
%\includegraphics[scale=0.53,clip]{R10_0J_sc_2019_ltop_kt-s_Tnvv_MSB_CMS14.pdf}
%
%\includegraphics[scale=0.53,clip]{C0_0J_sc_2019_ltop_kt-a_Tnvv_MSB_CMS14.pdf}
%\includegraphics[scale=0.53,clip]{R10_0J_sc_2019_ltop_kt-a_Tnvv_MSB_CMS14.pdf}
%
%\includegraphics[scale=0.53,clip]{C0_0J_sc_2019_ltop_kt-w_Tnvv_MSB_CMS14.pdf}
%\includegraphics[scale=0.53,clip]{R10_0J_sc_2019_ltop_kt-w_Tnvv_MSB_CMS14.pdf}
%\caption{$\Delta Y$-dependence of the $\varphi$-averaged cross section, $C_0$ (left), 
%  and of the $R_{10} \equiv C_{1}/C_{0}$ azimuthal correlation (right),
%  for the inclusive Higgs-jet hadroproduction in the large top-mass limit, 
%  for the three considered $p_T$-ranges and for $\sqrt{s} = 14$ TeV.
%  Shaded bands give the combined effect of the scale variation 
%  with the uncertainty coming from the phase-space numerical integration.}
%\label{fig:C0-Rnm_ltop}
%\end{figure}

\begin{figure}[hpt]
\centering
\includegraphics[scale=0.505,clip]{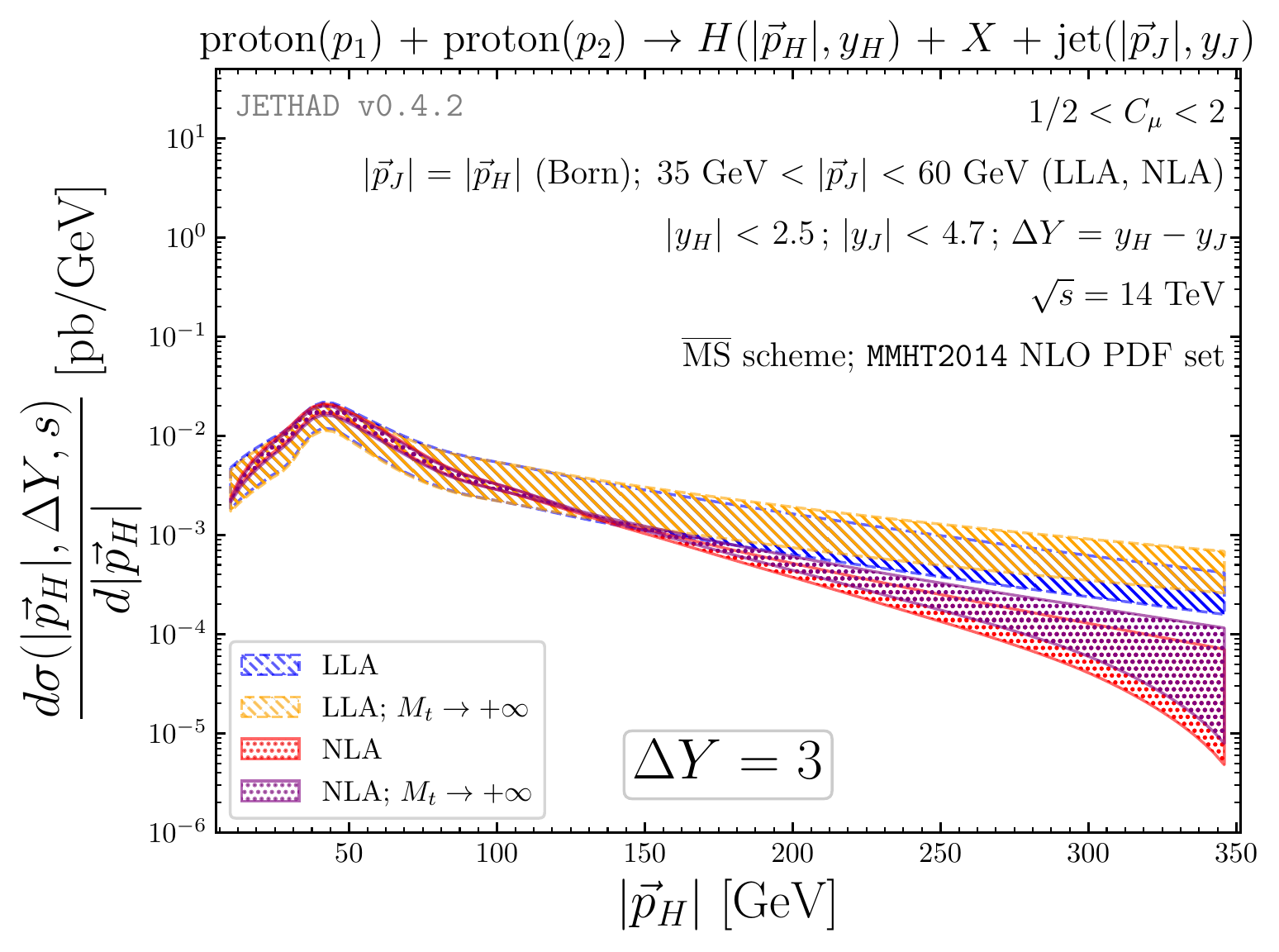}
\includegraphics[scale=0.505,clip]{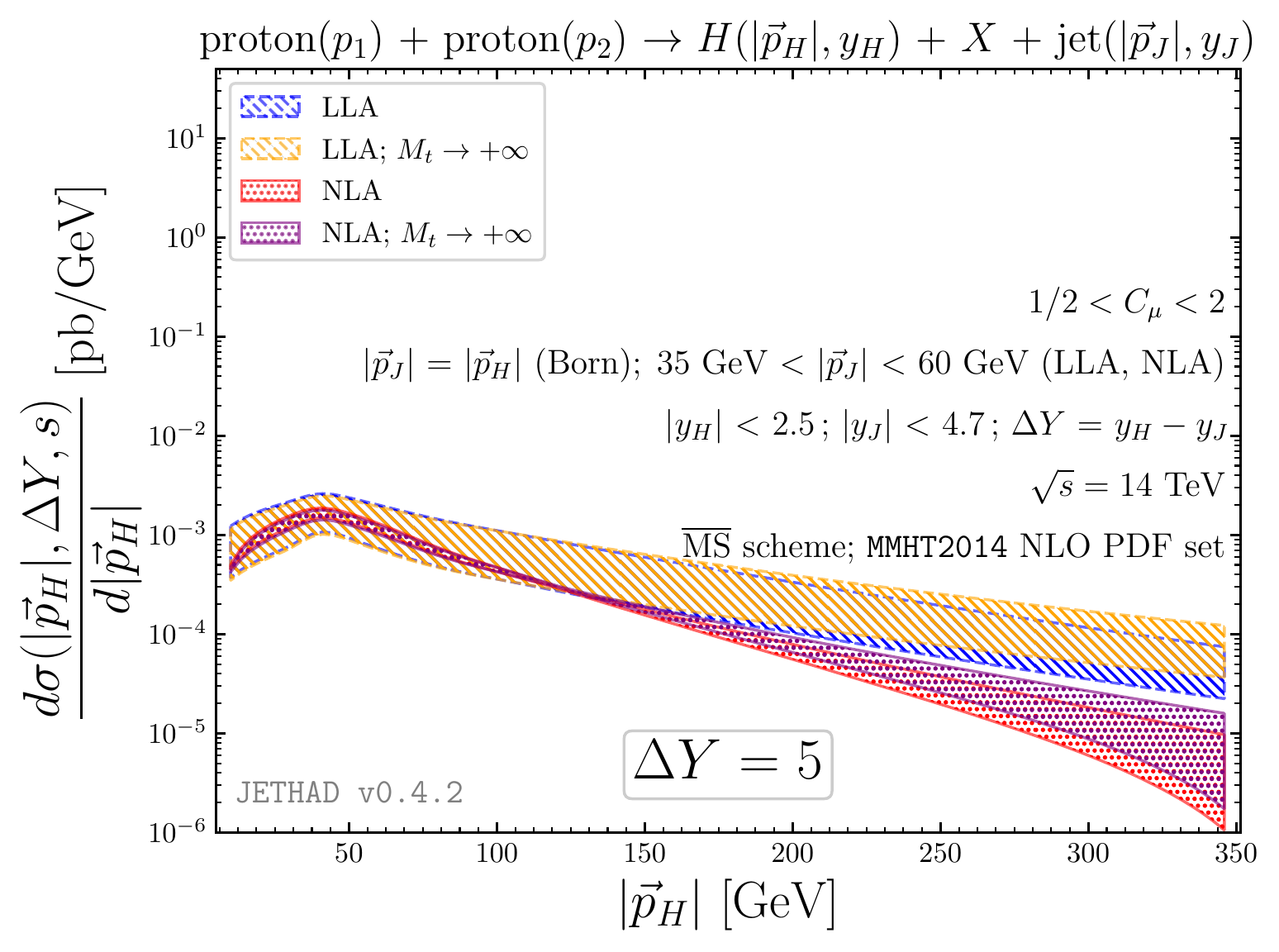}
\caption{$p_T$-dependence of the cross section for the inclusive Higgs-jet
  hadroproduction in the large top-mass limit, 
  for 35 GeV $< p_T <$ 60 GeV, $\sqrt{s} = 14$ TeV and for
  $\Delta Y = 3, 5$.
  %Shaded bands give the combined effect of the scale variation with the uncertainty coming from the phase-space numerical integration
  }
\label{fig:pT_kt-w_ltop}
\end{figure}

%===============================
\subsection{Numerical specifics and uncertainty estimation}
%===============================
\label{numerical_strategy}
All the numerical studies were completed making use the previous mentioned tool {\tt JETHAD}~\cite{Celiberto:2020wpk}. An auxiliary, independent
\textsc{Mathematica} interface allowed us to test the numerical reliability
of our results. Quark and gluon PDFs were calculated through the \textsc{MMHT2014} NLO PDF
set~\cite{Harland-Lang:2014zoa} as provided by the LHAPDFv6.2.1
interpolator~\cite{Buckley:2014ana}, whereas we selected a two-loop
running coupling setup with $\alpha_s\left(M_Z\right)=0.11707$ and with
dynamic-flavor threshold.

All the relevant sources of numerical uncertainty, coming from the
multidimensional integration over the final-state phase space and from the one-dimensional integral over the longitudinal momentum fraction $\zeta$ in the NLO impact factor
corrections (Eqs.~(\ref{cH1}) and~(\ref{cJ1})), were directly estimated by the \textsc{Jethad} integration tools.

Furthermore, we gauged the effect of concurrently varying the renormalization
scales ($\mu_{R_{1,2,c}}$) and the factorization ones ($\mu_{F_{1,2}}$) of them
around their \emph{natural} values in the range 1/2 to two. 
%
%The parameter
%$C_{\mu}$ entering the inset of panels in
%Figs.~\ref{fig:C0_kt-asw}, \ref{fig:Rnm_kt-s}, \ref{fig:Rnm_kt-a},
%\ref{fig:Rnm_kt-w} and \ref{fig:pT_kt-w} gives the ratio
%\begin{equation}
%\label{Cmu}
%C_\mu = 
%\frac{\mu_{{R,F}_1}}{M_{H,\perp}} = 
%\frac{\mu_{{R,F}_2}}{|\vec p_J|} = 
%\frac{\mu_{R_c}}{\sqrt{M_{H,\perp} |\vec p_J|}} \, .
%\end{equation}
%===============================
\subsection{Summary}
\label{conclusions}
%===============================
We have proposed the inclusive hadroproduction of a Higgs boson and of a jet
featuring high transverse momenta and separated by a large rapidity distance as another diffractive semi-hard channel to test the BFKL resummation. At variance with previously discussed analyses, the Higgs + jet production channel exhibit quite a fair stability under higher-order corrections, so that the renormalization scale needs not to be too large as for other processes where BLM optimization had to be used.
The obtained results for the distributions differential in the Higgs
transverse momentum, provided an evidence that a high-energy treatment is valid
and can be afforded in the region where Higgs $p_T$ and the jet one are of the
same order. Beyond that, the description for Higgs momentum distribution
should rely on many-sided resummation formalism unifying different approaches.
A possible extension of this work consists in the full NLA BFKL analysis,
including the NLO jet impact factor, with a realistic implementation of the
jet selection function, and the NLO forward-Higgs impact factor\footnote{See Appendix~\ref{A.3}.}.
%===============================
\section{Bottom-flavored inclusive emissions}
\label{sec:bflavour}
\begin{figure*}[hpt]
	\begin{center}
	\includegraphics[width=0.4\textwidth]{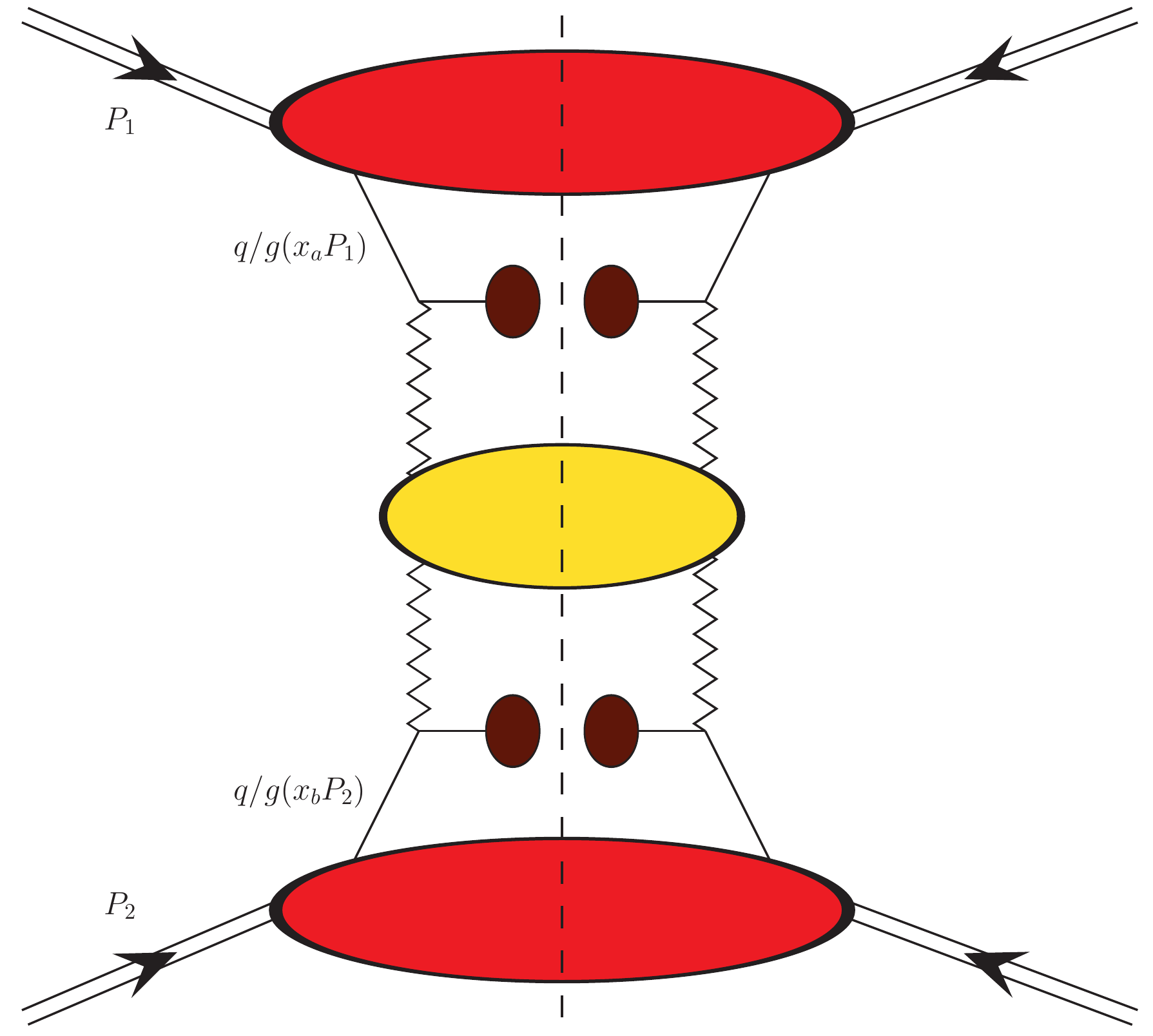}
	\hspace{0.25cm}
	\includegraphics[width=0.4\textwidth]{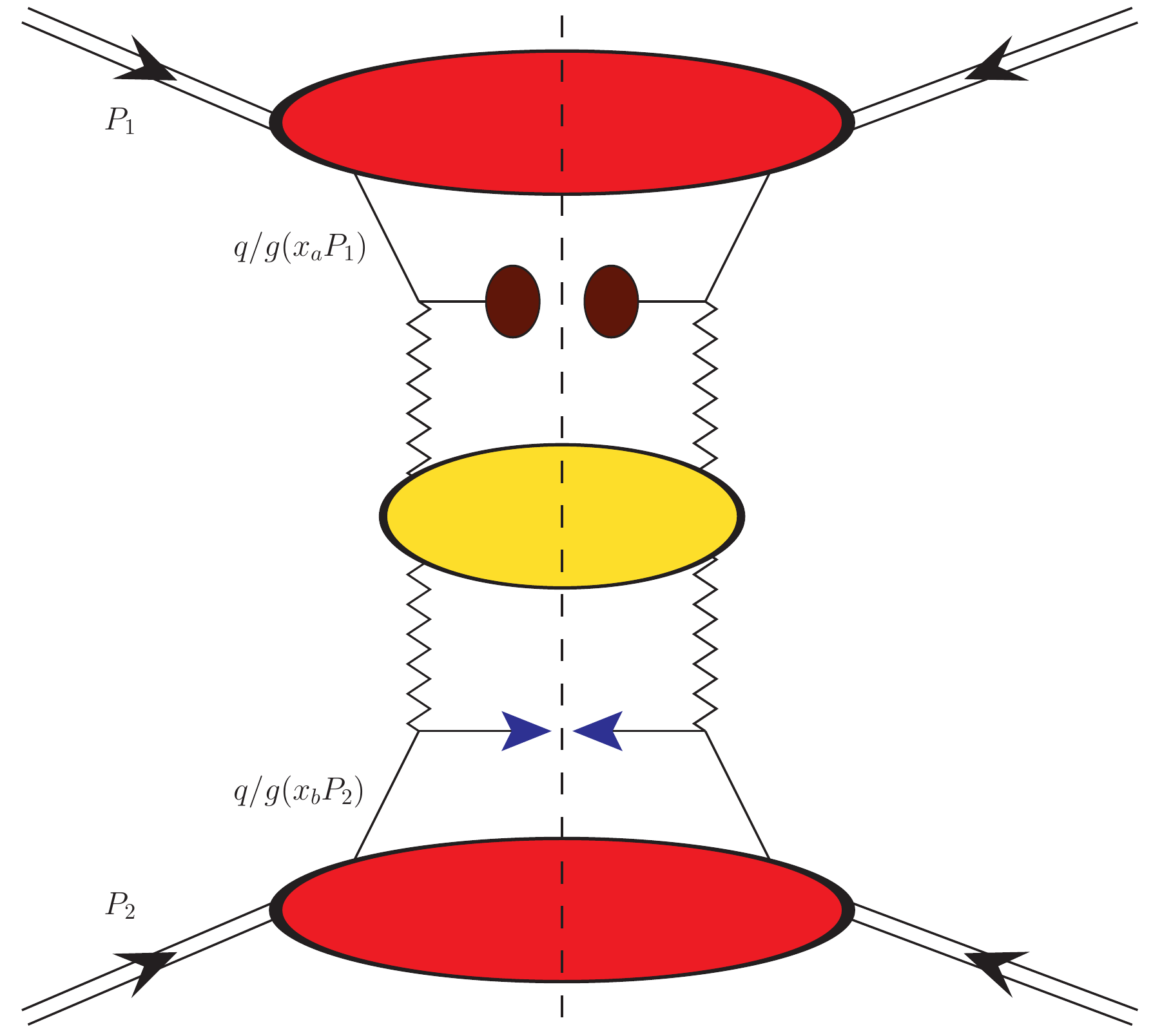}
	\\ \vspace{0.25cm}
	\hspace{-0.50cm}
	a) Double $H_b$ \hspace{5.50cm}
	b) $H_b$ $+$ jet
	\caption{Schematic representation of the two inclusive processes under investigation.}
	\label{fig:process}
	\end{center}
\end{figure*}

%===============================
In this section, we present the inclusive semi-hard production, in proton-proton collisions, of two bottomflavored hadrons, as well as of a single bottom-flavored hadron accompanied by a light jet (depicted in Fig.~\ref{fig:process}), as novel probes for investigating the stabilization effects of the high-energy resummation under NLO corrections.
%==============================================
\subsection{Cross section in the NLA BFKL}
%===============================================
The final state configurations of the inclusive processes under consideration is
schematically represented in ~Fig.~\ref{fig:process}, where a \textit{b}-hadron $(p_1,y_1)$ is emitted along with another \textit{b}-hadron or a jet $(p_2,y_2)$, featuring a large rapidity separation, together with an undetected system of hadrons.
For the sake of definiteness, we will consider the case where the 
rapidity for first detected final state object $y_1$ is larger than the second one $y_2$, so that $\Delta Y\equiv y_1-y_2$ is always positive, which requires, for most considered
values of $\Delta Y$, that the first object is forward and the second is backward.

The final-state objects are also required to possess large transverse momenta,
$\vec p_1^2\sim \vec p_2^2 \gg \Lambda_{QCD}^2$.
The colliding protons' momenta $P_1$ and $P_2$ are taken as Sudakov basis vectors satisfying
$P^2_1= P^2_2=0$ and $2 (P_1\cdot P_2) = s$,  so that the momenta of detected
objects can be decomposed as
\begin{equation}\label{sudakov}
p_{1,2}= x_{1,2} P_{1,2}+ \frac{\vec p_{1,2}^2}{x_{1,2} s}P_{2,1}+p_{1,2\perp} \ , \quad
p_{1,2\perp}^2=-\vec p_{1,2}^2
\end{equation}

The parton momentum fractions $x_{1,2}$ in
the large rapidity limit are connected to the respective rapidities through the relation
$y_{1,2}=\pm\frac{1}{2}\ln\frac{x_{1,2}^2 s}
{\vec p_{1,2}^2}$, 
so that $dy_{1,2}=\pm\frac{dx_{1,2}}{x_{1,2}}$,
and $\Delta Y=y_1-y_2=\ln\frac{x_1x_2 s}{|\vec p_1||\vec p_2|}$, here the
spatial part of the four-vector $p_{1\parallel}$ being taken positive.

In QCD collinear factorization, the processes can be viewed as
started by two protons each emitting one parton, according to its parton distribution function
(PDF) and ended with the detected $b$-hadron in the final state according to that its fragmentation functions (FFs). Within the collinear factorization LO cross section is given as a convolution of the (PDFs) and (FFs) with the subsequent partonic hard scattering as follows
\begin{equation}
\begin{split}
\frac{d\sigma^{pp\rightarrow H_b H_b}}{dx_Hdx_Jd^2k_Hd^2k_J}
=\sum_{r,s=q,{\bar q},g}\int_0^1 dx_1 \int_0^1 dx_2\ f_r\left(x_1,\mu_F\right)
\ f_s\left(x_2,\mu_F\right)\\
\times\int_{x_1}^1\frac{d\beta_1}{\beta_1}\int_{x_2}^1\frac{d\beta_2}{\beta_2}D^{H_b}_{r}\left(\frac{x_1}{\beta_1}\right)D^{H_b}_{s}\left(\frac{x_2}{\beta_2}\right)
\frac{d{\hat\sigma}_{r,s}\left(\hat s,\mu_F\right)}
{dx_1dx_2d\beta_1d\beta_2d^2\vec p_1d^2\vec p_2}\;,
\end{split}
\end{equation}
for the $H_{b}$ channel, and similarly 
\begin{equation}
\begin{split}
\frac{d\sigma^{pp\rightarrow H_b jet}}{dx_Hdx_Jd^2k_Hd^2k_J}
=\sum_{r,s=q,{\bar q},g}\int_0^1 dx_1 \int_0^1 dx_2\ f_r\left(x_1,\mu_F\right)
\ f_s\left(x_2,\mu_F\right)\\
\times\int_{x_1}^1\frac{d\beta_1}{\beta_1}D^{H_b}_{r}\left(\frac{x_1}{\beta_1}\right) 
\frac{d{\hat\sigma}_{r,s}\left(\hat s,\mu_F\right)}
{dx_1dx_2d\beta_1d\beta_2d^2\vec p_1d^2\vec p_2}\;.
\end{split}
\end{equation}
in the $H_{b}$ plus jet channel.

Here the $r, s$ indices specify the parton types 
(quarks $q = u, d, s, c, b$;
antiquarks $\bar q = \bar u, \bar d, \bar s, \bar c, \bar b$; 
or gluon $g$), $f_{r,s}\left(x, \mu_F \right)$ and $D^{H_b}_{r,s}\left(x/\beta, \mu_F \right)$ denote the initial proton PDFs and the final detected baryon FFs, respectively; 
$x_{1,2}$ are the longitudinal fractions of the partons involved in the hard
subprocess, while $\mu_F$ is the factorization scale;
$d\hat\sigma_{r,s}\left(\hat s \right)$ is
the partonic cross section and $\hat s \equiv x_1x_2s$ is the squared
center-of-mass energy of the parton-parton collision subprocess.

In the BFKL approach the cross section of the hard subprocesses can be presented as the Fourier sum of the azimuthal coefficients ${\cal C}_n$, 
having so
\begin{equation}
\frac{d\sigma}
{dy_1dy_2\, d|\vec p_1| \, d|\vec p_2|d\phi_1 d\phi_2}
=\frac{1}{(2\pi)^2}\left[{\cal C}_0+\sum_{n=1}^\infty  2\cos (n\phi )\,
{\cal C}_n\right]\, ,
\end{equation}
where $\phi=\phi_1-\phi_2-\pi$, with $\phi_{1,2}$ the baryon/jet 
azimuthal angles, while $y_{1,2}$ and $\vec p_{1,2}$ are their
rapidities and transverse momenta, respectively. 
The $\phi$-averaged cross section ${\cal C}_0$ 
and the other coefficients ${\cal C}_{n\neq 0}$ are given
by
\begin{equation}\nonumber
{\cal C}_n \equiv \int_0^{2\pi}d\phi_1\int_0^{2\pi}d\phi_2\,
\cos[n(\phi_1-\phi_2-\pi)] \,
\frac{d\sigma}{dy_1dy_2\, d|\vec p_1| \, d|\vec p_2|d\phi_1 d\phi_2}\;
\end{equation}
\begin{equation}\nonumber
= \frac{e^{\Delta Y}}{s}
\int_{-\infty}^{+\infty} d\nu \, \left(\frac{x_1 x_2 s}{s_0}
\right)^{\bar \alpha_s(\mu_R)\left\{\chi(n,\nu)+\bar\alpha_s(\mu_R)
	\left[\bar\chi(n,\nu)+\frac{\beta_0}{8 N_c}\chi(n,\nu)\left[-\chi(n,\nu)
	+\frac{10}{3}+2\ln\left(\frac{\mu_R^2}{\sqrt{\vec p_1^2\vec p_2^2}}\right)\right]\right]\right\}}
\end{equation}
\begin{equation}\nonumber
\times \alpha_s^2(\mu_R) c_1(n,\nu,|\vec p_1|, x_1)
[c_2(n,\nu,|\vec p_2|,x_2)]^*\,
\end{equation}
\begin{equation}\label{Cm}%\nonumber
\times \left\{1
+\alpha_s(\mu_R)\left[\frac{c_1^{(1)}(n,\nu,|\vec p_1|,
	x_1)}{c_1(n,\nu,|\vec p_1|, x_1)}
+\left[\frac{c_2^{(1)}(n,\nu,|\vec p_2|, x_2)}{c_2(n,\nu,|\vec p_2|,
	x_2)}\right]^*\right]\right.
\end{equation}
\begin{equation}\nonumber
\left. + \bar\alpha_s^2(\mu_R) \ln\left(\frac{x_1 x_2 s}{s_0}\right)
\frac{\beta_0}{4 N_c}\chi(n,\nu)f(\nu)\right\}\;.
\end{equation}
\\
\subsection{Phenomenological analysis}
\label{heavy_phenomenology}
%===============================
We give results in the full NLA of cross sections, azimuthal correlations and double differential distributions in the transverse momenta of final-state particles, the latter is proposed as a common basis to investigate the interplay of distinct resummations mechanisms. In our analysis, the light-flavored jet is always tagged in its typical CMS ranges~\cite{Khachatryan:2016udy}, \emph{i.e.} $|y_J | < 4.7$ and 35 GeV $< p_J <$ 60 GeV
For our hadron, we admit a tagging of $b$-hadrons on a slightly wider range, $|y_H|<$  2.4 and 20 GeV $< p_H <$ 60 GeV. For the non-perturbative ingredients in our formalism, we employed the MMHT14 PDF set, and for the fragmentation of our hadrons we use the KKSS07 NLO FFs. 
\subsubsection{Observables:}
\begin{itemize}
	\item \textit{$\Delta Y$-distribution}:
	The first observable we take into our account is the $\phi$-averaged contribution to the cross section, also known as $\DY$-distribution or simply $C_0$:
	\begin{equation}
	\label{DY_distribution}
	C_0 =
	\int_{y_1^{\rm min}}^{y_1^{\rm max}} d y_1
	\int_{y_2^{\rm min}}^{y_2^{\rm max}} d y_2
	\int_{p_1^{\rm min}}^{p_1^{\rm max}} d |\vec p_1|
	\int_{p_2^{\rm min}}^{p_2^{\rm max}} d |\vec p_2|
	\, \,
	\delta (\DY - (y_1 - y_2))
	\, \,
	{\cal C}_0
	%\Bigm \lvert_{y_2 = y_1 - Y}
	\, .
	\end{equation}
	
	\item \textit{Azimuthal correlations}:
	Analogously to $C_0$, we define the phase-space integrated higher azimuthal coefficients, $C_{n \neq 0}$, to build their ratios
	\begin{equation}
	\label{Rnm}
	R_{nm} \equiv \frac{C_n}{C_m} =
	\frac{
		\int_{y_1^{\rm min}}^{y_1^{\rm max}} d y_1
		\int_{y_2^{\rm min}}^{y_2^{\rm max}} d y_2
		\int_{p_1^{\rm min}}^{p_1^{\rm max}} d |\vec p_1|
		\int_{p_2^{\rm min}}^{p_2^{\rm max}} d |\vec p_2|
		\, \,
		\delta (\DY - (y_1 - y_2))
		\, \,
		{\cal C}_n
	}
	{
		\int_{y_1^{\rm min}}^{y_1^{\rm max}} d y_1
		\int_{y_2^{\rm min}}^{y_2^{\rm max}} d y_2
		\int_{p_1^{\rm min}}^{p_1^{\rm max}} d |\vec p_1|
		\int_{p_2^{\rm min}}^{p_2^{\rm max}} d |\vec p_2|
		\, \,
		\delta (\Delta Y - (y_1 - y_2))
		\, \,
		{\cal C}_m
	}
	\, .
	\end{equation}
	The $R_{n0}$ ratios determine the values of the mean cosines $\langle \cos n \varphi \rangle$, while the ones without zero indices represent ratios of correlations. We study the behavior of the $R_{nm}$ moments as functions of $\DY$ and in the kinematic ranges defined above.
	
	\item \textit{Double differential $p_T$-distribution}:
	The transverse-momentum double differential cross section given as
	\begin{equation}
	\label{bhd_pT_distribution}
	\frac{d \sigma(|\vec p_{1,2}|, \Delta Y, s)}{d |\vec p_1| d |\vec p_2| d \Delta Y} =
	\int_{y^{\rm min}_1}^{y^{\rm max}_1} \drv y_1
	\int_{y^{\rm min}_2}^{y^{\rm max}_2} \drv y_2
	\; \delta (\Delta Y - (y_1 - y_2))
	\; {\cal C}_0\left(|\vec p_1|, |\vec p_2|, y_1, y_2 \right)
	\, .
	\end{equation}
	 
\end{itemize}

%==============================================
%\subsection{Results and discussion}
%===============================================
%-----------------------------------------
\subsection{Results and discussion}
\label{sec:bhad-results}
\begin{figure}[h]
	%\centering
	
	\includegraphics[scale=0.53,clip]{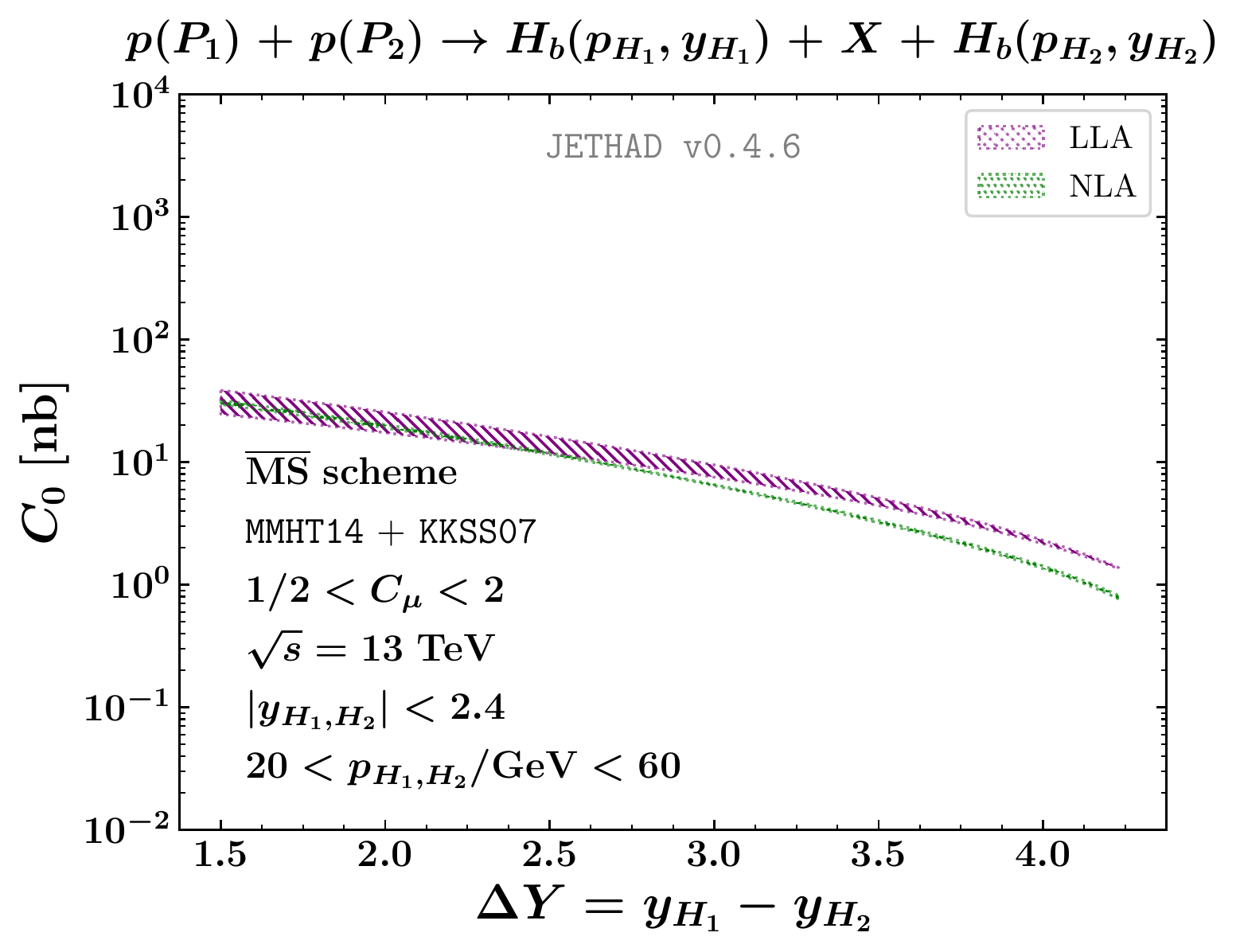}
	%   \hspace{0.05cm}
	\includegraphics[scale=0.53,clip]{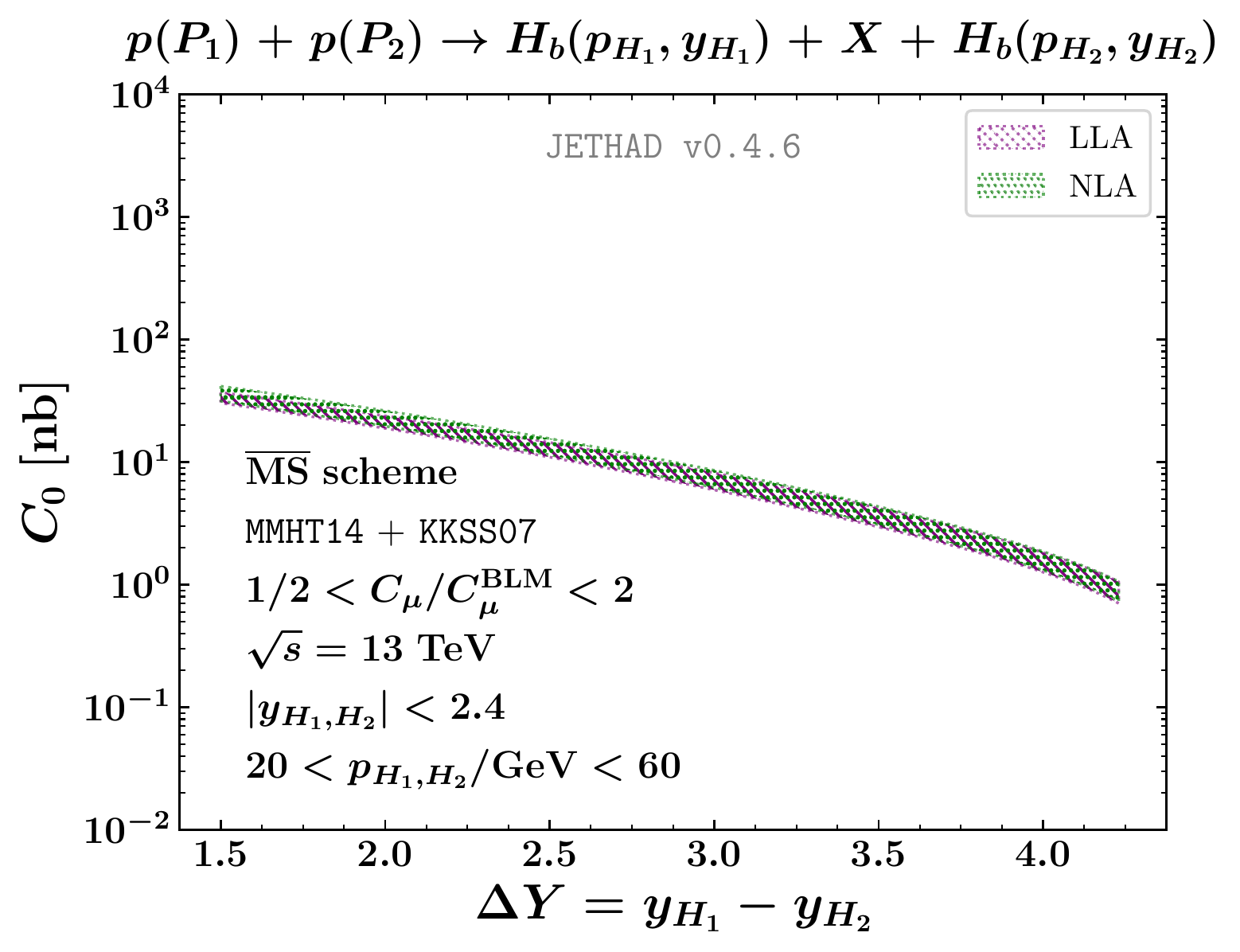}
	
	\includegraphics[scale=0.53,clip]{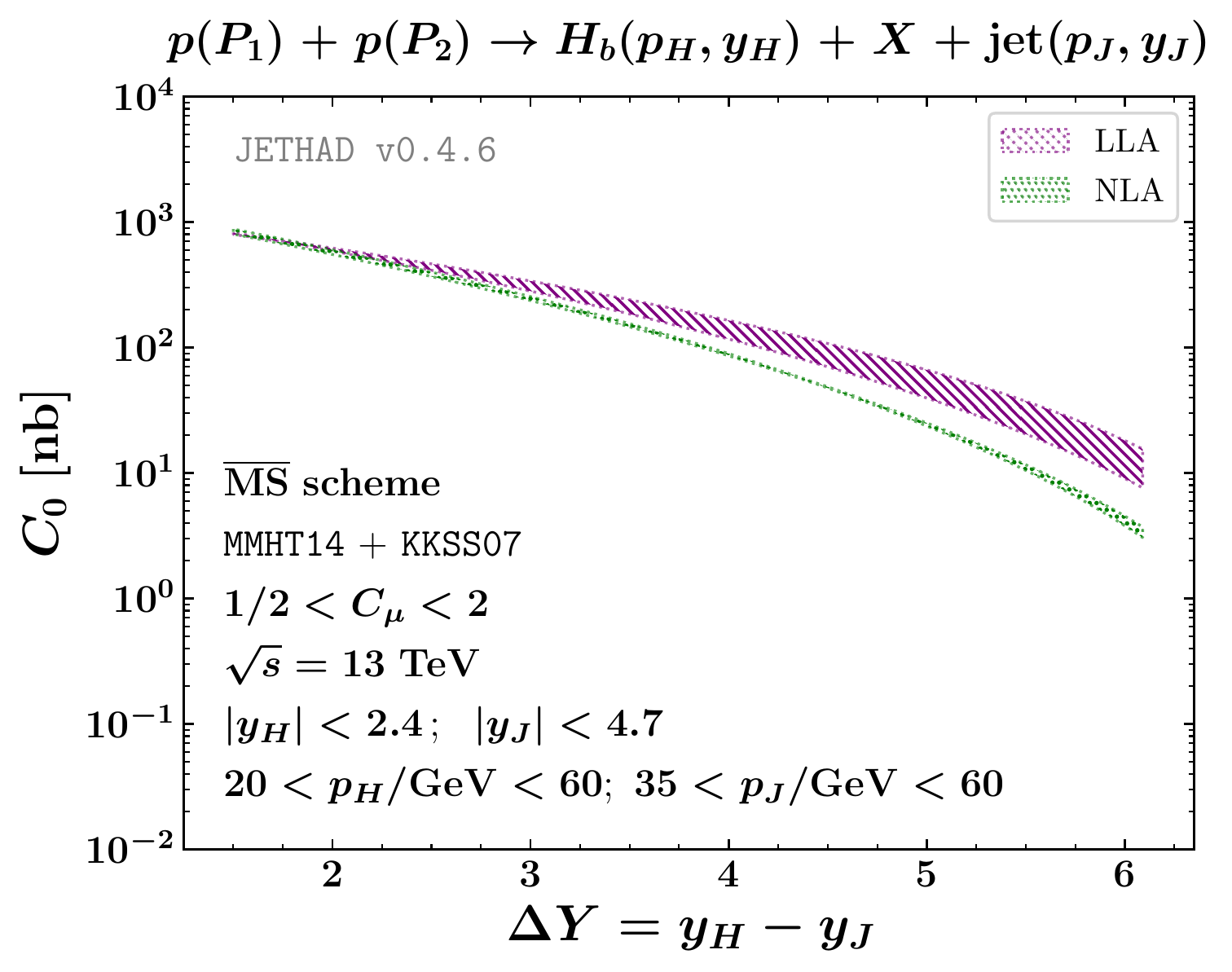}
	\hspace{0.10cm}
	\includegraphics[scale=0.53,clip]{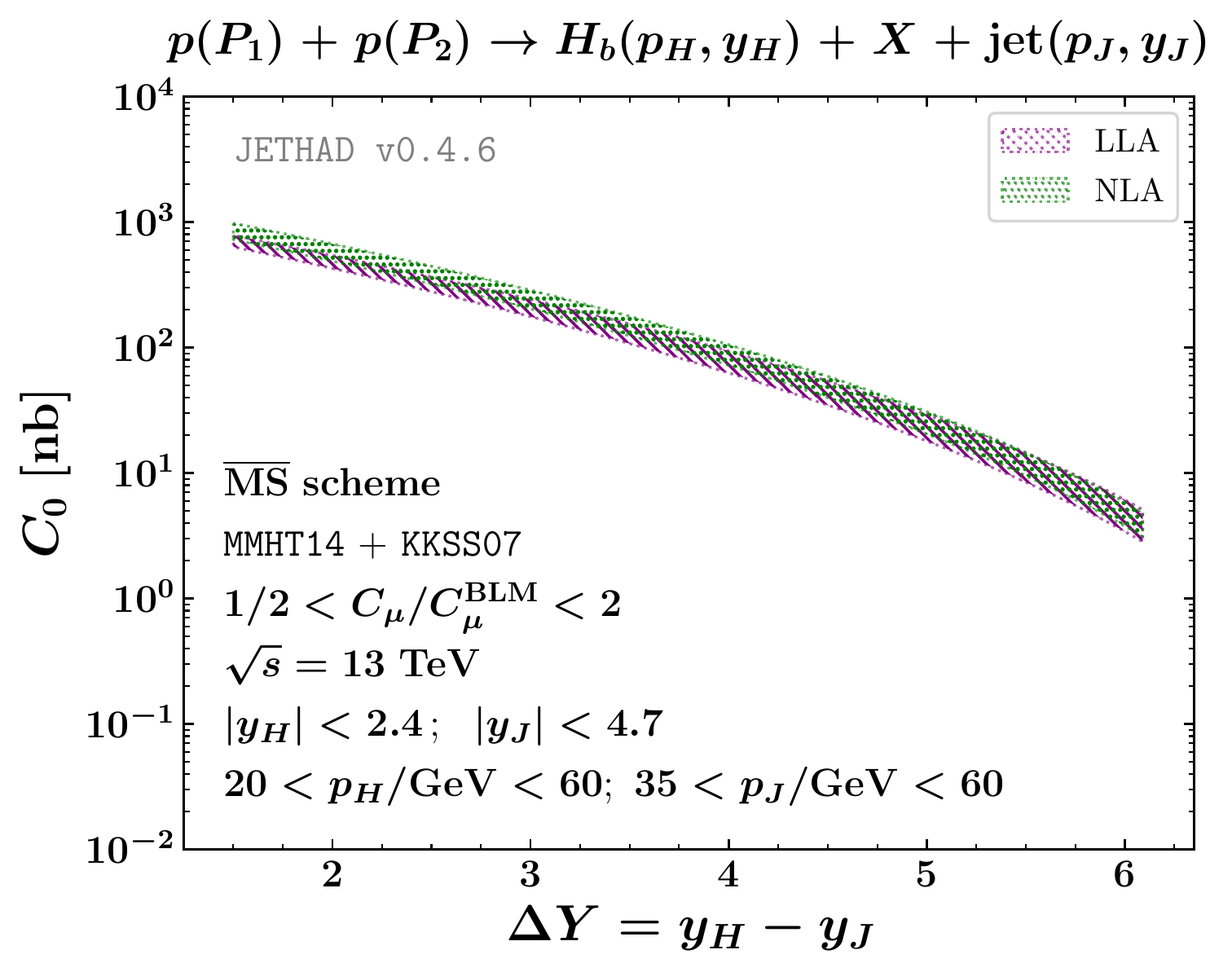}
	
	\caption{$\Delta Y$-shape of $C_0$ in the double $H_b$ (upper) and in the $H_b$~$+$~jet channel (lower), at natural (left) and BLM-optimized scales (right), and for $\sqrt{s} = 13$ TeV 
		%Text boxes inside panels show transverse-momentum and rapidity ranges. Uncertainty bands embody the combined effect of scale variation and phase-space multi-dimensional integration
		.}
	\label{fig:C0}
\end{figure}

%-----------------------------------------
In Fig.~\ref{fig:C0} we report in the upper panel the $C_0$ results for the double $H_b$ production, and for the $H_b$ + jet reaction the $C_0$ results are presented in the lower one. Predictions using natural scales present in the left panel of the figure, while the BLM results shown in the right one. Here the decreasing behavior of both LLA and NLA with the increasing intervals of rapidity is a usual BFKL pattern observed in previous semi-hard processes. Although the resummation of high-energy logs leads to a growth with energy of the pure hard cross section, the total effect is downtrend by the convoluted PDFs and FFs. We observe that NLA uncertainty bands are almost contained inside LLA one at BLM scale, and decouple form each other at natural scale. This decoupling behavior is a  manifestation of the fact that NLA series is stable under varying scale. Moreover in the case of double $H_b$ the LLA bands tend to shrink, while getting wider in $H_b + $ jet one.

\begin{figure}[h]
	%	\centering
	
	\includegraphics[scale=0.53,clip]{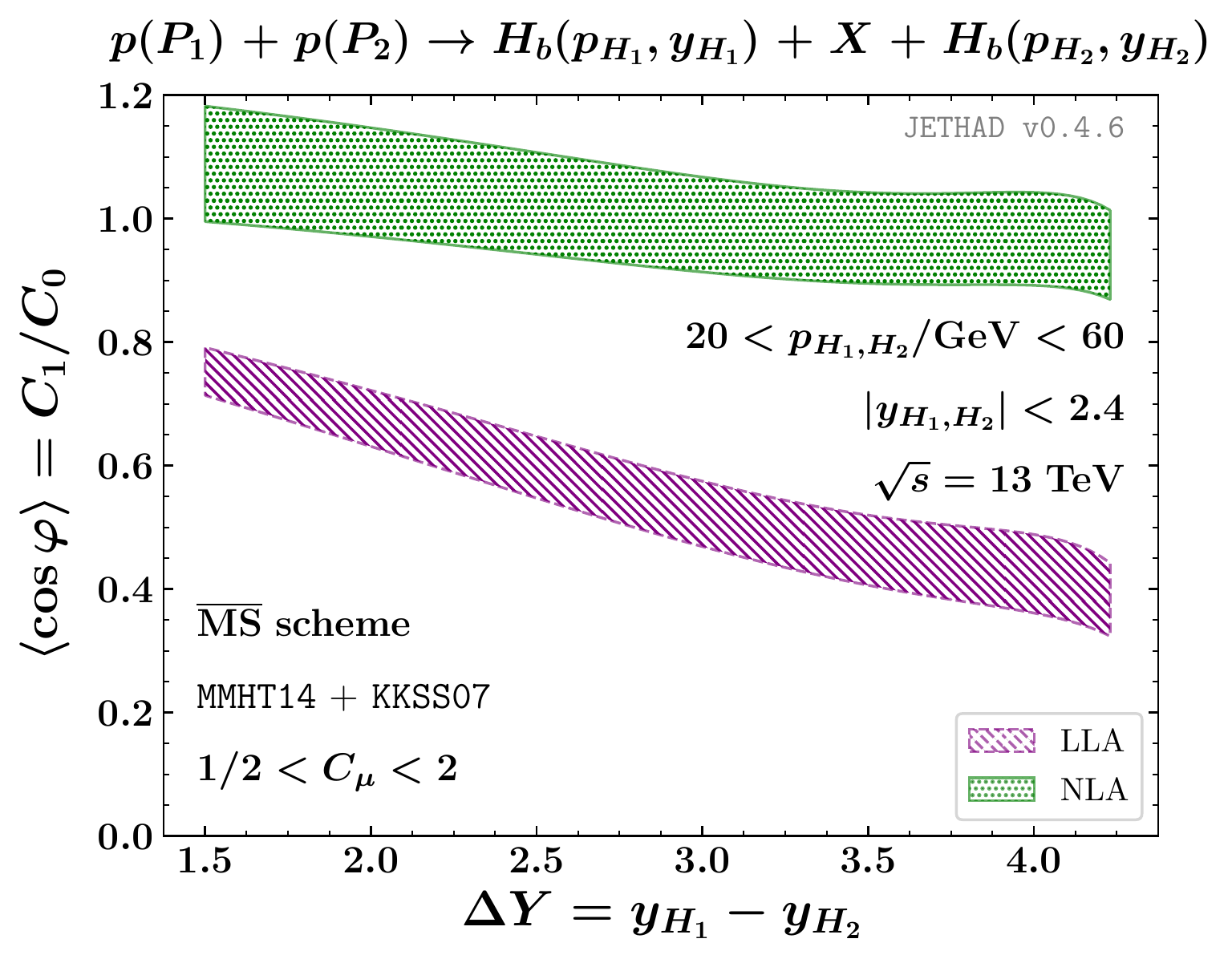}
	\includegraphics[scale=0.53,clip]{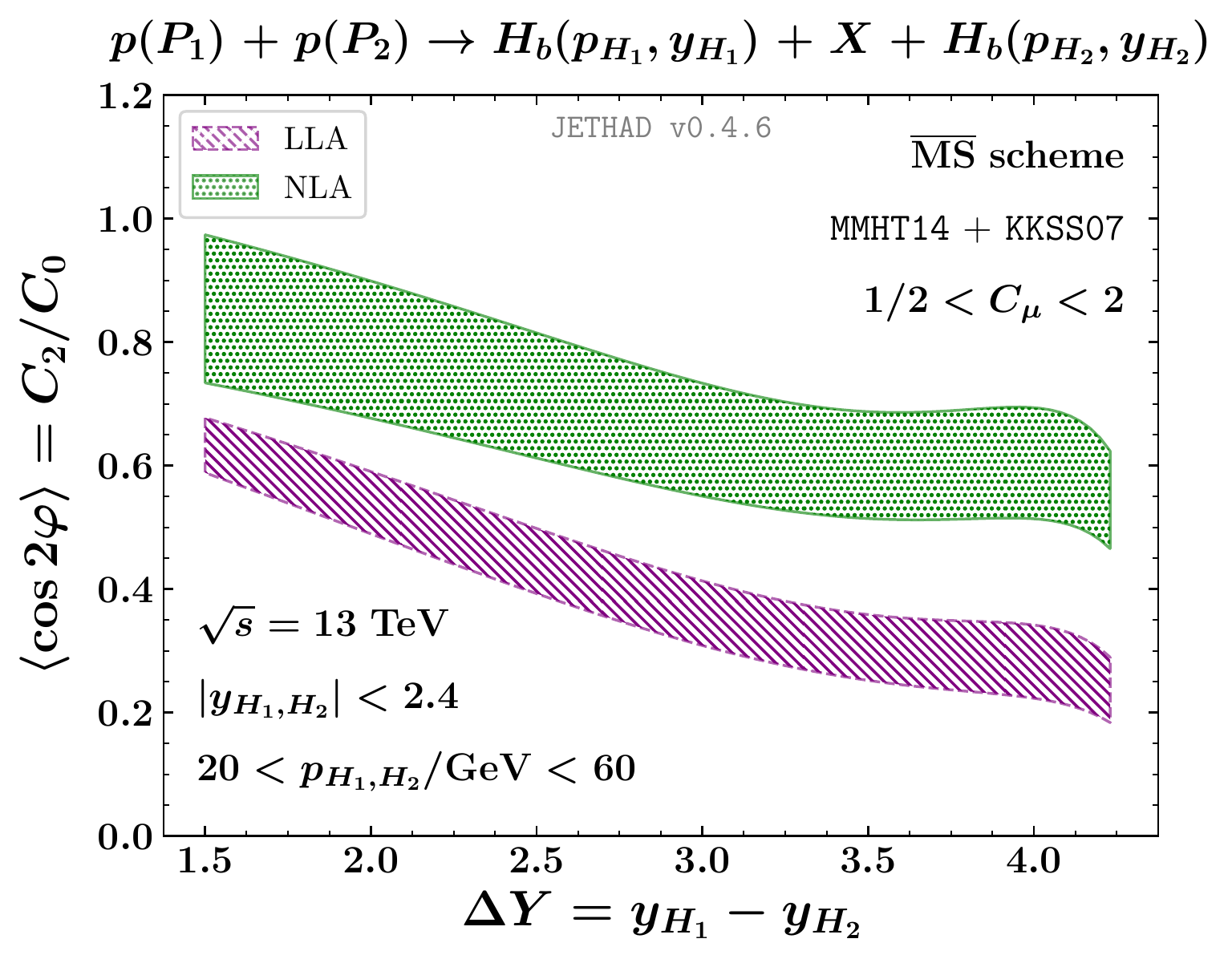}
	\hspace{0.10cm}
	\includegraphics[scale=0.53,clip]{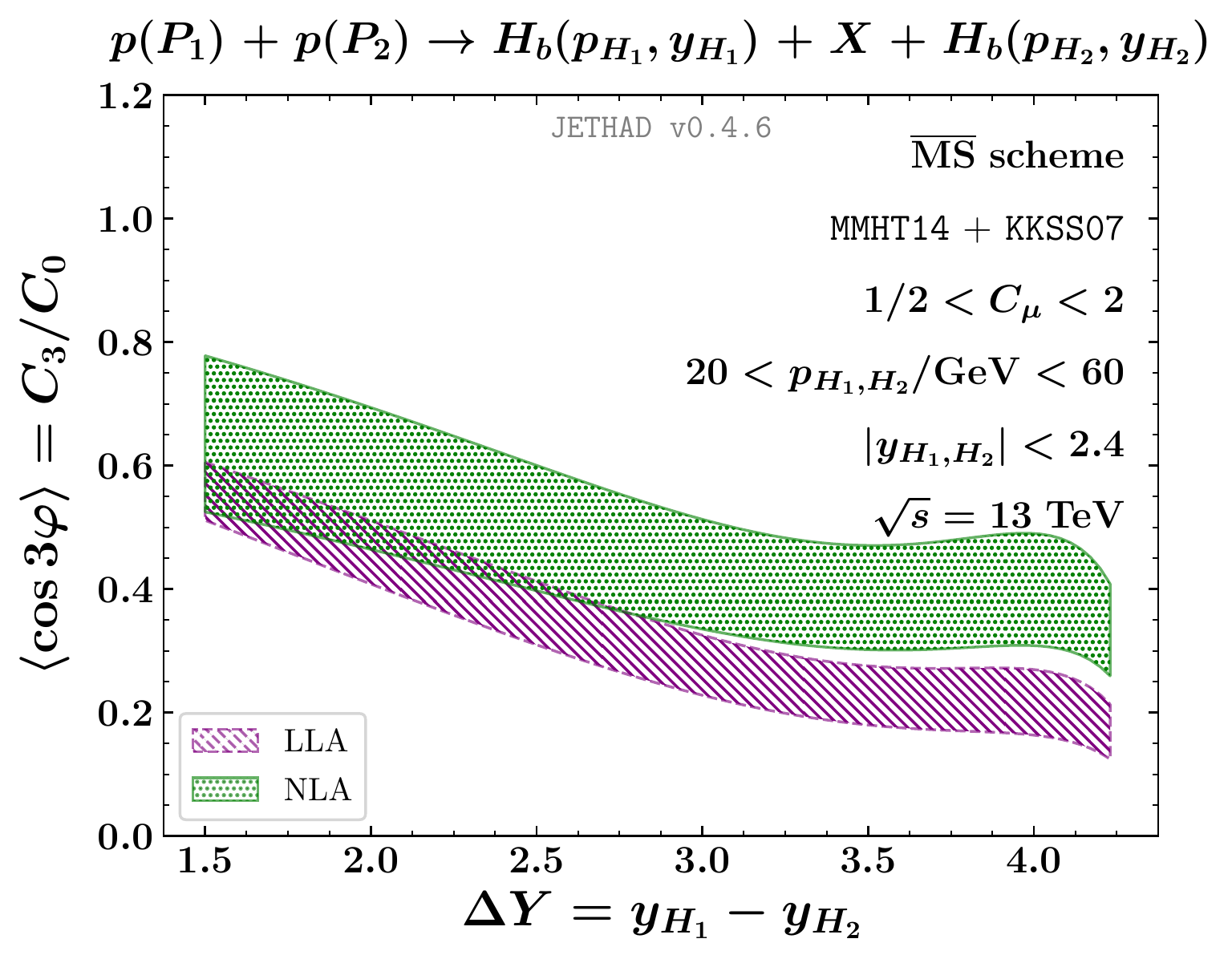}
	\includegraphics[scale=0.53,clip]{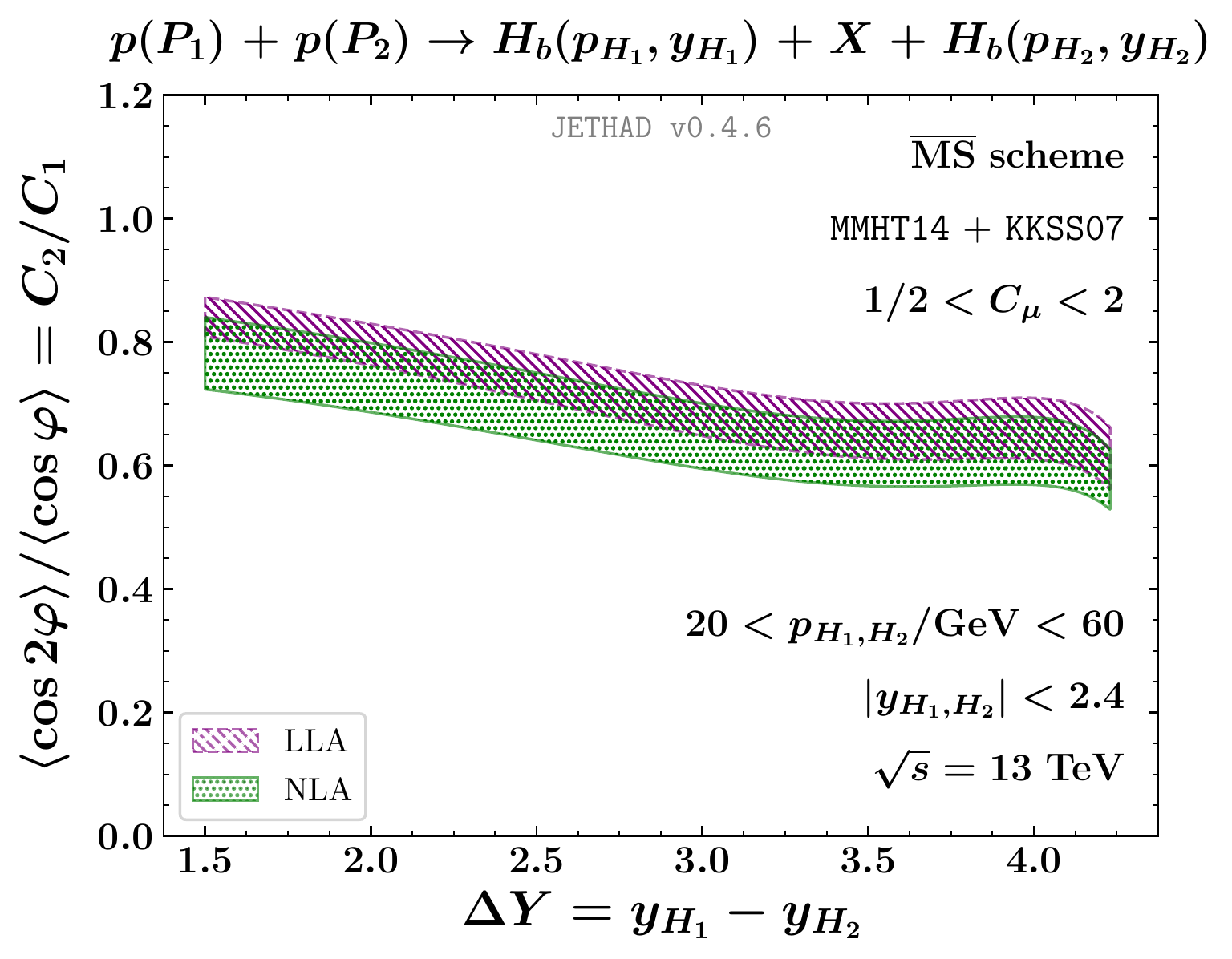}
	
	\caption{$\Delta Y$-shape of azimuthal correlations, $R_{nm} \equiv C_{n}/C_{m}$, in the double $H_b$ channel, at natural scales, and for $\sqrt{s} = 13$ TeV 
		%Text boxes inside panels show transverse-momentum and rapidity ranges. Uncertainty bands embody the combined effect of scale variation and phase-space multi-dimensional integration
		.}
	\label{fig:Rnm_HbHb_NS}
\end{figure}

\begin{figure}[h]
	%	\centering
	
	\includegraphics[scale=0.53,clip]{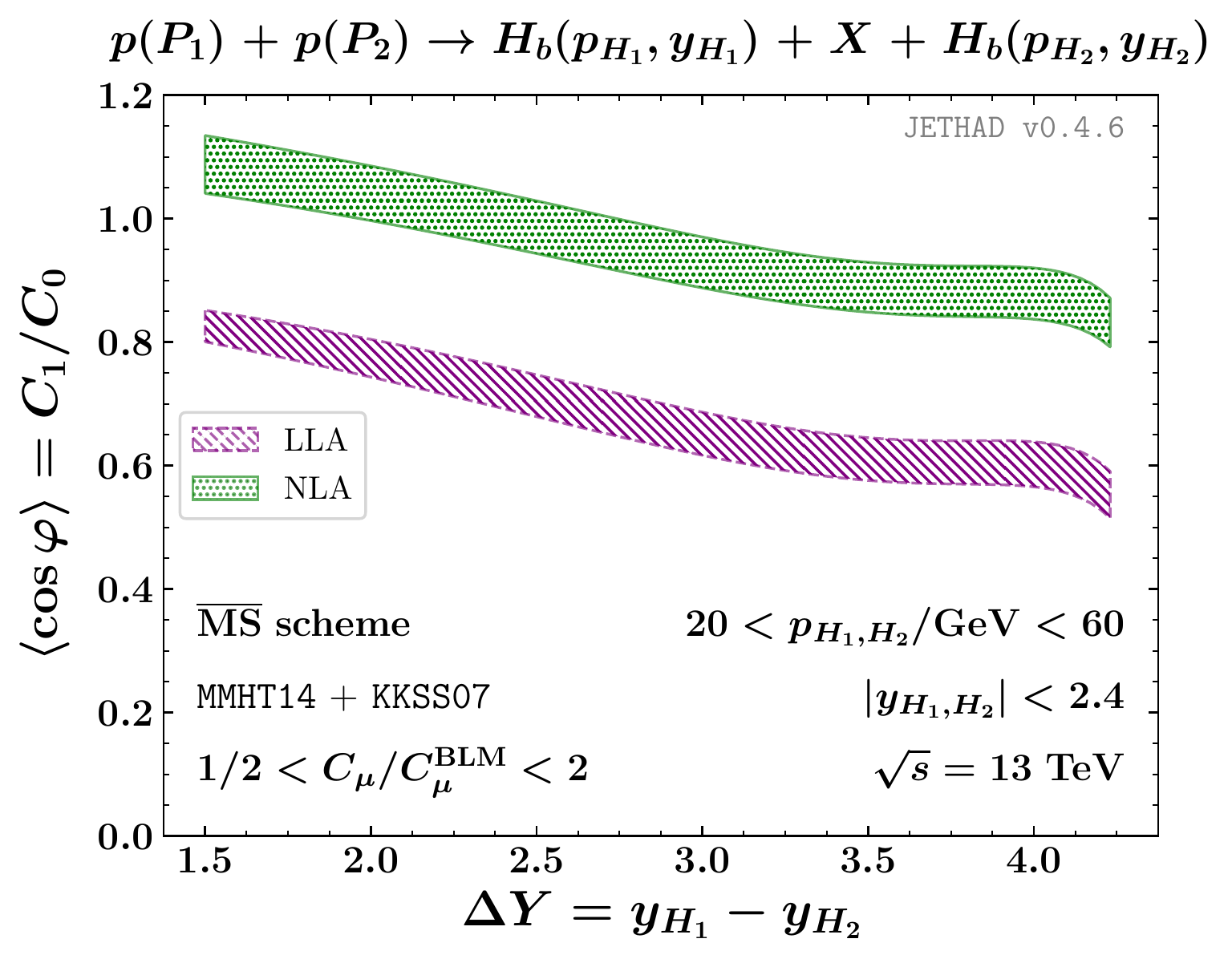}
	\includegraphics[scale=0.53,clip]{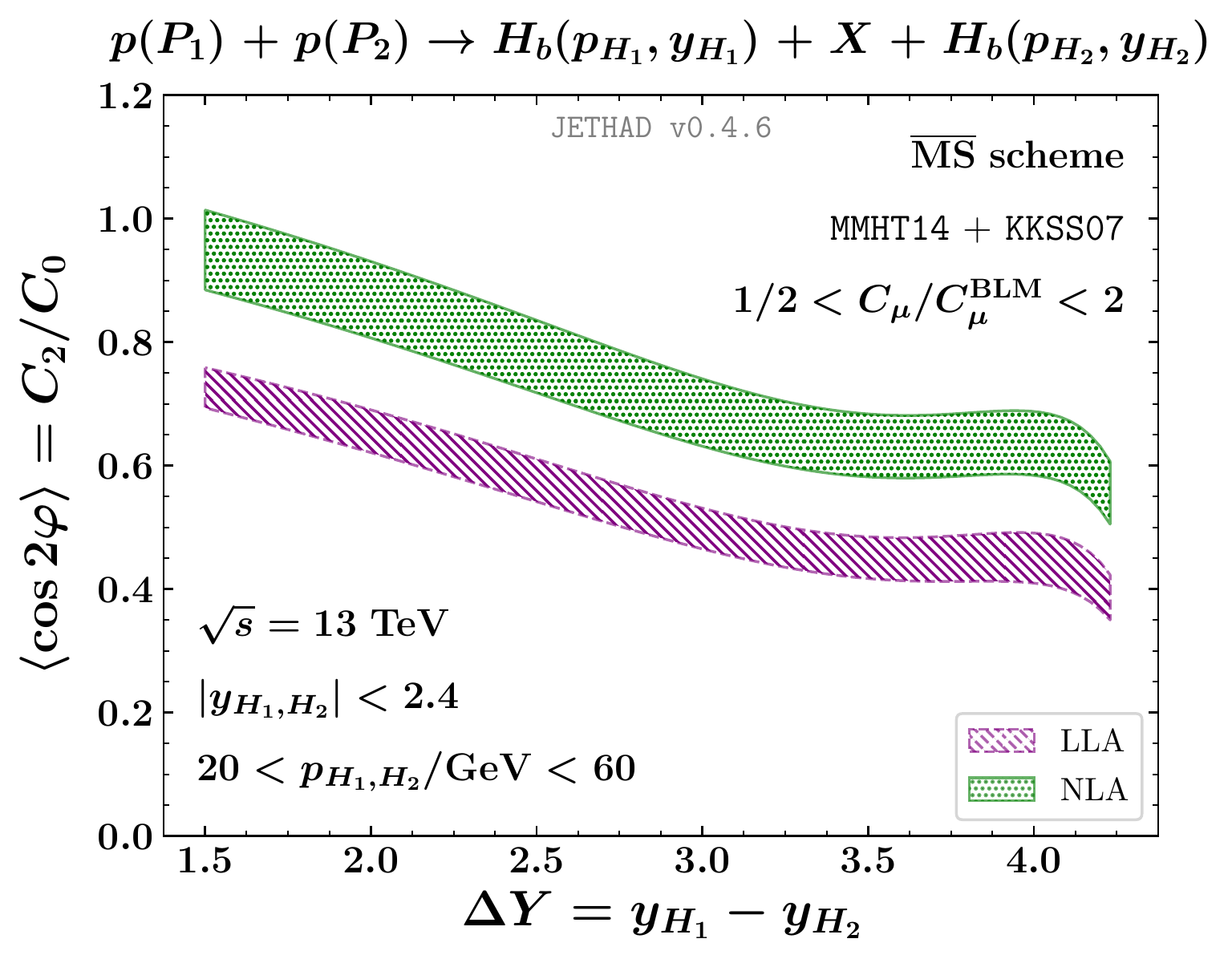}
	\hspace{0.10cm}
	\includegraphics[scale=0.53,clip]{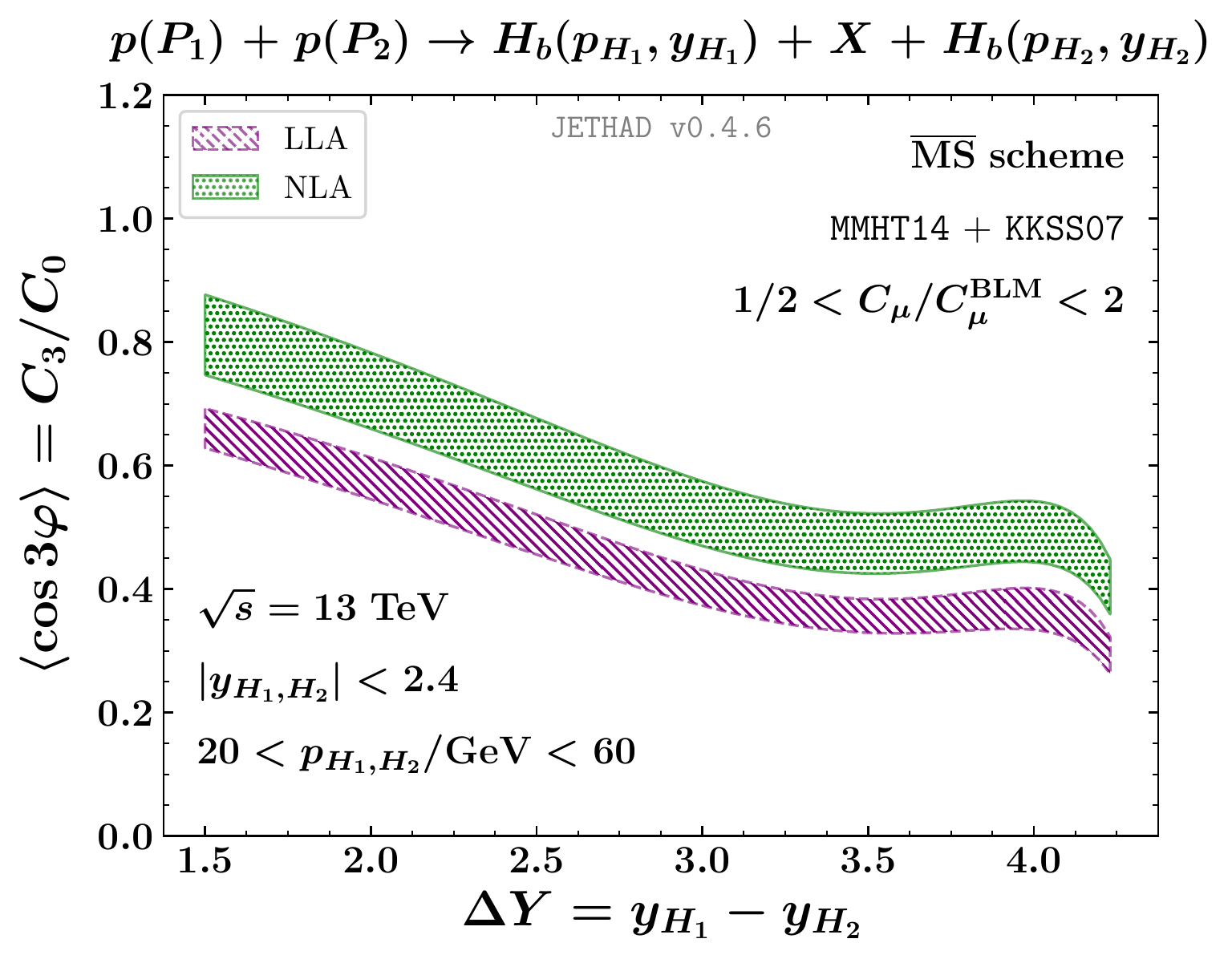}
	\includegraphics[scale=0.53,clip]{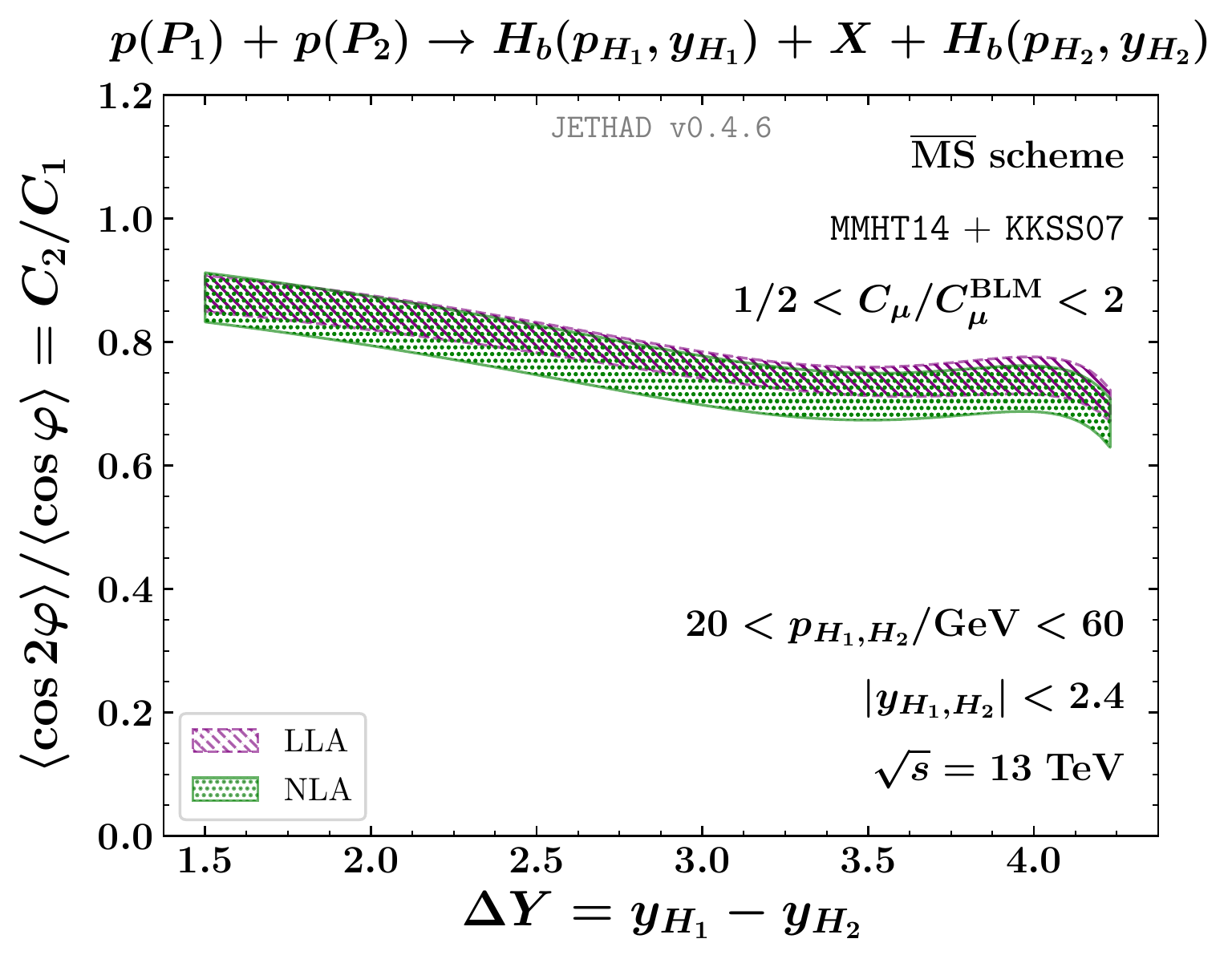}
	
	\caption{$\Delta Y$-shape of azimuthal correlations, $R_{nm} \equiv C_{n}/C_{m}$, in the double $H_b$ channel, at BLM scales, and for $\sqrt{s} = 13$ TeV 
		%Text boxes inside panels show transverse-momentum and rapidity ranges. Uncertainty bands embody the combined effect of scale variation and phase-space multi-dimensional integration
		.}
	\label{fig:Rnm_HbHb_BLM}
\end{figure}

In Figs.~\ref{fig:Rnm_HbHb_NS} and~\ref{fig:Rnm_HbHb_BLM} we present the  $R_{nm}\equiv C_n/C_m$ moments results for the double $H_b$ probe channel at natural and BLM scales, respectively. The notable observation here is that the natural scale predictions are closely in shape to the corresponding optimized (BLM) ones. The reason of the increased stability for $R_{n0}\equiv C_n/C_0$, is due to the relatively small uncertainty on $C_0$.

\begin{figure}[h]
	%\centering
	\includegraphics[scale=0.53,clip]{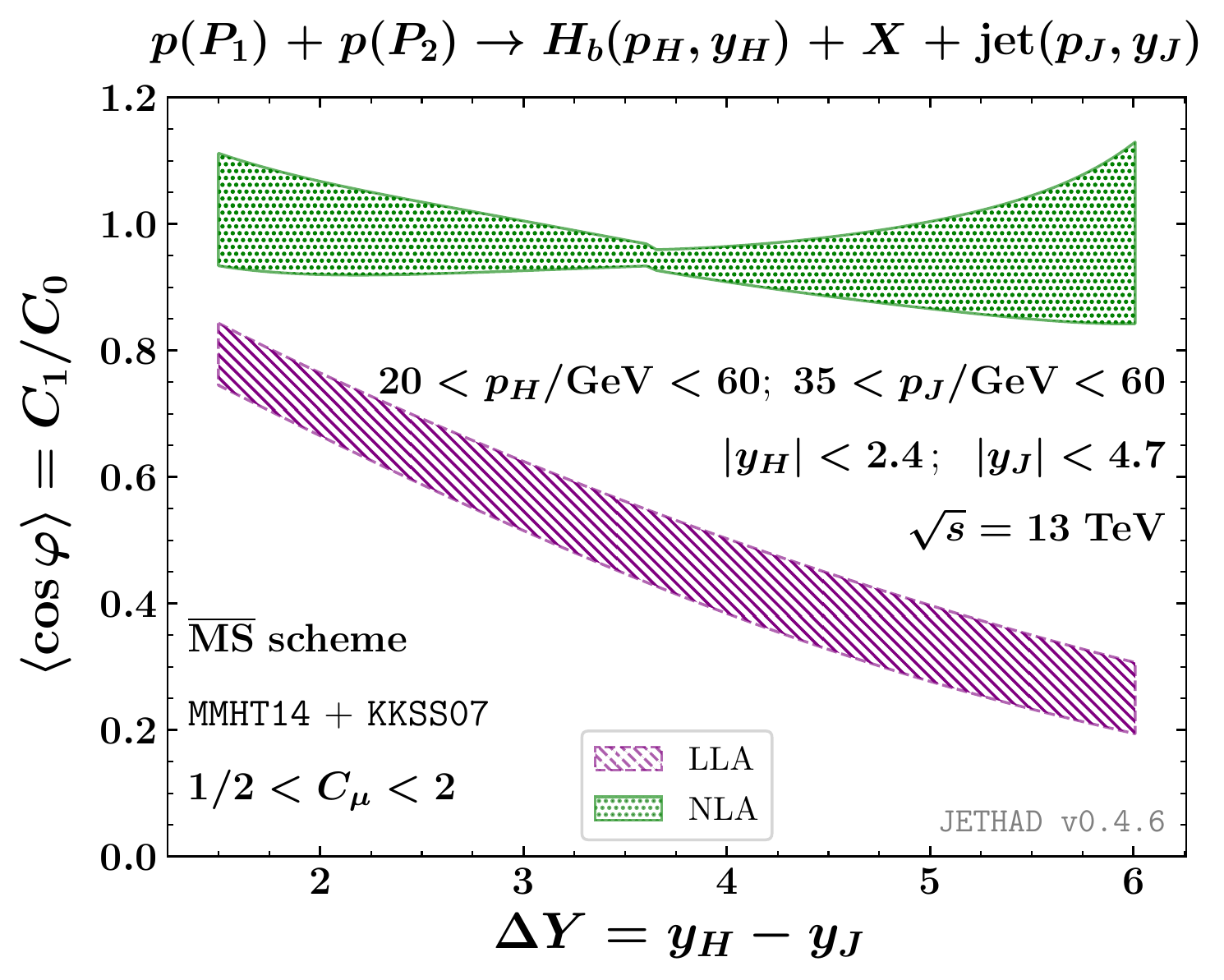}
	\includegraphics[scale=0.53,clip]{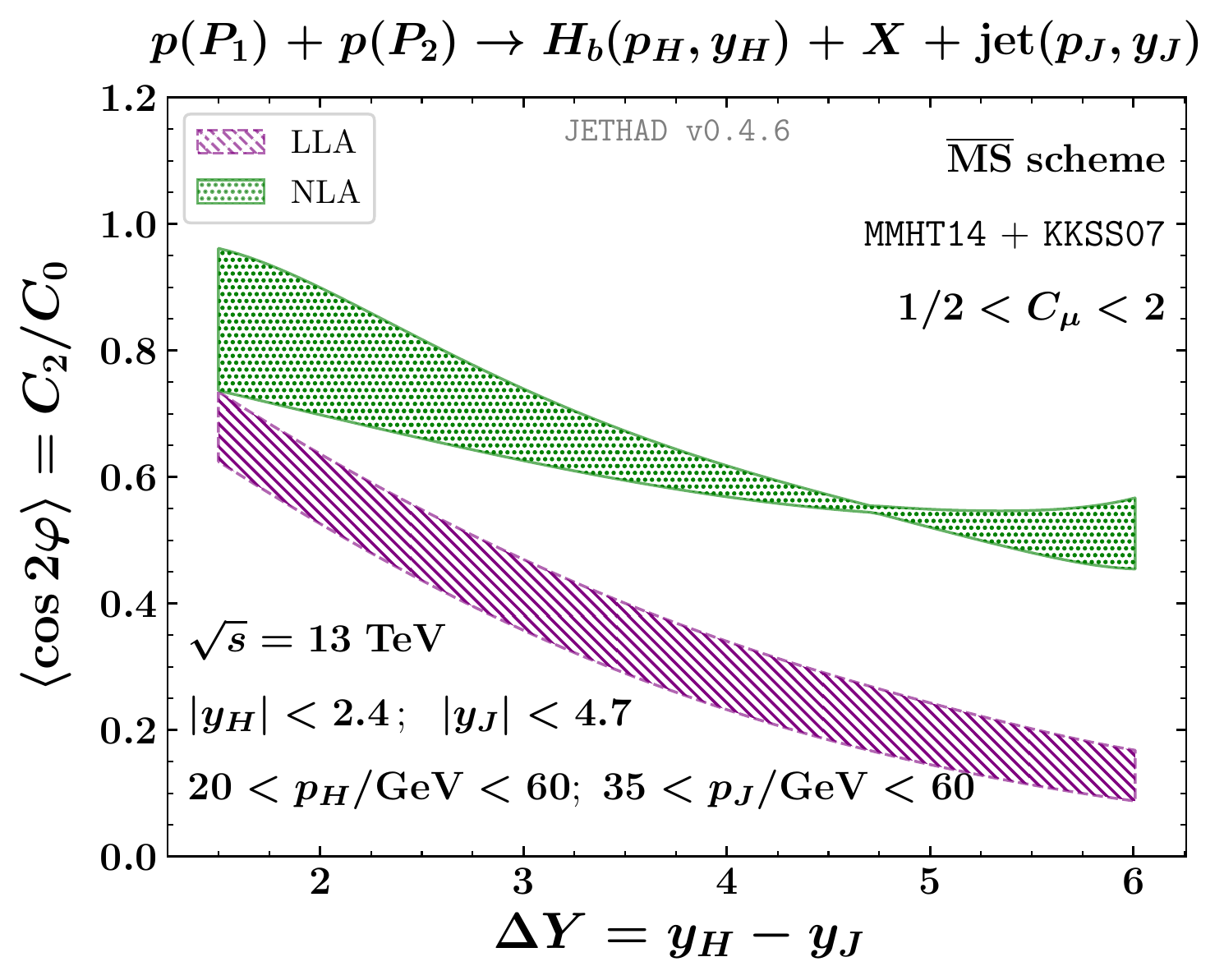}
	\hspace{0.10cm}
	\includegraphics[scale=0.53,clip]{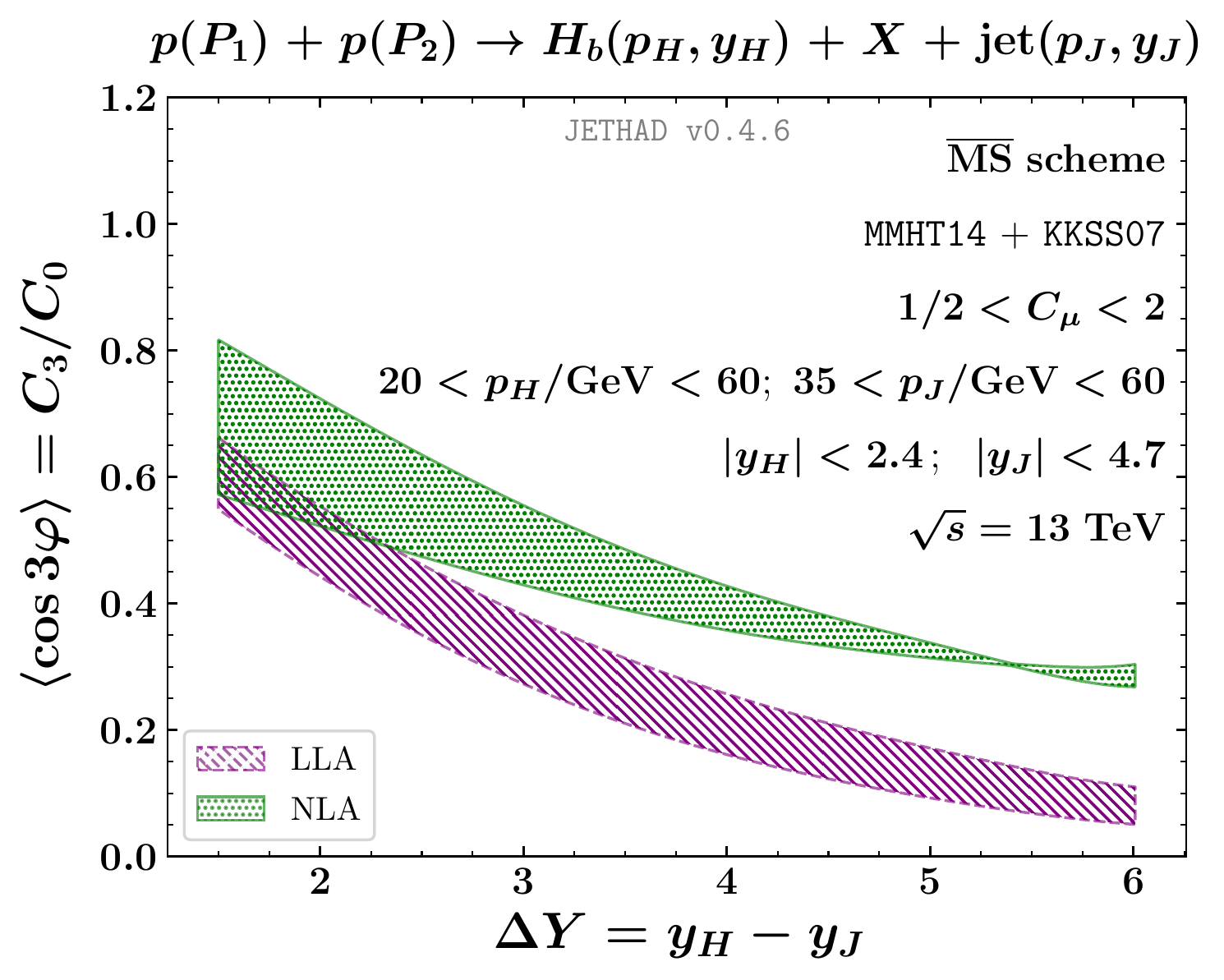}
	\includegraphics[scale=0.53,clip]{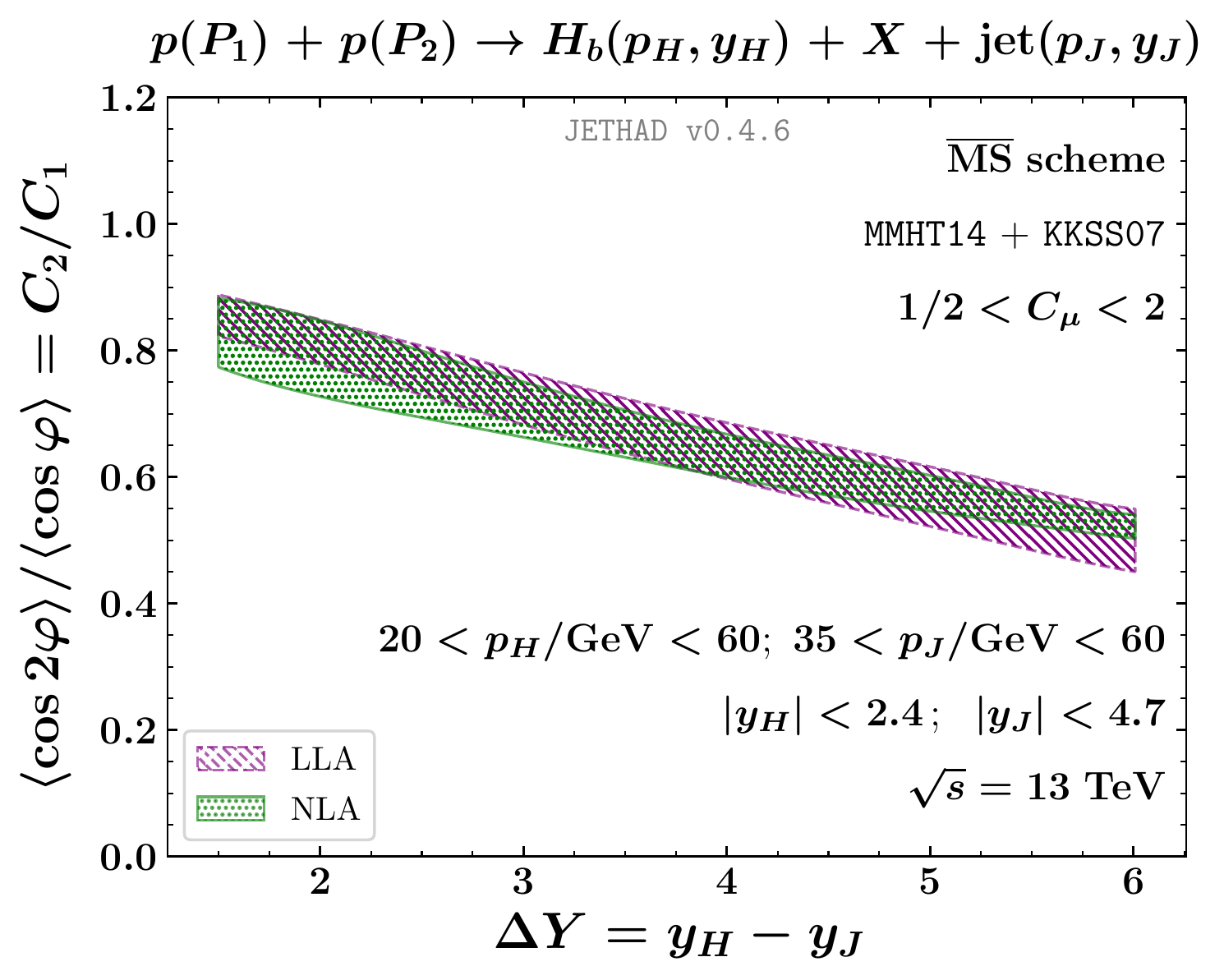}
	\caption{$\Delta Y$-shape of azimuthal correlations, $R_{nm} \equiv C_{n}/C_{m}$, in the $H_b$~$+$~jet channel, at natural scales, and for $\sqrt{s} = 13$ TeV
		% Text boxes inside panels show transverse-momentum and rapidity ranges. Uncertainty bands embody the combined effect of scale variation and phase-space multi-dimensional integration
		.}
	\label{fig:Rnm_HbJ_NS}
\end{figure}

\begin{figure}[h]
	%	\centering
	\includegraphics[scale=0.53,clip]{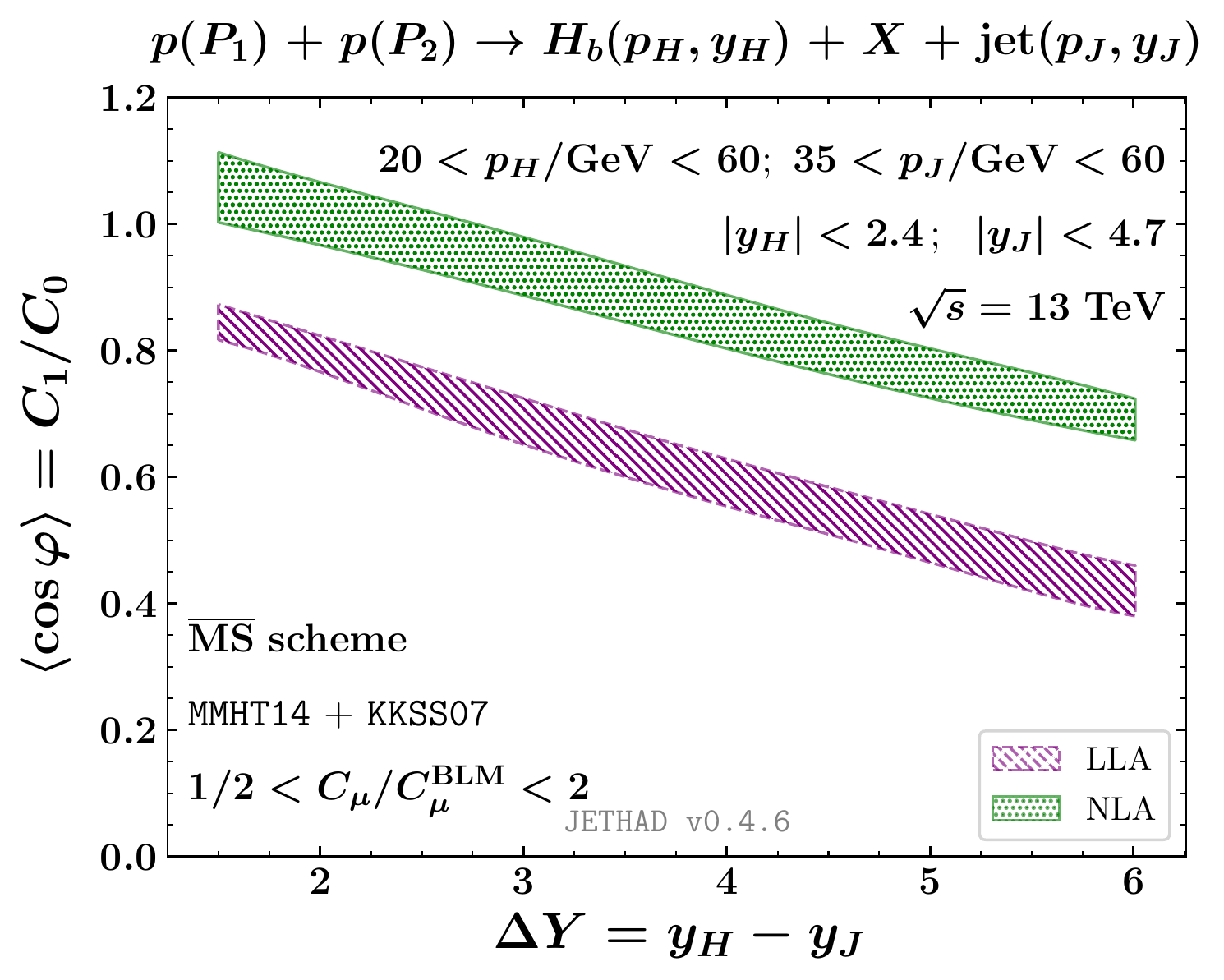}
	\includegraphics[scale=0.53,clip]{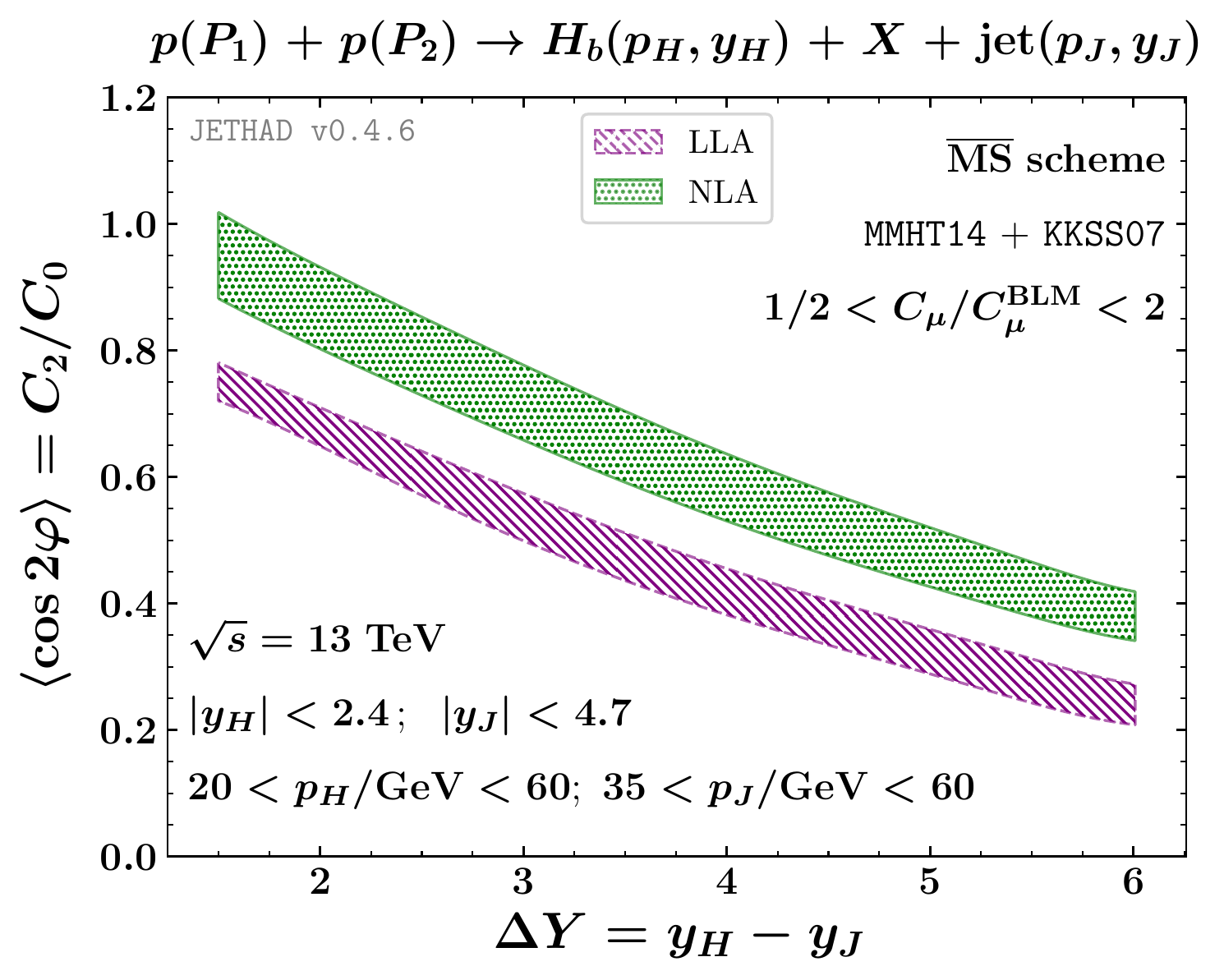}
	\hspace{0.10cm}
	\includegraphics[scale=0.53,clip]{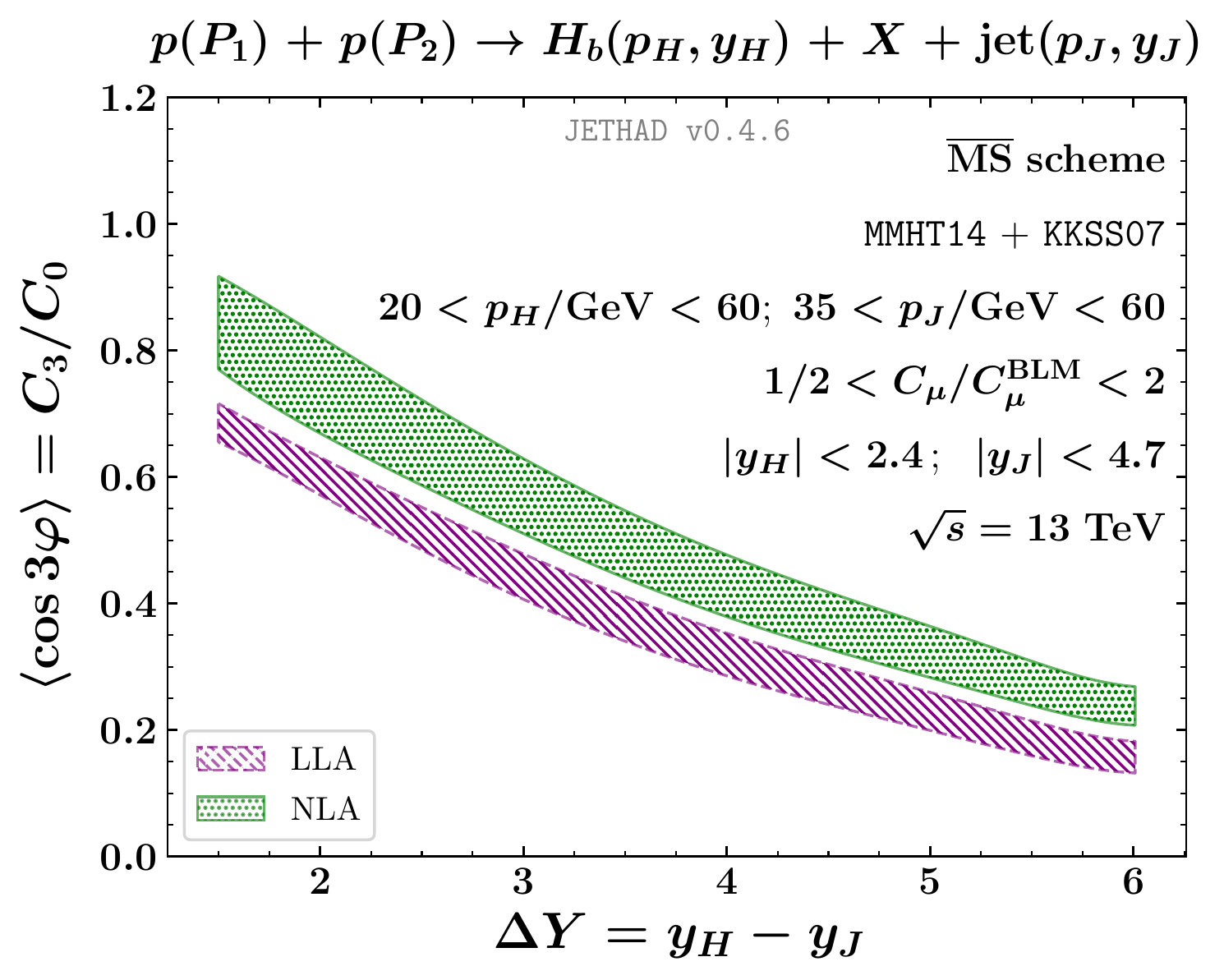}
	\includegraphics[scale=0.53,clip]{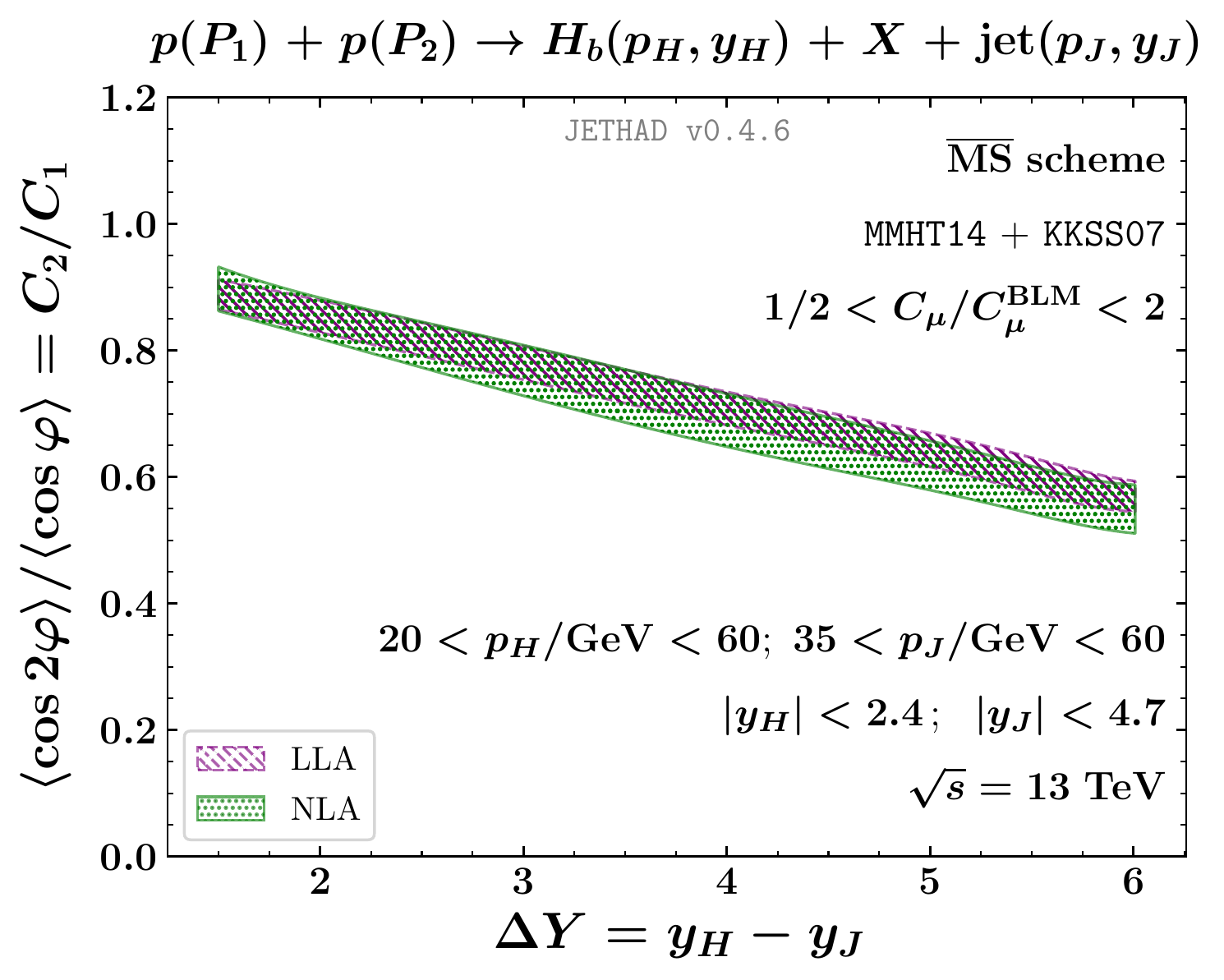}
	\caption{$\Delta Y$-shape of azimuthal correlations, $R_{nm} \equiv C_{n}/C_{m}$, in the $H_b$~$+$~jet channel, at BLM scales, and for $\sqrt{s} = 13$ TeV 
		%Text boxes inside panels show transverse-momentum and rapidity ranges. Uncertainty bands embody the combined effect of scale variation and phase-space multi-dimensional integration
		.}
	\label{fig:Rnm_HbJ_BLM}
\end{figure}

In Figs.~\ref{fig:Rnm_HbJ_NS} we present results for azimuthal ratios in the $H_b +$ jet reaction at natural scales. Here, the NLA $R_{n0}$ correlations appear to be strongly sensitive to scale variation\footnote{In semi-hard processes with emitted jet, such as Mueller–Navelet dijet or Hadron + jet, whenever instabilities emerging at natural scales are so strong, these probes cannot be studied at natural scales.}. Corresponding results for $R_{n0}$ moments at BLM scale (shown in Fig~\ref{fig:Rnm_HbJ_BLM}) have a similar shape to predictions of the double $H_b$ channel. 

\begin{figure}[h]
	%	\centering
	
	\includegraphics[scale=0.41,clip]{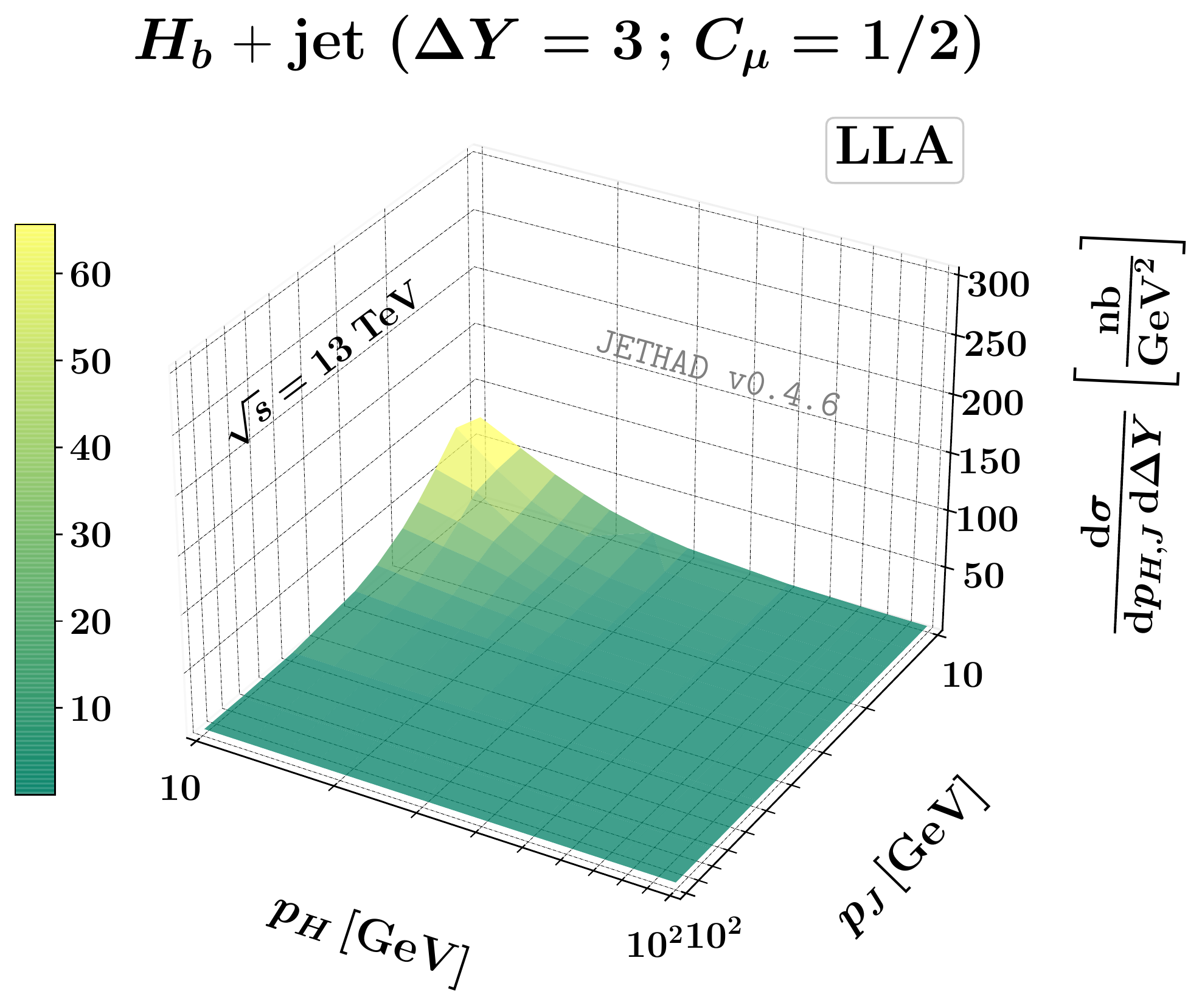}
	\hspace{0.25cm}
	\includegraphics[scale=0.41,clip]{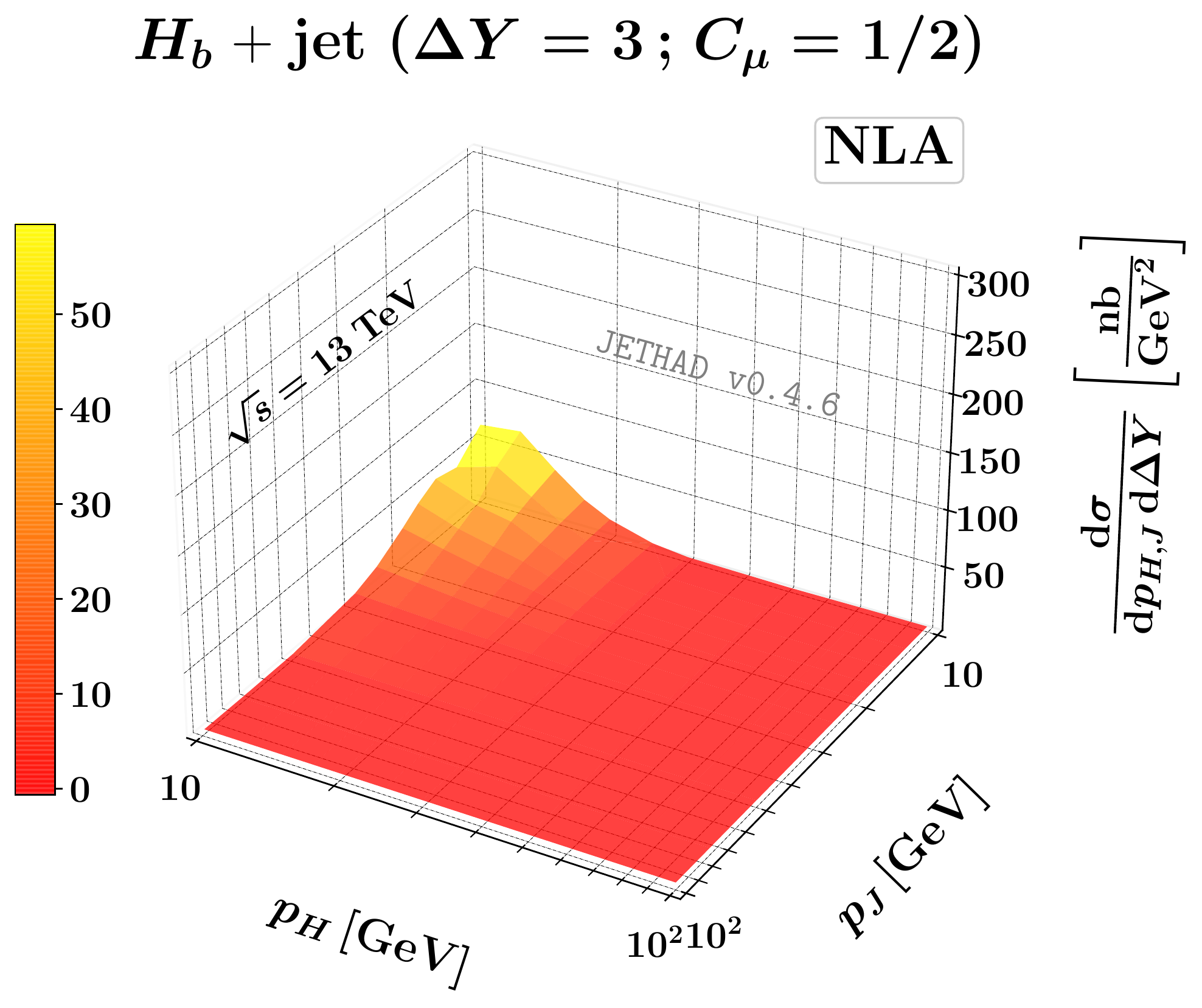}
	\vspace{0.15cm}
	
	\includegraphics[scale=0.41,clip]{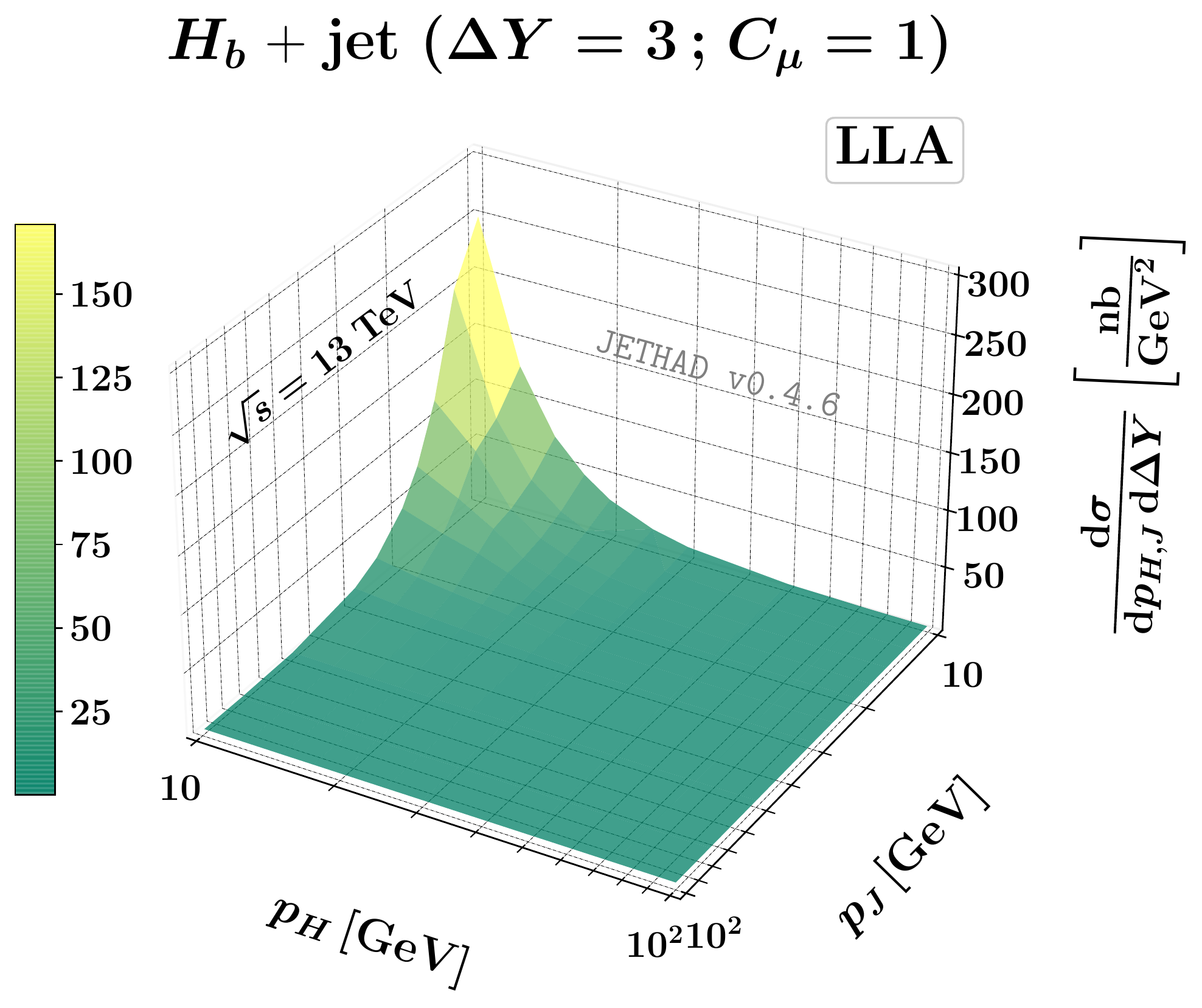}
	\hspace{0.25cm}
	\includegraphics[scale=0.41,clip]{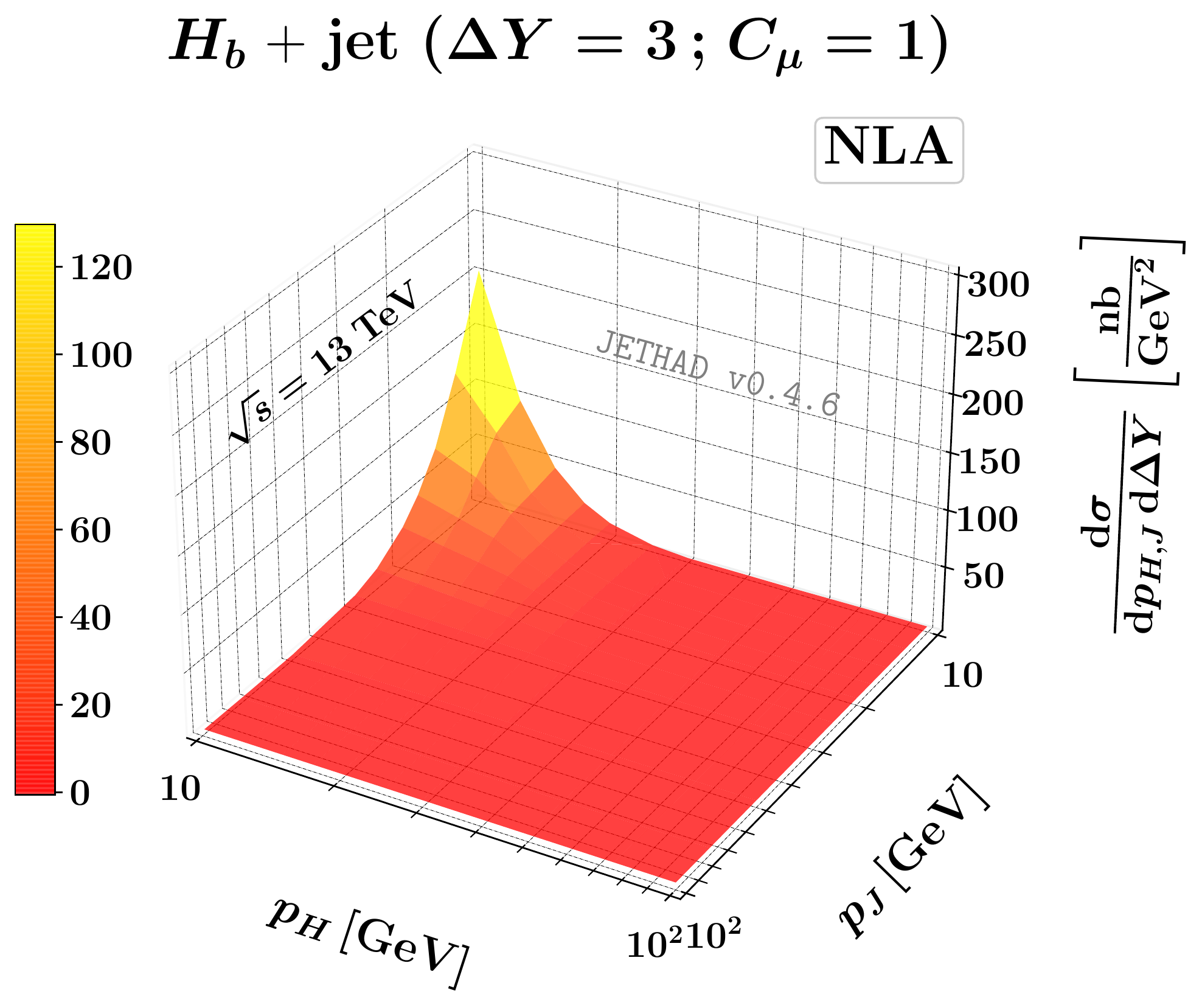}
	\vspace{0.15cm}
	
	\includegraphics[scale=0.41,clip]{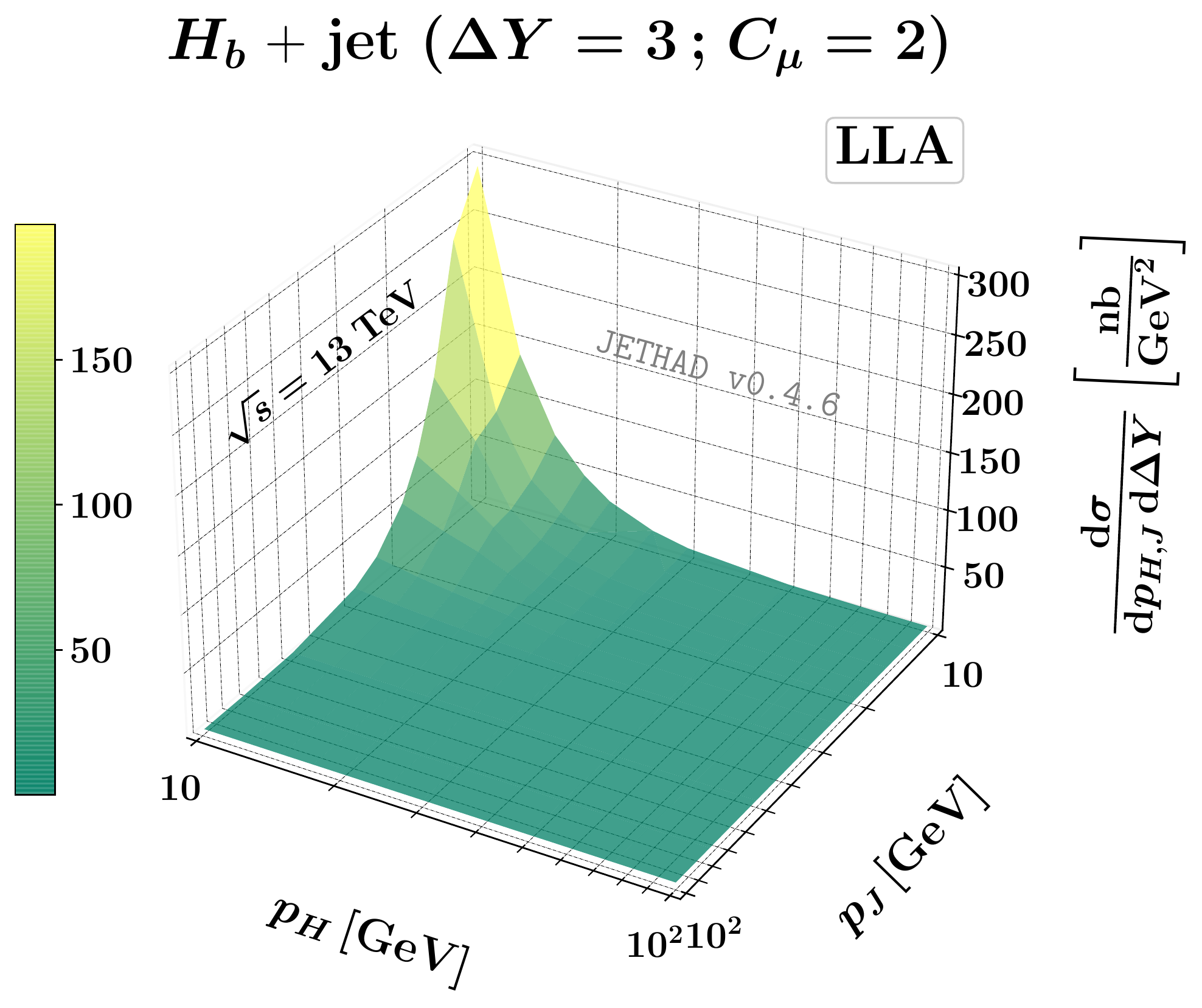}
	\hspace{0.25cm}
	\includegraphics[scale=0.41,clip]{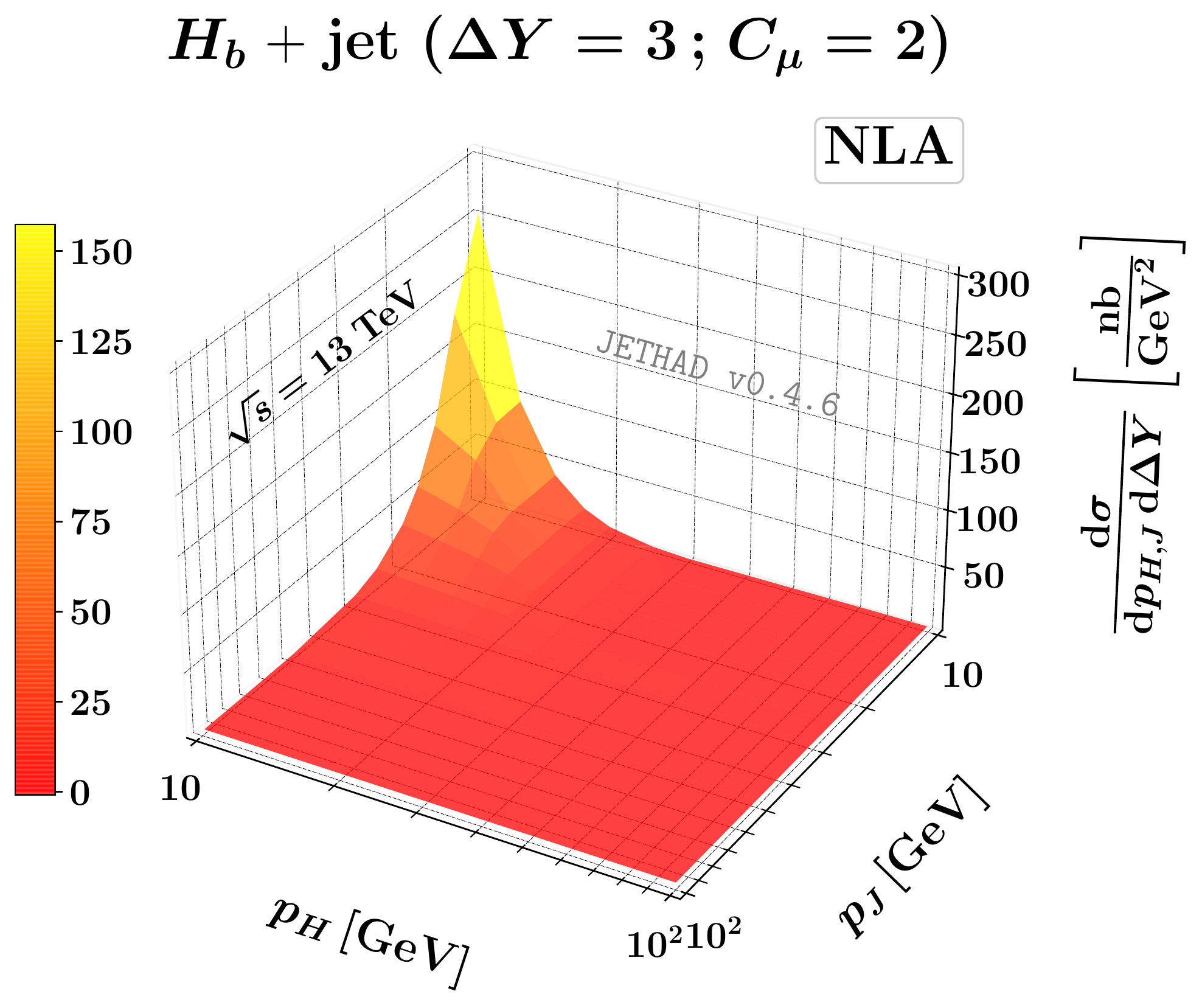}
	\vspace{0.15cm}
	
	\caption{Double differential $p_T$-distribution for the $H_b$~$+$~jet channel at $\DY=3$, $\sqrt{s} = 13$ TeV, and in the LLA (left) and NLA (right) resummation accuracy %Calculations are done at natural scales, and the $C_\mu$ parameter is in the range 1/2 to 2 (from top to bottom)
		.}
	\label{fig:Y3-2pT0}
\end{figure}

\begin{figure}[h]
	%\centering
	
	\includegraphics[scale=0.41,clip]{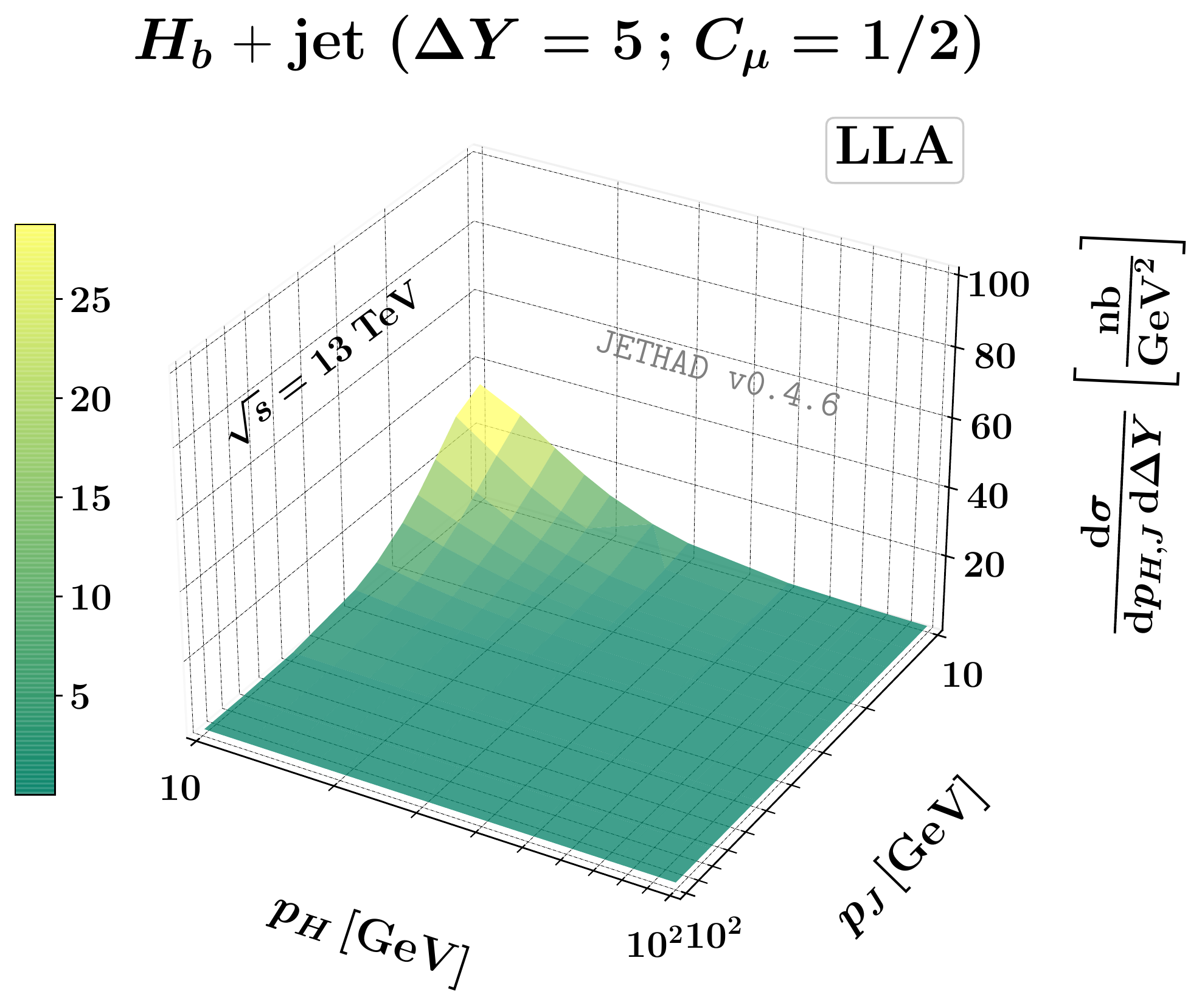}
	\hspace{0.25cm}
	\includegraphics[scale=0.41,clip]{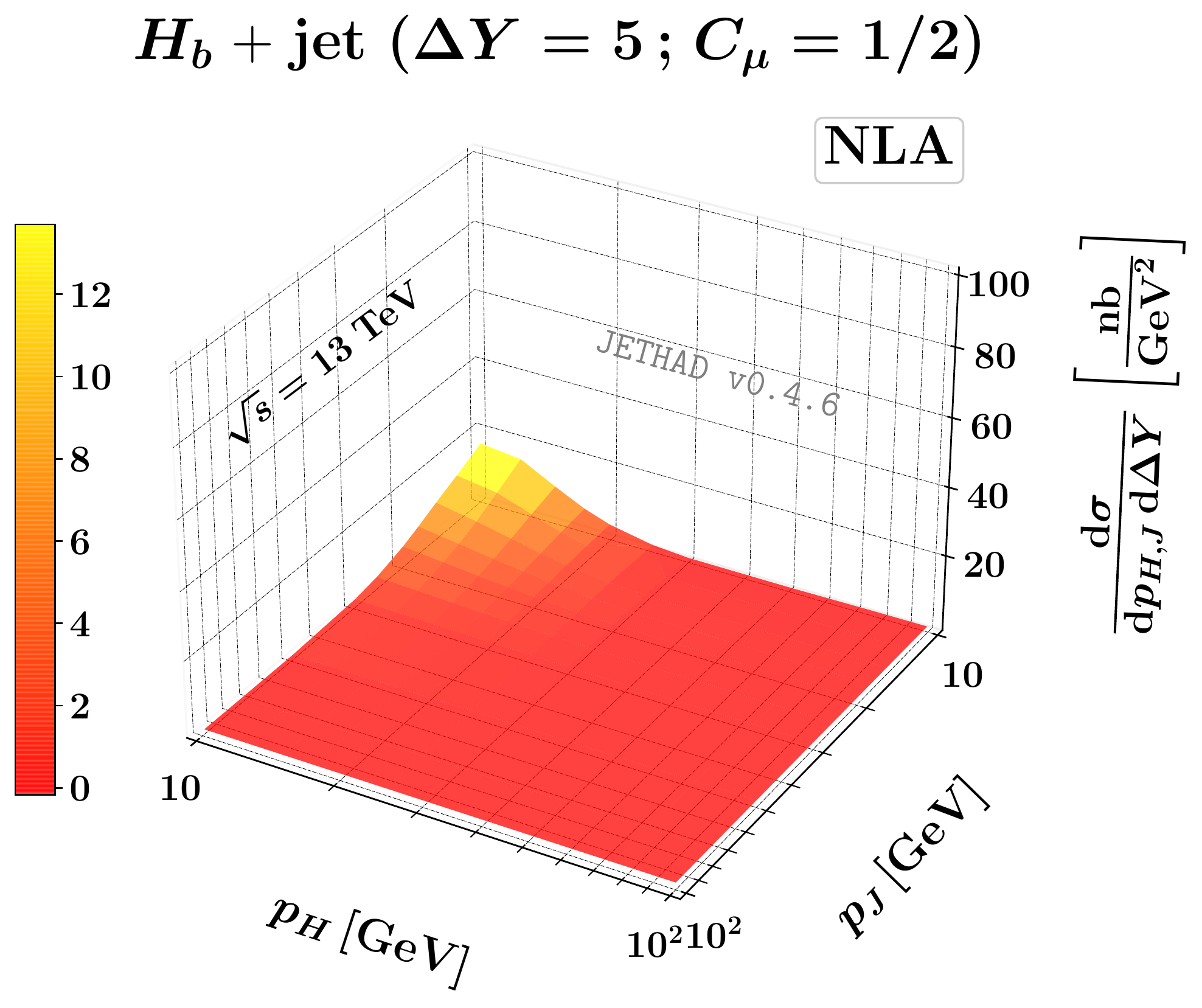}
	\vspace{0.15cm}
	
	\includegraphics[scale=0.41,clip]{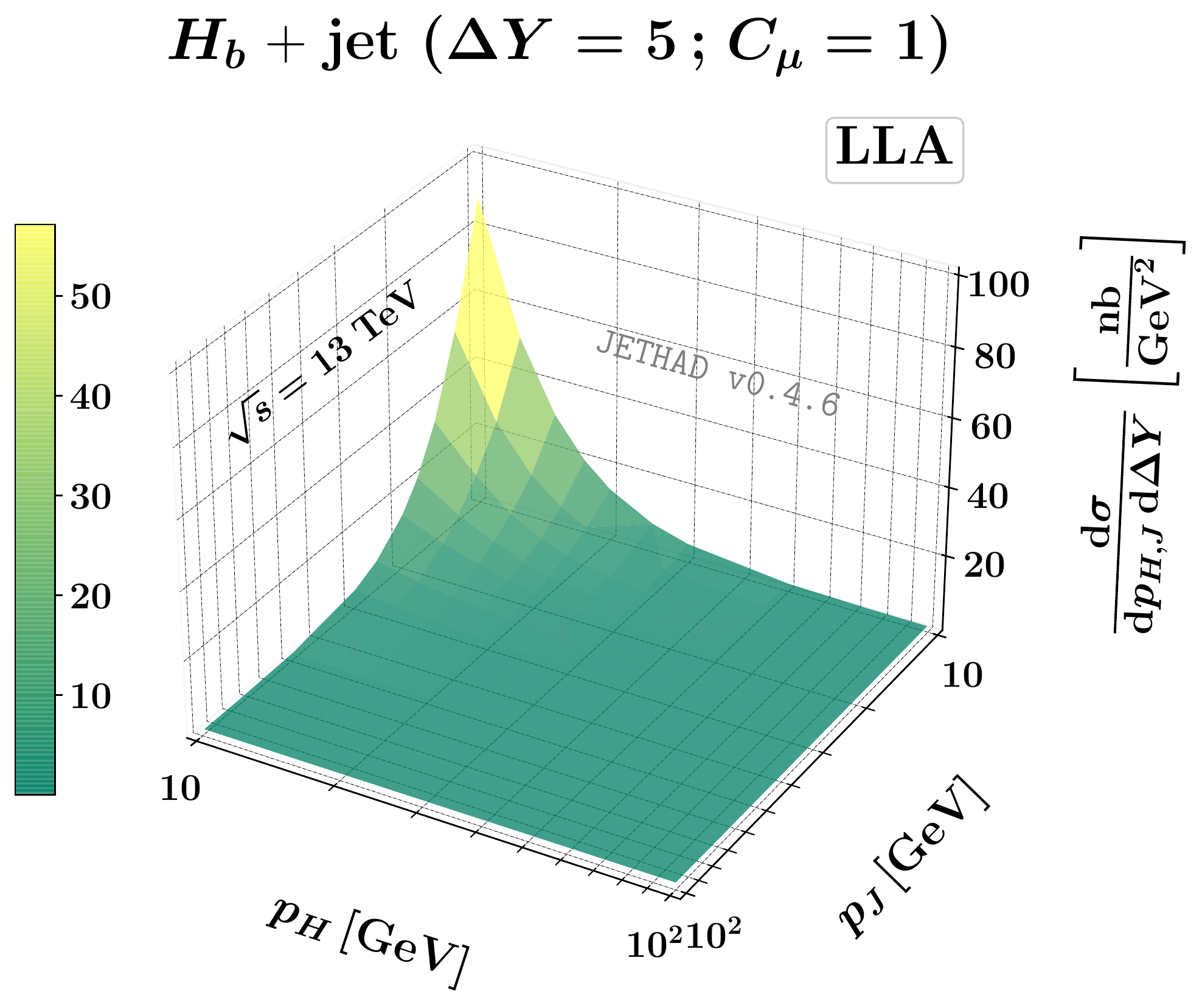}
	\hspace{0.25cm}
	\includegraphics[scale=0.41,clip]{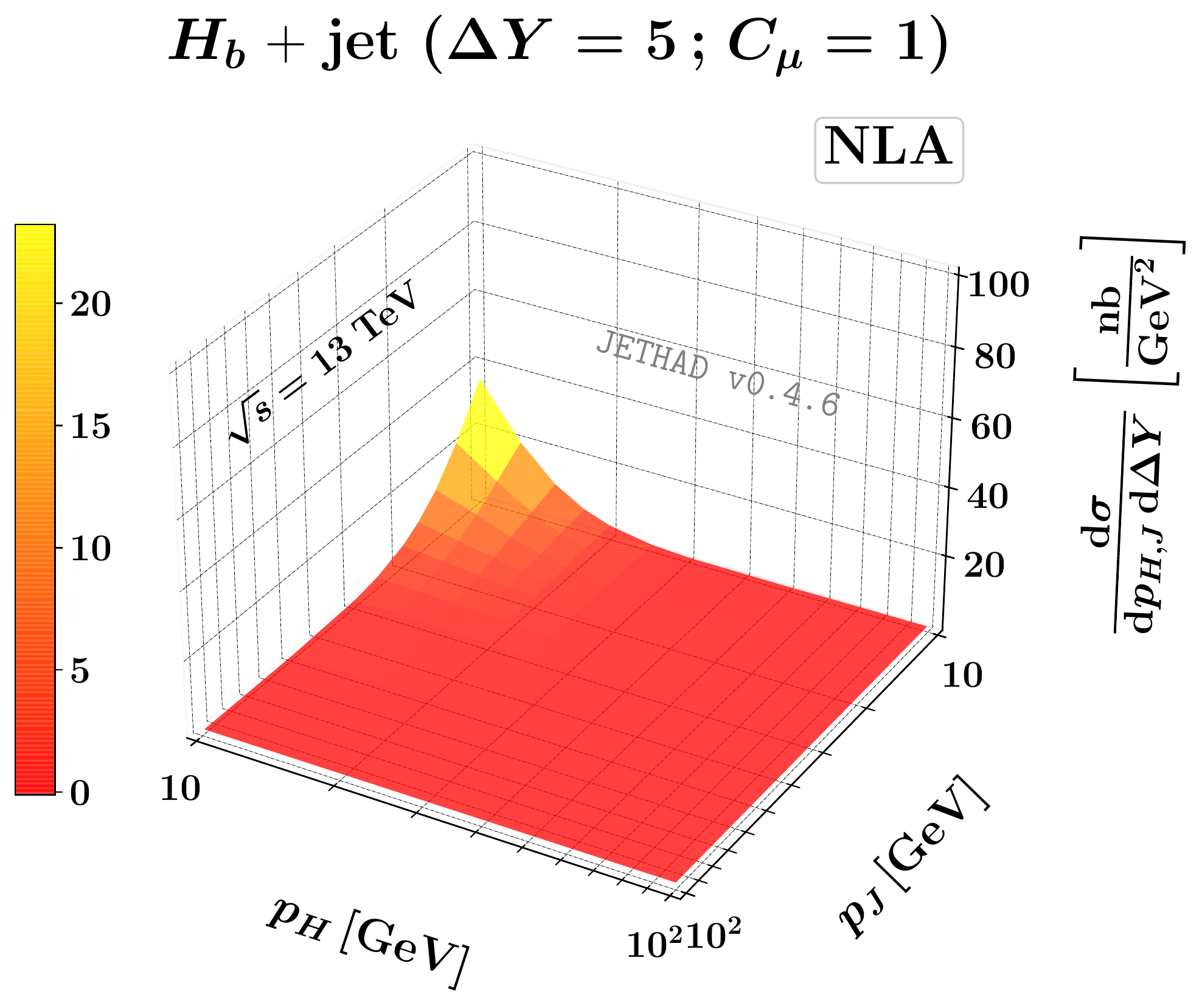}
	\vspace{0.15cm}
	
	\includegraphics[scale=0.41,clip]{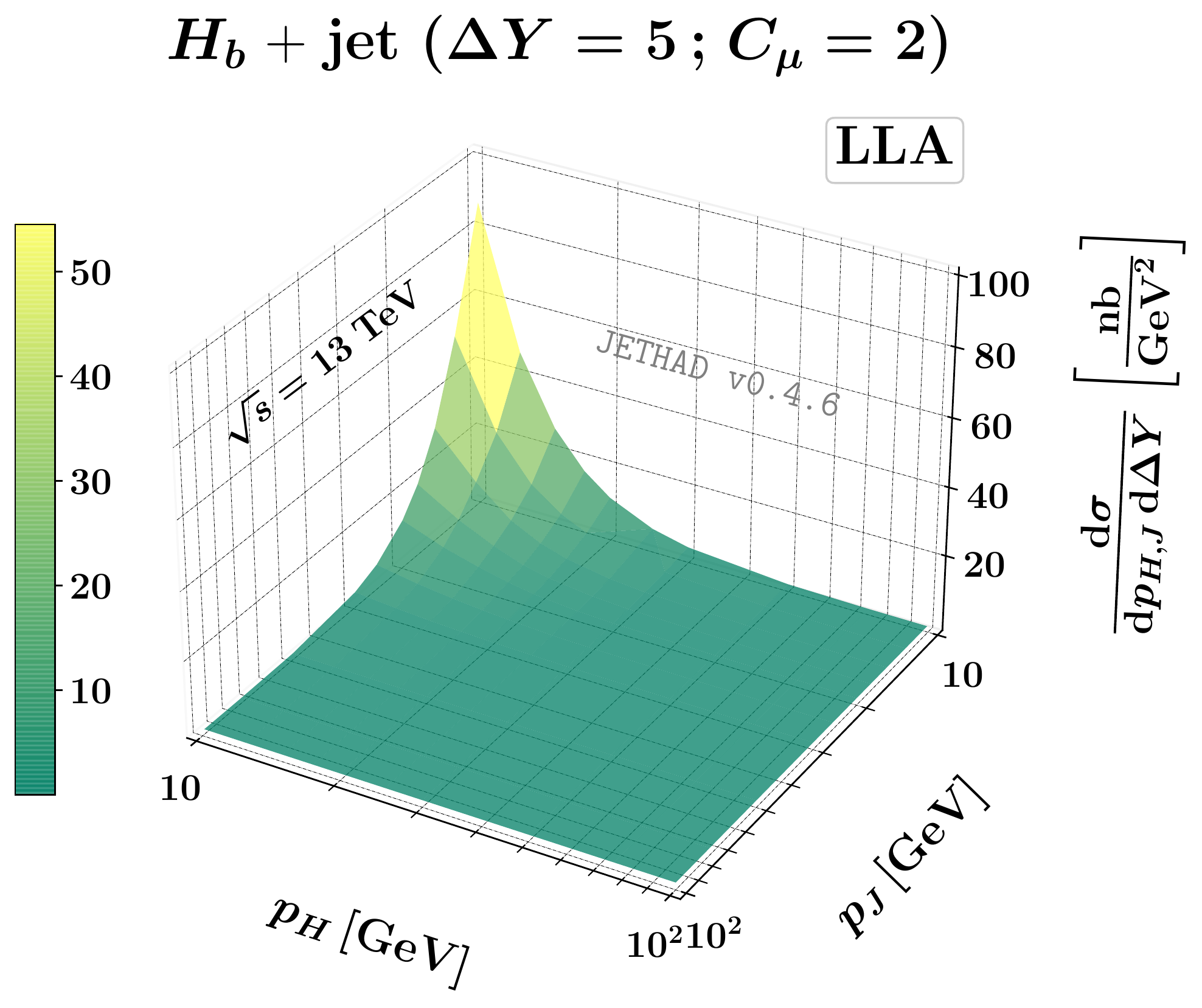}
	\hspace{0.25cm}
	\includegraphics[scale=0.41,clip]{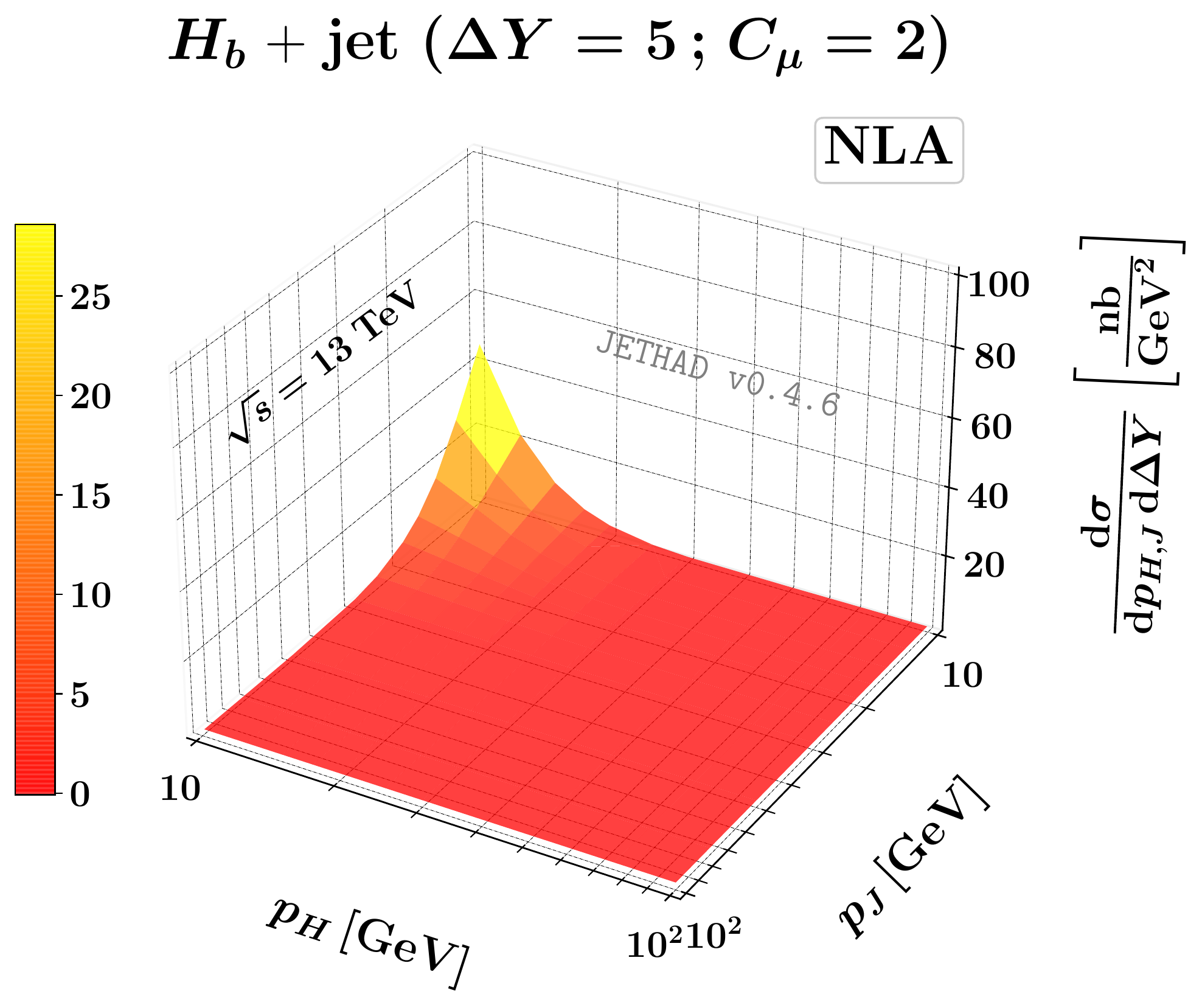}
	\vspace{0.15cm}
	
	\caption{Double differential $p_T$-distribution for the $H_b$~$+$~jet channel at $\DY=5$, $\sqrt{s} = 13$ TeV, and in the LLA (left) and NLA (right) resummation accuracy 
		%Calculations are done at natural scales, and the $C_\mu$ parameter is in the range 1/2 to 2 (from top to bottom)
		.}
	\label{fig:Y5-2pT0}
\end{figure}

In Figs~\ref{fig:Y3-2pT0} and~\ref{fig:Y5-2pT0} we report $p_T$-distributions results  in $H_b +$ jet channel at $\Delta Y = $ 3 and 5, without   employing any scale optimization. We observe that results fall rapidly at larger values of transverse momenta, $ |\vec{p}_H |$ and $|\vec{p}_J |$. Here NLA predictions (right panels ) are lower than LLA ones (left panels) as expected.  The scale variation effect shows pertinent with respect to what happens for the rapidity distribution and azimuthal correlations.

Tables\tref{tab:Y3-2pT0} and\tref{tab:Y5-2pT0} give us explicit quantitative information, by presenting numerical values of our distributions for a representative sample of $(|\vec p_H|, |\vec p_J|)$ pairs.
We observe that moving away from the symmetric $p_T$-region, $|\vec p_H| \simeq |\vec p_J|$, the sensitivity on scale variation of all the predictions increases.  
Furthermore, for almost all the considered $p_T$-pairs in Tables\tref{tab:Y3-2pT0} and\tref{tab:Y5-2pT0}, LLA results fall off when the $C_\mu$ scale parameter increase, while NLA prediction tend to oscillate around $C_\mu = 1$, which appears to act as a critical point for them.

This clearly indicates that our distributions are more stable on scale variation when higher-order corrections are included.
At the same time for $\Delta Y = 3$~(Table\tref{tab:Y3-2pT0}), their sensitivity on $C_\mu$ is almost of the same order (up to 45\%) for both LLA and NLA cases, while it is roughly halved when passing from LLA (up to 50\%) to NLA (up to 25\%) for $\Delta Y = 5$~(Table\tref{tab:Y5-2pT0}). 
This reflects the fact that the stabilizing effect of higher-order corrections is more pronounced when we go through the BFKL-sensitive region, \emph{i.e.} when $\Delta Y$ grows.  
Finally, we observe that our distributions are much smaller when $|\vec p_H| > |\vec p_J|$ than when $|\vec p_H| < |\vec p_J|$. Actually, it becomes more and more difficult to produce a $b$-flavored bound state than a light jet when the transverse momentum grows.

 \begin{table}[h]
 	\centering
 	\caption{Representative values of the double differential $p_T$-distribution [nb/GeV$^2$] for the $H_b$~$+$~jet channel, at $\DY=3$ and $\sqrt{s} = 13$ TeV.}
 	\label{tab:Y3-2pT0}
 	\scriptsize
 	\begin{tabular}{r|r|llllll}
 		%\hline\noalign{\smallskip}
 		\toprule
 		$|\vec p_H|$ [GeV] &
 		$|\vec p_J|$ [GeV] &
 		$\tarr c \rm LLA \\ C_\mu = 1/2 \earr$ & 
 		$\tarr c \rm LLA \\ C_\mu = 1 \earr$ &
 		$\tarr c \rm LLA \\ C_\mu = 2 \earr$ &
 		$\tarr c \rm NLA \\ C_\mu = 1/2 \earr$ &
 		$\tarr c \rm NLA \\ C_\mu = 1 \earr$ & 
 		$\tarr c \rm NLA \\ C_\mu = 2 \earr$ \\
 		\midrule
 		12.5 & 12.5 &  54.505(11) & 110.673(20) & 119.483(22) & 59.87(26) & 96.16(11) & 107.02(21) \\
 		20 & 20 &  9.5523(18) & 10.8676(20) & 9.9592(17) & 9.679(11) & 9.845(16) & 9.604(28) \\
 		20 & 30 &  4.2318(11) & 4.7487(10) & 4.5908(10) & 3.516(11) & 3.564(12) & 3.816(13) \\
 		30 & 20 &  2.45567(87) & 2.35844(71) & 1.97517(80) &  1.404(19) & 1.028(19) & 0.851(20) \\       
 		30 & 30 &  1.31402(33) & 1.22764(28) & 1.03082(27) & 1.2051(14) & 1.0995(28) & 1.0184(38) \\
 		30 & 50 &  0.37493(11) & 0.357939(83) & 0.31849(10) & 0.2780(11) & 0.2648(12) & 0.2706(12) \\
 		50 & 30 &  0.20170(11) & 0.166163(72) & 0.129175(83) & 0.0481(18) & 0.0326(14) & 0.0309(14) \\
 		50 & 50 &  0.079078(22) & 0.064566(20) & 0.050412(21) & 0.06287(15) & 0.05511(21) & 0.04978(25) \\
 		75 & 75 &  0.0069684(20) & 0.0053265(11) & 0.0039949(12) & 0.00489(19) & 0.004310(22) & 0.003885(23) \\
 		\bottomrule
 	\end{tabular}
 \end{table}
 
 \begin{table}[h]
 	\centering
 	\caption{Representative values of the double differential $p_T$-distribution [nb/GeV$^2$] for the $H_b$~$+$~jet channel, at $\DY=5$ and $\sqrt{s} = 13$ TeV.}
 	\label{tab:Y5-2pT0}
 	\scriptsize
 	\begin{tabular}{r|r|llllll}
 		%\hline\noalign{\smallskip}
 		\toprule
 		$|\vec p_H|$ [GeV] &
 		$|\vec p_J|$ [GeV] &
 		$\tarr c \rm LLA \\ C_\mu = 1/2 \earr$ & 
 		$\tarr c \rm LLA \\ C_\mu = 1 \earr$ &
 		$\tarr c \rm LLA \\ C_\mu = 2 \earr$ &
 		$\tarr c \rm NLA \\ C_\mu = 1/2 \earr$ &
 		$\tarr c \rm NLA \\ C_\mu = 1 \earr$ & 
 		$\tarr c \rm NLA \\ C_\mu = 2 \earr$ \\
 		\midrule
 		12.5 & 12.5 &  23.0020(29) & 36.3534(77) & 33.2496(52) & 11.979(44) & 15.952(19) & 18.320(25) \\
 		20 & 20 &  3.01358(59) & 2.82293(18) & 2.25985(22) & 1.4016(17) & 1.3654(18) & 1.4063(26) \\
 		20 & 30 &  1.02761(16) & 0.91503(13) & 0.74331(11) & 0.44181(85) & 0.43683(93) & 0.47148(95) \\
 		30 & 20 &  0.89192(16) & 0.721321(96) & 0.530710(87) &  0.2406(21) & 0.2053(20) & 0.2079(21) \\       
 		30 & 30 &  0.322595(57) & 0.256804(29) & 0.19209(3) & 0.13595(17) & 0.12690(23) & 0.12645(29) \\
 		30 & 50 &  0.0671327(66) & 0.052709(11) & 0.0402854(42) & 0.024583(87) & 0.024557(86) & 0.026155(76) \\
 		50 & 30 &  0.056804(16) & 0.040257(10) & 0.0277934(41) & 0.00681(16) & 0.00685(15) & 0.00820(15) \\
 		50 & 50 &  0.0136641(35) & 0.009805(20) & 0.0069481(15) & 0.004873(15) & 0.004708(15) & 0.004682(14) \\
 		75 & 75 &  0.000847907(94) & 0.00057898(48) & 0.000397990(32) & 0.0002501(15) & 0.0002620(12) & 0.0002661(10) \\
 		\bottomrule
 	\end{tabular}
 \end{table}
%===============================
\subsection{Summary}
In this proposed probes, motivated by looking for signals of stabilization of the high-energy resummation under higher-order corrections and under scale variation, we found that these effects are present and allow for the description or BFKL-sensitive observables at natural scales, moreover, study azimuthal moments at natural scales also when jet emissions are allowed, which is a novel feature which corroborates the statement that heavy-flavored emissions of bound states act as fair stabilizers of the high-energy series.
%===============================
\end{spacing}

\chapter{Conclusions and Outlook}\label{chapter4}
%===============================
\section{Conclusions} 
%===============================
\begin{spacing}{1.5}
%Tracing the path toward performing precision calculations via BFKL resummation of High-energy, 
%in this thesis we tracked and discussed the phenomenological analyses for distinct inclusive processes highlighting the issue of instabilities under higher-order corrections and energy-scales variations, that would aborted any possibility to investigate semi-hard reactions with high-precision at natural energy-scales. At the same time we presented a new proposed reactions acted as fair stabilizers of the high-energy series. 
We pursued the goal of following the phenomenological paths toward hunting signals of stabilization of the high-energy resummation under higher-order corrections and energy-scale variation, in order to perform a precision calculations via BFKL formalism. The BFKL theoretical key ingredients have been introduced in chapter~\ref{chapter1}.
Chapter~\ref{chapter2} was devoted to describe a proposed candidate probe of BFKL dynamics
at the LHC in the channel of inclusive production of an identified
charged light hadron and a jet, separated by a large rapidity gap. We have given some general arguments that this reaction, though being a simple hybridization of two already well studied ones, presents some own features which can make it worthy of consideration in future
analyses at the LHC. In view of that, we gave some theoretical predictions, with
next-to-leading accuracy, for the cross section averaged over the azimuthal-angle between the identified jet and hadron and for ratios of the azimuthal coefficients, employing the BLM optimization approach for fixing the renormalization scale $\mu_R$. 
The trends observed in the distributions over the rapidity interval between
the jet and the hadron are not different from the cases of Mueller-Navelet
jets~(see,~Refs.\cite{Colferai:2010wu,Caporale:2012ih,Ducloue:2013hia,Ducloue:2013bva,Caporale:2013uva,Caporale:2014gpa,Colferai:2015zfa,Caporale:2015uva,Ducloue:2015jba,Celiberto:2015yba,Celiberto:2015mpa,Celiberto:2016ygs,Celiberto:2016vva,Caporale:2018qnm}), and hadron-hadron~\cite{Celiberto_2016,Celiberto_2017,Celiberto_2017s}, when the jet is detected by CMS, and due to the different nature of the objects produced in this channel, asymmetric cuts were naturally considered, thus enhancing the BFKL effects, whereas some new features have appeared when the jet is seen by CASTOR, which deserve further investigation.  

We have followed in chapter~\ref{chapter3} the new direction toward hunting signals for the stability of the BFKL series at natural scales; section~\ref{sec:higgs-jet} was dictated to investigate the inclusive hadroproduction of a Higgs boson and of a jet featuring high transverse momenta and separated by a large rapidity distance. In this process the large transverse mass of the emitted Higgs boson suppresses the higher-order corrections, thus affording us a peerless opportunity to study energy scales around their natural values. Furthermore, we have extended this study to distributions differential in the Higgs transverse
momentum, providing evidence that a high-energy treatment is valid and can be afforded in the region where Higgs $p_T$ and the jet one are of the similar size, and beyond, the description for Higgs momentum distribution would be relied on a multi-lateral resummation formalism unifying different approaches. The stability effects that allow for the description or BFKL-sensitive observables at natural scales have been spotted in section~\ref{sec:bflavour}, where we have proposed the inclusive emission, in proton-proton collisions, of a forward bottom-flavored hadron accompanied by another backward bottom-flavored hadron or a backward light-flavored jet in semi-hard regimes that can be studied at current LHC energies. In the $H_b$ + jet channel it was possible to study azimuthal moments at natural scales
although previous studies on Mueller–Navelet dijet or ligther-hadron + jet production have shown that instabilities emerge when jet emissions are studied at natural scales, so strong to prevent any realistic analysis. All that would support the recent assumption, that heavy-flavored emissions of bound states act as fair stabilizers of the high-energy series. 
Here, the stability of our predictions motivates our interest in proposing the hybrid high-energy and collinear factorization as an additional tool to improve the fixed-order description. Moreover we believe that future, exhaustive studies of the inclusive exclusive observables, such as the double differential transverse-momentum distributions, would benefit from the inclusion of high-energy effects in a unified formalism where distinct resummations are concurrently embodied.
%===============================
\section{Outlook}
%===============================
% New BFKL-sensitive observables and more exclusive final-state reactions are needed, 
Future studies are needed to unveil the connection between the sensitivity of some observable, such as $R_{n0}$ ratio, on scale variation and other potential sources of uncertainty, as the jet algorithm selection. For the Higgs + jet program, an obvious extension consists in the full NLA BFKL analysis,
including a NLO jet impact factor and the NLO Higgs impact factor, the latter one was calculated quite recently in the large top-mass
limit, and a full calculation in the so-called ($\nu, n$)-representation (projected on the eigenfunctions of the BFKL kernel)
which can be easily implemented in numerical calculations, is underway.
The next step in our program of investigating semi-hard phenomenology relies on a two-fold strategy:
first, we plan to compare observables sensitive to heavy-flavor production in regimes where
either the light-flavored quarks and gluons are accounted for by proton collinear parton
densities (PDFs) or in case when all flavors are present in the initial state and are taken massless, and possibly do a match between the two
descriptions. The inclusion of quarkonium production channels will certainly enrich our phenomenology. The second step is extending our studies on heavy flavor by considering wider kinematic ranges. 
\end{spacing}
\begin{appendices}
	\renewcommand{\thesubsection}{\Alph{subsection}}
	\chapter{Forward Jet, Hadron, and Higgs impact factors}\label{A}
	In this Appendix the expressions for the LO and NLO corrections to both jet and an identified hadron impact factors are given. 
%=========================================================
	\section{Jet impact factor}\label{A.1}
%=========================================================
In this subsection the expressions for the LO forward jet impact factor and the NLO correction in the small-cone limit are given, (for more details about the latter see Ref.~\cite{Ivanov:2012ms})\\
For the jet vertex at LO it reads:
\begin{equation}
c_{1}(n,\nu,|\overrightarrow{k}|,x)=2\sqrt{\frac{C_{F}}{C_{A}}}(\overrightarrow{k}^{2})^{-i\nu-1/2}\left \{  \frac{C_{A}}{C_{F}}f_{g}(x)+\sum_{a=q,\bar{q}} f_{a}(x) \right \},
\end{equation}
where the functions $f(x)$ are the parton distribution functions (PDFs), and $C_{A}=N$, $C_{F}=(N^{2}-1/2N)$.
\begin{equation}
c_{2}(n,\nu,|\overrightarrow{k}|,x)=\bigg[c_{1}(n,\nu,|\overrightarrow{k}|,x)\bigg]^{*}.
\end{equation}
The NLO correction to the forward jet impact factor in the small-cone limit\\ $c_{2}^{(1)}(n,\nu,|\overrightarrow{k}|,x)$ has following expression
\begin{equation}\label{a3}
\begin{split}
c_{1}^{(1)}(n,\nu,|\overrightarrow{k}|,x)=\frac{1}{\pi}\sqrt{\frac{C_{F}}{C_{A}}}(\overrightarrow{k})^{i\nu-1/2}\int^{1}_{x}\frac{d\zeta}{\zeta}\zeta^{-\bar{\alpha}_{s}(\mu_{R})\chi(n, \nu)}\Bigg\{\sum_{\alpha=q,\bar{q}}f_{a}(\frac{x}{\zeta})\\
\times
\bigg[\bigg(P_{qq}(\zeta)+\frac{C_{A}}{C_{F}}P_{gq}(\zeta)\bigg)\ln\frac{\overrightarrow{k}^{2}}{\mu_{F}^{2}}-2\zeta^{-2\gamma}\ln R(P_{qq}(\zeta)+P_{gq}(\zeta))-\frac{\beta_{0}}{2}\ln\frac{\overrightarrow{k}^{2}}{\mu_{R}^{2}}\delta(1-\zeta)\\
+C_{A}\delta(1-\zeta)\bigg(\chi(n,\gamma)\ln\frac{s_{0}}{\overrightarrow{k}^{2}}+\frac{85}{18}+\frac{\pi^{2}}{2}+\frac{1}{2}\bigg(\psi^{\prime}(1+\gamma+\frac{n}{2})-\psi^{\prime}(\frac{n}{2}-\gamma)-\chi^{2}(n, \gamma)\bigg)\bigg)\\
+(1+\zeta^{2})\bigg[C_{A}\bigg(\frac{1+\zeta^{-2\gamma}\chi(n, \gamma)}{2(1-\zeta)_{+}}-\zeta^{-2\gamma}\bigg(\frac{\ln(1-\zeta)}{1-\zeta}\bigg)_{+}\bigg)+(C_{F}-\frac{C_{A}}{2})\bigg[\frac{\bar{\zeta}}{\zeta^{2}}I_{2}-\frac{2\ln\zeta}{1-\zeta}\\
+2\bigg(\frac{\ln(1-\zeta)}{1-\zeta}\bigg)_{+}\bigg]\bigg]+\delta(1-\zeta)\big(C_{F}(3\ln2-\frac{\pi^{2}}{3}-\frac{9}{2})-\frac{5n_{f}}{9}\big)\\
C_{A}\zeta+C_{F}\bar{\zeta}+\frac{1+\bar{\zeta}^{2}}{\zeta}\bigg(C_{A}\frac{\bar{\zeta}}{\zeta}I_{1}+2C_{A}\ln\frac{\bar{\zeta}}{\zeta}+C_{F}\zeta^{-2\gamma}(\chi(n, \gamma)-2\ln\bar{\zeta})\bigg)\bigg]+f_{g}\bigg(\frac{x}{\zeta}\bigg)\frac{C_{A}}{C_{F}}\\
\times \bigg[\bigg(P_{gg}(\zeta)+2n_{f}\frac{C_{A}}{C_{F}}P_{qg}(\zeta)\bigg)\ln\frac{\overrightarrow{k}^{2}}{\mu^{2}_{F}}-2\zeta^{-2\gamma}\ln R(P_{gg}(\zeta)+2n_{f}P_{qg}(\zeta))-\frac{\beta_{0}}{2}\ln\frac{\overrightarrow{k}^{2}}{4\mu_{R}^{2}}\delta(1-\zeta)\\
+C_{A}\delta(1-\zeta)\bigg(\chi(n,\gamma)\ln\frac{s_{0}}{\overrightarrow{k}^{2}}+\frac{1}{12}+\frac{\pi^{2}}{6}+\frac{1}{2}\bigg(\psi^{\prime}(1+\gamma+\frac{n}{2})-\psi^{\prime}(\frac{n}{2}-\gamma)-\chi^{2}(n, \gamma)\bigg)\bigg)\\
+2C_{A}(1-\zeta^{-2\gamma})\bigg(\bigg(\frac{1}{\zeta}-2+\zeta\bar{\zeta}\bigg)\ln\bar{\zeta}+\frac{(1-\zeta)}{1-\zeta}\bigg)\\
+C_{A}\bigg[\frac{1}{\zeta}+\frac{1}{(1-\zeta)_{+}}-2\zeta\bar{\zeta}\bigg]\bigg((1+\zeta^{-2\gamma})\chi(n, \gamma)-2\ln\zeta+\frac{\bar{\zeta}^{2}}{\zeta^{2}}I_{2}\bigg)\\
+n_{f}\bigg[2\zeta\bar{\zeta}\frac{C_{F}}{C_{A}}+(\zeta^{2}+\bar{\zeta}^{2})\bigg(\frac{C_{F}}{C_{A}}\chi(n, \gamma)+\frac{\bar{\zeta}}{\zeta}I_{3}\bigg)-\frac{1}{12}\delta(1-\zeta)\bigg]\bigg]\Bigg\},
\end{split}
\end{equation}
\begin{equation}
c_{2}^{(1)}(n,\nu,|\overrightarrow{k}|,x)=\bigg[c_{1}^{(1)}(n,\nu,|\overrightarrow{k}|,x)\bigg]
\end{equation}
where $\bar{\zeta}=1-\zeta$, $\gamma=i\nu-1/2$, and $P_{ij}(\zeta)$ are the LO DGLAP kernels defined as follows:\\
$$P_{qq}(\zeta)=C_{F}\bigg(\frac{1+\zeta^{2}}{(1-\zeta)_{+}}+\frac{3}{2}\delta(1-\zeta)\bigg),$$
$$P_{gq}(\zeta)=C_{F}\frac{1+(1-\zeta)^{2}}{\zeta},$$
$$P_{gg}(\zeta)=2C_{A}\bigg\{\frac{1}{(1-\zeta)_{+}}+\frac{1}{\zeta}-2+\zeta(1-\zeta)+\bigg(\frac{11}{6}C_{A}-\frac{n_{f}}{3}\bigg)\delta(1-\zeta)\bigg\},$$
$$P_{qg}(\zeta)=T_{R}\bigg(\zeta^{2}+(1-\zeta)^{2}\bigg),$$
where $C_{F}$ is the Casimir operator associated with gluon emission from a quark, $C_{F}=(N^{2}_{c}-1)/2N_{c}$ and $T_{R} = 1/2$ is the colour factor associated with the splitting of a gluon into a quark-antiquark pair.\\
For the $I_{1,2,3}$ functions we have the results:
\begin{multline}
I_{2}=\frac{\zeta^{2}}{\bar{\zeta^{2}}}\bigg\{\zeta\left(\frac{2{F_{1}(1,1+\gamma-\frac{n}{2},2+\gamma-\frac{n}{2},\zeta)}}{\frac{n}{2}-\gamma-1}-\frac{2F_{1}(1,1+\gamma+\frac{n}{2},2+\gamma+\frac{n}{2},\zeta)}{\frac{n}{2}+\gamma+1}\right)
\\
+\zeta^{-2\gamma}\left(\frac{2{F_{1}(1,-\gamma-\frac{n}{2},1-\gamma-\frac{n}{2},\zeta)}}{\frac{n}{2}+\gamma}-\frac{2F_{1}(1,-\gamma+\frac{n}{2},1-\gamma+\frac{n}{2},\zeta)}{\frac{n}{2}-\gamma}\right)\\
+(1+\zeta^{-2\gamma})(\chi(n, \gamma)-2\ln\bar{\zeta})+2\ln\zeta\bigg\},
\end{multline}
\begin{equation}
I_{1}=\frac{\bar{\zeta}}{2\zeta}I_{2}+\frac{\zeta}{\bar{\zeta}}\left[\ln\zeta+\frac{1-\zeta^{-2\gamma}}{2}(\chi(n, \gamma)-2\ln\bar{\zeta})\right],
\end{equation}
\begin{equation}
I_{3}=\frac{\bar{\zeta}}{2\zeta}I_{2}-\frac{\zeta}{\bar{\zeta}}\left[\ln\zeta+\frac{1-\zeta^{-2\gamma}}{2}(\chi(n, \gamma)-2\ln\bar{\zeta})\right].
\end{equation}
7)
Using the following property of the hypergeometric function,
\begin{equation}
2F_{1}(1,a+1,\zeta)=a\sum_{n=0}^{\infty}\frac{(a)_{n}}{n!}\left[\psi(n+1)-\psi(a+n)-\ln\bar{\zeta}\right],
\end{equation}
one can easily see that $\zeta\rightarrow1$
\begin{equation}
I_{2}=O(\ln\bar{\zeta}), \qquad I_{1}=O(\ln\bar{\zeta}), \qquad I_{3}=O(\ln\bar{\zeta}),
\end{equation}
which implies that the integral over $\zeta$ in (\ref{a3}) is convergent on the upper limit.
%=========================================================
\section{Hadron impact factor}\label{A.2}
%=========================================================
Here the expressions for the LO and NLO to the identified hadron impact factor in $\nu$ -representation is given (see Ref. \cite{Ivanov:2012iv} for further details).\\
The LO impact factor case, written in expression contains the PDFs of the gluon and of the different quark/anti-quark flavors in the proton, and the FFs of the detected hadron, reads:
\begin{multline}
c_{1}(n,\nu,|\overrightarrow{k}|,\alpha)=2\sqrt{\frac{C_{F}}{C_{A}}}(\overrightarrow{k}_{1}^{2})^{i\nu-1/2}\int_{\alpha}^{1}\frac{dx}{x}\bigg(\frac{x}{\alpha}\bigg)^{2i\nu-1}\\
\times\left \{  \frac{C_{A}}{C_{F}}f_{g}(x)D_{g}^{h}(\frac{\alpha}{x})+\sum_{a=q,\bar{q}} f_{a}(x)D_{a}^{h}(\frac{\alpha}{x}) \right \},
\end{multline}
where $D^{h}_{i}$ are parton fragmentation functions.
\begin{equation}
c_{2}(n,\nu,|\overrightarrow{k}|,\alpha)=\bigg[c_{1}(n,\nu,|\overrightarrow{k}|,\alpha)\bigg]^{*}
\end{equation}
The NLO impact factor corrections in $(\nu, n)$-representation $c_{1}^{(1)}(n,\nu,|\overrightarrow{k}|,\alpha)$, have the form
\begin{multline}
c_{1}^{(1)}(n,\nu,|\overrightarrow{k}|,\alpha)=2\sqrt{\frac{C_{F}}{C_{A}}}(\overrightarrow{k}^{2})^{i\nu-1/2}\frac{1}{2\pi}\int_{\alpha}^{1}\frac{dx}{x}\int_{\frac{\alpha}{x}}^{1}\frac{d\zeta}{\zeta}\left(\frac{x\zeta}{\alpha}\right)^{2i\nu-1}\\
\times \bigg\{\frac{C_{A}}{C_{F}}f_{g}(x)D_{g}^{h}(\frac{\alpha}{x\zeta})C_{gg}(x, \zeta)+\sum_{a=q,\bar{q}} f_{a}(x)D_{a}^{h}(\frac{\alpha}{x\zeta})C_{qq}(x, \zeta)\\
+D_{g}^{h}(\frac{\alpha}{x\zeta})\sum_{a=q,\bar{q}} f_{a}(x)C_{qg}(x, \zeta)+\frac{C_{A}}{C_{F}}f_{g}(x)\sum_{a=q,\bar{q}}D_{g}^{h}(\frac{\alpha}{x\zeta})C_{gq}(x, \zeta)\bigg\}.
\end{multline}
The expressions for the coefficient functions $C_{gg}, C_{qq}, C_{qg} and C_{gq}$, are given in Ref \cite{Ivanov:2012iv}.
 
 %=========================================================
 \section{Higgs impact factor}\label{A.3}
 %=========================================================
 \begin{figure*}[hbt!]\label{F1}
 	\begin{center}
 		\begin{tikzpicture}
 		\begin{feynman}[/tikzfeynman/large]	
 		\vertex (a);
 		\vertex [below right=of a] (b);
 		\vertex [below right=of b] (e);
 		\vertex [right=of e] (f);
 		
 		\vertex [right=of f] (F);
 		
 		\vertex [above right=of F] (B);
 		\vertex [above right=of B] (A);	
 		\vertex [below =of b] (c);
 		\vertex [below =of B] (C);
 		\vertex [below =of c] (d){\(\nu,b\)};
 		\vertex [below =of C] (D);	
 		
 		\diagram* [large] {
 			(a)  -- [gluon, momentum={\(k=x_{1}p_{1}\)},edge label'=\(\epsilon^{\mu}\)] (b) -- [fermion, momentum=\(l+k\)] (e) -- [fermion,momentum=\(l-q\)] (c)  -- [fermion,momentum=\(l\)] (b) ,
 			(e) --[red, charged scalar,momentum'=\(k+q\)](f),
 			%				(x) -- [momentum=\(q\)] (y),
 			(F) --[fermion](B) --[fermion](C) --[fermion](F)--[red,charged scalar](f),
 			%				(D) -- [rmomentum] (C),
 			(A)  -- [gluon] (B)
 		};
 		\vertex [above=0.4em of a] {\(\mu,a\)};
 		\vertex [below left=1.2em of c] {\(\epsilon^{\nu}=p^{\nu}_{2}/s\)};
 		\vertex [above right=1.5em of d] (x);
 		\vertex [below =0.2em of x] (y);
 		\diagram*{
 			%		(x)--[momentum=\(q\)](y),	
 		};		
 		%% Find the midpoint, which is halfway between f and F.
 		\coordinate (midpoint) at ($(f)!0.05!(F)$);
 		%% Draw a line starting 3 units above the midpoint, and ending 3 units below
 		%% the midpoint.
 		\draw [dashed] ($(midpoint) + (0, 3)$) -- ($(midpoint) + (0, -3)$);
 		\coordinate (point) at ($(d)!0.05!(c)$);
 		\draw[decoration = {zigzag,segment length = 3mm, amplitude = 1mm},decorate] (d)--(c);
 		\coordinate (point) at ($(d)!0.05!(c)$);
 		\draw[decoration = {zigzag,segment length = 3mm, amplitude = 1mm},decorate] (D)--(C);
 		\coordinate (xpoint) at ($(x)!0.5!(y)$);
 		\draw[->] (x)-- node[label=right:$q$] {}($(xpoint) + (0, +1)$);
 		\end{feynman}
 		\end{tikzpicture}
 		\caption{Optical theorem: Representative Feynman diagram for the process $gg^{*}\longrightarrow H$.}	
 	\end{center}
 \end{figure*}
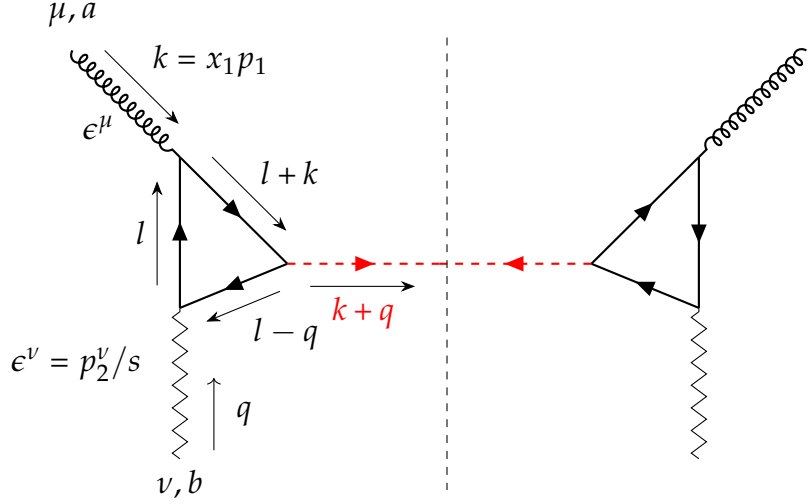
  We can define LO impact factor for the production of the Higgs in gluon-gluon fusion, as follow:
 \begin{equation}\label{38}
 V^{(0)}_{gg\rightarrow H}(\overrightarrow{q})=\sum_{\left\{f\right\}}\int  \frac{ds_{gH}}{2\pi}\overline{\lvert\mathcal{M}\rvert^{2}}dP.S.^{(f)}.
 \end{equation} 
 The integration here is over the standard phase space $dP.S.^{(f)}$ of an
 intermediate state $\{f\}$ as well as over its squared invariant mass $s_{gR}$. The LO impact factors in our case take contribution only from a one-particle intermediate state in the gluon-Reggeon collision, and then the phase space integration simply gives
 \begin{equation}\label{39}
 P.S^{(1)}=\int\frac{d^{4}P_{H}}{(2\pi)^{3}}\delta(P^{2}_{H}-M^{2}_{H})(2\pi)^{4}\delta(k+q-P_{H})= 2\pi\delta(s_{gR}-M^{2}_{H}),
 \end{equation} 
 where $P_{H}$ is the Higgs boson momentum, using this result we end up with  
 \begin{equation}\label{40}
 V^{(0)}_{gg\rightarrow H}(\overrightarrow{q})= \frac{\alpha^{2}_{s}}{v^{2}}\frac{1}{128\pi^{2}\sqrt{N^{2}_{c}-1}}|{{q}}^{2}_{\perp}||\mathcal{F}(|{{q}}^{2}_{\perp}|)|^{2},
 \end{equation}
 satisfying $V^{(0)}_{gg\rightarrow H}(\overrightarrow{q})|_{\overrightarrow{q}=0}\longrightarrow 0$, which will guarantee the infra-red finiteness of the BFKL amplitude. Applying the large top mass limit ($I_{1}\rightarrow \frac{1}{3}$), our LO impact factor reads
 \begin{equation}\label{41}
 V^{(0)}_{gg\rightarrow H}(\overrightarrow{q})= \frac{\alpha^{2}_{s}}{v^{2}}\frac{1}{72\pi^{2}\sqrt{N^{2}_{c}-1}}|{{q}}^{2}_{\perp}|,
 \end{equation}
 which is in agreement with what has been computed in section B of Ref \cite{DelDuca:2003ba}.\\
 Now after inclusion of PDF of the gluon, Eq.(\ref{40}) becomes
 \begin{equation}\label{42}
 V^{(0)}_{gg\rightarrow H}(\overrightarrow{q})= \frac{\alpha^{2}_{s}}{v^{2}}\frac{|\mathcal{F}(|{{q}}^{2}_{\perp}|)|^{2}}{128\pi^{2}\sqrt{N^{2}_{c}-1}}|{{q}}^{2}_{\perp}|\int_{0}^{1}dx_{1} f_{g}(x_{1})
 \end{equation}
 In order to establish the proper normalization for our impact factor, we insert into
 (\ref{42}) a delta function depends on the produced Higgs boson transverse
 momentum $\overrightarrow{P_{H}}$, then the LO result for the impact factor reads
 \begin{equation}\label{43}
 \frac{dV^{(0)}_{gg\rightarrow H}(\overrightarrow{q})}{\overrightarrow{q}^{2}}=
 \frac{\alpha^{2}_{s}}{v^{2}}\frac{|\mathcal{F}(|{{q}}^{2}_{\perp}|)|^{2}}{128\pi^{2}\sqrt{N^{2}_{c}-1}}|{{q}}^{2}_{\perp}|\int_{0}^{1}dx_{1} f_{g}(x_{1}) \frac{d^{2}\overrightarrow{P_{H}}}{\overrightarrow{P_{H}}^{2}}\delta^{(2)}(\overrightarrow{P_{H}}-\overrightarrow{q}).
 \end{equation}
 This will simplify the calculation of the impact factors when we project them into the so called ($\nu,n$)-representation,  i.e. the eigenfunctions
 of LO BFKL kernel. The transfer to ($\nu,n$)-space is done as follows:
 \begin{equation}\label{44}
 dV^{(0)}_{gg\rightarrow H}(\nu,n)=\int d^{2}\overrightarrow{q}\frac{dV^{(0)}_{gg\rightarrow H}(\overrightarrow{q})}{\overrightarrow{q}^{2}}\frac{\big(\overrightarrow{q}^{2}\big)^{i\nu-1/2}}{\pi\sqrt{2}}e^{in\phi}.
 \end{equation}
 Using (\ref{43}) and absorbing our inserted delta function we get the following differential expression for our LO impact factor in the azimuth symmetric $n=0$ sector
 \begin{equation}\label{45}
 \frac{dV^{(0)}_{gg\rightarrow H}(\nu,n=0)}{dx_{1}d^{2}\overrightarrow{P_{H}}}= \frac{\alpha^{2}_{s}}{v^{2}}\frac{|\mathcal{F}(|{{q}}^{2}_{\perp}|)|^{2}}{128\pi^{3}\sqrt{2(N^{2}_{c}-1)}} \big(\overrightarrow{P_{H}}^{2}\big)^{i\nu-1/2}f_{g}(x_{1}).
 \end{equation}

 \section*{Higher-order corrections}
Going to higher order, we have to calculate the virtual and real emission corrections, which both of them contain infrared divergences canceling each other, moreover, the real corrections could contain a collinear divergences, which can be handled by taken into account the Altarelli–Parisi splitting functions up to the appropriate order. The next-to-leading order calculations involve the computation of two-loop triangular diagrams with massive quarks, which complicate the process, but fortunately, the large quarks mass limit $(m_{t}\rightarrow \infty)$ allows us to replace the coupling of Higgs with the two gluon via the quarks loop by an effective vertex, which is help to reduce the number of loops appear in any given diagram by one, moreover it can serves as a valuable check of the complete two-loop calculations for the gluonic radiative corrections. The coupling vertex between the two gluons and the produced Higgs can be obtained from the following effective Lagrangian\footnote{Which can be derived from the top-quark contribution to gluon self-energy by taking the derivative of the gluon propagator with respect to the bare quark mass.},
\begin{equation}\label{eff}
\mathcal{L}_{eff}=-\frac{g_{H}}{4}H G^{a}_{\mu\nu}G^{\mu\nu}_{a},
\end{equation} 
where $G^{a}_{\mu\nu} $ is the gluon field strength tensor , $H$ is the Higgs filed, and the effective coupling $g_H$ given by 
$$g_H=\frac{\alpha_s}{3\pi v}\left(1+\alpha_s\frac{11}{4\pi}\right).$$
Using this effective Lagrangian we can generate all the building blocks vertices which involve  the two, three and four gluons with the Higgs.
 \section*{1-loop Virtual corrections to $gg^{*}\rightarrow H$}
 Before we proceed to the virtual correction, it will be useful to rewrite our previous LO (Born)-amplitude at large quarks mass limit in $(d=4-2\epsilon)$ dimensions, considering this, there is only one extra $\epsilon$ term appears due to averaging over the different possible polarization and color states of the incoming gluons,
 \[
 \mathcal{M}^{(B)}_{gg^{*}\rightarrow H}=ig_H \delta ^{a b} \left(g^{\mu  \nu } (k_1\cdot q)-k_1^\nu q^\mu\right)\frac{k_2^{\nu }}{s}\epsilon_{\mu}(k_1),
 \]
 \begin{equation}\label{eq1}
 |\mathcal{M}^{(B)}_{gg^{*}\rightarrow H}|^{2}=\frac{\alpha_{s}^{2}}{72\pi^{2}}\frac{|{{q}}^{2}_{\perp}|}{v^{2}}\frac{1}{(1-\epsilon).}
 \end{equation}
 \begin{figure*}[hbt!]\label{F1}
 	\begin{center}
 		\includegraphics[scale=0.9]{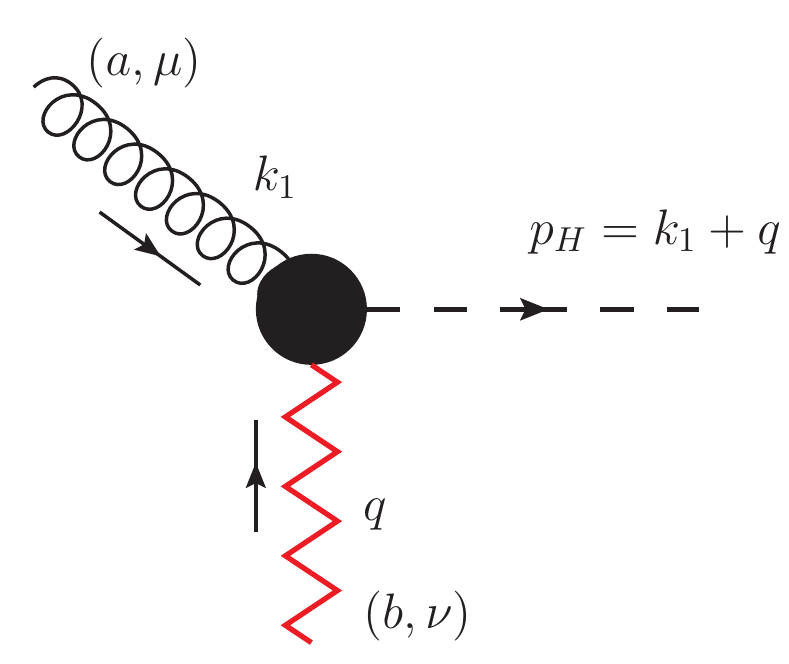}
 	\end{center}
 	\caption{Diagrammatic representation for $gg^{*}\rightarrow H$ tree level amplitude at HEFT}
 \end{figure*} 
 There are several diagrams appearing at one-loop, few of them giving null contributions, and the only non-zero contribution to $gg^*\rightarrow H$, comes from six different diagrams shown in Fig.\ref{F2}. These diagram could be classified according to the type of the correction into three contributed groups of diagrams, the first group represents the one-gluon exchange contribution, the second and third contributions are those from the Riggezed gluon loop correction and the two-gluon exchange diagrams respectively.
 
 \begin{figure*}[hbt!]\label{F2}
 	\begin{center}
 		\includegraphics[scale=0.3]{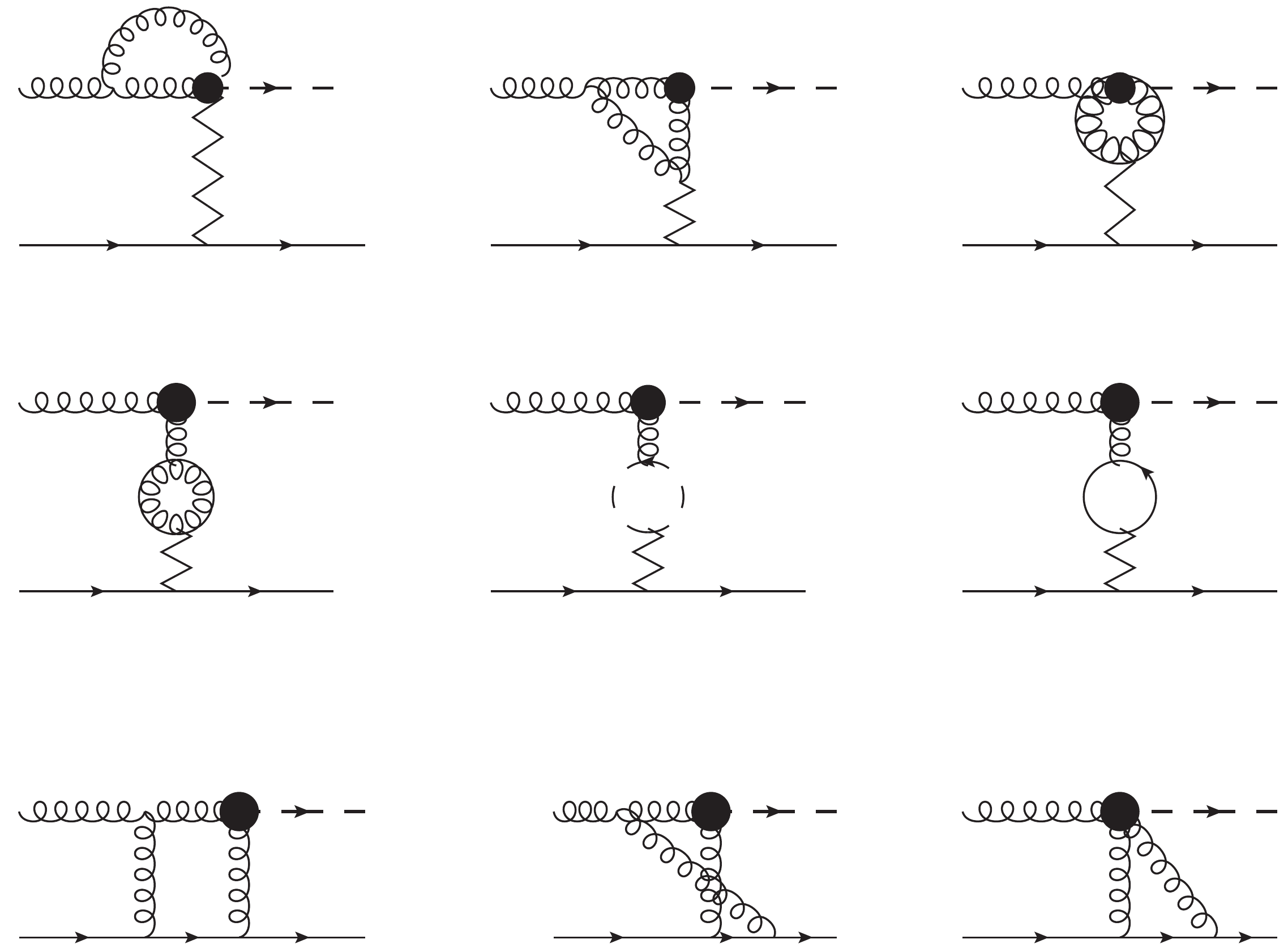}
 	\end{center}
 	\caption{Feynman diagrams giving non-zero contribution to $gg^{*}\rightarrow H$ at one-loop.}
 \end{figure*} 
\section*{Real correction}
The real correction contains two different  contribution comes from both gluon initiated processes, and only one initial quark process contribute to the total real correction.
we can simply read the Feynman amplitudes of our processes one by one directly from their corresponding diagrams. 
\section*{Quark initiated process}
\begin{figure*}[hbt!]
\begin{center}
	\includegraphics[scale=0.7]{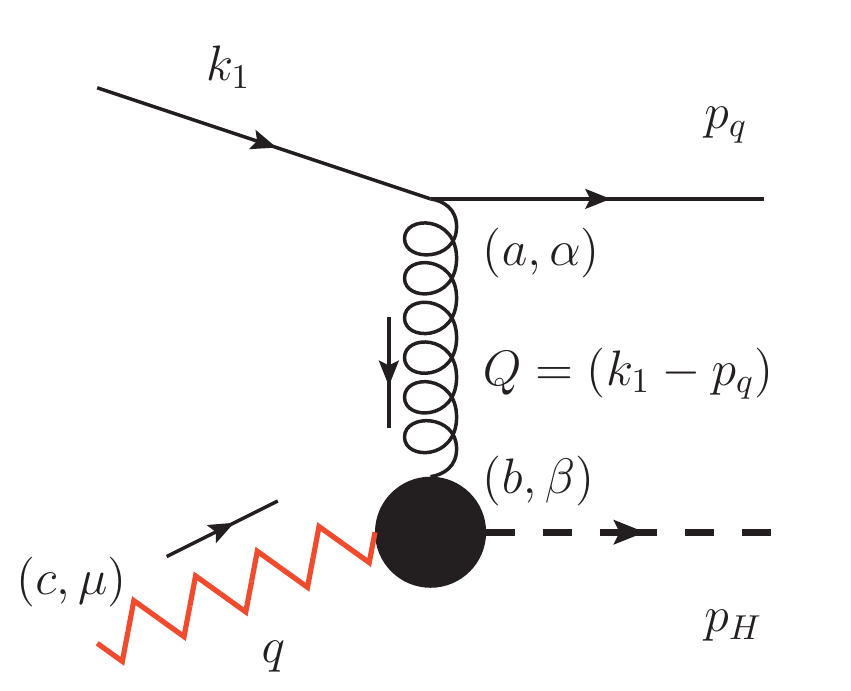}
	\caption{Feynman diagram for $qg^*\rightarrow Hq$ process.}
\end{center}
\end{figure*} 
\section*{Gluon initiated process}
\begin{figure*}[hbt!]
	\begin{center}
		\includegraphics[scale=0.5]{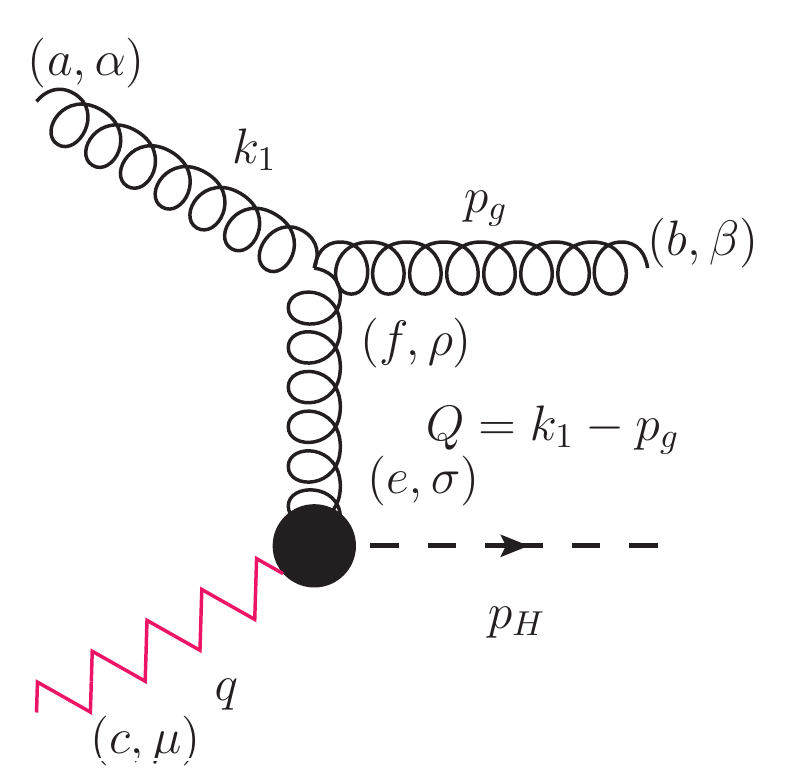}
		\includegraphics[scale=0.5]{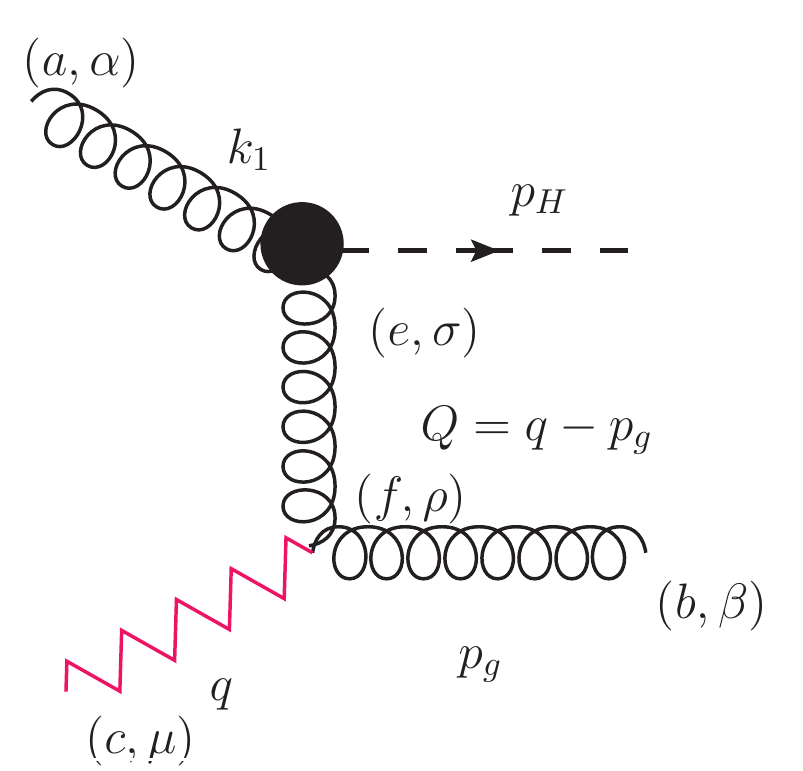}\\
		\includegraphics[scale=0.5]{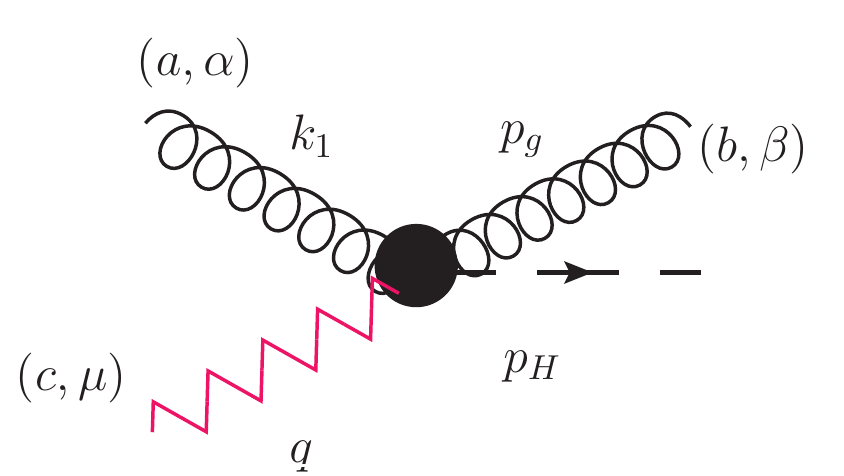}
		\includegraphics[scale=0.5]{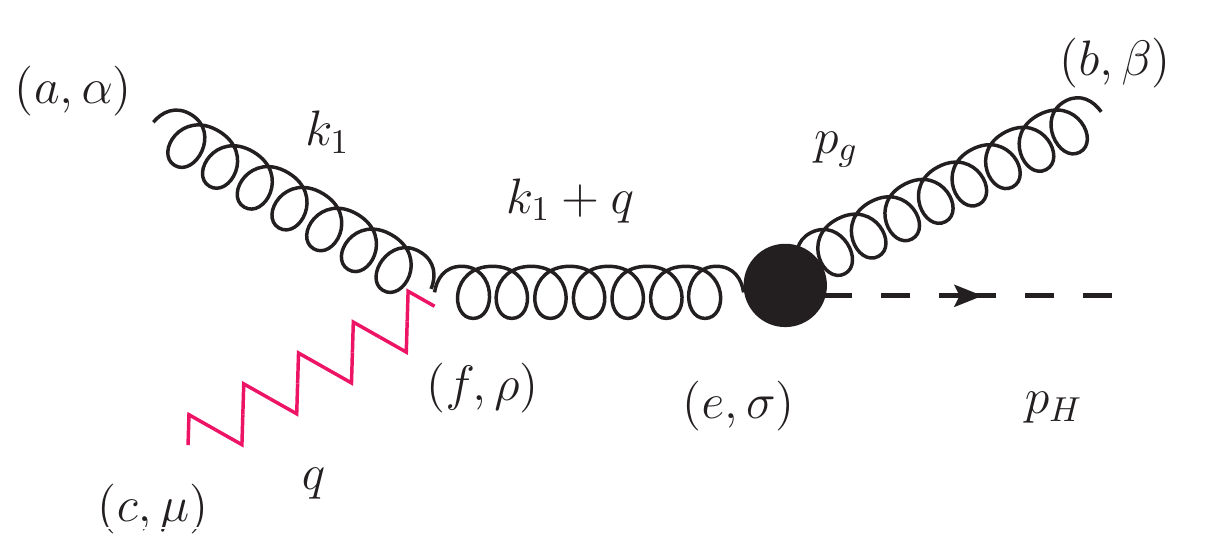}
		\caption{Feynman diagram for $gg^*\rightarrow Hq$ process.}
	\end{center}
\end{figure*} 
 
\end{appendices}

\cleardoublepage
\phantomsection
\addcontentsline{toc}{chapter}{Bibliography}
\bibliographystyle{SP_bibstyle}
\bibliography{References.bib}  
 
\end{document}